# Key Event Receipt Infrastructure (KERI) Design [1][2]

Samuel M. Smith Ph.D.

Abstract—An identity system-based secure overlay for the Internet is presented. This includes a primary root-of-trust in self-certifying identifiers. It presents a formalism for Autonomic Identifiers (AIDs) and Autonomic Namespaces (ANs). They are part of an Autonomic Identity System (AIS). This system uses the design principle of minimally sufficient means to provide a candidate trust-spanning layer for the internet. Associated with this system is a decentralized key management infrastructure (DKMI). The primary root-of-trust are self-certifying identifiers that are strongly bound at issuance to a cryptographic signing (public, private) key-pair(s). These are self-contained until/unless control needs to be transferred to a new key-pair. In that event, an append-only chained key-event log of signed transfer statements provides end-verifiable control provenance. This makes intervening operational infrastructure replaceable because the event logs may be served up by any infrastructure, including ambient infrastructure. End-verifiable logs on ambient infrastructure enable ambient verifiability (verifiable by anyone, anywhere, at any time).

The primary key management operation is key rotation (transference) via a novel key pre-rotation scheme. Two primary trust modalities motivated the design, these are a direct (one-to-one) mode and an indirect (one-to-any) mode. The direct mode depends on append-only verifiable data structures called key event logs (KELs). The identity controller establishes control via verified signatures of the controlling key-pair(s) on each event. The indirect mode depends on witnessed key event receipt logs (KERL) as a secondary root-of-trust for validating events. This gives rise to the acronym KERI for key event receipt infrastructure. The indirect mode extends that trust basis with witnessed key event receipt logs (KERL) for validating events. But for short we call both types of logs key event logs or KELs. The security and accountability guarantees of indirect mode are provided by KAWA or KERI's Algorithm for Witness Agreement (KAWA) among a set of witnesses.

The KAWA approach may be much more performant and scalable than more complex approaches that depend on a total ordering distributed consensus ledger. Nevertheless, KERI may employ a distributed consensus ledger when other considerations make it the best choice. The KERI approach to DKMI allows for a more granular composition. Moreover, because KERI is event streamed, it enables DKMI to operate in-stride with data events streaming applications such as Web 3.0, IoT, and others where performance and scalability are more important. The core KERI engine is identifier-independent. This makes KERI a candidate for a universal portable DKMI.

## 1 Introduction

The major motivation for this work is to provide a *secure decentralized* foundation of trust for the Internet as a *trustable spanning layer* [7; 24; 42]. A major flaw in the design of the Internet

1. https://arxiv.org/abs/1907.02143
2. https://keri.one

Protocol was that it had no security layer. There is no built-in mechanism for security. Anyone can forge an IP (Internet Protocol) packet. Specifically, the IP packet header includes a source address field to indicate the IP address of the device that sent the packet [81]. Because the source address may be forged, a recipient may not know if the packet was sent by an imposter. This means that security mechanisms for the Internet must be overlaid (bolted-on).

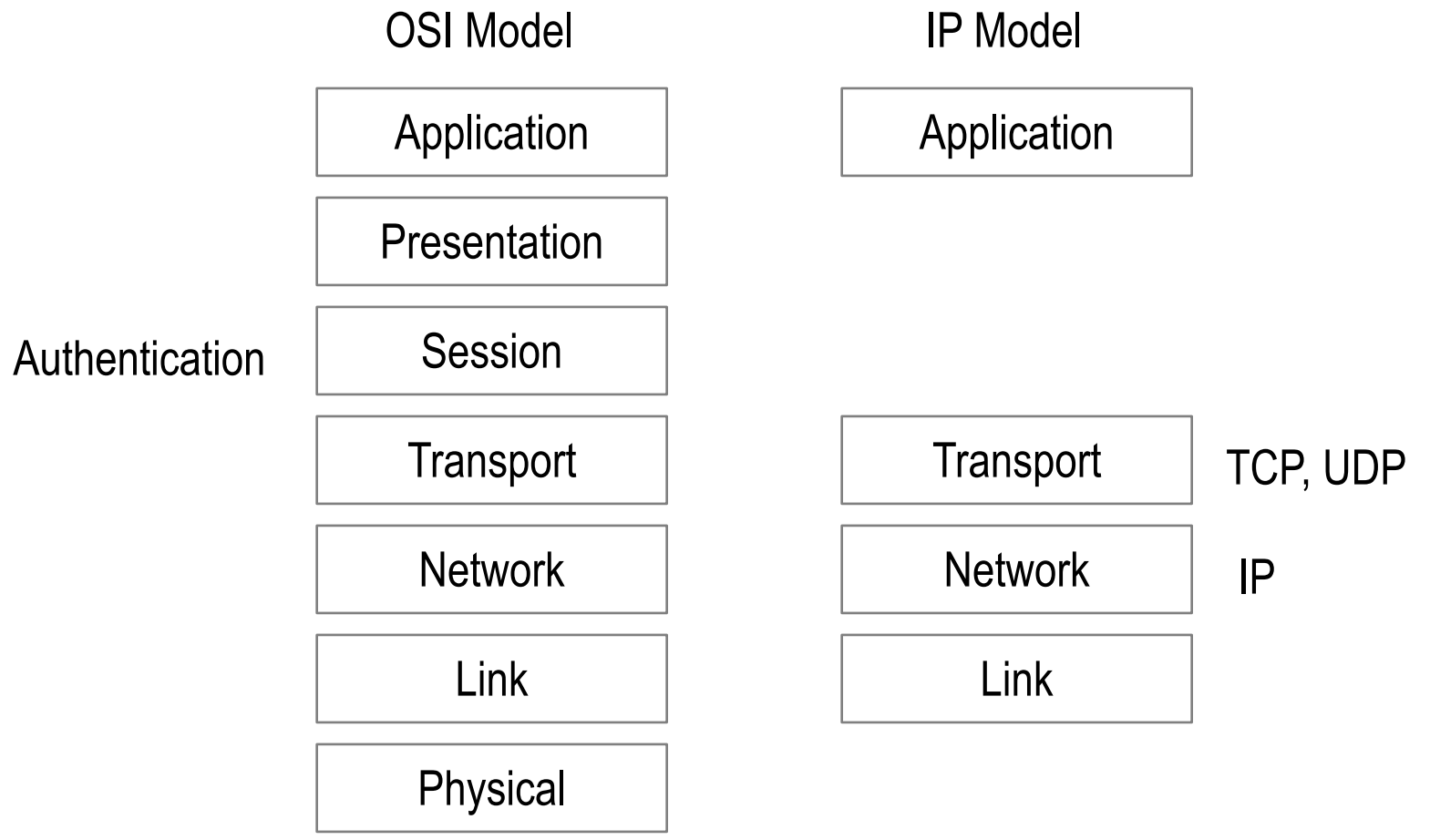

Figure 1.1. OSI vs IP Protocol Stack. IP stack is missing both the session and presentation layers that provide security (authentication). Without a built-in security layer, IP security must be provided by bolt-on identity system security overlays.

A typical Internet security overlay is based on some form of an identity system that leverages the properties of asymmetric public-key cryptography (PKI) [117]. The purpose of the overlay is to establish the authenticity of the message payload in an IP Packet. The following diagram shows the basic concept.

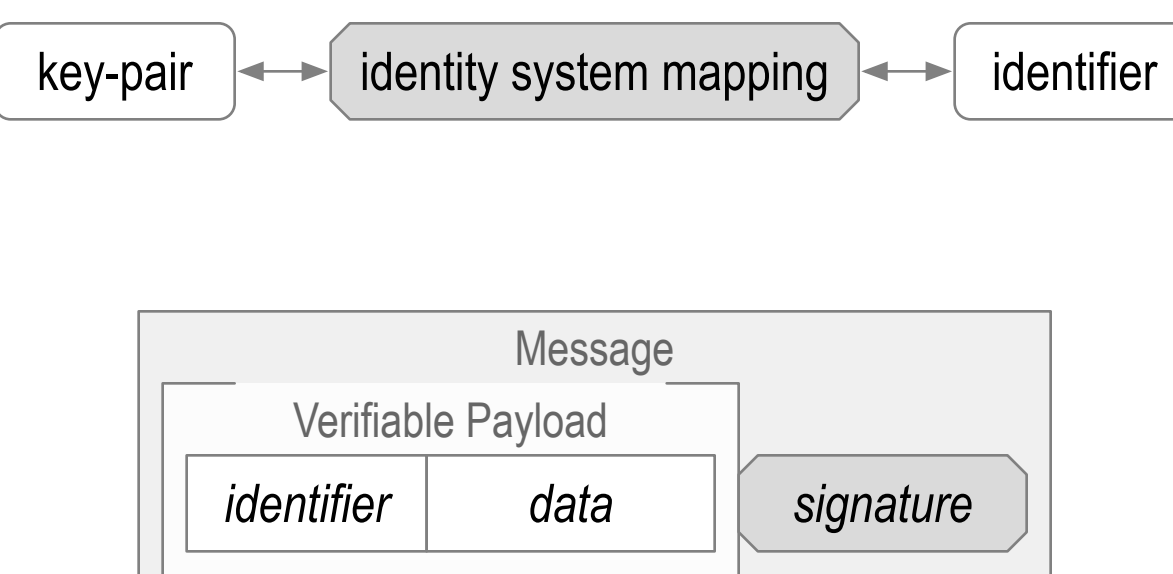

Figure 1.2. Identity System Security Overlay. The authenticity of a verifiable packet payload is derived from a mapping between an asymmetric key pair and an identifier. The attached digital signature exclusively (non-repudiably) associates the verifiable payload to the holder of the private key from the key-pair. The overlay's security is contingent on the mapping's security.

In general, with such a security overlay, the message payload inside an IP packet includes a unique identifier provided by the identity system that is exclusive to the sender of the packet. The payload is signed with a digital signature generated by the private key of a (public, private) key-pair. This signature is attached to the packet. The identity system provides a directory that links or maps the public key from the (public, private) key pair to the sender identifier. A valid signature verifies that the packet came from the holder of the private key. The mapping between identifier and public key establishes that the message came from the holder. This approach assumes that only a valid sender of the packet with payload has access to the private key, i.e. the sender is the exclusive holder or controller of the private key. As a result, the recipient of the packet may use the identity system to look up the corresponding public key given the unique sender identifier in the packet payload and then, with that public key, cryptographically verify the signature attached to the payload. Only the holder of the private key may generate such a verifiable payload. Without access to the private key, an imposter may not forge a verifiable payload. Indeed, a more accurate label for this approach would be an *identifier* system security overlay because identity is inimically bound to verifiable control over the identifier.

This security mechanism ostensibly establishes that the payload in the packet came from the correct source. Ostensibly, because the strength of this establishment depends on the strength of the security of the mapping between identifier and key-pair. Indeed, the overlay's security is contingent on the mapping's security.

In addition, when ordering is important, other contents of the payload, such as sequence numbers, may be used to protect against replay of the same payload in a different packet. Notably, securely establishing the authenticity of the payload via a signature is orthogonal to ensuring that the contents of the payload are confidential via encryption. Encryption is an additional task. Confidentiality alone may be insufficient when the payload is from the wrong source.

## 1.1 Binding

The essential feature of an identity (identifier) system security overlay is that it binds together controllers, identifiers, and key pairs. A sender *controller* is exclusively bound to the public key of a (public, private) key pair. The public key is exclusively bound to the unique identifier. The sender controller is also exclusively bound to the unique identifier. The strength of such an identifier system-based security overlay is derived from the security supporting these bindings.

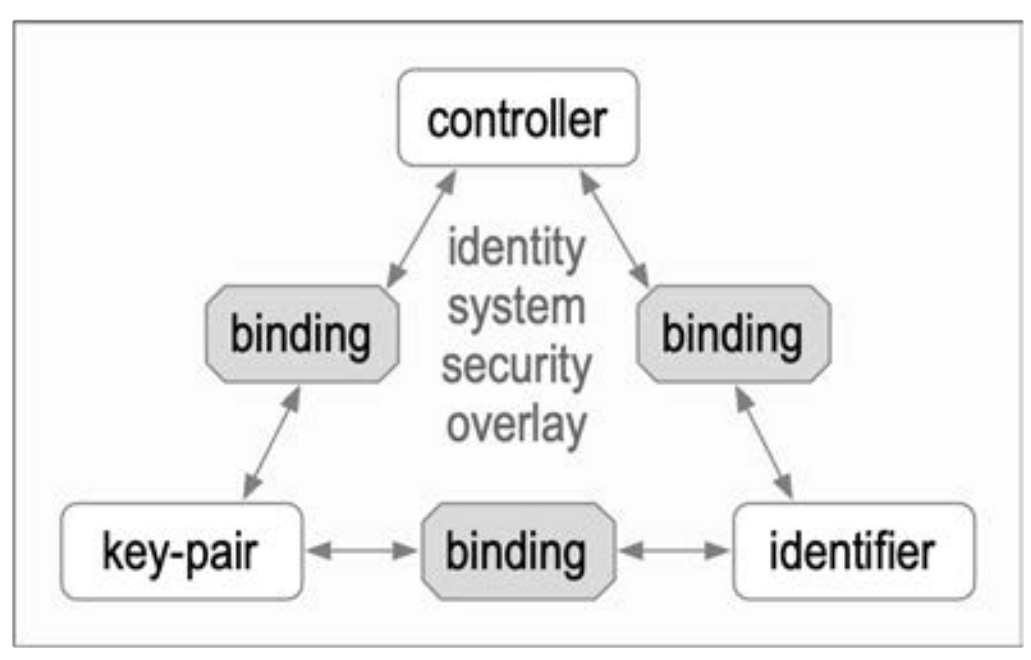

Identifier Issuance

Figure 1.3. Identity system as security overlay. The mechanism provides secure bindings between controlling entities, asymmetric key pairs, and identifiers.

To elaborate, typically, the associated identity systems are administrative in nature. This means the identifiers sourced from such an identity system are ultimately controlled by the associated administrator (administrative organization) of the identity (identifier) system. From a security perspective, the most important function of the administrator is to authoritatively and usually exclusively permit some other entity to use the identifier. We call the permitted use by an entity of an identifier a *binding* between the entity and the identifier. The permitted entity we call the *controller* (or, for short, the user). In this sense, the administrator binds that *controller* (user) to an identifier. This binding enables the controller to use the identifier to make mappings to resources such as IP addresses. The administrator also binds the public key from a cryptographic (public, private) key pair to this entity-identifier binding. The controller holds the private key and may prove its permitted use of the identifier by signing a statement (packet) with the private key. By virtue of holding the private key, the controller is bound to the key pair.

When the identifiers are sourced from the administrator, then their use is contingent on the administrator's continuing permission (usually by paying rent). The exclusivity and security of the bindings between controller and identifier and the key-pair and identifier are based on trust in the administrator and, by association, trust in the ostensibly secure operation of essential computing infrastructure to sustain and manage control over the bindings. By *essential*, we mean necessary to the security function of the overlay. Thus, a controller of an identifier sourced from such an administrative identity system is merely a renter of the identifier and must also rely on the operational security of the administrator's essential computing infrastructure. Consequently, any value the controller builds in its identifiers is contingent on both continuing permission from the administrator and trust in the administrator's essential operational infrastructure.

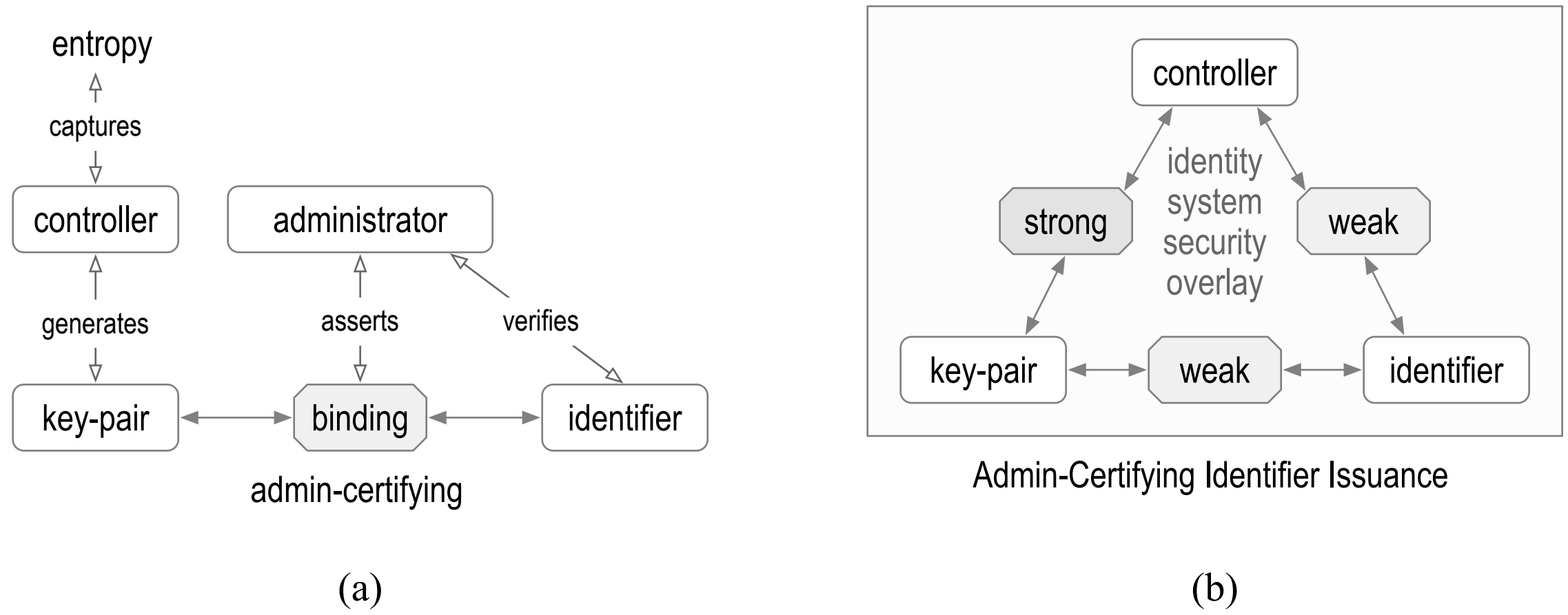

Figure 1.4. Administrative identifier issuance and binding. The controller captures entropy to generate a key pair. Administrator issues identifier and asserts bindings between key-pair and identifier and controller and identifier. (a) The strength of administrative certified bindings between key-pair and identifier and controller and identifier depends on the security of administrative infrastructure supporting its assertion. (b) Resultant weak (insecure) bindings between controller and identifier and key-pair and identifier.

In summary, Internet security mechanisms by-in-large use some form of identity system security overlay. The most well-known example is the Internet Domain Name System (DNS) with Certificate Authorities (CAs) [40; 54]. DNS binds controllers to domain name identifiers. DNS also maps domain names such as "mail.example.com" to IP addresses such as "54.85.132.205". A CA binds controllers with domain names to public keys (certificates) from (public, private) key pairs. A DNS-based URL namespace maps URL identifiers to a resource [148]. Besides DNS-CA, other identifier based security mechanisms are in wide use. These include web platform access control systems that use usernames (identifiers) and passwords and OAuth/OpenID Connect [113]. These other identifier based security mechanisms systems typically leverage DNS/CA.

Because DNS was bolted onto IP and CA was later bolted onto DNS, they are not well integrated from a security perspective. This has resulted in well-known exploits [69; 74]. In order to better mitigate exploits, various security mechanisms have been further bolted on such as pinning, notaries, and DNSSEC [44; 52; 55; 71; 76; 126; 128]. All these require trust in administratively operated essential computing infrastructure and, as a result, suffer from related security vulnerabilities. More recently Certificate Transparency (RFC6962) has had some success at mitigating exploits of the DNS/CA security overlay [41; 57; 70; 95–97]. DNS servers that employ certificate transparency maintain a global immutable append-only audit log of certificate authority activities (issuance and revocation events). The log may detect inconsistent behavior by a given CA and thereby trigger prophylactic measures by clients of servers using the associated domain names. In spite of the mitigation potential of certificate transparency, the DNS-CA security overlay continues to suffer from security exploits that fall into the class of attacks called

DNS hijacking. [4; 69; 140]. A related vulnerability is BGP hijacking using AS path poisoning to spoof the domain verification process by CAs [28–30; 66]. These both enable an attacker to induce a CA to issue a bogus but verifiable certificate. At the time of the referenced study, a majority of the internet domains were vulnerable to this attack. These sorts of attacks exploit the weak binding between key-pair and identifier from trusted issuers that is inherent to the DNS-CA system.

In general, the principal weakness of current internet security overlays is that establishing the binding between key-pair and identifier via certificates relies on essential computing infrastructure, which is inherently vulnerable to faulty operational administration of that infrastructure. The design goal of KERI is a security overlay that does not rely on essential operational infrastructure, thereby removing that weakness.

## 1.2 Trust Basis and Trust Domain

In security systems design, a *root-of-trust* is some component or process of the system upon which other components or processes are reliant. The *root-of-trust* has trustworthy security properties that provide a foundation of trust that other components or processes in the system may rely on [119]. A system may have multiple *roots-of-trust*. Some roots may be considered primary and others secondary depending on the degree of reliance on the given root-of-trust. We define *primary* and *secondary roots-of-trust* as follows:

A *primary* root-of-trust is *irreplaceable*. (1.1)

This means that the component that serves as the primary root-of-trust is an essential component in the system. It alone provides trust that is uniquely necessary to the system.

A *secondary* root-of-trust is *replaceable*. (1.2)

This means that the component that serves as the secondary root-of-trust is not an essential component in the system. It provides trust that may also be provided redundantly by another component. A secondary root-of-trust may serve an essential role, but the component itself is not uniquely necessary to establish trust in the system. In this case, at least one component must serve in the role, but replacement of that component does not remove trust.

Exploitable weaknesses in a *primary root-of-trust* may destroy the security of the associated system. Thus, good security design starts with selecting the root or roots-of-trust and includes making trade-offs between cost, convenience, and security of different roots-of-trust as integrated with the rest of the security system. Often, the term root-of-trust is narrowly applied to some hardware device that is sufficiently secure that it may always be trusted. To be clear, in this work, we generalize the concept of *root-of-trust* to include all the operational elements that may be trusted even when not very trustable or trustworthy. These all contribute to what we call the *trust basis*. In an Internet identity system security overlay, we define the *trust basis* thusly:

A *trust basis* binds controllers, identifiers, and key-pairs. (1.3)

A trust basis may include various roots-of-trust both primary and secondary, as well as other operational components or processes. When the *trust basis* includes administratively managed essential operational infrastructure, then the security of the bindings may be dependent on the operational virtue of the administrator to support and maintain those bindings. In that case, the administrator may be considered a *root-of-trust* for the bindings and hence control of the identifier. This includes the policies and procedures that create and establish control over bindings and those that manage or govern the associated infrastructure.

Each *trust basis* has associated with it what we call a *trust domain*. A trust domain is the ecosystem of interactions between entities using the internet that relies on a given trust basis to imbue trust in those interactions. The trustability of the trust domain (ecosystem) depends on its trust basis. We define the *trust domain* thusly:

A *trust domain* is an ecosystem of interactions that relies on a trust basis. (1.4)

For example, the Facebook ecosystem of social interactions is a *trust-domain* that relies on Facebook's identity system of usernames and passwords as its trust basis. The following diagram illustrates the relationship between trust basis and trust domain in an identifier-based security overlay.

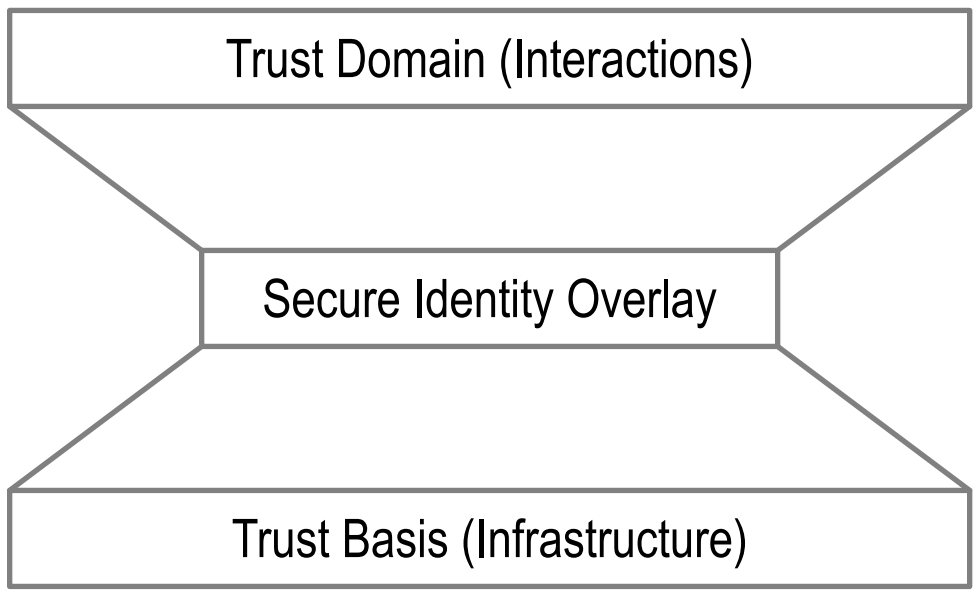

Figure 1.5. Relationship between Trust Basis and Domain

The essential role of the *trust basis* is to enable the trustworthy establishment of control over an identifier. A common misconception is that administrative issuance of identifiers is inherently trustworthy. Careful analysis of the process of bootstrapping from roots-of-trust to trustworthy internet interactions reveals that this process is often exploitable (see for example the exposition above for the DNS/CA system). As mentioned previously, the primary weakness of many administrative approaches to establishing or proving control over an identifier depends on a root-of-trust in the administrator's management of its computing infrastructure. To elaborate, the goal of a security overlay is to eventually enable a controller to make digitally cryptographically verifiable statements referencing an identifier using universally unique cryptographic digital signatures based on a key-pair comprised of a public key and a private key (or key pairs when multi-signature) . But this first requires that one trust the binding between that identifier and the cryptographic key-pair. Usually, this binding, especially when the identifier is human-meaningful is self-asserted by the administrator and therefore has a primary root-of-trust in the administrator's management or governance of its computing infrastructure [142; 155]. Thus, trust in such identifiers issued by the administrator requires trust in the administrator, and when that binding is issued in a digital way, trust in the computing infrastructure that is used to issue that binding. Moreover, the binding may be based on a manual establishment process that may require direct physical interaction with the administrative entity, such as visiting the office of the administrator to obtain some form of hardcopy credential that proves a given key pair is bound to an identifier issued by the administrator. Once the binding between an administratively issued identifier and a digital signing key pair is established, then cryptographically secure signed internet interactions may occur. Thus, the provenance of trust in an administrative entity is crucial. Trustable administrative entities must go to great lengths to establish and maintain a credible provenance of trust.

When the mechanisms used to establish control over identifier bindings are not interoperable in any material way between trust bases, then the establishment of secure control over identifier bindings may be *trust-basis dependent*. This may be especially true of administrative identifiers. In this case, identifiers are effectively *locked* to a given *trust basis* and thereby its associated *trust domain* by virtue of the dependence on the administrator's operational infrastructure used to ensure a trust-ably secure binding. In other words, the binding and, hence, identifier is not portable to other trust bases. As a result, any value attached to the identifier is not portable to other trust domains. Moreover, different trust bases that rely on widely differing mechanisms for establishing or proofing the validity of a binding may place a large burden on any controller (user) that wishes to interoperate concurrently with others in more than one *trust domain* because the controller has to support the different establishment mechanisms for each trust basis. This may include redundant resource instances with redundant non-interoperable identifiers on different operational infrastructures, such as both a FaceBook and YouTube URL for the same video.

To restate, one of the main problems with the vast majority of identity-based security mechanisms arises when the bindings between controller, identifier, and key pair are both administratively permissioned and locked to the associated trust basis. A controller's interactions using that identifier are thereby locked to the associated trust domain. Merely providing interoperability between trust bases (and hence domains) does not fix the administrative permissioning problem. Merely removing the administrative permissioning problem does not fix the trust domain locking problem. Consequently, one goal of this work is to design a system that provides secure interoperable bindings (hence identifiers) that are both self-administering and portable between trust bases and their trust domains. The goal is to create bindings and identifiers that are not locked to a given trust basis/domain but allow the controller to move identifiers between interoperable trust bases/domains. We believe this means that interoperable trust bases must share a common mechanism for establishing control over those bindings that thereby span both the supporting trust bases and the associated applied trust domains. This mechanism provides the foundation for a trust-spanning layer protocol.

## 1.3 Decentralization

Our definition of *decentralization* (*centralization*) is about control, not spatial distribution. In our definition, *decentralized* is not necessarily the same as *distributed*. By *distributed*, we mean that activity happens at more than one site. Thus *decentralization* is about *control* and *distribution* is about *place*. To elaborate, when we refer to decentralized infrastructure, we mean infrastructure under decentralized (centralized) control, no matter its spatial distribution. Thus, *decentralized infrastructure* is infrastructure sourced or controlled by more than one *entity*. *Entities* are not limited to natural persons but may include groups, organizations, software agents, things, and even data items. This control may lie on a scale from highly *centralized* to highly *decentralized*. A *centralized* administratively managed identity system may be controlled by a single governing organization. A governing organization might also be hierarchical in nature, with multiple subordinate organizations that operate under the auspices of the next higher-level organization. The associated operational infrastructure might itself be highly spatially distributed despite being under highly centralized control or vice-versa. For example, although DNS is administered by a single organization, IANA, the operational infrastructure is distributed worldwide [77].

Much of the operation of internet infrastructure is inherently decentralized, but control over the value transported across this infrastructure may be much less so. Centralized value capture systems concentrate value and power and thereby provide strong temptations for malicious administrators to wield that concentrated power to extract value from participants. We believe that decentralization of value transfer is essential to building trust. Consequently, a key component for a decentralized foundation of trust is an interoperable decentralized identity (identifier) system [10; 132; 133; 152].

Within a centralized identity/identifier system used as a security overlay, one entity, the administrator, controls all the identifiers. In contrast, within a decentralized system, disparate entities each control some of the identifiers but in an interoperable way. Each entity may have a set of identifiers (namespaces) that it controls but are still recognized by the other entities. In other words, each entity may be sovereign over a set of identifiers. In this case, where an identifier refers to its controlling entity then that entity in that sense is *self-sovereign* over its own identifier and hence the identity associated with that identifier. Ideally, a *self-sovereign identifier* is one that the user issues and controls without deference to or permission from any other administrative organization. The associated identity system is also, in that sense, *self-administering*. This property of decentralized, *self-administering* identity systems has given rise to the term *self-sovereign identity* (SSI) [10; 127; 133; 152]. A fully *self-sovereign identity* system would be highly decentralized.

To summarize, two disadvantages to participants in a centralized identity-system-based security overlay are as follows:

- participants require permission from the administrator often at some cost. The administrator may censure a participant.
- participants must trust the operational infrastructure (trust basis) of the administrator. The value produced by the participant may be locked to the infrastructure and/or subject to exploitation due to administrative vulnerability.

In either case, the participant is not ultimately in control of the value the participant brings to the associated trust domain. Whereas decentralization of the trust basis enables a more competitive eco-system with portability between infrastructure providers where the choice and means to port identifiers are under participant control [129; 134].

# 2 Self-Certifying Identifiers

We believe fully decentralized and trustable control over identifiers is best enabled by a trust basis with a primary root-of-trust that is based on the properties of a *self-certifying* identifier [68; 90; 102; 103]. This concept was first described in the 1990s as a way of creating secure identifiers under decentralized control. Applications include file systems and internet URLs that did not depend on centralized DNS certificate authorities for secure control establishment. However, the concept never achieved meaningful adoption when it was first introduced. Recent concerns over both the increasing centralization of the internet and its attendant security vulnerabilities now make the *self-certifying identifier* concept much more attractive. We proposed using *self-certifying identifiers*, called cryptonyms, for decentralized or self-sovereign identity in early 2015 [132; 133]. More recently, some variations of identifiers that follow the emerging W3C Decentralized Identifier (DID) standard are self-certifying [151]. Also, more recently, the Trusted Computing Group (TCG) uses the terms "implicit identity" and "embedded certificate authority" to describe the process whereby device identifiers are automatically generated by the associated computing device [143]. This mechanism is called the Device Identifier Composition Engine (DICE). It generates DICE Compound Device Identifiers (DCI). These are compatible with the IEEE 802.1AR-2018 Secure Device ID (DevID) identifier standard and the IETF RATS standards [3; 118; 124; 154]. They also provide a type of *self-certifying* identifier.

## 2.1 Basic Concept

The simplest form of a *self-certifying* identifier includes either the public key or a unique fingerprint of the public key as a *prefix* in the identifier. Because the public key is included in the identifier, the identifier is strongly and securely bound to the (public, private) key pair. This cryptographic secure binding makes the identifier *self-certifying* [68; 90; 102; 103].

This *self-certifying* property is the critical feature needed to establish a secure, decentralized, portable identity system. The holder of the private key is in sole *control* of the identifier because only that holder can sign statements that may be cryptographically verified with the public key bound to that identifier. The holder of the private key is the controller and, by virtue of that con-trol, is implicitly bound to the identifier. Only the holder of the private key (controller) may successfully respond to any challenge to prove control over the identifier by signing the challenge response with the private key.

A *self-certifying identifier* cryptographically binds an identifier to a key pair. (2.1)

The primary root-of-trust in a self-certifying identifier comes from the properties of asymmetric public key cryptography (public-private key pairs)[117]. To elaborate, the private key is derived from a random seed represented as a very large random number. The degree or strength of randomness determines how difficult it would be for someone else to reproduce the same large random number. This is called collision resistance. Shannon's information theory uses the term *entropy* to describe the degree of unpredictability of a message [78]. Entropy is measured in bits.

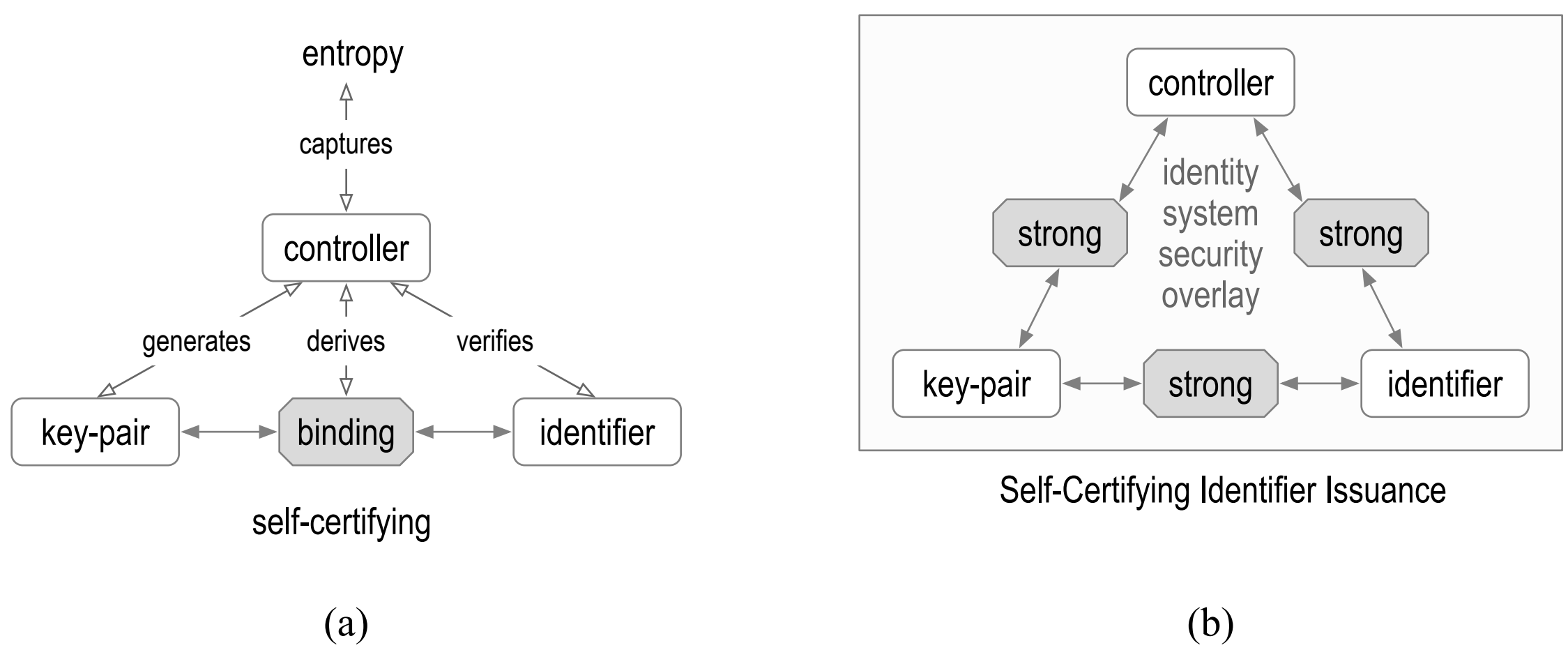

Figure 2.1. Self-certifying identifier issuance and binding. The controller captures entropy to generate key-pair and derive identifiers. Self-certifying identifiers provide strong cryptographic binding between key-pair and identifier. (a) The strength of self-certified binding between key-pair and identifier is cryptographic. (b) Resultant strong (secure) bindings between controller and identifier and key-pair and identifier.

Likewise, the randomness of a number or message can measured by the number of bits of entropy in the number. A cryptographic strength random number may have as many bits of entropy as the number of bits in the number. An $N$-bit long base-2 random number has $2^N$ different possible values. A random seed with $N$ bits of entropy means that there are $2^N$ possible values of that seed.

The highest level of cryptographic security is called information-theoretic security [79]. A special case of information-theoretic security is called perfect security. Perfect security means that the cipher text provides no information about the key [79]. A crypto-system with this level of security cannot be broken algorithmically. It must be broken by brute force, if at all. Brute force means that in order to guarantee success, the adversary must search every combination of the seed/key.

For cryptosystems with perfect security, the fundamental parameter is the number of bits of entropy needed to resist any practical brute-force attack. In other words, when a large random number is used as the seed/key that is the input to a crypto-system with perfect security, the question to be answered is how large does the random number need to be to withstand a brute force attack against the output of the crypto-system? Assuming non-quantum computers, the conventional wisdom is that, for systems with perfect security, the seed/key input needs to have on the order of 128 bits (16 bytes) of entropy to practically withstand any possible brute force at-tack for a given output. For other cryptosystems, the size of the seed/key may need to be much larger. Given that with perfect security, no other information is available to an attacker, the attacker may need to try every possible value as an input before finding the correct one that results in the desired output. Thus, with N bits of entropy in an N-bit random number, the number of attempts or trials that the attacker would have to make will be on the order of $2^N$. Theoretically, quantum computers, using the asymptotically optimal Grover's Algorithm, may be able to brute force a given output of a black box crypto-system where the input seed/key has *N* bits of entropy in as little as $2^{N/2}$ trials [145]. Thus, once practical quantum computers exist, for algorithms that are vulnerable to Grover's attack, the size of *N* needed to achieve cryptographic strength would increase from 128 to 256.

Just to understand how hard it would be to brute force attack a system with a cryptographic strength of 128 bits, given available computing power, let's suppose an adversary had access to some multiple of all the world's supercomputing resources. The aggregate performance of the top 500 supercomputers combined is estimated to be less than 2 exaflops [145]. One exaflop is on the order of one quadrillion operations per second [63]. A quadrillion is approximately $2^{50}$ = 1,125,899,906,842,624. Because a single CPU core is limited to about 4 billion operations per second the most powerful supercomputers must run many many cores in parallel. Suppose an adversary could use over a million ($2^{20}$ = 1,048,576) exaflop capable supercomputers. The adversary could then try $2^{50} * 2^{20} = 2^{70}$ values per second (assuming that each try only took one operation). There are about $3600 * 24 * 365 = 313,536,000 = 2^{log2313536000} = 2^{24.91} \sim= 2^{25}$ seconds in a year. Thus, this set of over one million exascale supercomputers could try about $2^{50+20+25} = 2^{95}$ values per year. With 128 bits of entropy in the input seed/key, the adversary would need on the order of $2^{128-95} = 2^{33} = 8,589,934,592$ years to guess the right value. This also assumes that the value gained by breaking the crypto-system is worth the expense of that much computing power. Consequently, a crypto-system with $N = 128$ bits of cryptographic strength is practically impossible to break (pre-quantum). The term for this is computationally infeasible. To restate, a cryptographic strength of 128 bits is strong enough that no adversary may break it by brute force attack with any practically imaginable amount of (current pre-quantum) computational resources, i.e., it is considered computationally infeasible. Successful attacks must instead exploit some other vulnerability that exposes the private key, such as poor private key management, socially engineered disclosure of private keys, or a side-channel weakness in the digital signing compute infrastructure [22; 130; 137].

Cryptographic strength random numbers may be generated in various ways. A common software-based approach is a cryptographically secure pseudo-random number generator (CSPRNG) [46]. More sophisticated approaches use special-purpose hardware such as trusted platform modules [146]. In any case, inexpensive methods to collect sufficient *entropy* to generate a cryptographic strength random number are readily available. This means that it is entirely practical for any entity to capture sufficient entropy to generate a cryptographically strong random seed. Moreover, once an entity captures that entropy in a seed, no other entity may practically reproduce that seed; the likelihood is infinitesimally small. This makes the original capturing entity the sole holder, possessor, or controller over that random seed. It is thereby the sole sovereign or the sole authority over that seed. The holder of that seed obtained control of that seed by virtue of capturing entropy without deference, permission, or dependence on any other entity or authority. This is *true* decentralized control.

## 2.2 Generating Self-Certifying Identifiers

Given current computing power, each cryptographic operation in the process of generating the self-certifying identifier should maintain no less than 128 bits of cryptographic strength. Each operation is a type of cryptographic *one-way function* [112]. A *one-way* function is relatively easy to compute going in one direction but extremely hard to reverse (invert) [111]. A one-way function with 128 bits of security would take on the order of $2^{128}$ operations to invert.

ECC scalar multiplication is a type of one-way function. Public key generation in elliptic curve cryptography (ECC) uses elliptic curve scalar multiplication as a one-way function to generate a public key from a private key. To restate, public key generation is a type of one-way function. Take Ed25519 keys, for example. They are relatively strong for their length and take relatively little computation to generate [1; 26; 49]. The following description uses Ed25519 as an example, but other ECC keys like ECDSA have similar characteristics [35; 58; 107; 125]. An Ed25519 private key may be represented by a 256-bit long (32-byte) number. of length (not strength) 256 bits (32 bytes). The cryptographic strength of a scalar multiplication compliant with the Ed25519 scheme using a 256-bit long scalar is no less than 128 bits [26; 89]. The reason the cryptographic strength with a 256-bit scalar multiplication is still only 128 bits is that well-known attack algorithms can invert an N-bit ECC scalar multiplication in $2^{N/2}$ independent trials. This means that the size of the underlying field should be roughly twice the security parameter. Thus for 128 bits of security, the private key must be at least 256 bits in length [27; 53; 59; 75]. An ECC public key is actually a point on the elliptic curve with both an $x$ and $y$ coordinate. Each coordinate is a 32-byte number, for a total of 64 bytes. Because, however, the curve itself is known, only the $x$ coordinate and a sign bit that indicates which side of the $x$-axis the $y$ coordinate lies on are needed to recover the $y$ coordinate. Thus, public keys are only 32 bytes when stored ($x$ coordinate and sign bit).

Given a 128-bit random seed, the next step in generating a self-certifying identifier is to create an asymmetric (public, private) key pair. If the required length of the private key is 256 bits (as is the case for Ed25519), then we first need to stretch the random seed from 128 bits (16 bytes) to 256 bits (32 bytes). This stretched seed then becomes an input to the Ed25519 key generation algorithm [49]. Key stretching is another type of one-way function [92]. A good algorithm for key stretching is Argon2 [15; 114]. Once stretched, this 256-bit seed becomes the private key. Using the LibSodium code library, which provides the associated ECC operations, we can create the associated public key. Under the hood, the Ed25519 code hashes the private key/seed before performing the scalar multiplication on the results to compute the public key. But this hashed value is never exposed to the user. The resultant 32-byte public key may then be encoded and used to generate the self-certifying identifier.

Cryptographic operations produce large binary numbers. The binary number format is not very useful for identifiers in most applications. Identifiers are better represented as strings of characters. Thus, in order to practically create an identifier from a public key, we first need to convert it to a string of characters. A highly interoperable and relatively compact encoding is RFC-4648 Base-64 (URL safe) [87]. Base-64 encodes every 3 bytes of a binary number into 4 ASCII characters. When the *N-byte binary number is not an exact multiple of 3 bytes,* there will be either 1 or 2 bytes of pad characters added to the end of the Base-64 encoding. A 32-byte public key would encode to 44 Base-64 characters with one trailing pad character inclusive. For example, consider the following 32-byte private key represented as a 64-character hex string:

```
0caac9c64711f66e6ed71b37dc5e69c5124fe93ee12446e1a47ad4b650dd861d
```

This is encoded as the following 44-character Base-64 string (includes one trailing pad character):

```
DKrJxkcR9m5u1xs33F5pxRJP6T7hJEbhpHrUtlDdhh0=
```

### 2.2.1 Derivation Code

To properly extract and use the public key embedded in a self-certifying identifier, we need to know the cryptographic signing scheme used by the key pair. In this case, we need to know that the key pair follows Ed25519 and is used for signing. This information would also allow us to infer the length of the public key. This information consists of the cipher suite and operation. In general, this defines the process used to derive the self-certifying identifier. This derivation information must either be assumed or included in the identifier. One way to include this very compactly in the identifier is to replace the pad character with a special character that encodes the derivation process. Let's call this the *derivation code*. Because this derivation information is needed to correctly parse the encoded public key and the convention is to parse from left to right, we prepend the *derivation code* to the public key and delete the pad character. The result is still 44 characters in length. For example, suppose that the 44-character Base-64 with trailing pad character for the public key is as follows:

`F5pxRJP6THrUtlDdhh07hJEDKrJxkcR9m5u1xs33bhp=`

If B is the value of the derivation code then the resultant self-contained string is as follows:

`BF5pxRJP6THrUtlDdhh07hJEDKrJxkcR9m5u1xs33bhp`

Now, the derivation code is part of the identifier. This binds both the public key and its derivation process to the identifier. Because the string is valid Base-64, it may be converted to binary before being parsed with binary operations, or it may be parsed before conversion. If parsed before conversion, the derivation code character must be extracted from the front of the string. The first character tells how to parse the remaining characters, including the length. This allows compact but parseable concatenation of cryptographic material. One pad character must be re-appended before converting the remaining characters that comprise the encoded public key back to binary. Once converted, the binary version of the public key may be used in cryptographic operations. This approach to compact encoding of cryptographic material given rise to the CESR (Compact Event Streaming Representation) protocol. Proposed sets of derivation codes for KERI are provided in Section 14.2. Each is either 1, 2 or 4 characters long to replace 1, 2 or 0 pad characters respectively. This ensures that each self-contained identifier string is compliant with Base-64 specification which must be a multiple of 4 base-64 characters. This allows conversion before parsing, which may be more efficient. This approach to encoding the derivation process is similar to that followed by the multi-codec standard but is more compact because it efficiently exploits pad bytes and only needs to support cryptographic material in KERI events [105]. Furthermore, a KERI design goal is to support cryptographic agility for all the cryptographic material in its events. A compact derivation code makes supporting this degree of granular cryptographic agility much more efficient. The derivation codes are not only used for self-certifying identifier prefixes but also for keys, signatures, and digests. For example, the Base-64 representation of the private key from the end of the Section 2.2 above may be encoded with a derivation code as follows:

`ADKrJxkcR9m5u1xs33F5pxRJP6T7hJEbhpHrUtlDdhh0`

This enables the use of derivation codes for efficient representation of cryptographic material throughout the key management infrastructure used to support KERI.

### 2.2.2 Inception Statement

The practical use of a self-certifying identifier may require some initial configuration data. We call this the inception data, and it is formally represented in a signed inception statement. The inception data must include the public key and the identifier derivation from that public key and may include other configuration data. The identifier derivation may be simply represented by the derivation code. A statement that includes the inception data with an attached signature made with the private key comprises a cryptographic commitment to the derivation and configuration of the identifier that may be cryptographically verified by any entity that receives it. It is completely self-contained. No additional infrastructure is needed or, more importantly, must be trust-ed in order to verify the derivation and initial configuration (inception) of the identifier. The initial trust basis for the identifier is simply the signed inception statement. A diagram of a basic inception statement is shown below:

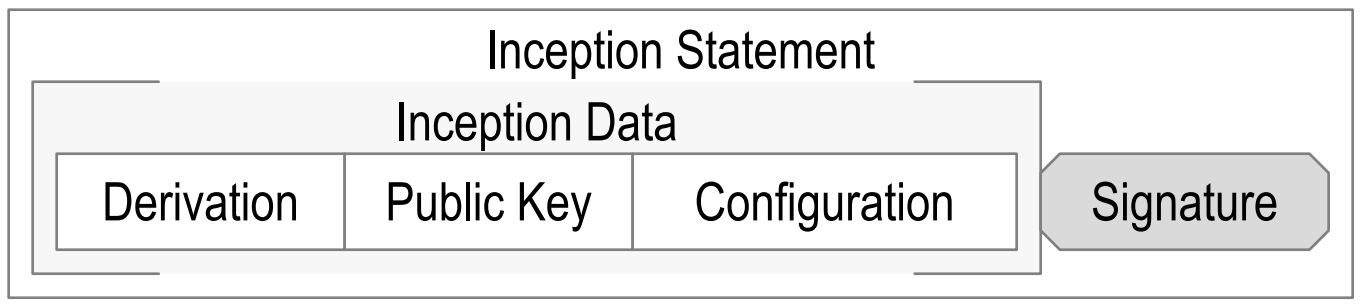

Figure 2.2. Inception Statement

### 2.2.3 Transferability

Unfortunately, over time, the private key controlling a self-certifying identifier may be exposed to potential compromise. In this case, best practices for key management indicate that the (public, private) key pair needs to be changed. This is called key rotation. Key rotation effectively transfers control of an identifier to a different key pair. The transfer may be performed with a signed transfer statement. This makes a cryptographically verifiable commitment to the transfer. For a self-certifying identifier, however, after the transfer, the new controlling key pair is no longer uniquely bound to the identifier. Nonetheless, its new binding is still cryptographically verifiable, given one has a copy of the transfer statement. The log of transfer statements is *end-verifiable*. This means that the log may be verified by any end user that receives a copy. No trust in intervening infrastructure is needed to verify the log and validate the chain of transfers and thereby establish the current control authority. Because any copy of the record or log of transfer statements is sufficient, any infrastructure providing a copy is replaceable by any other infrastructure that provides a copy. Any infrastructure may do. Therefore, the infrastructure used to maintain a log of transfer statements is merely a *secondary root-of-trust* for control establishment over the identifier. This enables the use of *ambient infrastructure* to provide a copy of the log. The combination of *end verifiable* logs served by *ambient infrastructure* enables *ambient verifiability*, that is, anyone can verify anywhere at any time. This approach exhibits some of the features of certificate transparency and key transparency with end-verifiable event logs but differs in that each identifier has its own chain of events that is rooted in a self-certifying identifier [93; 97].

This presents two different fundamental classes of self-certifiable identifiers. They are *transferable* (rotatable) and *non-transferable* (non-rotatable). The transferability class of the identifier may be declared in either the derivation code (see Section 14.2) or the inception statement's configuration data. When the private key for a non-transferable identifier becomes exposed to potential compromise, then the identifier must be abandoned by the controller as it is no longer secure. Whereas when the private key for a transferable identifier becomes exposed to potential compromise, then control over the identifier may be transferred to a new key pair to maintain security. Many applications of self-certifying identifiers only require temporary use, after which the identifier is abandoned. These are called *ephemeral* identifiers. Other applications may only attach a limited amount of value to the identifier such that replacing the identifier is not onerous. Because a non-transferable (ephemeral) identifier is not recoverable in the event of compromise, the only recourse is to replace the identifier with another identifier. In some applications, this may be preferable, given the comparable simplicity of maintaining the key state. In either of these cases, a non-transferable self-certifying identifier is sufficient. In other applications, control over the identifier and its attached value must persist over time. In this case, a transferable self-certifying identifier is required. This persistent control property gives rise to a new type of identifier, an *unbounded term* identifier. In public key cryptography it is common to use bothg short term and long term public keys as identifiers. But even long term public keys must be eventually abandoned due to exposure. This means any value or reputation binding to that public key as an identifier must eventually be lost. Whereas an unbounded term identifier need never abandon its bound value or reputation. It need never be abandoned. This is one of the novel unique properties that KERI provides.

The early implementations of self-certifying identifiers, as well as the more recent secure device ID implementations, are, in general, non-transferable. The main innovation of KERI is that it provides a universal decentralized mechanism that supports both non-transferable and, more importantly, transferable (persistent) self-certifying identifiers.

## 2.3 Types of Self-Certifying Identifiers

### 2.3.1 Basic

A basic self-certifying identifier includes a *prefix* that is composed of a Base-64 (URL safe) derivation code prepended to Base-64 encoding of a public digital signing key. The derivation code indicates that the derivation is basic, as well as the cipher-suite for the signing scheme. Abstractly, the prefix may be represented as follows:

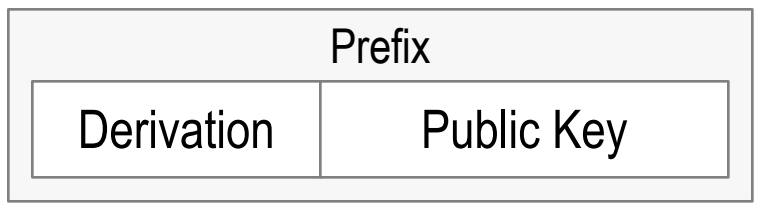

Figure 2.3. Basic Prefix.

Including the derivation code in the prefix binds the derivation process along with the public key to the resultant identifier. An example of the prefix with a one-character derivation code and a 32-byte public key encoded into a 44-character Based-64 string follows:

`BDKrJxkcR9m5u1xs33F5pxRJP6T7hJEbhpHrUtlDdhh0`.

A diagram showing the derivation from a random seed to the public key in the prefix for a basic self-certifying identifier follows:

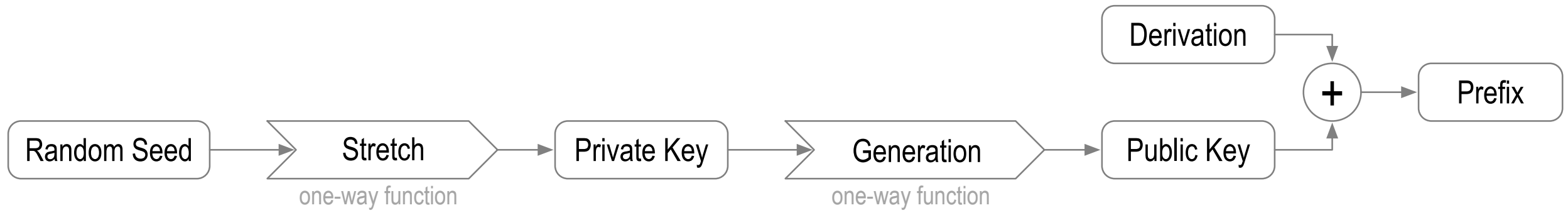

Figure 2.4. Basic Prefix Derivation.

This derivation process was described in detail in Section 2.2.1 above. The basic type is the simplest to parse. It's easy to extract the derivation and public key directly from the identifier prefix. Moreover, when the derivation code also indicates that the identifier is non-transferable, then the inception configuration data may be empty. This means there must never be any transfer statements, and everything needed to verify the signature on any statement is contained in the identifier prefix. This means that there is no need for an inception statement. This provides a very compact representation and trust basis for ephemeral self-certifying identifiers. Furthermore, when non-transferable, a convention may be adopted that allows repurposing of the signing key pair to also be used as an encryption key pair.

For example, an Ed25519 signing key pair may be converted to an X25519 encryption key pair because the two curves are birationally equivalent [56; 149]. The public key from the X25519 key pair may be used either for anonymous asymmetric public encryption or exchanged with the public key from another party to create a Diffie-Hellman shared symmetric encryption key. This would allow bootstrapping a secure authenticated (signed) and encrypted communications channel between any two parties with only a non-transferable basic self-certifying identifi-er from each party. No other infrastructure is needed. Once the secure channel is bootstrapped, the two parties may thereby exchange non-transferable identifiers.

### 2.3.2 Self-Addressing

When there is a need for incepting configuration data, then the inception statement must not be empty. In some cases, it would be useful to be able to strongly bind the inception statement to the identifier and not merely rely on a cryptographically verifiable inception statement for the associated identifier. This binding between inception data and private key can be created by replacing the public key in the identifier prefix with a cryptographic strength content digest (hash) of the inception statement (that includes the public key). A type of one-way function produces a content digest (hash). Given that the hashing function has sufficiently strong collision resistance then it uniquely securely cryptographically binds its output digest to the contents of the inception data. Furthermore, given that the inception data includes both the derivation code and the public key, then, the digest output is bound to the public key, which in turn is bound to the private key. This assumes that the full derivation process, including the hashing step, is known by any verifier. In general, the process of successively chaining sufficiently strong one-way functions may maintain the cryptographic strength of the binding backward along the whole chain from the final output to the originating private seed/key. With reference to the example in Section 2.2 , a 32 byte Blake2b, Blake2s, or Blake3 hash function has 128 bits of cryptographic strength [31; 33; 121]. Blake2b is relatively fast for its length and strength, Blake2s works on limited-resource hardware, and Blake3 is even faster [31; 32]. These functions maintain 128-bit strength with a 32-byte digest.

Using a content digest binds the inception statement to the identifier and makes it easy to use the identifier prefix to retrieve the inception statement from a content addressable database. This requires that the database store the signature with the inception data so that a reader may cryptographically verify that the inception data is authoritative. Without the signature, the digest is repudiable by the controller. This approach also enables content or document-specific but self-certifying identifiers. This provides a content-addressable database mechanism to enforce non-repudiable unique attribution to the content creator as controller. This binds the content to the creator in the content address. Different creators mean different addresses. This makes confidential (encrypted) content more usable as the content address is bound to the controller from whom a decryption key must be obtained.

Abstractly, the prefix for a content self-addressing self-certifying identifier may be represented as follows:

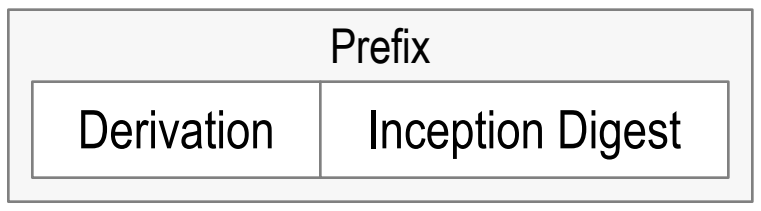

Figure 2.5. Self-Addressing Prefix.

An example of the prefix with a one-character derivation code and a 32-byte hash encoded into 44 character Base-64 string follows:

`EXq5YqaL6L48pf0fu7IUhL0JRaU2_RxFP0AL43wYn148.`

A diagram showing the derivation from a random seed to a hash of the inception data in the prefix for a self-addressing self-certifying identifier follows:

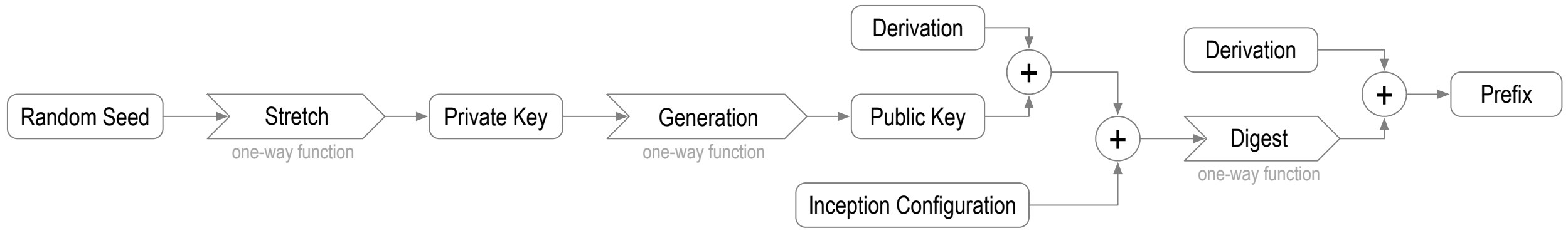

Figure 2.6. Self-Addressing Prefix Derivation. The public key has a derivation distinct from the digest's derivation.

### 2.3.3 Multi-Sig Self-Addressing

Adding a content digest at the end of the identifier prefix derivation opens up other derivation possibilities for a self-certifying identifier. For example, the inception statement need not merely contain one public key but could include a set of public keys and a threshold such that multiple signatures meeting a threshold are required for verification. Multiple signatures may significantly increase the security of the establishment of control authority over the identifier derived from the inception content. This assumes that the multiple signature (multi-sig) scheme used in the derivation is known by any verifier.

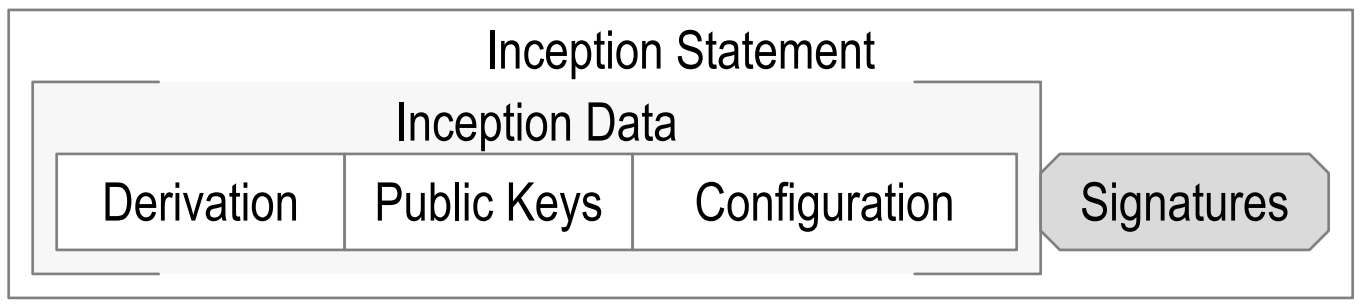

Figure 2.7. Multiple Public Key Inception Statement

Abstractly, the prefix for a multi-sig content self-addressing self-certifying identifier differs from a single signature in the derivation code specifies a multi-sig scheme with inception data that includes multiple public keys appended together (not just one). The prefix may be represented as follows:

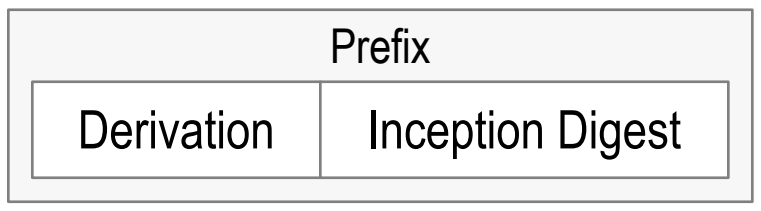

Figure 2.8. Multi-Sig Self-Addressing Prefix.

A diagram showing the derivation from multiple random seeds to a hash of the inception data in the prefix for a multi-sig self-addressing self-certifying identifier follows:

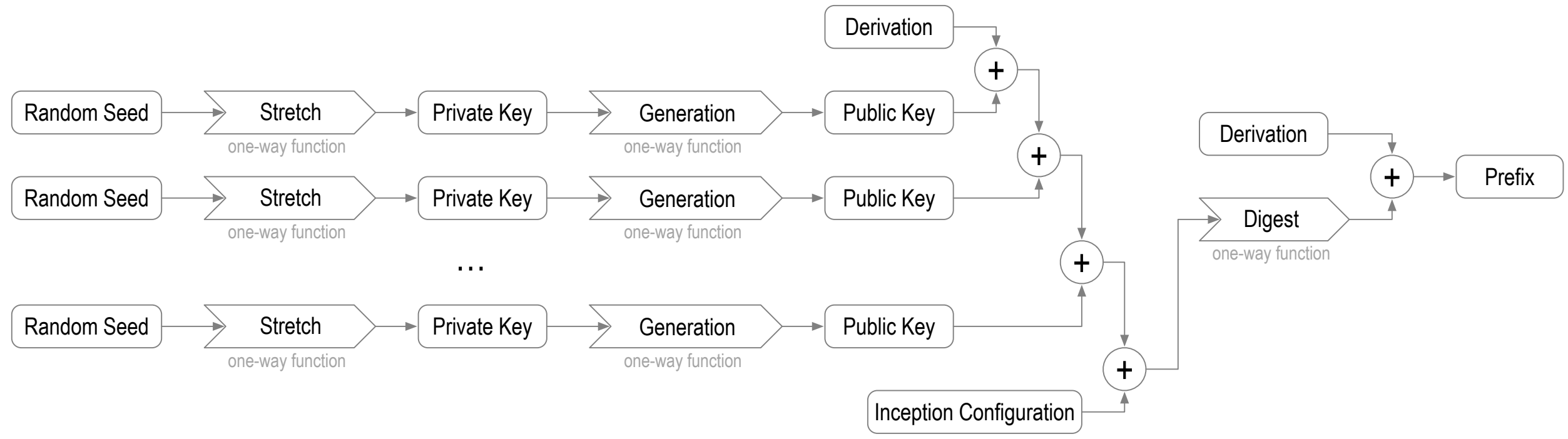

Figure 2.9. Multi-Sig Self-Addressing Prefix Derivation. The public keys have a derivation distinct from the digest's derivation.

The inception configuration data block shown in the figure could also include the value and type of threshold of the number of signatures from the multi-sig set of public keys that are needed for satisfaction.

### 2.3.4 Delegated Self-Addressing

An important use case for self-certifying identifiers is delegated identifiers, where a delegating identifier authorizes the delegated identifier to perform some subset of activities. A delegated identifier prefix may be created by including, at the very least, the delegating identifier prefix in the delegated prefix inception configuration data. This may also be generalized to include other essential delegating configuration data. This inclusion of the delegator's identifier cryptographically binds the delegating prefix to the delegated prefix and other delegating configurations, as well as the delegated prefix's controlling key pairs and other configuration data. The prefix may be represented as follows:

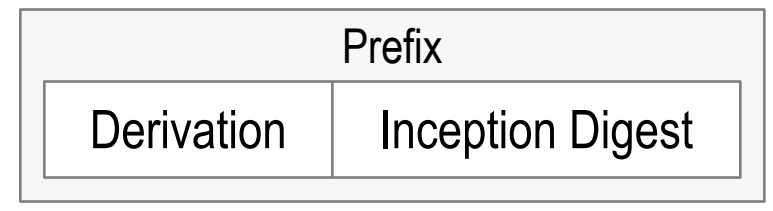

Figure 2.10. Delegated Self-Addressing Prefix.

A diagram showing the derivation to a hash of the inception data in the prefix for a delegated self-addressing self-certifying identifier follows:

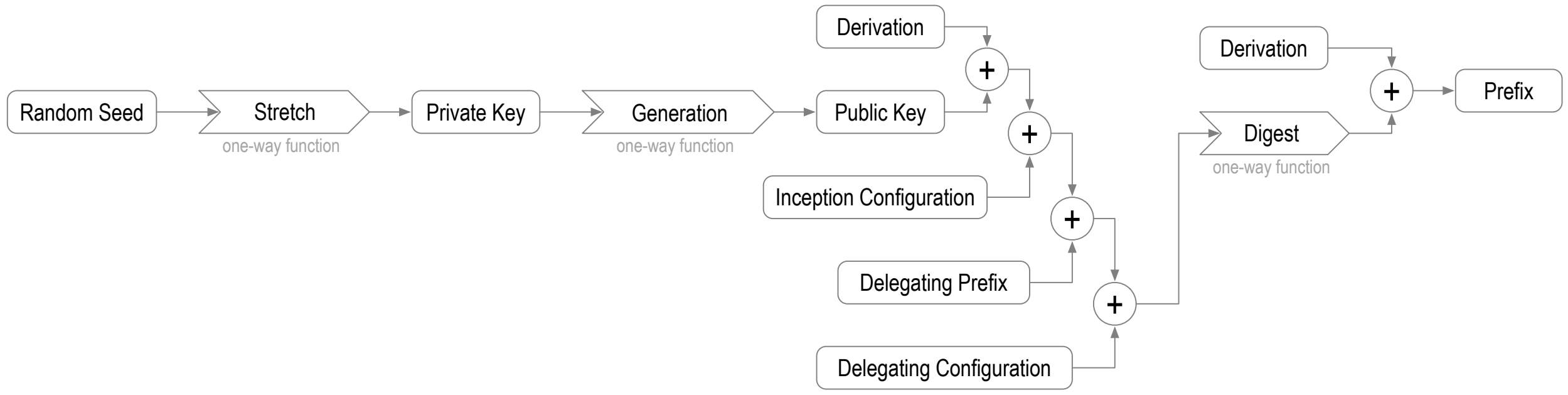

Figure 2.11. Delegated Self-Addressing Prefix Derivation. The public key has a derivation distinct from the digest's derivation.

A delegated prefix may also be made multi-sig by including a set of multi-sig public keys.

### 2.3.5 Self-Signing

While using a content digest in the prefix of a self-addressing self-certifying identifier may bind the identifier prefix to the inception data, it still requires an attached signature for verification. If instead of a content digest (hash), a signature of the inception data is used in the prefix, then no attached signature is required. Signatures not only provide collision-resistant content hashes, but the signature is non-repudiable. This gives the identifier self-contained verifiability that the controller (holder of the private signing key) created the signature. In other words, this makes the inception data both self-addressing for content addressable storage and self-signing. This approach may be particularly useful for legal documents or other digital media where the contents of the document (or a digest) are included in the inception configuration data. This makes the content addressable identifier strongly bound to the verifiable signature on the document, which establishes control over the document at the time of its entry into the database. The drawback of this approach is that for the same level of cryptographic strength, the signature is typically twice as long as the digest. This is the case for Ed25519 used in the example in Section 2.2 above, where signatures are 64 bytes long. This may or may not be a net disadvantage, depending on the application. Removing the burden of managing attached signatures might simplify some content-addressable storage systems. Furthermore, including the signature in the identifier prefix makes the inception statement more compact overall but may make other statements longer because the identifier is longer. Some content-addressable document storage applications, however, do not need many statements, so identifier length may not contribute much to total storage.

Abstractly the prefix for a content self-addressing self-certifying identifier may be represented as follows:

| Prefix | |
|---|---|
| Derivation | Inception Signature |

Figure 2.12. Self-Signing Prefix.

Signatures that are 64 bytes long encode to an 88 character Base-64 string that includes 2 pad characters. This means the derivation code, in this case, is two Base-64 characters long, not one. An example of the prefix with a two-character derivation code and a 64-byte signature encoded as an 88-character Base-64 string follows:

```
0BYbLIOlRx8xh5taCxW-_aCBoPboLAZjK5-
d1DP4OZ9PWn13BpPCe12ZFVZfFlSsM3Pv-zljbsJnR6Adz7iE5ZAw.
```

A diagram showing the derivation from a random seed to a hash of the inception data in the prefix for a self-addressing self-certifying identifier follows:

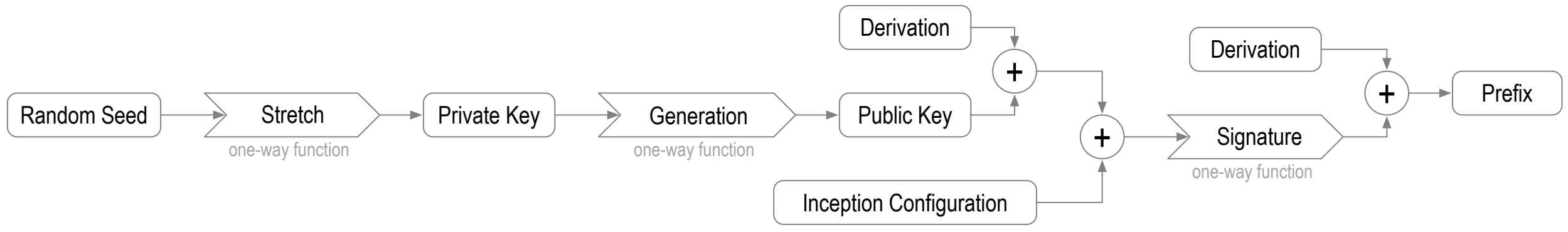

Figure 2.13. Self-Signing Prefix Derivation.

### 2.3.5.1 Generalized

Any unique chain of cryptographic one-way functions of sufficient cryptographic strength may be used to securely generate an associated type of self-certifying identifier. The derivation information must properly indicate the full chain of one-way functions used in the derivation. Given such a derivation and any other content included in the derivation, any recipient of the identifier may verify that the identifier was derived from a given private key(s) and thereby enable the recipient of any associated statement with attached signature(s) to verify the statement as authoritative for the identifier. KERI's security properties and support may apply to these generalized self-certifying identifiers. This enables universal extensibility for future types of self-certifying identifiers.

The prefix for a generalized self-certifying identifier may be composed of a derivation and the resulting derivative. This may be abstractly represented as follows:

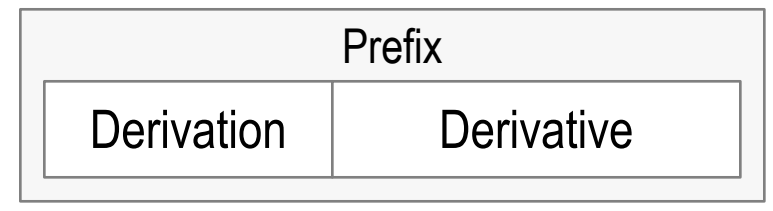

Figure 2.14. Generalized Prefix.

A diagram showing the derivation from a random seed to some number of one-way functions to the final derived self-certifying identifier follows:

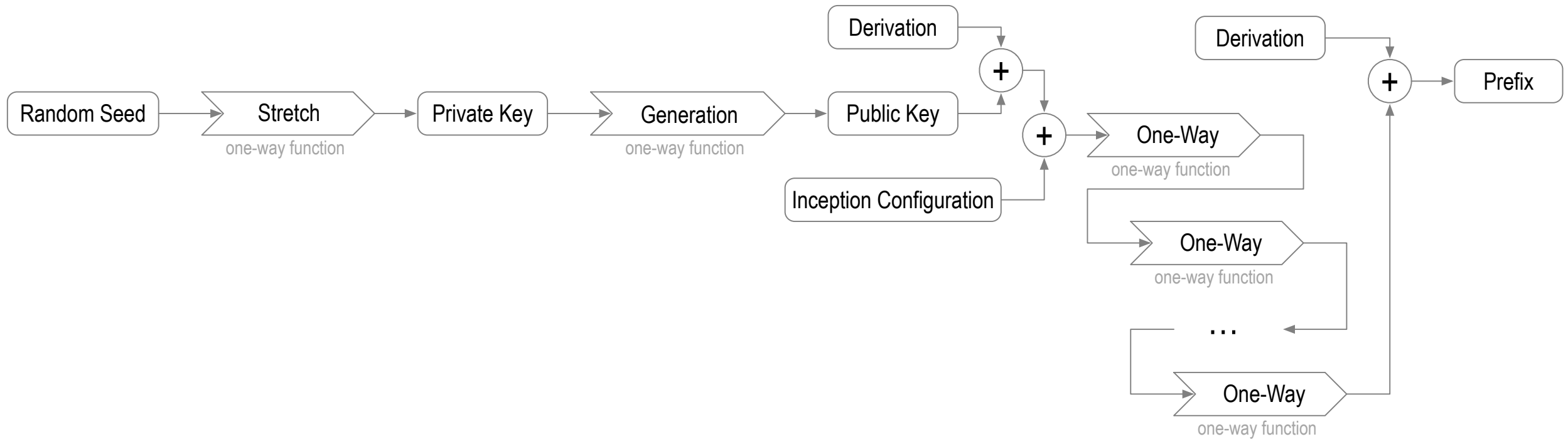

Figure 2.15. Generalized Prefix Derivation.

## 2.4 Cryptographic Trust Basis

The design of our portable decentralized identity system security overlay may be best understood through three important properties of its trust basis. These are *roots-of-trust*, *sources-of-truth*, and *loci-of-contro*l.

### 2.4.1 Roots-of-Trust

A *root-of-trus*t was defined previously. The primary *root-of-trust* for a self-certifying identifier is cryptographic in that the derivation process at inception makes a unique binding between one or more signing key pairs and the identifier prefix. This unique binding is established with one or more chained one-way functions that bind the entropy captured by the controller to the identifier prefix. Only the controller may universally uniquely issue the identifier associated with that entropy. The binding at issuance is cryptographically verifiable and does not depend on any operational infrastructure for its initial establishment. This provides for a totally decentralized root-of-trust for the secure issuance of identifiers.

A cryptographic root-of-trust in the binding between a digital signing key pair and an identifier provides a verifiable trust property. A digitally signed verifiable statement provides two additional verifiable trust properties, these are integrity and non-repudiability [110]. Integrity means the statement has not been tampered with since the signature was created. One can trust the statement was made as received. Non-repudiability means that no entity besides the private key holder (controller) could have created the signature. This means that the holder may not repudiate the statement while maintaining that they have sole possession of the private key. This means that key possession is crucial to maintaining the root-of-trust.

What a digital signature does not provide is a basis for trusting the facts of the signed statement or, equivalent the veracity of *what was said.* Informally, this means that it does not provide trust in *what was said* but only in *who said it* or equivalently the authenticity of *what was said*. Nonetheless, trust in *who said it* is useful because non-repudiation enables *consistent attribution*. Consistent attribution to *who said it* is a necessary condition for maintaining the possibility of trust (veracity) in *what was said* when that trust (veracity) is based on the reputation, credibility, or authoritativeness of *who said it.* With consistent attribution, one may trust that the controller made a set of statements. If those statements are also consistent in their content, then there is thereby the possibility of trust in the veracity. On the other hand, if the statements are inconsistent in their contents, then the non-repudiable source of that statement (the controller) is thereby proven to be inconsistent. Inconsistency is a basis for distrust. Another term for inconsistency is duplicity. Duplicity exhibits itself as inconsistent statements about some fact. To summarize, consistent attribution is a basis for assessing the veracity of *what was said* because *who said it* is not known to be untrustworthy. One may then build on that basis over time a stronger likelihood of trust by maintaining histories of statements and verifying them for consistency in order to detect duplicity.

### 2.4.2 Sources-of-Truth

A *source-of-truth* is someone or something that may make authoritative statements. One very useful approach to information systems design is to establish a single or authoritative *source-of-truth* for any item of information in the system [8; 94; 131]. In the case of an identity (identifier) system security overlay, we wish to establish a cryptographically verifiable source-of-truth about an identifier. Because at issuance, self-certifying identifiers make a universally unique cryptographically strong binding between the identifier and a key pair, there may be no other verifiable source-of-truth besides the controller who created the key pair and thereby holds the private key.

To restate, the process of creating a self-certifying identifier means that only the controller may make verifiably authentic control statements about the identifier. The identifier is thereby bound to the controller. This makes the controller the only authoritative or single source-of-truth for the identifier.

To elaborate, at inception, self-certifying identifiers have a single root-of-trust and a single source-of-truth that are cryptographically bound together. This binding creates what we call the *root-authority* for the self-certifying identifier. Because *root-authority* uniquely binds an entity, the controller, to a private key(s) by which that authority is exercised, therefore without loss of generality, we may refer to *root-authority* as either the power of control and/or the entity that exercises that power of control. This results in the following definition:

A *root-authority* has total and exclusive control over a self-certifying identifier. (2.2)

This exclusive total control is imbued in the *root-authority* at the inception of a self-certifying identifier by virtue of the process of creation from captured entropy. Exercise of this root-authority is first exhibited at issuance of the identifier. Root-authority may later be transferred, delegated, or attenuated by the exercise of that root-authority via cryptographically verifiable statements to that effect.

In contrast, in an administrative identity system, by virtue of its designation as administrator, the administrator is/holds the root-authority over identifiers it issues and not the holders of the private keys that the administrator binds to those identifiers. Thus, the security of an administrative binding is comparatively weak because it is a function of the security of the administrator's processes and not cryptography.

### 2.4.3 Loci-of-Control

In psychology, the term *loci-of-control* refers to the scope of a person's life over which they feel they have control. The root meaning of the word loci or locus means place. The term *loci-of-control* refers to the extent to which control over one's life either is driven by external forces or arises internally from the exercise of one's agency [2; 100]. We apply the term *loci-of-control* to an identifier system to answer the question of *"who controls what?"*. Does the control over bindings between identifiers, controllers, and key-pairs arise externally from an administrator or does it arise internally from the controller itself? One way of analyzing the degree of decentralization in an identity system is to categorize the relative scope of external vs internal *locus-of-control* over the bindings.

With self-certifying identifiers, the singular root-of-trust and source-of-truth, that support the its control-authority enable consistent attribution of verifiable authoritative control statements about an identifier. Given that there is only one source-of-truth, that source of truth may exclusively determine the sequence of control statements and any dependencies between control statements. Control over the sequencing means that the type of infrastructure needed to support and maintain the history of control statements may be simplified because any identifier's history of control statements may be ordered independently of any other identifier's history. Recognition of this property directly leads to KERI's approach. Primarily, control statements are used to establish and maintain control over the identifier. This includes the following types of statements:

- Control over the creation or issuance of an identifier. This may include designations of support infrastructure or pre-commitments to future transfers of control.
- Control over operations on an identifier.

Control operation statements may include the following:

- Transfer of control (non-revocable). This means transfer of control of the identifier to a different key-pair or set of key-pairs and different signing schemes such as multi-signature. This is also known as rotation. A transfer/rotation operation revokes the current set of key pairs and replaces them with a new set. This may also include transfer designations of support infrastructure
- Delegation of identifier (revocable). This creates a new identifier with its own set of keys and authorizes (revocable) some degree of control authority.

A notional diagram of the composition of these control statements follows:

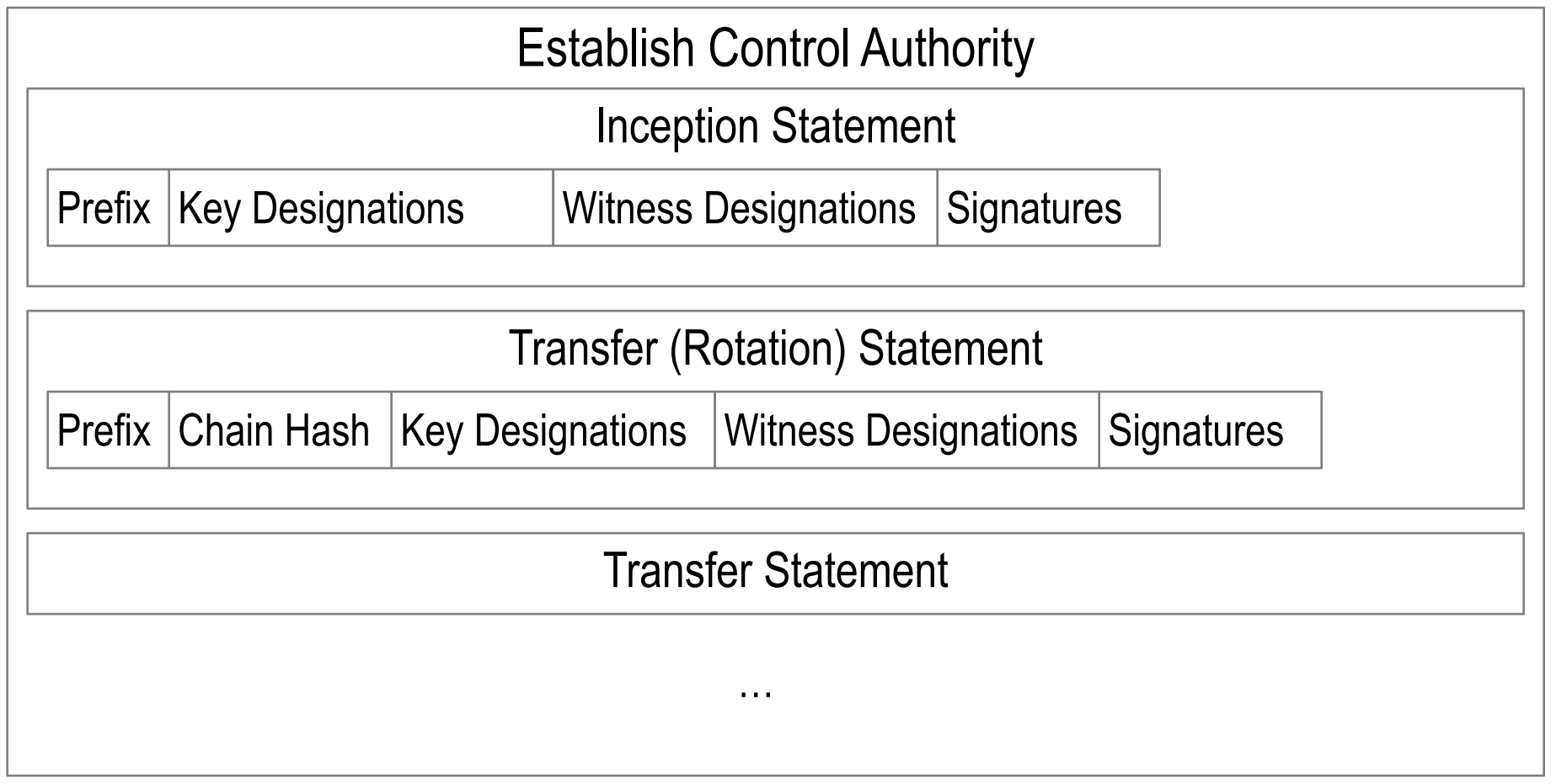

Figure 2.16. Control Establishment Statements

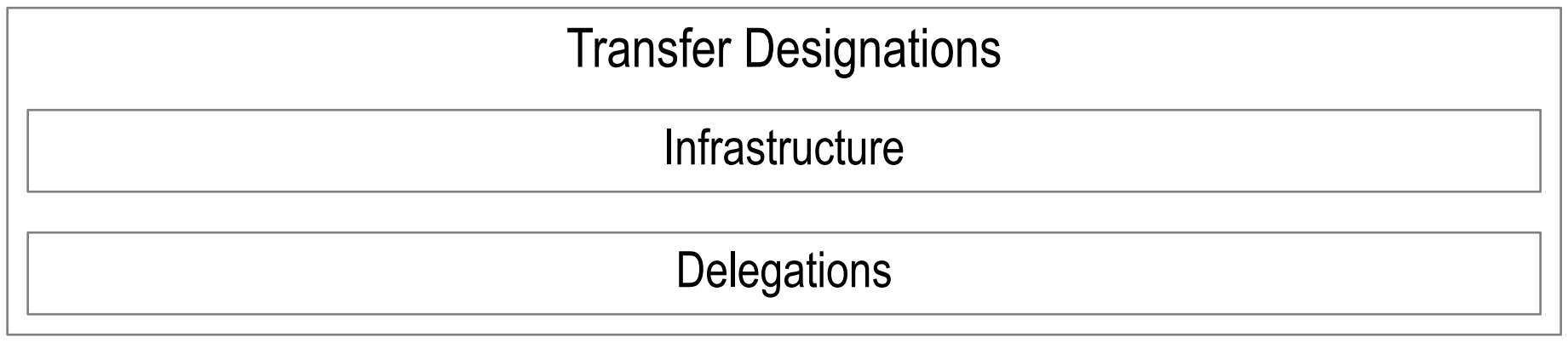

Figure 2.17. Transfer Designation

Once current control authority over an identifier has been established, that current control authority may make additional statements for purposes unrelated to establishing and/or maintaining its control authority. These statements may include authorizations of encryption keys for communication, endpoints for routing communications, service endpoints for other resources, or signed transactions for doing business. These are diagramed below:

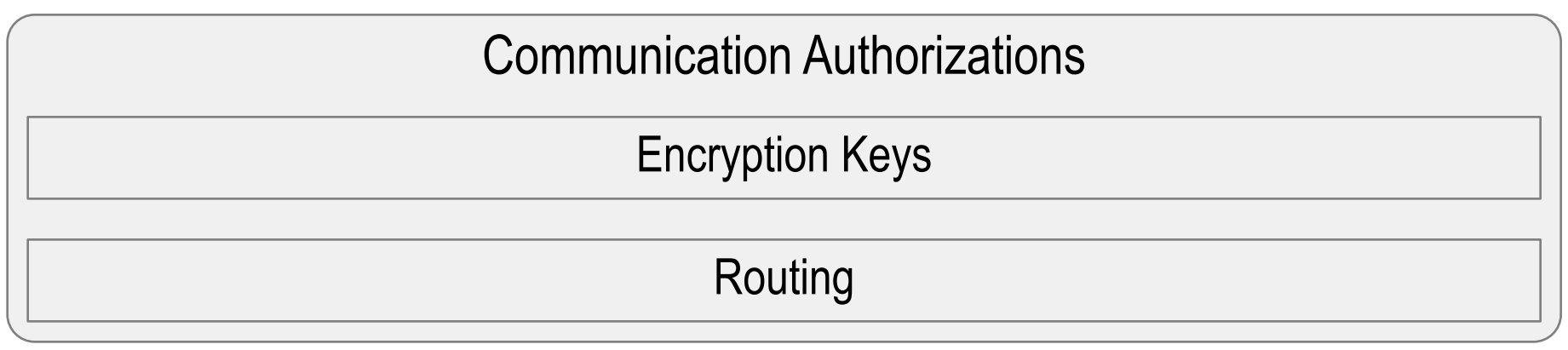

Figure 2.18. Communications Authorizations

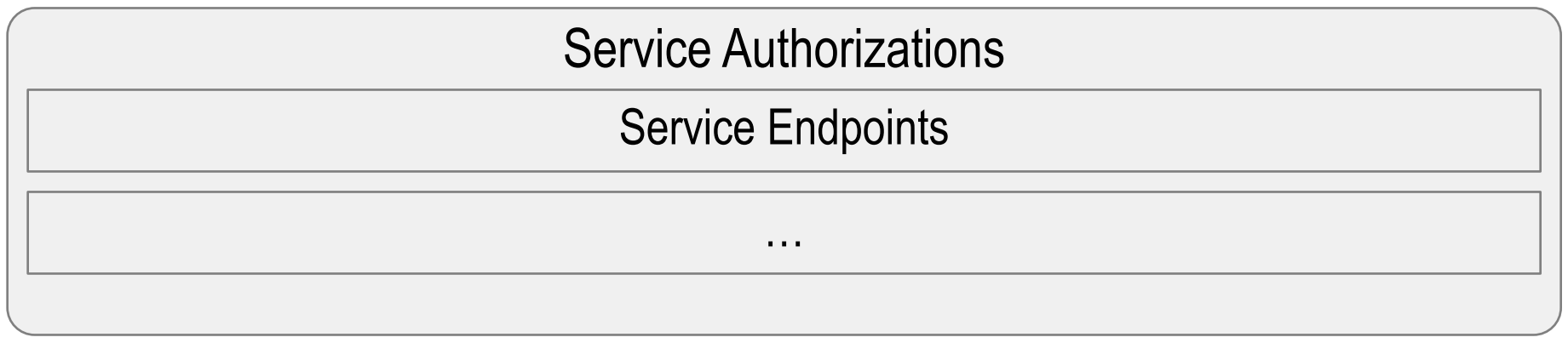

Figure 2.19. Service Authorizations

# 3 Autonomic Identifier (AID)

Although SSI is becoming recognized as an important new approach to identity, different implementations of SSI vary in the degree of decentralization and security. As a result there is a need to define more precisely a normalized theoretical model that best represents the ideals of self-sovereign identity. To avoid confusion with other variants, we have chosen to use the terms *Autonomic Identity System* (AIS) and *Autonomic Identifier* (AID) to name this more precise model. The roots of the word *autonomic* are the same as the word autonomous, which comes from the Greek words auto (self) and nomos (law), which together mean literally, self-rule or when applied as an adjective, self-ruling, self-governing, or self-regulating [18; 21]. The word *autonomic* is most commonly used to refer to the body's *autonomic nervous system* that controls the circulatory, digestive, respiratory, and other systems that act unconsciously, i.e., are self-regulating without conscious oversight [20]. Likewise, the term *autonomic computing (AC)* means some type of self-managing software capability such as self-configuring, self-securing, or self-scaling [6; 19; 51].

In that vein, we can say that an *autonomic identifier* is one that has *self-managing* or *self-governing* capabilities. This more fully evokes the meaning of *self-sovereign* but with the addition of self-management. The most important *self-management* capability is *self-certification* which, as we have shown, leads to *self-administration*. We defined *self-certification* above in Section 2.2 but it bears generalization here. As defined above, self-certifying identifiers include a prefix that uniquely (strongly) binds the identifier to the incepting controlling key pair at issuance. The self-certifying property is inherently completely decentralized in the broadest sense of the word. This *prefix* enables that identifier to be part of a namespace governed by that controller of the key pair.

## 3.1 Autonomic Namespace (AN)

A *namespace* groups symbols or identifiers for a set of related objects [108]. In an identity system, an identifier can be generalized as belonging to a *namespace* that provides a systematic way of organizing related identifiers with their resources and attributes.

To elaborate, a *namespace* employs some scheme for assigning identifiers to the elements of the namespace. A simple name-spacing scheme uses a prefix or prefixes in a hierarchical fashion to compose identifiers. The following is an example of a namespace scheme for addresses within the USA that uses a hierarchy of prefixes:

$$\texttt{state.county.city.zip.street.number.} \tag{3.1}$$

An example element in this namespace may be identified with the following:

$$\texttt{utah.wasatch.heber.84032.main.150S.} \tag{3.2}$$

where each prefix location has been replaced with the actual value of the element of the address. Namespaces provide a systematic way of organizing related elements and are widely used in computing.

We formally define an *autonomic namespace* (*AN*) as a namespace with a *self-certifying prefix*. A *self-certifying prefix* provides cryptographic verification of root control authority over its namespace [68; 90; 102; 103]. To clarify, each identifier from an autonomic namespace includes as a prefix an identifier that either is the public key or is uniquely cryptographically derived from the public key of the incepting (public, private) key pair. The associated private key may then be used by the controller to authoritatively (nonrepudiably) sign statements that authenticate itself and authorize use of the identifier. These statements include responses to challenges to prove control over the identifier. It is the essential enabling property of identifiers in a fully decentralized self-sovereign identity system. Thus *self-certification* enables both *self-authentication* and *self-authorization* capabilities as well as *self-management of* cryptographic signing keypairs. Together, these properties make the namespace *self-administering* .

An *autonomic namespace* is *self-certifying* and *self-administrating*. (3.3)

The *self-management* capabilities of *autonomic namespaces* best exemplify the touted *self-governing* property associated with self-sovereign identity systems [10; 45; 127; 132; 133; 136]. All derived *autonomic identifiers* (AIDs) in the same *autonomic namespace* (AN) share the same *root-of-trust*, *source-of-truth,* and *locus-of-control* (RSL). The governance of the namespace is, therefore, unified into one entity, that is, the controller who is/holds the *root authority* over the namespace. This unification of control authority simplifies the security system design and analysis. The associated implications lead to the KERI design.

The primary purpose of an autonomic identity system (AIS) is to enable any entity to establish control over an identifier namespace in an independent, interoperable, and portable way. This approach builds on the idea of an identity (identifier) meta-system [37; 38; 153] that enables interoperability between systems of identity (identifiers) that not only expose a unified interface, but add decentralized control over their identifiers. This best enables portability, not just interoperability. Given portability in an identity (identifier) meta-system system, *transitive trust* can occur, that is, the transfer of trust between contexts or domains. Because *transitive trust* facilitates the transfer of other types of value, a portable decentralized identity meta-system, e.g., an autonomic identifier system, enables an identity meta-platform for commerce [134].

## 3.2 Namespace Syntax

There are many name-spacing conventions for the syntax of hierarchically organizing the identifiers in the namespace. Of those conventions, the most widely supported is the Uniform Resource Identifier (URI) of which URLs (Uniform Resource Locators) and Uniform Resource Name (URN) are special cases (IETF RFC-3986) standard [147; 148]. Given its almost universal popularity, we adopt its hierarchical path, query, and fragment syntax for name-spacing autonomic identifiers (AIDs). In our national definition, the identifier starts with the characters *aid* to indicate the scheme. This is followed by a colon, and then the Base-64 encoded prefix as defined above. The prefix includes the derivation code and the derived self-certifying identifier. Other colon-delimited prefix options may follow. This is then followed by the URI components. More formally the syntax is as follows:

$$\texttt{aid:}\textit{prefix}\texttt{[:}\textit{options}\texttt{][/}\textit{path}\texttt{][?}\textit{query}\texttt{][\#}\textit{fragment}\texttt{]} \tag{3.4}$$

A notional example of an AID using URI components is as follows:

```
aid:BXq5YqaL6L48pf0fu7IUhL0JRaU2_RxFP0AL43wYn148/path/to/
resource?name=secure#really
```

This is not a complete specification, but it is sufficient to illustrate the concept. The prefix provides a means of authentication. The other components provide a means of accessing a resource. Any signed statement referencing an AID may be verified as authoritative and may serve as authorization.

### 3.2.1 Properties

An AID has several useful properties.

- Universally Unique Identifier (UUID). By virtue of its prefix, an AID is universally unique thus it can be used instead of a UUID as defined by RFC 4122 [5]. UUIDs enable distributed applications to create unique identifiers without a central identifier server. Prefixed name-spacing allows for sorting and searching properties such as time order, lexical order, nesting, etc.
- Mini language for performing operations on resources (ReST). The URI components are compatible with the tooling that has been developed for ReST interfaces. This fosters adoption in micro-service and serverless applications but with security built-in not bolted on.
- Hierarchically deterministic derived self-certifying identifier. Using the URI components, one can easily specify hierarchically deterministic derivation paths for identifiers that enable the creation of derived (public, private) key pairs without having to store private keys but merely re-derive them from the identifier and the root private key. This helps with key management. An example follows:

```
aid:selfcertroot:/path/to/data?key-path=parent/child/child/child
```

These features make AIDs a candidate for a universal identifier scheme to be used throughout a computing system. This body of work will explore in more depth the design of *autonomic namespaces* and their associated security and operational features. There is some similarity between the use of the cryptographic identifier prefix in KERI and public encryption keys in the *key transparency* project [93]. The similarity is that in KERI, a basic self-certifying identifier is a public signing key with a derivation code, whereas key transparency is a verifiable directory of public keys, primarily public encryption keys, although signing keys may be used. There are several other important differences. In KERI, identifier prefixes may be any of the self-certifying identifier types, not merely public keys. Furthermore, the key events in KERI are separable for each identifier prefix. This enables support for GDPR’s right to be forgotten [67]. KERI uses pre-rotation for key rotation (see later). KERI’s key event logs allow anchoring of issuance events. Finally, the identifier derivation code enables self-contained crypto agility. In summary, KERI is a full-featured DKMI, whereas key transparency is simply a verifiable public key directory.

## 3.3 Unified Identifier Model

In general, we want to have a theory or model of how to design identifier systems with desirable properties. Historically identifier systems has been designed without any unifying theory to guide the designers. This means that many identifier systems are largely bespoke to a given application. This may be problematic especially when the identifier system does not have a well defined approach to security. More recently, explicitly characterizing identifier properties has gained some recognition as a prerequisite to identifier system design. For example, the well known Zooko's triangle or so called tri-lemma is based on recognizing a conflict between three different desirable properties for an identifier [155]. The three properties are human meaningful, secure, and decentralized. The tri-lemma states that an identifier may have any two of the three properties by not all three. Some have claimed to have solved the tri-lemma by using a hybrid identifier system where human meaningful but insecure identifiers may be registered on a distributed consensus ledger. The registration process itself uses identifiers that are secure and decentralized but not human meaningful [142]. The fact that not one identifier but a combination of identifiers is needed to solve the tri-lemma hints that the identifier model therein employed is incomplete.

The autonomic identifier (AID) model, however, provides a complete unified model of identifiers (digital). This model takes a security first approach. The model is based on the recognition that all other properties of an identifier are of little value if the identifier is not securely controlled. In this sense the AID model provides a formal theory of identifiers. All identifiers maybe classified and described within the framework of this theory. This theory provides a principled approach to identifier system design and may be used to inform and guide the design and development of any concrete implementation of an identifier system.

The function of KERI is to establish control authority over an identifier (AID) in a cryptographically secure and verifiable way. Once that control authority has been established, a verifier may then securely verify any associated authorizations issued by the controller of the AID. These authorizations have the form of a signed authorization statement where the statement typically includes the AID under which the authorization is issued. A verifier may then verify the authorization by verifying the attached signature using the keys that were authoritative at the time the authorization was issued. These authorizations are secure to the extent that the established control authority is secure. The authorizations inherit their security from their associated AID. The trust domain of an AID is the set of statements that may be securely verified as being issued by the controller of that AID (along with any associated transactions or interactions). Authorizations then inhabit and are protected by this trust domain. In other words the authorizations are under the aegis of the AID (or equivalently the controller of the AID) that issued them.

### 3.3.1 Legitimized Human Meaningful Identifiers (LIDs)

Given an AID we may now unify the other desirable property of an identifier, that is human meaningfulness. The cryptographic material in an AID makes it decidedly non-human meaningful. In contrast, examples of human meaningfulness might be easily recognized words or names or more relevant some efficiently hierarchically organized composition of characters. Human meaningfulness has two limiting characteristics. The first is scarcity, this exhibits itself in various undesirable ways such as name squatting, or race conditions to register or otherwise assert control. More importantly, there is no inherent security property of a human meaningful identifier. This makes them insecure by default. Nevertheless, given the trust domain of an AID any human meaningful identifier may be uniquely authorized, sanctioned, or legitimized within that trust domain. Because only the controller of the AID may issue verifiable authorizations associated with that AID, that controller alone may authorize the use of any human meaningful identifier under the aegis of its trust domain. The associated authorization statements legitimize the use of that human meaningful identifier within the trust domain. This gives rise to a new class of identifiers, we call a member of this class a legitimized human meaningful identifier and use the acronym LID. The important property of an LID is its verifiability with respect to a given AID. The pair forms a new identifier couplet that we may represent as follows:

$$aid|lid \tag{3.5}$$

where $|$ represents that the succeeding *lid* is authorized within the trust domain of the preceding *aid*. We can diagram this relationship as follows:

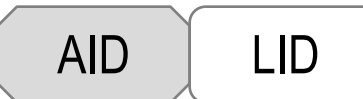

Figure 3.1. AID|LID couplet

This coupling is a special type of name spacing. For example suppose the *lid* is a library Dewey Decimal code for a book such as:

`625.127C125r` (3.6)

Further suppose that the aid prefix is given by:

`EXq5YqaL6L48pf0fu7IUhL0JRaU2_RxFP0AL43wYn148` (3.7)

then full *aid|lid* couplet may be expressed as follows:

`EXq5YqaL6L48pf0fu7IUhL0JRaU2_RxFP0AL43wYn148|625.127C125r` (3.8)

The trust domain of an AID provides a context in which to interpret the appearance of any LID. The AID is implied by the context. This means that the AID may not have to be prepended or appear with the LID. This allows the human meaningfulness of the LID to exhibit itself without being encumbered by the AID. Any verifier of the LID, however, knows from the given context how to cryptographically verify the legitimacy of the LID. Thus any existing human meaningful identifier may be converted to a verifiable LID by association with an AID's trust domain. This is shown below:

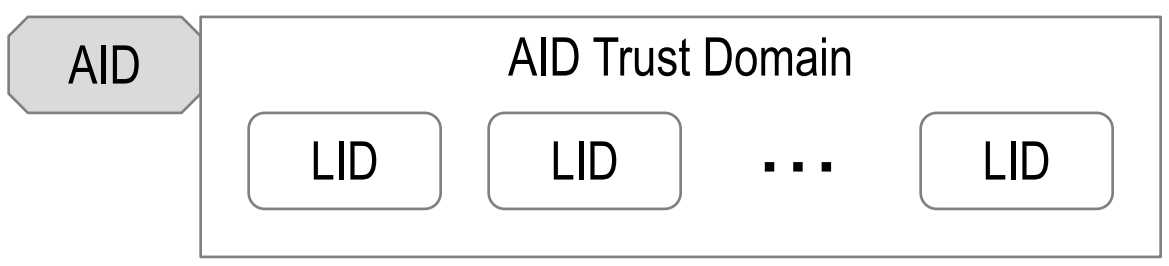

Figure 3.2. Multiple LIDs within the trust domain of an AID.

To elaborate, this model of an *aid|lid* couplet unifies all desirable identifier properties into one identifier system model. The *aid* part provides the security infrastructure while *lid* part provides the application-specific human meaningfulness. The connection between the two is provided by a verifiable legitimizing authorization represented by the |. With this model, there is no longer any global scarcity associated with a given LID because each AID may have its own copy of the LID. Scarcity only occurs within the trust domain of each AID but that is completely managed by the controller of that trust domain (AID).

To further explicate this concept, we may characterize any *aid|lid* couplet as consisting of two classes of identifiers, namely, *primary* and *secondary*. *Primary* identifiers are AIDs and are self-certifying to a cryptographic root-of-trust. *Secondary* identifiers are LIDs and are not self-certifying but are secured under the aegis of an associated *primary* identifier's trust domain.

All other identifiers used in a system may be grouped into a *tertiary* class of identifiers that has no external security properties but may be useful within the confines a local application. Identifiers that appear within an application user interface or other local context but have no explicit verifiable connection to a root-of-trust are *tertiary*. To clarify, *tertiary* identifiers may only be used internally whereas because *primary* and *secondary* identifiers are both secured via a verifiable authorization with respect to the root-of-trust of the *primary,* both may be used externally. A *primary* may appear alone but a *secondary* must either appear within the known context of the authorizing *primary's* trust domain or as a couplet with the *primary* itself.

This unified model may now be used to guide the design of any identifier system. The design of the *primary* AIDs may be tuned to the specific security, performance, and governance properties of the application. The design of the *secondary* LIDs may be tuned to provide any other desirable human meaningful properties of the identifier.

# 4 DISTRIBUTED LEDGERS

The primary alternative to autonomic namespace identifiers for decentralized key management infrastructure are identifiers registered on a distributed consensus algorithm-based ledger that provides the primary root-of-trust and source-of-truth for associated key management operations. In other words, the ledger is the trust basis for a decentralized key management infrastructure (DKMI). There are many types of distributed consensus algorithms with various properties. One useful property of many distributed consensus algorithms is a total (global) ordering of events from multiple sources. This allows all transactions on the associated ledger to have a unique ordering with respect to one another. In the case of key management, for example, the total ordering property makes it easy to establish the ordering of key inception and rotation events. A distributed consensus ledger, however, may require a significant amount of infrastructure that must be set up, operated, and maintained. Typically infrastructure that depends on distributed consensus ledgers must make trade-offs between cost, throughput, and latency. As a result, the infrastructure may not be as scalable or performant as infrastructure that does not depend at all on a distributed consensus ledger or that minimizes that dependency.

## 4.1 Hybrid Trust Basis

Practically speaking many so-called SSI systems provide some degree of self-sovereignty over identifiers if not complete self-sovereignty. They are more-or-less decentralized. A popular approach to SSI is to use a shared distributed consensus ledger to provide the bindings between controller, identifier, and key-pair(s). As mentioned above, distributed consensus ledgers can be implemented in ways that are highly resistant to attack. The governance of these ledgers can be highly decentralized. The trust basis involves two different primary roots-of-trust. The first-root of-trust is the ledger itself because transactions on the ledger are authoritative for the identity system. The first root-of-trust is therefore operational infrastructure, albeit highly resistant to exploit. The second root-of-trust is the user-created address (identifier) used to access the ledger. This identifier is derived from a public key of a user-created key-pair The second root of trust is therefore cryptographic and is a limited form of a self-certifying identifier described above in Section 2.#.2. A shared distributed consensus ledger is a logically centralized construct but whose validator nodes cooperate (i.e. come to consensus) to write replicas of the ledger. Ledgers may be governed in a more-or-less decentralized manner. Some ledgers are permission-less with respect to validator nodes and some are permissioned. Permissioned ledgers may have either private or public governance. Typically the ledgers are open for use by any participant who pays the cost of use. Even though, on a permission-less ledger, the ledger is renting space to participants (users) there is no way for the ledger nodes as a whole to censor which users may submit and pay for transactions. Public permissioned ledgers may provide checks and balances on the ability of nodes to censor users.

## 4.2 Ledger Registration

In general, the registration process is as follows: In order to enter transactions on a ledger, each participating user first creates a (public, private) key pair. The public key is used to create a special case identifier that enables access by the user to enter transactions on the ledger. For example the “Bitcoin address” or “Ethereum address” identifier is derived from this public key. The user proves control of the public key as ledger access identifier by signing transactions with the associated private key. Given the user now has access, the user may now register identifiers on the ledger that belong to an associated ledger-based identity system. Typically such an identity system uses an incepting transaction on the ledger to register a binding between a controller, identifier, and one or more key-pairs. The registered identifier and registered key-pair may be different from the ledger access identifier (public key) used to initiate the registration transac-tion. Registrations, however, makes an implicit binding between the controller of the ledger access identifier and the ledger registered identifier. There may be two (public, private) key-pairs involved in this registration process. The first to control the registration transaction and the second to prove control over the registered identifier. As a special case, they may be the same key-pair. When they are distinct, the first controller to register the second will be authoritative. This creates a potential race condition. When they are the same the registration transaction effectively transfers control of the access/registered identifier to the new key-pair bound in the registration transaction. The following diagram illustrates this process.

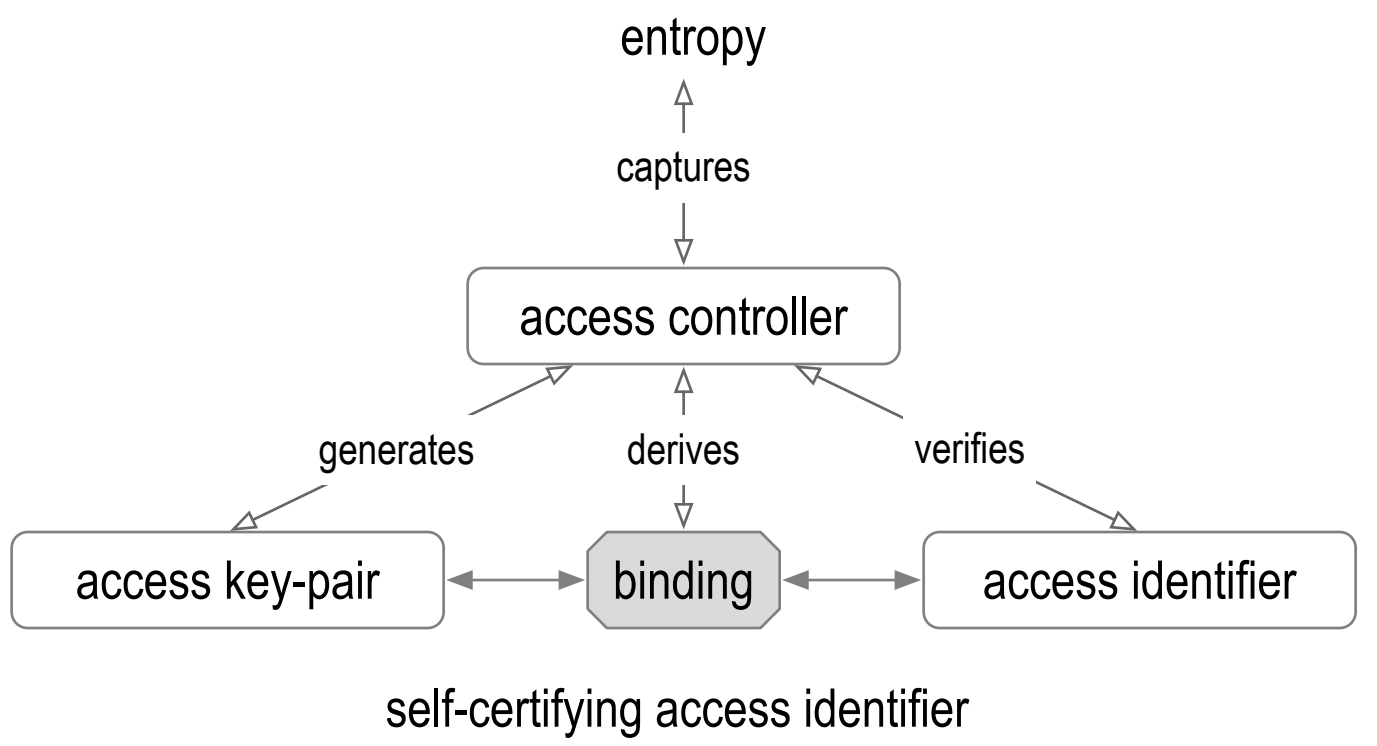

Figure 4.1. Ledger Access Identifier Binding. User (controller) generated access identifier key-pair that is used to create identifier registration transaction on ledger.

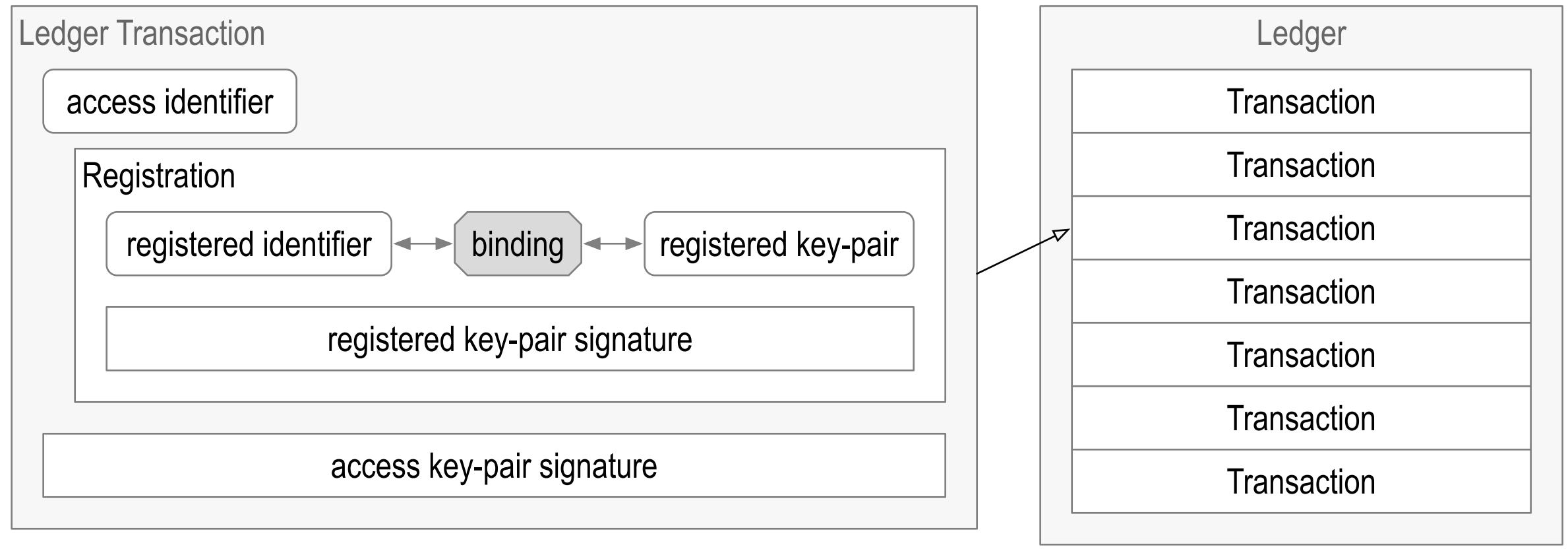

Figure 4.2. Ledger Registered Identifier. Registration transaction creates binding on ledger between a registered identifier and its key-pair.

To reiterate, a form of secure decentralized registration is enabled by allowing each user to create their own access key-pair where the access identifier is derived from that key-pair. The purpose of this identifier is to enable the user to interact with the ledger. The user proves control of its access identifier by signing with the associated private key. This identifier is usually different from the identifier registered on the ledger. As a result, there are two roots-of-trust in this approach. The first is the ledger which requires trust in its operational infrastructure and governance. The second is the user's ledger access identifier that is the public key of the user-selected public/private key pair. This is cryptographic. The binding between the registered identifier and the user control over it is based on both roots-of-trust. As such the trust basis is a hybrid of cryptographic and operational roots-of-trust. The important insight is that although the registration process is decentralized the trust basis of that registration includes as a primary root-of-trust the ledger and its associated transaction security mechanisms. As such the registered identifier is locked to the ledger and any associated trustable interactions are thereby locked to the ledger's trust domain. Therefore the registered identifier per se may not be necessarily portable to other trust domains.

In contrast, autonomic identifiers, are fully portable across trust bases and domains. A ledger could be used with an autonomic identifier to store a copy of the control transfer log mentioned above. In this case the identifier is not locked to that ledger and as a result the ledger is merely a secondary root-of-trust. The KERI design provides an identity system security overlay with full portability of identifiers by using a trust basis that relies on a cryptographic primary root-of-trust and only uses operational infrastructure as a secondary root-of-trust.

# 5 Trust Spanning Layer

The projection of any entity onto the internet with respect to any other entity is just the data stream communicated between the entities. Indeed an entity and its projected data stream are made indistinguishable by virtue of the internet as a digital communications medium. This is why in the limit a data item is a valid entity with respect to an identity system. One purpose of those projected data streams is to enable entities to interact remotely. These interactions may have value and therefore need to be trustworthy. The problem with the internet is that we do not have a human basis for building trust in those projections [45; 153]. We, therefore, need another basis for trust. But as mentioned in the introduction, the internet was not designed with a built-in security mechanism to provide that basis of trust. To reiterate, this work seeks to provide an internet-appropriate basis for trust by way of a universal security overlay.

## 5.1 Unifying Trust Bases and Trust Domains

The ideal would be a universal trust basis that resulted in a single trust domain for all internet interactions. Absent such an ideal a more practical approach is a *trustable spanning layer* for internet interactions [7; 23; 24; 42]. The essential feature of a *spanning layer* is that it provides a single interface through which all the supporting infrastructure below it and all the supported applications above it may interoperate. In terms of security, a trustable spanning provides a single interoperable mechanism or protocol through which all spanned trust domains are supported by all spanned trust bases. This single mechanism simplifies the interaction between the participants in the spanned trust domains. It must provide secure portability between trust bases for the binding between the identifier and its control well as control of the mapping between the identifier and associated resources. Associated resources include not only IP addresses but any supporting associated infrastructure, services, and especially content. Convenient portability effectively unifies all spanned trust domains into one and subsumes all trust bases into one. We believe autonomic namespaces with ambient verifiability are the best candidate to provide that spanning layer protocol.

## 5.2 Background

The concept of a spanning layer was formulated to describe the hourglass shape of the dependencies between internet protocols [7; 23; 24; 42]. The following diagram illustrates this perspective:

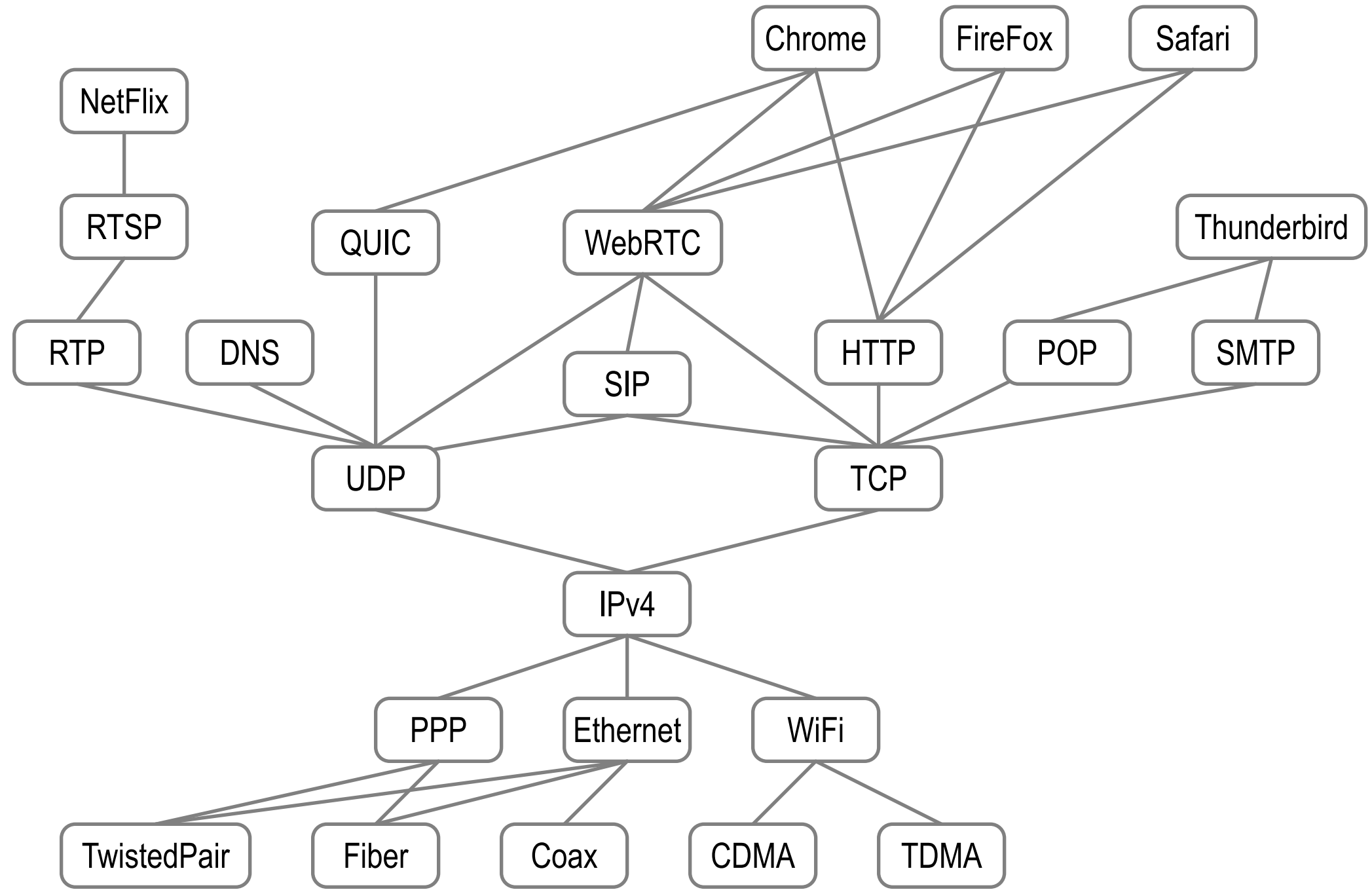

Figure 5.1. IP Hourglass Shape

As the diagram above shows, the IP protocol is the one common protocol for all supporting protocols below it and all applying protocols above it. In other words it *spans* both its support and its applications. It is therefore the *spanning layer* of the internet [7; 23; 24; 42]. This can be represented more abstractly as a set of layers in an hourglass where the narrow waist of the hourglass is the IP layer.

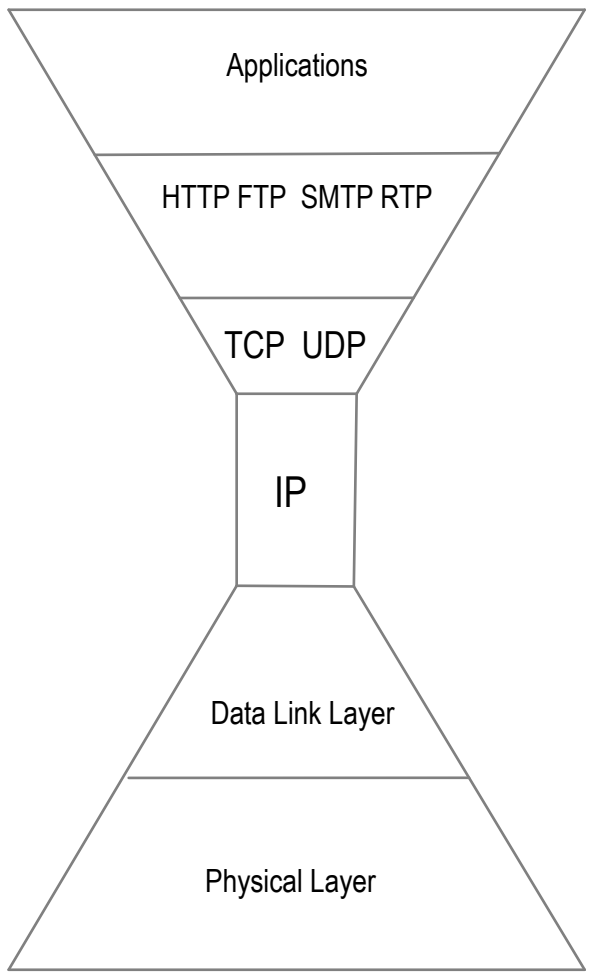

Figure 5.2. Abstract Hourglass

The realization that IP is both the spanning layer for the internet and has no built in security mechanism makes it obvious why security on the internet is broken; its spanning layer is insecure (not trustable)! One solution to this security problem would be to add a trustable security overlay to IP that becomes a trust spanning layer for the internet. The design of such a spanning layer may be informed by the principles in the hourglass theorem [23; 24].

The hourglass theorem characterizes the features of a spanning layer [23; 24]. Each layer in a layered system design may span multiple supports in the layer beneath it and multiple applications in layer above it. The hourglass theorem defines *weakness* in a layer as providing less functionality. The simplicity, generality, resource limiting, and deployment scalability of *weaker* lay-ers tend to gain broader adoption vis-a-vis the complexity of *stronger* layers. To elaborate weakness gives rise to the following properties in a layer:

- *Simplicity* means that there one way to access (orthogonality) any service or resource through the spanning layer.
- *Generality* means the spanning layer has the richest set of possible applications without increasing its logical strength.
- *Resource limiting* means the spanning layer abstracts and limits resource access by supported applications.
- *Deployment scalability* means widespread adoption though minimally essential functionality.

A *minimally sufficient* spanning layer is the weakest possible layer that still supports the necessary applications. Its an optimization. The hourglass theorem posits that the design of the Internet as a layered stack of weak layers led to its universal adoption. The following diagram illustrates the optimization:

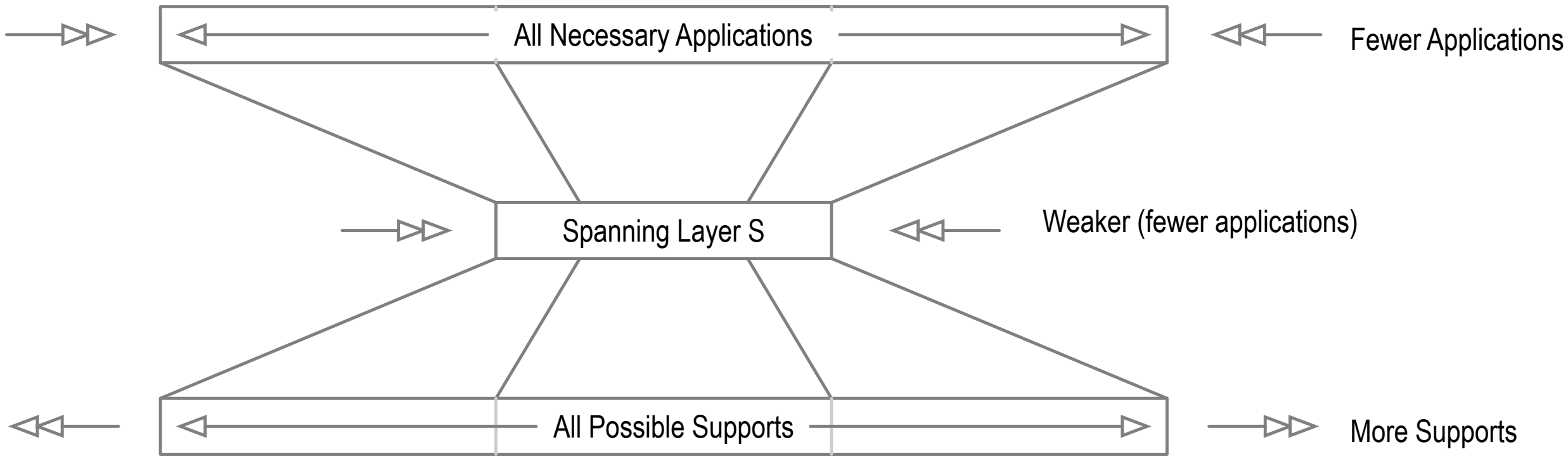

Figure 5.3.  Optimal Spanning Layer

We believe that autonomic identifiers AIDs with KERI form an Autonomic Identity System (AIS) that is a viable candidate for a minimally sufficient trust spanning layer for the internet. Its self-certifiable primary root-of-trust is simple and requires no operational infrastructure to issue identifiers. As will be seen below, KERI provides minimal secondary root-of-trust to support transfer of control to new key-pairs. Because an identity system security overlay necessarily uses protocols above the IP layer, AIS cannot span the internet at the IP layer but must span somewhere above it. This gives rise to a double waisted or waist and neck shape where the security overlay is the neck. This is shown in the following diagram.

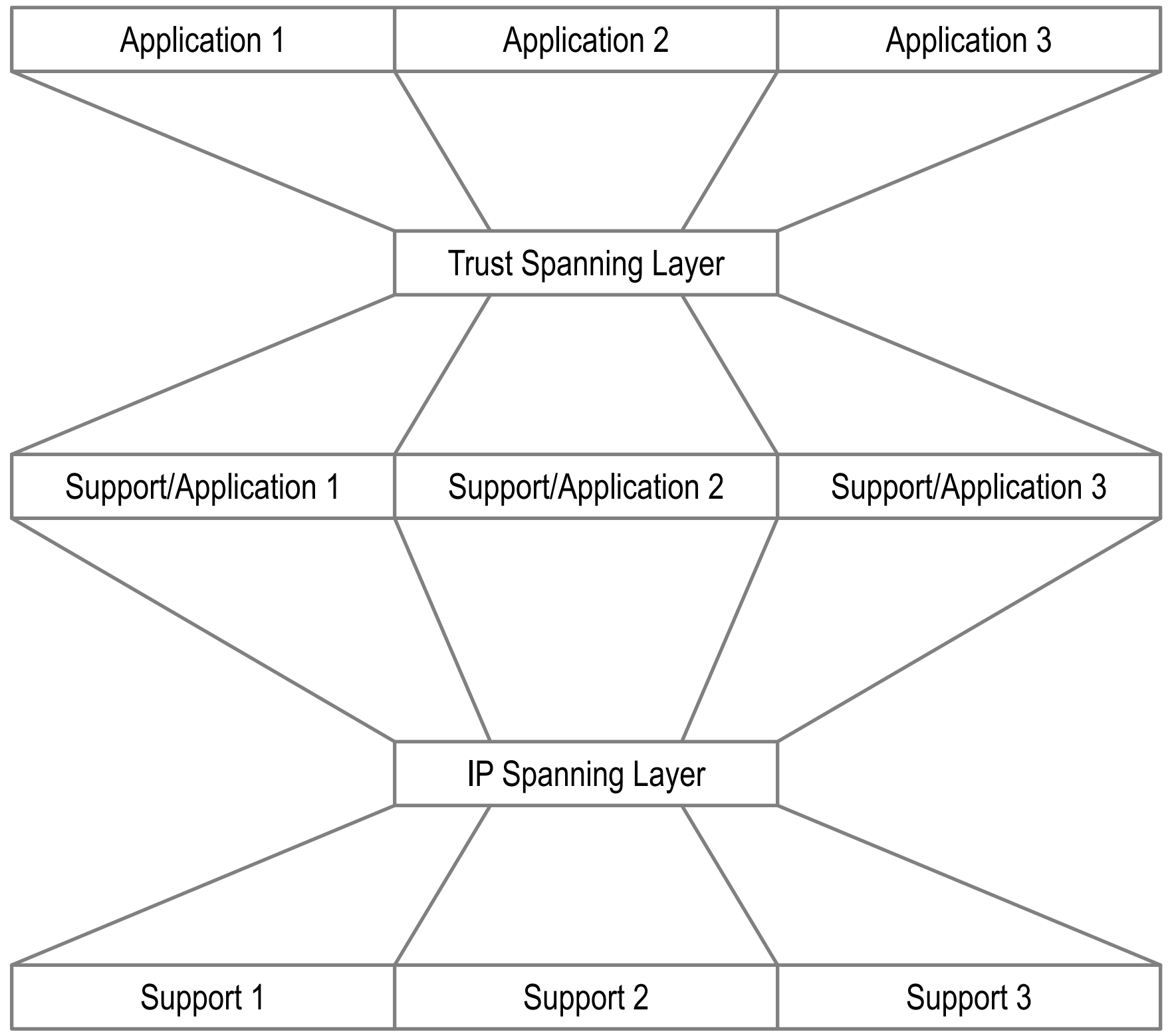

Figure 5.4. Trust Spanning Layer as Secure Neck of the Internet

In this case the supports are trust bases or platforms that interoperate using the standard set of KERI statements. Portability between application trust domains is thereby enabled. This is shown in the following diagram.

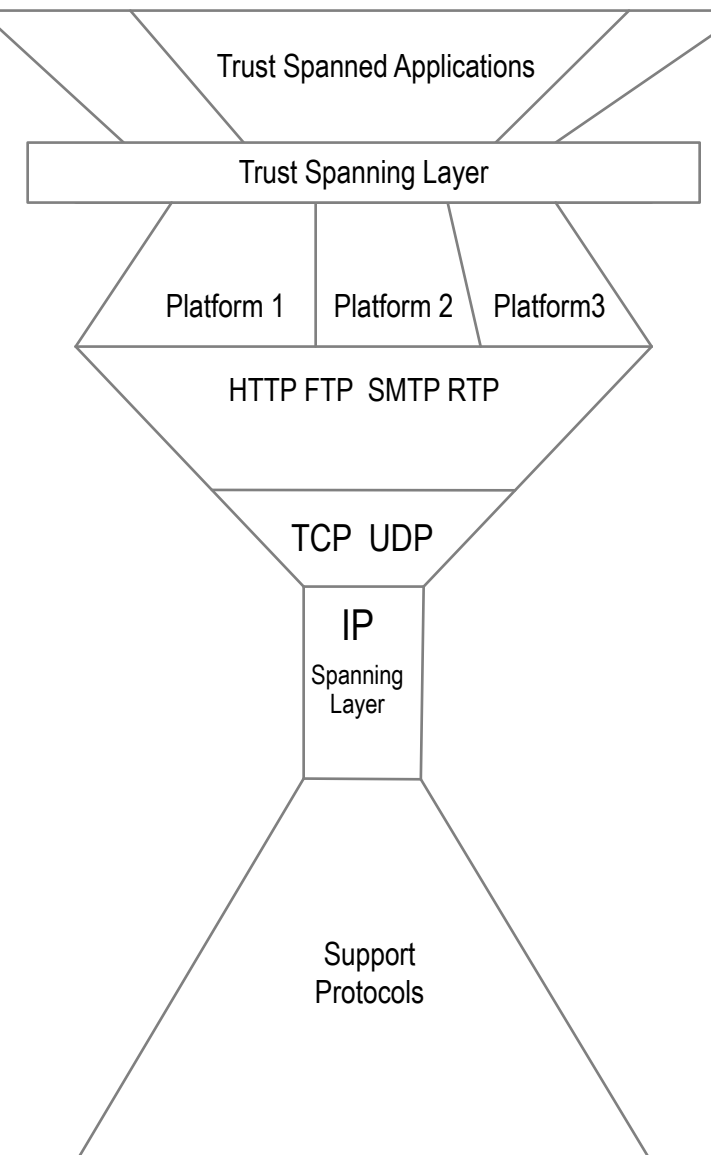

Figure 5.5. Trust Spanning Layer

## 5.3 Secure Identifier to IP Address Mapping

Given that an Autonomic Identity System (AIS) may serve as a trust spanning layer for Internet, what then becomes the role of DNS? Initially DNS can act as a layer below AIS with secure authorizations that map AIDs to DNS identifiers. Eventually however AIS can replace DNS by mapping AIDs directly to IP addresses. Recall that the cryptographic root-of-trust means that a signed statement may service as a cryptographically verifiable authorization. In DNS the mapping between identifiers and IP addresses happens in a zone file. As discussed above the security weaknesses of the DNS system arise from its dependence on the proper administrative management of operational infrastructure. The provenance of the mapping is subject to vulnerabilities in the operational infrastructure. Whereas an AIS may provide cryptographically verifiable provenance back to the self-certifying root-of-trust.

Although not provided here, an extension of this work would to be define a standard discovery and resolution system for AIS (KERI) that could replace DNS. This would provide a decentralized secure universal open standard overlay that would allow any controller (user) to map identifiers of its own choosing to computing devices with associated IP addresses. The value inherent in the identifiers would therefore be solely under the control of the user. Although IP addresses are administratively allocated, the original intent of the internet design was that IP addresses world be freely and non-discriminately allocated to computing devices based on geography. Unfortunately, the original allocation of IPv4 addresses reserved large blocks for some entities that created a premature artificial shortage thereby making them not so free [81; 82]. The switchover to IPv6 will restore the essentially free non-discriminatory nature of IP address allocation [83; 84]. This switchover is also being accelerated by the deployment of 5G networks [34]. The transition to IPv6 provides a uniquely historic opportunity to "fix" the trust layer of the internet. IPv6 addresses are 128 bits long with a 64 bit prefix and a 64 bit suffix. The 64 bit host suffix allocates 64 bits of address space to each local network thereby giving each local subnet more address space than the entire IPv4 global address space [83]. This enables usage, service, or data flow specific addressing per device in each local address space. This means that a secure direct identifier overlay to IPv6 that bypasses rent-seeking DNS is not only possible but imminently practical.

As previously mentioned, an identity system security overlay binds controllers, identifiers, and key-pairs. These bindings enable the system to thereby securely map the identifiers to IP addresses of associated devices. With an AIS a controller may prove via a verifiable digital signature the mapping between its identifiers and the IP addresses of its devices. It makes secure establishment of control authority happen first not after the fact. Given that a controller is able to independently control the mapping of its identifiers to IP addresses, the controller may assign at will those identifiers to devices with IP addresses and then reassign them as needed. The controller may then also move at will the resources or content associated with those identifiers. The URI portion of an AID may then be mapped transparently between devices without breaking any links. This provides controller (user) defined indirection of links (identifiers) to content and hence value. Thus content and value in a sense become autonomic as well!

# 6 Key Management

## 6.1 Minimally Sufficient Means

In our opinion, the term *minimally sufficient means* better captures the meaning of weakness used in the hourglass theorem for layered protocol design [23; 24]. We have long used minimally sufficient means as a design aesthetic [45; 136]. This design aesthetic lead us to design this proposed approach to a trust spanning layer (security overlay) for the internet. Likewise this design aesthetic leads to more scalable and performant Decentralized Key Management Infrastructure (DKMI) that does not require a totally ordered distributed consensus ledger, but may still use one as a secondary root-of-trust or source-of-truth if at all. It leverages the fact that only the holder of the private key may create and order events that perform verifiable operations on the keys. Thus a secondary root-of-trust and source-of-truth merely needs to witness events and their ordering not provide the ordering. As long as one complete verifiable copy of the event history is preserved the provenance of control authority may be established. The main goal of KERI is to provide a standard mechanism for provenancing the sequences of controlling key-pairs for transferable self-certifying identifiers. This mechanism is a witnessed (signed) log of key events receipts that provides a secondary root-of-trust and source of truth to enable end-verifiability of the current control authority over the identifier.

For decentralized identity systems based on self-certifying identifiers, protection and management of the associated private keys is essential. Because the controlling entity holds the private key(s) the primary burden of management falls on that entity or its assigns. The security of the identity is a function of the security of the management infrastructure. As mentioned above, unlike a centralized or administrative identity system where a central administrative entity controls all the identifiers, a decentralized identity system may have a multitude of controlling entities each controlling one or more identifiers. Some of these entities may not have the resources or the expertise to design, build, and maintain secure key management infrastructure. Consequently there is a need for open interoperable decentralized key management infrastructure (DKMI). Moreover, some applications of decentralized identity may benefit from DKMI that is scalable and performant. Example applications include data streaming, supply chain, IoT (internet of things), and other applications where data provenance among multiple controlling entities is important and data processing is demanding. One design approach to composing scalable and performant infrastructure is to find minimally sufficient means for each of the key management tasks. This is a primary motivation for this work, that is, to identify the minimally sufficient means for essential key management tasks. This does not imply that other means might not be beneficial or best for a given application but that by first understanding minimally sufficient means an implementor might have at hand more design options that might be customized to better fit a broader class of applications.

The three main key management tasks are key reproduction, key recovery, and key rotation. We call these the three Rs of key management [136]. The focus of this work is key rotation which is usually the most difficult.

## 6.2 Reproduction

Key reproduction includes the creation and derivation of (public, private) key pairs and the associated tracking and storage of the private keys. A discussion of key reproduction is provided elsewhere. But in summary one method that simplifies key reproduction tasks is the use of hierarchically deterministic key derivation algorithms that produce ((HD keys) or keychains [9; 73; 91; 136]. An HD key pair is usually derived from a high entropy root or master private seed or key and some deterministic key derivation path. The key path may be public. This means that there is no need to store the derived HD private key which may be a security risk because the derived private key may be re-derived on demand from the root key and the public derivation path.

## 6.3 Recovery

Key recovery involves methods for securely backing up or distributing private keys such that they may be recovered in the event that the device holding the private key is lost or damaged. Key recovery approaches are discussed elsewhere [136].

## 6.4 Rotation

From a control perspective, key rotation is effectively transferring authoritative control (root authority) over the identifier from one set of key-pairs to another. Key rotation involves methods for securely revoking a key-pair and replacing it with a new key-pair. Revoke without replace may be accomplished by merely rotating to a null key. Thus rotation may be implemented as a primary operation and revocation only (without replace) may be implemented as a special case of rotation. Rotation to a null key means that no more operations using that identifier are permitted. The identifier is abandoned.

### 6.4.1 Purpose

In decentralized identity systems, key rotation is useful when the controller of a self-certifying identifier needs to maintain persistent control over that identifier indefinitely despite exploits of the private key(s). Otherwise in the event of exploit, the controller could just abandon the exploited identifier and create a new identifier with a new (public, private) key pair. The primary motivation for key rotation is to prevent, mitigate, or recover from an exploit of the private key due to exposure. The primary risk of exposure comes from use of the private key to sign statement. Creating a signature typically requires loading the private onto a computing device that then generates the signature. An attacker may gain access to the private key by remotely exploiting or physically capturing the computing device or intercepting the movement of the private key. In addition, over time advances in computing and cryptography may weaken the cryptographic strength of a given crypto-system. Best practice, therefore, is to enable the rotation (albeit infrequent) of a given (public, private) key pair to a new (public, private) key pair either using the same crypto-system or a stronger crypto-system.

### 6.4.2 Rotation History

With transferable self-certifying identifiers special semantics are applied to rotation. The self-certifying identifier derived from the public key is not changed, but merely the private key that is authoritative for signing statements is changed. Otherwise the identifier loses its value as an identifier. Consequently in order to verify a statement belonging to a transferable self-certifying identifier the verifier must know the key rotation history for that identifier. Recall, in a self-certifying identifier, its prefix derivation is bound though one-way functions to this original incepting public key. This original private key is used to cryptographically sign statements that prove control over the identifier at issuance. The original public key is used to cryptographically verify the associated signatures. This incepting key-pair is the means by which the root authority over the identifier may be exercised. Rotation of this key-pair to a new key-pair(s) is essentially transferring the means by which root authority may be exercised. A history of these rotation statements may be used to provenance the current root authoritative key-pair(s). This history forms a sequence of operations (events) that may be referred to as the root authority history. Other statements may delegate or attenuate the root authority but do not transfer or rotate the root authority. When we discuss rotation unless otherwise qualified we mean rotation of the root authority key-pairs.

To elaborate, each rotation operation expressed as a verifiable statement creates a new (public, private) key-pair(s). The first rotation operation must be signed at the very least with the original private key. Rotation does not change the identifier. It still references the original prefix derived from the public key. After rotation, however, authoritative statements are now signed with the new private key(s) and verified with the new public key(s). The original private key has been revoked and replaced with the new private key(s) specified in the rotation operation. The new public key(s) is included in the identifier's key rotation history. Validation of a statement first requires lookup and validation of the key rotation history. The final rotation entry provides the current key pair(s) used to sign and verify statements.

The key rotation history of digital signing keys used to control an identifier, provides the basis for authoritatively managing any other data affiliated with the identifier. In general, changes to the value of attributes associated with the identifier may be managed by verifiable signed asser-tions using the authoritative signing key-pair. Thus management of the signing key-pair enables management of affiliated data including other keys such as encryption keys.

### 6.4.3 Post Compromise Rotation

To clarify, regularly rotating the key bounds the risk of compromise resulting from exposure over time. This can be used proactively to upgrade the digital signature crypto-system to keep up with advances in computing. The more difficult problem to solve is rotation after a specific exploit may have already occurred (post-compromise). In this case, the exploiter may create a valid signed rotation operation to a key-pair under the exploiter's control prior to the original controller detecting the exploit. The exploiter could thereby either "capture" the identifier or create a conflict or race condition where two inconsistent but verifiable rotation events have been created. This work describes (in detail below) a protocol using *pre-rotation* of one-time first-time rotation keys that provides a simplified scalable performant solution to the problem of post-compromise secure rotation after an exploit may have occurred [136].

# 7 Nomenclature and Components

This system is designed to be compatible with performant data streaming or event sourcing applications in micro-services or serverless computing architectures. Consequently, each function is modularized into a composable functional primitive or elements. Components of the system may be built by the recursive composition of primitives. Thus system *components* may be comprised of combinations of these composable primitives.

The component descriptions below give rise to the acronym KERI for *Key Event Receipt Infrastructure*. This infrastructure includes the associated components described below, such as, *controllers, validators, verifiers, witnesses, watchers, jurors, judges, resolvers*, and their associated KELs, KERLs, and DELs [135]. This section also provides terminology definitions and symbolic representations of the various elements of the protocol messages.

## 7.1 One-way Function

As described above a *one-way* function is a cryptographic process that is easy to compute in one direction but impractical to invert or compute in the reverse direction [111; 112]. One-way functions provide essential capabilities in this protocol.

## 7.2 Large Integer Encoding

The *serialized encoding* standard used by this protocol is the RFC-4648 Base64 URL-Safe encoding standard [88]. This is used to encode prefixes, public keys, signatures, digests, and other elements that are derived from large integers (see below). They must be encoded in order to be consistently represented when stored, transmitted over networks, or included in serialized data structures that are signed or hashed. These are intended to be processed by computer algorithms and not transcribed by humans. Base64 uses a highly interoperable subset of the ascii character set. It is the most compact widely interoperable encoding.

## 7.3 Qualified Cryptographic Material

A *qualified cryptographic material item* is an expression of that material such as an identifier, public key, digest, or signature that attaches a specification of the material's derivation process to the material. A *qualified cryptographic material item* may be represented by a data structure or ordered tuple that includes both the cryptographic material and its derivation. A more compact form of such an item is a Base64 encoding with one or more prepended *derivation code* bytes. The context and location of the qualified cryptographic material determine both the type of derivation code used and how to interpret the derivation code. There are two significant locations.

The first is in a key event and the second is an attached signature to a key event. The attacked signatures may use a different derivation encoding that includes an index or offset into the list of current signing keys that facilitate matching signatures to the public key needed to verify the signature.

Examples of *qualified cryptographic material items* include as follows:

A *qualified identifier* is an *expression* of an identifier that attaches its *derivation process* information to the uniquely derived identifier material. The special term *prefix* or *identifier prefix* is used in this protocol to denote *qualified self-certifying identifiers*.

A *qualified public key* is an *expression of* a public key that attaches its *derivation process* information to the public key. This may be provided in compact form with a in Base64 with a prepended derivation code. The Base64 encoded format may be the same as the format used for the *basic* self-certifying identifier prefix type. Unless otherwise indicated all expressions of *public keys* in this protocol are *qualified*.

A *qualified digest* is an *expression* of a digest output from a cryptographic strength one-way hashing process that attaches its *derivation process* information to the digest. This may be provided in compact form with a in Base64 with a prepended derivation code. Unless otherwise indicated all representation of digests in this protocol are *qualified*.

A *qualified signature* is an *expression* of a an attached digital signature that prepends its *derivation process* information to the signature. This must include the signature scheme. This is different from a signature used in an identifier prefix. This may be provided in compact form with a in Base64 with a prepended derivation code. The derivation code also may include an index into the list of current signers (public keys) of the public key needed to verify the signature Unless otherwise indicated all expressions of *signatures* in this protocol are *qualified*.

A *hidden qualified cryptographic material item* is an *expression* of *qualified cryptographic material* whose derivation includes an additional one-way hashing step that produces a cryptographic strength digest of the material. The material is not provided explicitly in the *item*, but merely in the digest. In which case the actual material is hidden by the digest. This may be useful in making verifiable cryptographic commitments to material that may be disclosed in the future.

A *hidden public key* is a *qualified public key* where its *derivation* includes an additional one-way hashing step that produces a digest of the public key. In which case the actual public key is not provided explicitly in the *qualified public key*, but merely the digest, hence hidden. This may be useful in making verifiable cryptographic commitments to public keys that may be disclosed in the future.

A *hidden signing threshold and public key set* is a signing threshold specifying followed by a set of *qualified public keys* where the set's *derivation* includes an additional one-way hashing step that produces a digest of the serialized threshold specifier and set of qualified public keys. In which case the actual public keys are not provided explicitly, but merely the digest, hence hidden. This may be useful in making a compact verifiable cryptographic commitment to a set of public keys that may be disclosed in the future.

A *seal* is a *qualified digest* where its *derivation* code specifies what type of hashing function was used but does not include any other information about the associated data. The hashing step produces a digest of the serialized data that is referenced by the *seal*. The seal acts as an anchor of the data. To clarify, the actual data is not provided explicitly in the *seal*, but merely the digest of the serialized data, hence the data is hidden. This may be useful in making verifiable cryptographic commitments at the location of an event to data stored and/or disclosed elsewhere.

## 7.4 Self-certifying Identifier Prefix

A *self-certifying identifier prefix* or *identifier prefix* or for short *prefix* or *identifier* is a type of *qualified cryptographic material* that includes material universally uniquely derived though one or more one-way functions that includes one or more cryptographic digital signing (public, private) key-pairs. The *derivation process* cryptographically uniquely binds the resultant *self certifying identifier prefix* to those key-pairs. A *qualified prefix* attaches its *derivation process* information to its derived unique identifier material. A prefix's *derivation process specification* or *derivation* for short must indicate the cypher suite used for the digital signing scheme along with the other parts of the *derivation process*. A *derivation* may be represented as a data structure such as a mapping or an ordered tuple that includes elements for the raw public keys as well as the other parts of the derivation. A *derivation* may be compacted encoded as a character code compatible with a Base64 encoding. A *qualified prefix* therefore may be provided in compact form in Base64 with a prepended Base64 derivation code.

An *identifier prefix* is a type of *self certifying identifier* that is used to create a namespace of identifiers. The *identifier prefix* is bound via a derivation process to a set of incepting data. Unless otherwise indicated, the term *identifier prefix* may apply to any of the defined types of derivation processes from incepting data. This protocol description may use the simpler terms *identifier* or *prefix* to mean *self-certifying identifier prefix* unless the shorter expression would be ambiguous. Moreover, by virtue of its derivation process, the *prefix* is uniquely bound to one or more incepting or root (public, private) key-pairs. These key-pairs represents the root control authority over the *prefix* at its inception and issuance. They comprise the controlling set of keys for the prefix at issuance. Furthermore, all identifiers in the associated namespace of the *prefix* are controlled by the same controlling key-pairs as the *prefix*. Consequently the protocol need only consider the prefix when establishing control authority.

The generalized representation of a self-certifying identifier is composed of a derivation and a resultant unique derivative as follows:

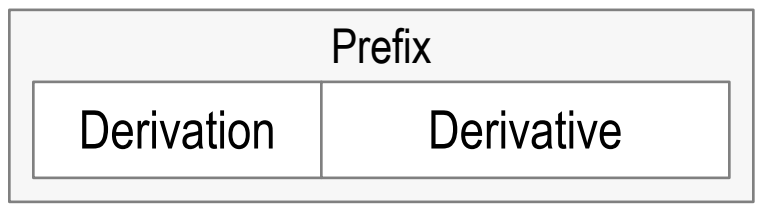

Figure 7.1. Generalized Prefix.

Together these may be encoded (BASE-64 URL Safe) into a single string of characters that is the prefix as follows:

```
BDKrJxkcR9m5u1xs33F5pxRJP6T7hJEbhpHrUtlDdhh0
```

Best practices cryptography limits the options that user may choose from for the various cryptographic operations, such as signing, encrypting, and hashing to a suite of balanced and tuned set of protocols, one for each operation. Each member of the set should be the *one and only one* best suited to that operation. This prevents the user from making bad choices. Many key-representation schemes allow the user the freedom to specify the features of the operation independently. This is usually a very bad idea [14; 138; 139]. Users should not be custom combining different features that are not part of a best practices cypher suite. Each custom configuration may be vulnerable to potential attack vectors. The suggested approach is to narrowly specify a single cypher suite family and version for each operation. If an exploit is discovered for a member of a suite and then fixed, the suite is updated wholly to a new version. The number of allowed cypher suites should be minimized to those essential for compatibility but no more. This approach increases expressive power because only one syntactic element is needed to specify a suite instead of a different element per feature.

The *derivation* element in the *prefix* specifies a cypher suite. An good example of this approach is the specification format used for the W3C DID (Decentralized Identifier) *type* field in the authentication section of a DID Document [151]. This is single string that includes a cypher suite family, operation, and version. For example the family = Ed25519, operation = verification, and version = 2018 may be expressed as follows:

$$\texttt{Ed25519VerificationKey2018} \tag{7.1}$$

Control authority in the form of digital signing key-pairs may be verified upon receipt of a signed inception statement. The statement includes the identifier's derivation process specification, the authoritative public key or keys, and any other associated configuration information. It is signed by the authoritative private keys.

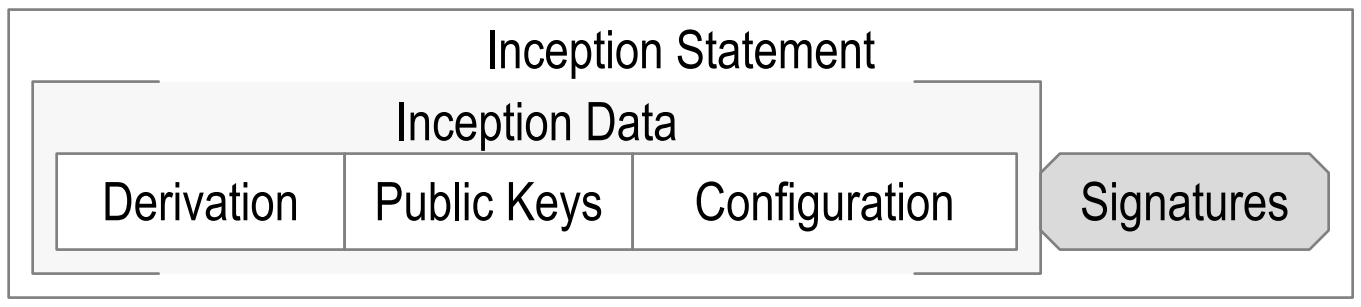

Figure 7.2. Verifiable Inception Statement

## 7.5 Key-Pair Label

A *key-pair labeling convention* makes event expressions both clearer and more compact. As a shorthand for its identifier prefix, each entity in an expression may be labeled by an uppercase letter symbol as an *alias* such as *A*, *B*, *C* etc. Furthermore each self-certifying identifier must have a least one initial controlling (public, private) key-pair bound to it at issuance (inception, origination). In the case when there is only one initial controlling key-pair, that key-pair may be represented with the same letter symbol as the entity's *alias* but where the public key uses uppercase and the private key uses lower case. For example, entity *A* with key-pair $(A, a)$ where *A* denotes the public key and *a* denotes the private key. Usually, the context determines which member of the key-pair is to be used, such as, the public key for verification or the private key for signing. In those cases, the uppercase letter alone may be sufficient to represent the appropriate member of the key pair. For example, *A* alone represents either or both members of key-pair $(A, a)$.

With a set of multiple controlling key-pairs (either successive or in concert), all members of the set can be put into a sequence. Each member may therefore be indexed with a superscript on the letter such as $(A^j, a^j)$ or individually $A^j$ and $a^j$. Indexing the key-pairs used to control an identifier may simplify the description of protocol events. More formally, an *indexed key-pair* belongs to an ordered sequence of (public, private) key-pairs used to control an identifier. This may be denoted with superscripted integer index *j,* such as, $(A^j, a^j)$ where $A^j$ and $a^j$ are the $j^{th}$ public and private keys respectively. With this convention, the zeroth index, $j = 0$, may be used for the initial (originating or incepting) key-pair for the associated self-certifying identifier prefix. For example, $(A^0, a^0)$, represents the initial (incepting, originating) key-pair bound to the entity labeled by alias *A* with identifier prefix controlled by private key $a^0$ and signatures verifiable with the public key, $A^0$. A sequence of indexed public keys may be used to represent the sequence of key-pairs as follows:

$$A^0, A^1, A^2, \ldots. \tag{7.2}$$

were $(A^0, a^0)$ is the initial key pair and $(A^1, a^1)$ for the next pair and so on.

When it is not important to specify the exact index of a key-pair but merely to indicate that it is the latest valid or authoritative key pair whatever its index then the superscripts may be left off. For example, $\left(A^1, a^2\right)$ becomes $\left(A, a\right)$.

Sometimes it may be helpful to denote the individual members of a class of similar entities. This may be represented by a numerically subscripted uppercase letter symbol where the letter symbol denotes the class and the numeric subscript denotes the member. Let C denote the class of controllers, then $C_0$ and $C_1$ respectively denote key-pairs from two different controllers.

Furthermore, it may be helpful to label each key-pair by the nature of its usage in an event with an upper case letter subscript. For example, given a controller labeled *C*, let *R* represent a key-pair used in a rotation event and *X* a key-pair used in an interaction event. These two key-pairs may be denoted respectively as $C_R$ and $C_X$.

Any two or all three of the conventions may be combined in a single label. These are a numeric superscript for event sequence order, a numeric subscript for members of a class, and a letter subscript for usage. For example, both a numeric superscript and numeric subscript may be used where the superscript indicates the event index and the subscript indicates a given controller from a class,, such as, $C_0^1$ and $C_1^1$. Alternatively, a letter or string subscript may be used to indicate the event usage, such as, $C_R^1$ and $C_X^2$. Both numeric and letter subscripts may be used to indicate both the entity class and event usage, such as, $C_{1R}^1$ and $C_{2R}^1$. When the numeric superscript for the event sequence is left off it may indicate the latest or current key-pair, such as, $C_{1R}$, and $C_{2R}$.

Because protocol expressions typically refer to a given controller, the usual case may leave off the class member numeric subscript. For example, given controller *C*, the sequence of usage labeled key-pairs for a sequence of events may be denoted as follows:

$$C_R^0, C_R^1, C_X^1, C_X^1, C_R^2, \ldots \tag{7.3}$$

Sometimes it may be helpful to indicate that the private key in a key-pair has become previously exposed as a result of it already having been used sometime in the past to sign a statement. This exposure may be represented by an over dot on the key-pair label, such as $\dot{C}_R^1$ and $\dot{C}_X^1$.

Likewise, it may be helpful to indicate that a given public key or set of public keys is hidden in a cryptographic digest of the key or set of keys. A hidden public key or set of keys may be represented by an underbar on the key pair label, such as, $\underline{C}_R^1$ and $\underline{C}_R^2$.

The former two conventions may be combined to represent exposed but originally hidden keys, such as, $\dot{\underline{C}}_R^1$ and $\dot{\underline{C}}_R^2$.

## 7.6 Serialized Data

A data structure with ordered fields may be represented by a tuple where the tuple is denoted with parentheses about a comma separated list of field names, such as, $\left(t, A, C\right)$. A serialized version of that data structure is denoted with angle brackets about a comma separated list of the field names, such as, $\left\langle t, A, C\right\rangle$. In general an ordered tuple may be converted to/from a data structure with labeled key value pairs. Ordering is important when serializing.

## 7.7 Signature

A digital *signature* is type of *qualified cryptographic material* that includes a string of characters that is produced by a cryptographic operation on a given string of characters (signed text) using the private or signing key from a (public, private) key pair. A *qualified signature* attaches its *derivation process* information to the digest. Its derivation includes the type of digital signing scheme. This may be provided in compact form in Base64 with a prepended derivation code. The derivation code may also include an index or offset into the list of current signing public keys that specifies which public key may be used to verify the signature.

The public or verifying key may be used in a related cryptographic operation on the given signed text to verify the validity of the signature. A digital signature has two important properties. The signature is unique to the key pair for the given signed text and the signature is non-repudiable by the signer, that is, only the holder of the private key can create the signature.

A digital signature operation may be represented with the lowercase greek sigma, $\sigma$ where the symbol of the signing key pair is provided as a subscript to the sigma and the serialized data or text is provided as the argument to the signing operation. For example, the digital signature of serialized data $\langle t, A, C \rangle$ for entity $A$ using key-pair $A^0$ may be denoted as follows:

$$\sigma_{A^0}\left(\langle t, A, C \rangle\right), \tag{7.4}$$

where the signing key is the private key from the key pair, $\left(A^0, a^0\right)$ controlled by $A$ and the signed text is the serialized data structure, $\langle t, A, C \rangle$. An equivalent form would use the lowercase symbol for the private key instead, as follows;

$$\sigma_{a^0}\left(\langle t, A, C \rangle\right). \tag{7.5}$$

But as stated above when there is no doubt as to the context, the uppercase symbol may be used to represent the appropriate member from the key pair. A message that includes both a serialized data structure and an attached signature may be denoted as follows:

$$\langle t, A, C \rangle \sigma_{A^0}\left(\langle t, A, C \rangle\right). \tag{7.6}$$

A message with two attached signatures, one each from entities $A$ and $B$ may be denoted as follows:

$$\langle t, A, C \rangle \sigma_{A^0}\left(\langle t, A, C \rangle\right) \sigma_{B^0}\left(\langle t, A, C \rangle\right), \tag{7.7}$$

where the signing key for entity $B$ is the private key from the key pair, $\left(B^0, b^0\right)$ and the other parts are as above. Without confusion, in the expression above the arguments to the signing operations may be redundant. Therefore a more compact form of the preceding expression may be denoted as follows:

$$\langle t, A, C \rangle \sigma_{A^0} \sigma_{B^0}. \tag{7.8}$$

Expressions eq. 7.7 and eq. 7.8 are equivalent. When the attached signatures use the latest key pair from both $A$ and and $B$ then an even more compact form of eq. 7.8 may be denoted as follows:

$$\langle t, A, C \rangle \sigma_A \sigma_B. \tag{7.9}$$

## 7.8 Transferable (Non-transferable) Identifier

A *transferable* (*non-transferable*) *identifier* allows (dis-allows) transfer of its control authority from the current set of controlling keys to a new (next or ensuing) set via a *rotation event* (see below). An *identifier* may be declared *non-transferable* at inception in its derivation code and/or in its *inception event* (see below). Derivation code declaration is only defined for basic self-cer-tifying identifiers. A *rotation event* (operation) on a *transferable* identifier may rotate to a null key thereby irreversibly converting it into a *non-transferable* identifier. Once an identifier becomes non-transferable no more events are allowed for that identifier. The identifier is effectively abandoned from the standpoint of KERI. By convention, when non-transferability of an identifier is declared in its derivation code then its authoritative (signing) key-pair may be converted to an encryption key-pair to enable a self-contained bootstrap to a secure communications channel using only the exchange of the non-transferable identifier [56; 149]. Although a non-transferable identifier is abandoned from the standpoint of KERI, it does not preclude a given application from employing the identifier. Its just that no more events within KERI are allowed on the identifier (see *event* definition below). An identifier declared at inception as non-transferable may have one and only one event, that is, the inception event. In this sense a non-transferable identifier at inception is pre-abandoned. These identifiers are typically meant to be used as ephemeral identifiers or identifiers where replacement of the identifier instead of key rotation is the preferred approach when the identifier becomes compromised.

## 7.9 Controller

A *controller* is a controlling entity of an *identifier*. At any point in time an *identifier* has at least one but may have more than one *controlling entity*. Let $L$ be the number of *controlling entities*. This set of *controlling entities* constitute the *controller*. All proper key management *events* on the *identifier* must include a *signature* from the sole *controlling entity* when there is only one member in the set of *controlling entities* or a least one *signature* from one of the *controlling entities* when there is more than one. This signatures may be expressed as a single collective signature when a collective signing scheme is used. Without loss of generality, when the context is unambiguous, the term *controller* may refer either to the whole set or a member of the set of *controlling entities*. Typically, when there is more than one *controlling entity*, control is established via $L$ signatures, one from each *entity* (*controller*). This is called *multi-signature* or *multi-sig* for short. Alternatively, with a $K$ of $L$ threshold control scheme, where $K \leq L$, control is established via any set of at least $K$ signatures each one from a subset of at least size $K$ of the $L$ *controllers*. A more sophisticated scheme may use fractional weighted multiple signatures. These thresholded multiple signatures may be expressed as a single collective threshold signature from an appropriate collective threshold signing scheme. The description of the ensuing protocol assumes the simplest case of individual not collective signatures but it is anticipated that the protocol may be extended to support collective multi-signature schemes.

## 7.10 Statement

A *statement* is any data that may be digitally signed. A *statement* may be used to perform one or more operations on an identifier or attributes or other data affiliated with that identifier. A *signed statement* includes a cryptographically verifiable signature. A *signed statement* is therefore a *verifiable statement*. In a given context, all statements may be assumed to be verifiable such that the qualifier *signed* or *verifiable* may be elided.

## 7.11 Message

A *message* as used by the protocol is a serialized data structure with one or more attached signatures. The data structure must be serialized in order to be digitally signed. A message is a well defined type of *signed* or *verifiable statement*.

## 7.12 Event

An *event* may be represented or conveyed by a *message* associated with the protocol, typically an operation associated with an *identifier*. *Event messages* are expressed as well defined *verifi-* *able statement*s. The protocol is primarily concerned with creating, transmitting, validating, and logging *events* as conveyed by *messages*. The *event data* may be represented by a serializable data structure that is serialized and then signed to create the event message. Given the one-to-one relationship between an *event* and its *message* they may be referred to interchangeably in context.

## 7.13 Version

The *version string* or *version* identifies (among other things) a given set of protocol specifications. Versioning enables interoperable extensibility of the protocol. Each *version* indicates a well defined feature set. The *version string* includes a *version code* with a *major* and *minor* version number. Small backward compatible changes in the protocol increment the *minor* version number whereas backward breaking changes increment the *major* version number. Each event includes the *version string*. The version string also includes the serialization coding (such as JSON, CBOR, MessagePack, etc) used to encode the associated event as well as the size of the serialized event. This allows a parser to determine how to parse the message and extract it from attached signatures. The *version string* also ends in a terminal delimiting character. This allows for future detectable changes in the version string format including length.

A compact example version string may be as follows:

`"KERI10CBOR0001c2"` (7.10)

where `KERI` is the identifier of KERI events, `1` is the hex major version code, `0` the hex minor version code, `CBOR`, is the code for the serialized encoding format of the event, and `0001c2` is the hex size of the serialized event. The version string provides a self-contained way of determining the encoding and length of the serialized event.

When using web protocols such as HTTP, one way to indicate encoding to to use the MIME type standard using structured syntax suffixes [65; 72]. Suggested mime types for KERI encodings would be `application/keri+json`, `application/keri+cbor`, `application/keri+binary`, and `application/keri+msgpack`.

## 7.14 Event Digest

A *key event message digest* or *digest* is a *qualified digest*. It is a type of *qualified cryptographic material* that includes a string of characters that is the output of cryptographic one-way hash function of the serialization of a *key event message*. A cryptographic digest is a space efficient but unique fingerprint of the content of some message. A *qualified digest* attaches its *derivation process* information to the digest. Its derivation includes the type of hashing function. This may be provided in compact form in Base64 with a prepended derivation code.

The important property of a cryptographic hash function is that the resultant digest is highly unique to the string of characters on which it was computed and is practically non-invertible. In other words, it is highly collision resistant. To compute a *digest*, a serialized string of characters encoded as a string of bits is provided as the argument to the hash function. An example of a suitable hashing function and digest format are Base-64 encodings of 256 bit (32 byte) character strings produced by the Blake3, Blake2b, or Blake2s hashing functions [17; 31; 33; 121]. These have a cryptographic strength of 128 bits. Blake3 is currently the most performant hashing function for digests of this length and strength [31]. A signed digest is a non-repudiable commitment to the content. Because digital signatures schemes include hash functions, a digital signature is simultaneously a digest and a non-repudiable commitment to the content. Event digests are fully qualified with a prepended derivation code that replaces the pad character(s).

## 7.15 Event Ilk

The *key event ilk* or *ilk* indicates the type (kind) of the key event to aid in its parsing. Suggested *ilk* values and events are as follows: `icp` for *inception* event, `rot` for *rotation* event, `ixn` for *interaction* event, `dip` for delegated *inception* event, `drt` for delegated *rotation* event, `rct` for event *receipt* message, and `vrc` for validator event receipt message. Specific applications may use other event *ilk* values.

## 7.16 Key Event

A *key event* is a special type of *event* associated with a given identifier that appears in a uniquely ordered sequence. The order of appearance of *key events* in the sequence is determined by the *controller*. A *key event* may be used to either manage the identifier's authoritative key-pair(s) or be managed by them. A proper *key event message* is verifiable in that it includes *signatures* from the authoritative key-pairs of the associated *identifier*. Proper key events may include several elements in a standard *key event header* such as *version*, *prefix*, *sequence number*, *digest*, and *ilk*. These are defined elsewhere. In addition to the *header*, a *key event* includes event specific *configuration* data. A *key event message* also includes an attached signature(s). An abstract representation follows:

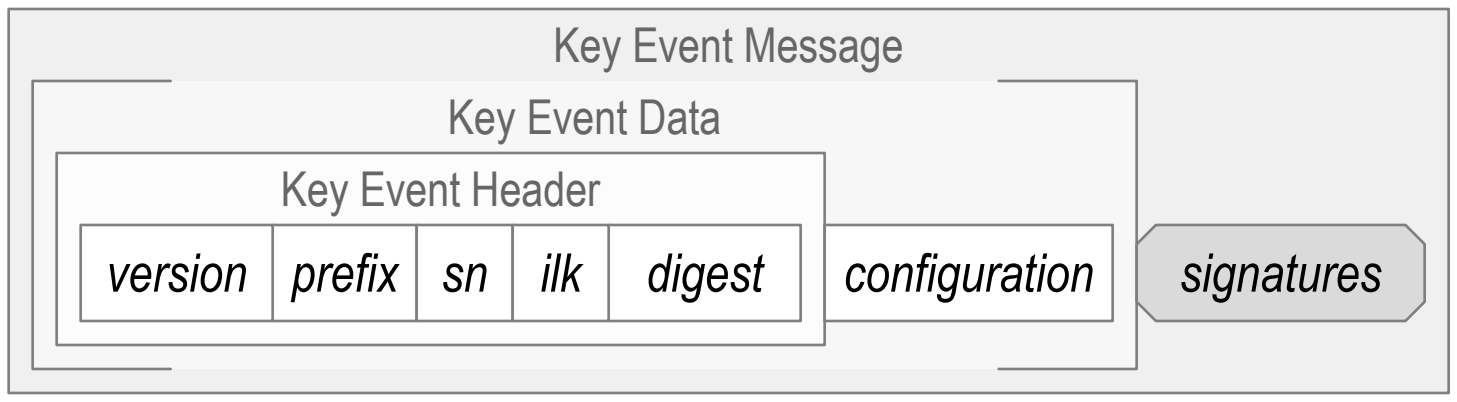

Figure 7.3. Key Event Message Detail

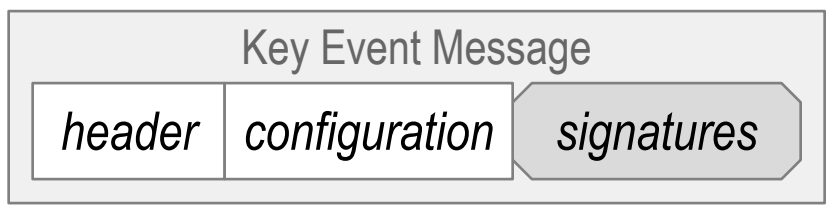

Figure 7.4. Key Event Message Simple

## 7.17 Key Event Sequence

A *key event sequence or history* is a uniquely ordered sequence of *key event messages*. The *key event messages* in the sequence are chained together in that each event message besides the first event includes a non-empty cryptographic strength content *digest* of the immediately preceding *key event* message. A chained sequence of event messages is shown below:

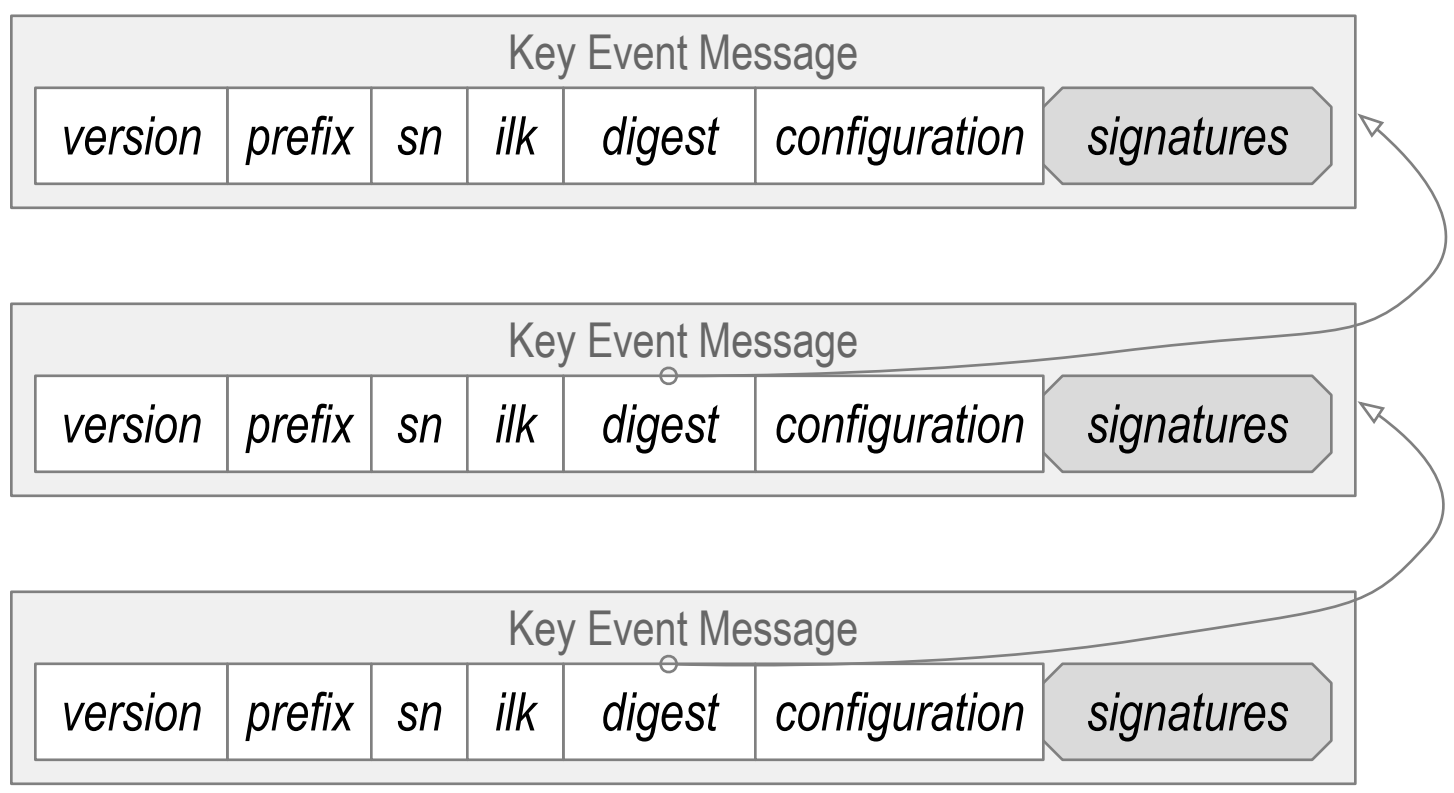

Figure 7.5. Chained Sequence of Key Event Messages

## 7.18 Key Event Sequence Number and Event Labeling

A *key event sequence number* or *sequence number (SN)* for each event in a key event sequence is represented by a unique monotonically increasing arbitrary precision non-negative integer. Arbitrary precision integers increase in length to accommodate bigger values so that they never wrap around [13]. There is wide support for arbitrary precision arithmetic in most programming languages [99]. Each arbitrary precision integer is represented internally by a data structure that is optimized for ranges of values. When serialized the number need only use as many characters as needed to represent the actual encoded value.

Should a *key event sequence number* ever reach an upper limit then the associated identifier would have to be abandoned. Recall the discussion above on the computational limits of attempting $2^{128}$ trials. Thus a practically infinite or inexhaustible upper limit would be the length of an encoded version of a 128 bit (16 byte) binary number. Pragmatically, a 64 bit (8 byte) binary number may be sufficient for many applications. Recall that there are $2^{25} \cdot 2^{20} = 2^{45}$ microseconds in a year. If a new key event were emitted every microsecond then a 64 bit sequence number would not reach its upper limit for $2^{64}$ - $2^{45} = 2^{19}$ = 524,000 years.

The purpose of a sequence number is to enable more convenient and performant secure processing of events. A sequence number provides a unique numerical index for each event in a sequence. This may be used to order sort, index, and organize events. Using indexes that are included in the signed content of an event message protects application programmer interface (API) calls that access or provide events by index. For transmission in a key event message, a 16 byte binary number encodes to 24 Base64 characters including 2 pad characters. Unfortunately, Base64 encodings do not preserve lexicographic so may be inappropriate for use in a database. Instead a hex encoding does preserve lexicographic ordering in a database where the indexes must be fixed length strings of characters instead of arbitrary precision integers.

Abstractly the symbol $t_k$ is used to represent the sequence number with a numeric subscript $k$ starting at zero to indicate the order of appearance of the sequence number , such as $t_k$ in the sequence $t_0, t_1, \ldots, t_n$. A simple format for $t_k$ is a counter with integer values starting at 0 that increment by 1 for each new event. In this case each sequence number value $t_k$ equals its numeric subscript $k$ , that is,

$$t_k = k\,, \tag{7.11}$$

and

$$t_0 = 0, t_1 = 1, \ldots, t_n = n\,. \tag{7.12}$$

The symbol, greek epsilon, $\varepsilon$ may be used to label a key event for an identifier known by its context. The label may be subscripted with the same $k$ from its sequence number as in $\varepsilon_k$, to represent its position in the event sequence, e.g., $\varepsilon_0$ for $k = 0$. A key event may be more fully represented in terms of its identifier by superscripting the event label with the identifier label, such as, $\varepsilon_0^A$ for the zeroth event for identifier $A$ or $\varepsilon_k^A$ for the $k^{th}$ event of identifier $A$. Using this convention, a signed event message for identifier $A$ might be represented as follows:

$$\varepsilon_0^A = \langle t, A, A^0, A^1 \rangle \sigma_{A^0}\,.$$

A *message digest* may be labeled with the lowercase greek eta, $\eta$. A digest of the serialized data structure $\langle t, A, A^0, A^1 \rangle$ may be denoted as:

$$\eta\left(\langle t, A, A^0, A^1 \rangle\right). \tag{7.13}$$

A digest of a full event message comprised of a serialized data structure and an attached signature may be denoted as follows:

$$\eta\left(\left\langle t, A, A^0, A^1 \right\rangle \sigma_{A^0}\right) = \eta\left(\varepsilon^A\right). \tag{7.14}$$

A digest's label may be subscripted with the sequence number index, *k*, of the message in which it appears such as $\eta_k$ or $\eta_k(\varepsilon)$. When a *digest* of some event message appears in a different event message, then the subscripts on the digest and event will be different. For example the digest of event $k-1$ appearing in event *k* may be labeled as follows :

$$\eta_k\left(\varepsilon_{k-1}\right). \tag{7.15}$$

When $k=1$ this gives $\eta_1\left(\varepsilon_0\right)$. In this protocol the message digest of a given message always appears in the immediately succeeding message. This means that a message digest label appearing in a message may unambiguously drop the label's argument as follows:

$$\eta_k\left(\varepsilon_{k-1}\right) = \eta_k. \tag{7.16}$$

When $k=1$ this gives $\eta_1\left(\varepsilon_0\right)=\eta_1$. Likewise a digest label may be superscripted with the identifier label, as in $\eta_k^A$, to indicate to which event stream the digest belongs.

## 7.19 Establishment Event

An *establishment event* is a class of of *key event* associated that is used to establish the current authoritative key-pairs for its identifier. The set of *establishment events* comprise an ordered subsequence of the associate full *key event* sequence. The primary purpose of an *establishment event* is to help establish the current control authority for an identifier and are typically related to key creation and rotation. The configuration data in an establishment event includes designations of supporting infrastructure for establishing and maintaining control authority over the identifier. This configuration data also declares or commits to the next (ensuing) set of authoritative keys as part of the *pre-rotation* scheme (described later). Types of *establishment events* include *inception* and *rotation events* (see below). A basic establishment event message is diagrammed below.

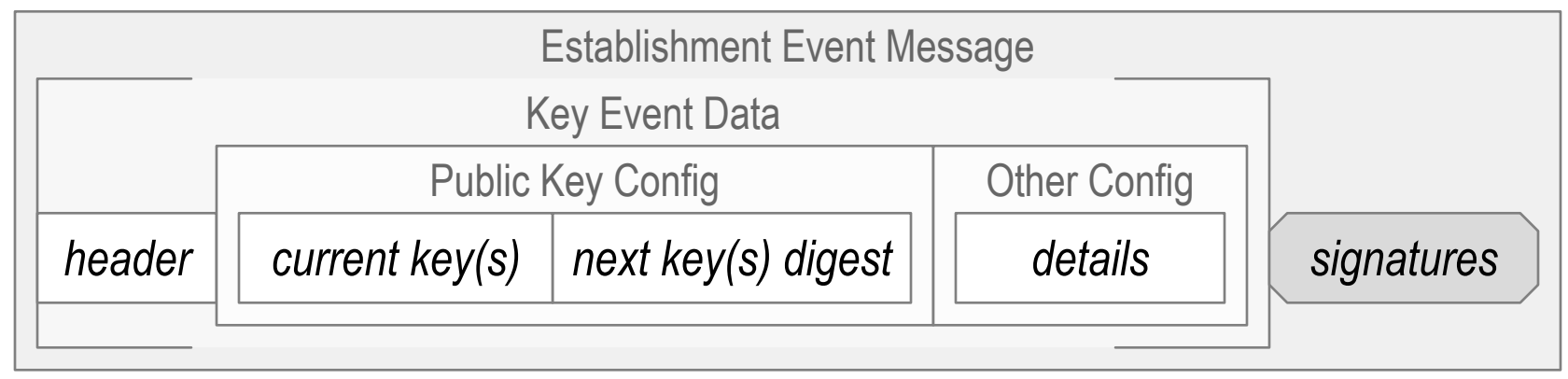

Figure 7.6. Basic Establishment Event Message.

## 7.20 Inception Event

An *inception event* is an *establishment key event* that represents the creation operation of an *identifier* including its derivation and its initial set of controlling keys as well as other inception or configuration data for supporting infrastructure There may be one and only one *inception event* operation performed on an *identifier*. An *inception event* is necessary to control *establishment*. An inception event configuration also declares or commits to the ensuing (next) set of authoritative keys as part of the *pre-rotation* scheme (described later). When that ensuing set of keys is null then the identifier is *non-transferable*. No more events for that identifier are allowed. A basic inception event message is diagrammed below.

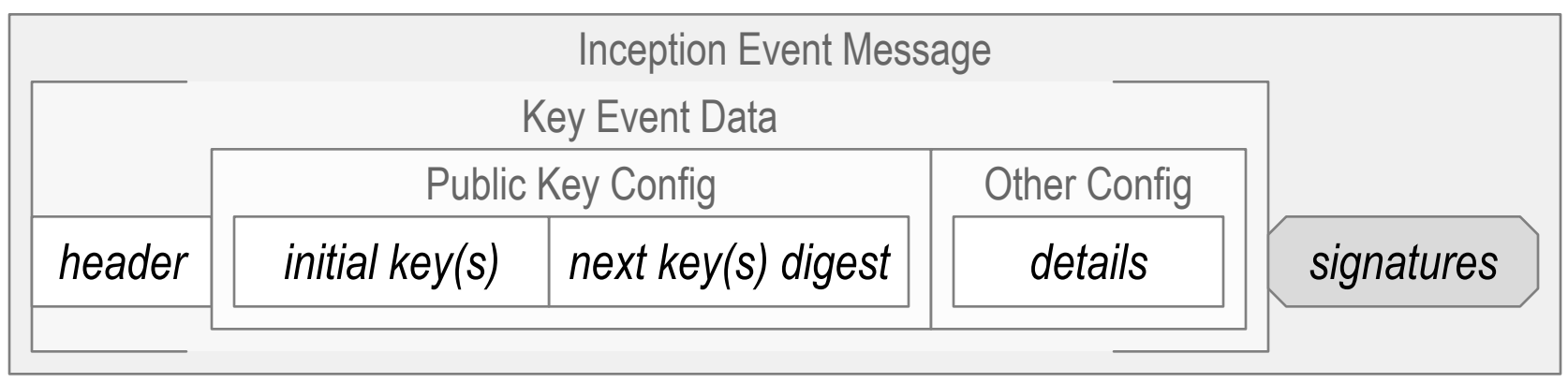

Figure 7.7. Basic Inception Event Message.

## 7.21 Rotation Event

A *rotation event* is an *establishment key event* that represents a rotation operation on an identifier that transfer control authority from the current set of controlling keys to new set. To clarify, rotation essentially transfers root control authority from one set of keys to a new set of keys. A *rotation event* is therefore an *establishment event*. A rotation operation can be viewed as a combination of a revocation followed by replacement of keys. Each rotation event declares or commits to the next (ensuing) set of authoritative keys as part of the *pre-rotation* scheme (defined later). When that next set of keys is null then the identifier becomes *non-transferable*. No more events for that identifier are allowed. A basic rotation event message is diagrammed below.

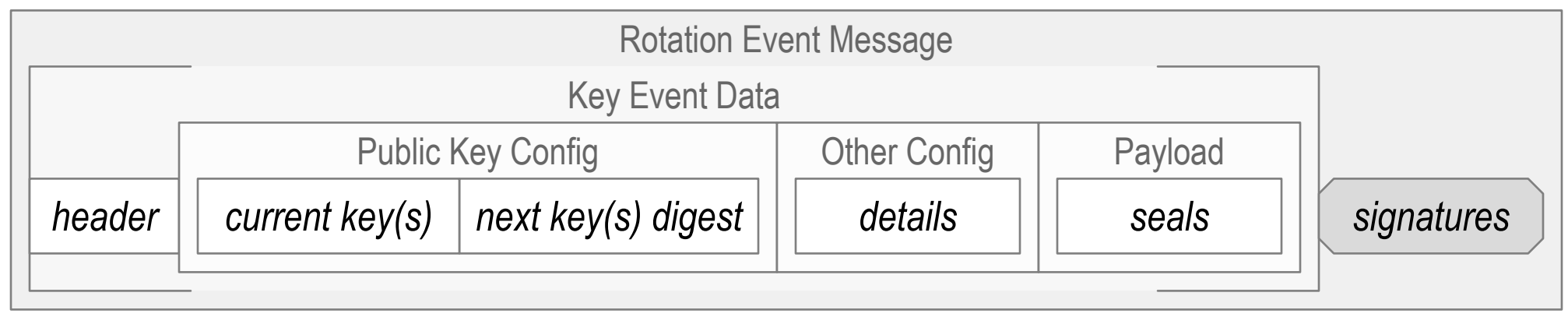

Figure 7.8. Basic Rotation Event Message.

To elaborate, for each *identifier (prefix)* there is a unique ordered sequence of chained *key event* messages that begins with one and only one *inception event* message. Within that sequence of key event messages is a subsequence of *establishment event* messages that begins with the *inception event message* and may be followed with one or more *rotation event messages*. This establishment subsequence may be called the *establishment* or *rotation (event) history* of the identifier. A valid key event sequence may contain only establishment events. In this case the establishment event subsequence equals the key event sequence. This is shown as follows:

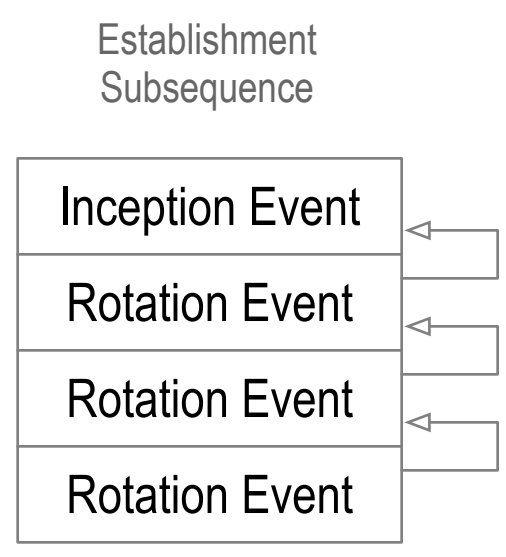

Figure 7.9. Establishment Key Event Sequence. Sequence composed exclusively of establishment events. Arrows represent digest chaining where event at arrow head of arrow is the content for the digest at arrow tail.

## 7.22 Non-Establishment Event

A *non-establishment event* is a type of *key event* that is interleaved into the key event history with the events from the *establishment event* subsequence (history) in order that the current control authority (root authoritative key-pairs) may be cryptographically verified at the point in the interleaved sequence order where the *non-establishment event* appears. An *non-establishment* event includes an event-specific *data payload*. A basic non-establishment event message is diagrammed below.

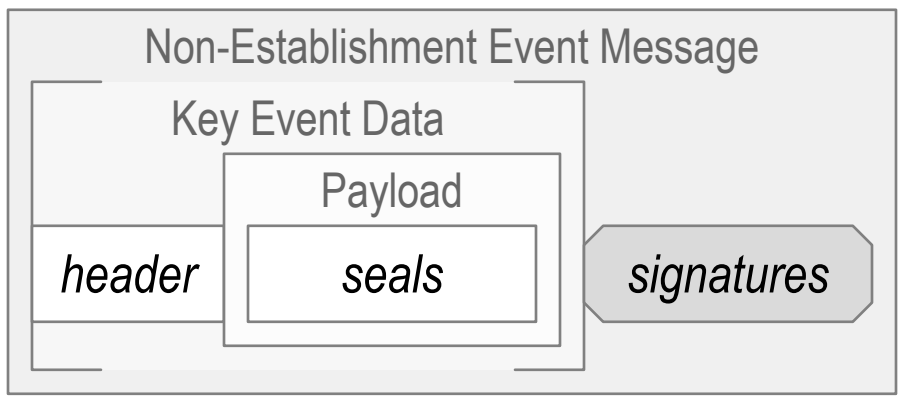

Figure 7.10. Basic Non-Establishment Event Message.

A *non-establishment* event does not declare *pre-rotated* keys (defined later) and is only signed by the current authoritative set of keys. Recall that a *key event history* is comprised of an ordered sequence of chained *key event messages*. The event messages are chained in that each event message in the sequence besides the *inception event* includes a cryptographic strength content digest of the immediately preceding event message in the sequence. The set of *establishment events* comprises a subsequence that is interleaved into the full key event sequence. The subsequence of *establishment events* preceding a *non-establishment event* may be used to establish current control authority for the *non-establishment event* thus binding the *non-establishment event* to its control authority. Specifically, the interleaving of *non-establishment events* with *establishment events* enables cryptographic verification of the authoritative key-pairs at the time of issuance of the *non-establishment event* where time of an event is measured by its order of appearance in the sequence. This provides an end-verifiable secure binding between the *non-establishment event* itself and the *key events* used to establish its control authority. Given verification of the authoritative key-pairs, the signatures on a *non-establishment key* event may then be verified as authoritative. Numerous *non-establishment events* may occur between *establishment (key rotation) events*. Although this exposes the signing keys through repeated use to sign the *non-establishment events*, the pre-rotated ensuing signing keys declared in the most recent *rotation event* are not exposed and may be used via a later rotation to recover from compromised signing keys. The data payload in a *non-establishment event* may not be ever used in the establishment of its own identifier's root control authority but may be used instead to make verifiable authoritative statements for other purposes. These might include authorizations of encryption keys, or communication routes or service endpoints and so forth. Transactions or workflows composed of *non-establishment events* are secured by virtue of being interleaved in the verifiable key event sequence with the verifiable authoritative *establishment* events.

## 7.23 Seal

A *seal* is a cryptographic commitment in the form of a cryptographic digest or hash tree root (Merkle root) that anchors arbitrary data or a tree of hashes of arbitrary data to a particular event in the key event sequence [104]. According to the dictionary, a *seal provides evidence of authenticity*. A key event sequence provides verifiable proof of current control authority at the location of each event in the key event sequence. In this sense, therefore, a *seal* included in an event provides proof of current control authority i.e. *authenticity* of the data anchored at the location of the seal in the event sequence. A *seal* is an ordered self-describing data structure. Abstractly this means each element of the seal has a tag or label that describes the associated element's value. So far there are four normative types of seals, these are *digest*, *root*, *event*, and *location* seals.

A *digest* seal includes a digest of external data. This minimal *seal* has an element whose label indicates that the value is a *digest*. The value is fully qualified Base64 with a prepended *derivation code* that indicates the type of hash algorithm used to create the digest.

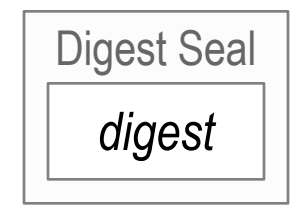

Figure 7.11. KERI Digest Seal.

A *root* seal provides the hash tree root of external data. This minimal *seal* has an element whose label indicates that the value is the *root* of a hash tree. The value is fully qualified Base64 with a prepended *derivation code* that indicates the type of hash algorithm used to create the hash root. In order to preclude second pre-image attacks, hash trees used for hash-tree roots in KERI seals must be sparse and of known depth similar to certificate transparency [47; 70; 95–97]. One simple way to indicate depth is that internal nodes in a sparse tree include a depth prefix that decrements with each level and must remain non-negative at a leaf [47].

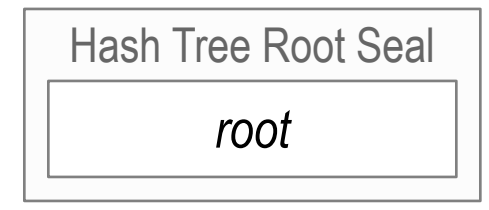

Figure 7.12. KERI Root Seal.

An *event* seal includes the *identifier prefix*, *sequence number*, and *digest* of an event in a key event log. The *prefix*, *sequence number*, and *digest* allow locating the event in an event log database. The digest also allows confirmation of the anchored event contents. An *event* seal anchors one event to another event. The two events may be either in the same key event sequence in two different key event sequences with different identifier prefixes. Thus a seal may provide a cryptographic commitment to some key event from some other key event.

Event Seal

| prefix | sn | digest |
|---|---|---|

Figure 7.13. KERI Event Seal.

An *event location* seal is similar to an *event* seal. A *location* seal includes the *prefix*, *sequence number*, *ilk*, and *prior digest* from an event. These four values together uniquely identify the location of an event in a key event log. A location event is useful when two seals in two different events cross-anchor each other. This provides a cross-reference of one event to another where the other event's digest must include the seal in the event contents so it cannot contain the first event's digest but the digest of the preceding event. To clarify, digest creation means that only one of the cross anchors can include a complete digest of the other event. The other cross anchor must use a unique subset of data such as the unique location of the event. The *ilk* is required in the location because of the special case of recovery where a rotation event supersedes an interaction event. This is described in detail later under-recovery. Location seals are also useful in external data that is anchored to an event log. The location seal allows the external data to include a reference to the event that is anchoring the external data's contents. Because the anchoring event includes a seal with the digest of the external data, it is another form of cross anchor.

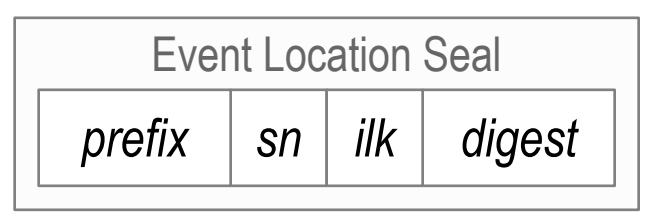

Figure 7.14. KERI Event Location Seal.

The data structure that provides the elements of a seal must have a canonical order so that it may be reproduced in a digest of elements of a event. Different types of serialization encodings may provide different types of ordered mapping data structures. One universal canonical ordering data structure is a list of lists (array or arrays) of (label, value) pairs. The order of appearance in each list of each (label, value) pair is standardized and may be used to produce a serialization of the associated values.

The interpretation of the data associated with the digest or hash tree root in the seal is independent of KERI. This allows KERI to be agnostic about anchored data semantics. Another way of saying this is that seals are data agnostic; they don't care about the semantics of its associated data. This better preserves privacy because the seal itself does not leak any information about the purpose or specific content of the associated data. Furthermore, because digests are a type of content address, they are self-discoverable. This means there is no need to provide any sort of context or content specific tag or label for the digests. Applications that use KERI may provide discovery of a digest via a hash table (mapping) whose indexes (hash keys) are the digests and the values in the table are the location of the digest in a specific event. To restate, the semantics of the digested data are not needed for discovery of the digest within a key event sequence.

To elaborate, the provider of the data understands the purpose and semantics and may disclose those as necessary, but the act of verifying authoritative control does not depend on the data semantics merely the inclusion of the seal in an event. It's up to the provider of the data to declare or disclose the semantics when used in an application. This may happen independently of verifying the authenticity of the data via the seal. This declaration may be provided by some external application programmer interface (API) that uses KERI. In this way, KERI provides support to applications that satisfies the spanning layer maxim of minimally sufficient means. Seals merely provide evidence of authenticity of the associated (anchored) data whatever that may be.

This approach follows the design principle of *context independent extensibility*. Because the seals are context agnostic, the context is external to KERI. Therefore the context extensibility is external to and hence independent of KERI. This is in contrast to context dependent extensibility or even independently extensible contexts that use extensible context mechanisms such as linked data or tag registries [98; 116; 120; 151]. Context independent extensibility means that KERI itself is not a locus of coordination between contexts for anchored data. This maximizes decentralization and portability. Extensibility is provided instead at the application layer above KERI though context specific external APIs that reference KERI *seals* in order to establish control authority and hence authenticity of the anchored (digested) data. Each API provides the context not KERI. This means that interoperability within KERI is focused solely on interoperability of control establishment. But that interoperability is total and complete and is not dependent on anchored data context. This approach further reflects KERI's minimally sufficient means design aesthetic.

## 7.24 Interaction Event

An *interaction event* is a type of *non-establishment key event* that is interleaved into the key event history with the events from the *establishment event* subsequence (history) in order that the current control authority (root authoritative key-pairs) may be cryptographically verified at the point in the interleaved sequence order where the *interaction event* appears.

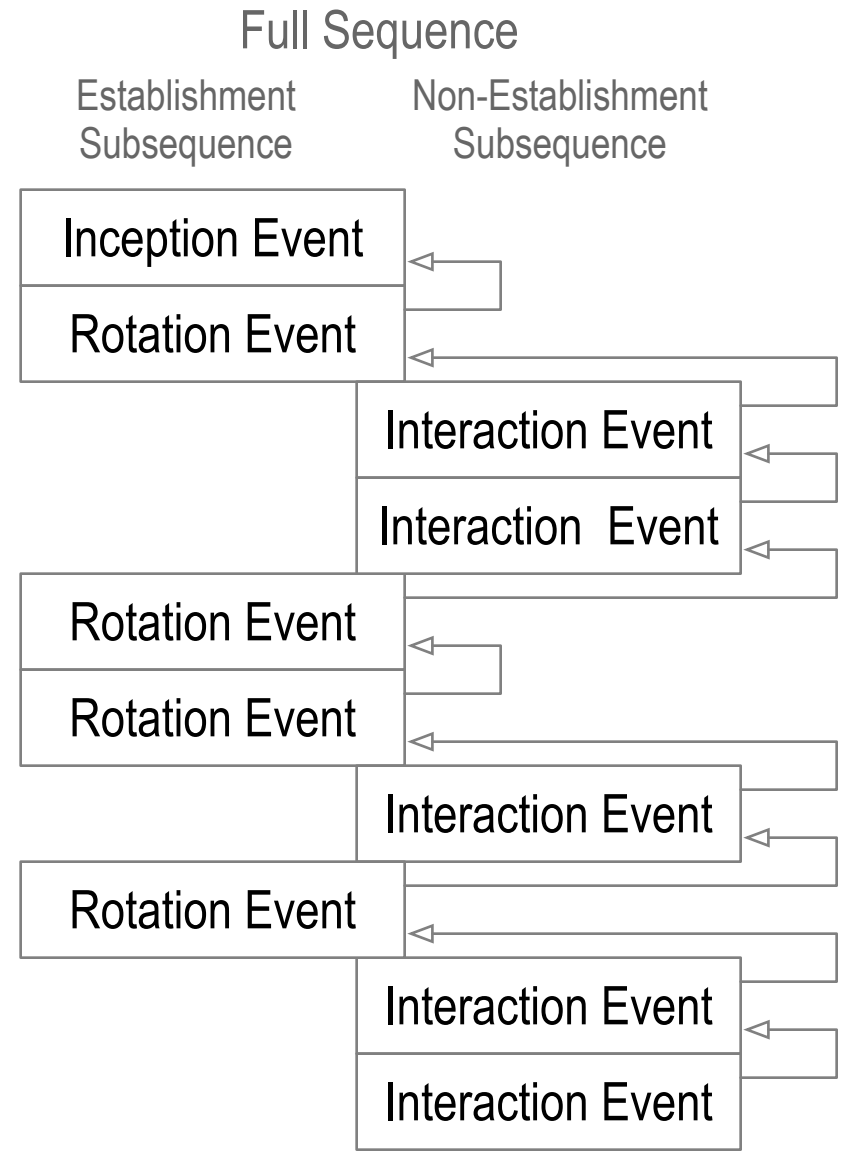

Figure 7.15. Key Event Sequence. Sequence composed of interleaved establishment (inception, rotation) and non-establishment (interaction) events. Arrows represent digest chaining where event at arrow head is the content for the digest at arrow tail.

An interaction event include an event specific interaction data payload in the form of an array of one or more seals. Each seal acts a an anchor that provides evidence of authenticity and proof of control authority for the associated data. Each entry in the payload array may be a digest of the data for a single interaction, or the Merkle root of a sparse hash tree of the data from a block of interactions. All the associated interaction data is thereby anchored to the event sequence at the location of the interaction event [104]. Applications that use KERI may provide discovery of a digest in an interaction event via a hash table (mapping) where the indexes (hash keys) to the table are the digests and the values in the table are the location of the digest at a specific event, at the offset of the entry in the seal array and where applicable the path to its leaf in the associated hash tree. A basic interaction event message is diagrammed below.

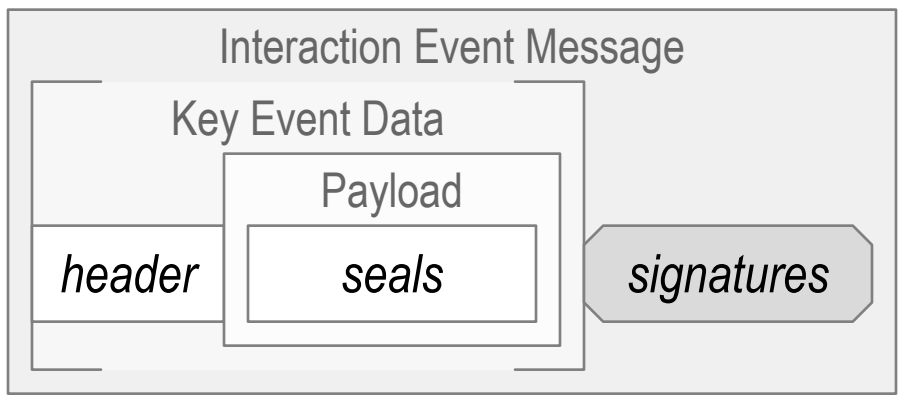

Figure 7.16. Basic Interaction Event Message.

## 7.25 Cooperative Delegation

A *delegation* or *identifier delegation operation* is provided by a pair of events. One event is the delegating event and the other event is the delegated event. This pairing of events is a somewhat novel approach to delegation in that the resultant delegation requires cooperation between the delegator and delegate. We call this *cooperative delegation*. In a *cooperative delegation*, a delegating identifier performs an establishment operation (inception or rotation) on a delegated identifier. A *delegating event* is a type of *event* that includes in its *data payload* an *event seal* of the delegated event that is the target the *delegation* operation. This *delegated event seal* includes a digest of the *delegated event*.

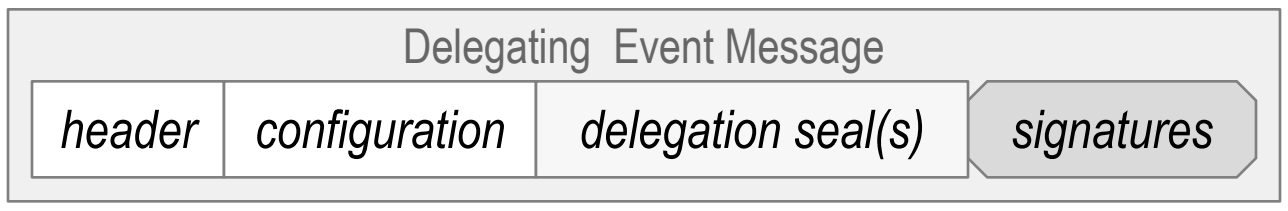

Figure 7.17. Delegating Event.

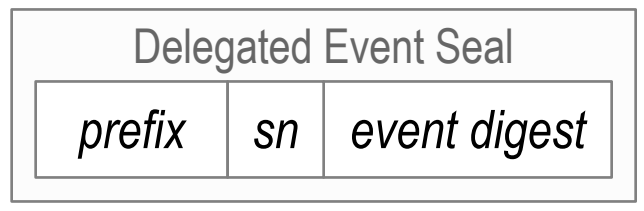

Figure 7.18. Delegated Event Seal.

Likewise, the inception event of the *delegatee's* KEL includes the *delegator's* AID. This binds the inception and any later establishment events in the *delegatee's* KEL to a unique *delegator*. A validator must be given or find the delegating seal in the *delegator's* KEL before the event may be accepted as valid. The pair of bindings (delegation seal in delegator's KEL, delegator's AID in delegatee's inception event) make the delegation cooperative. Both must participate. As will be seen later, this cooperation adds an additional layer of security to the delegatee's KEL and provides a way to recover from prerotated key compromise.

ToDo XXXX diagram of delegated inception event

Figure 7.19. Delegated Event.

Because the delegating event payload is a list, a single *delegating event* may perform multiple delegation operations, one per set of *delegation seal*s.

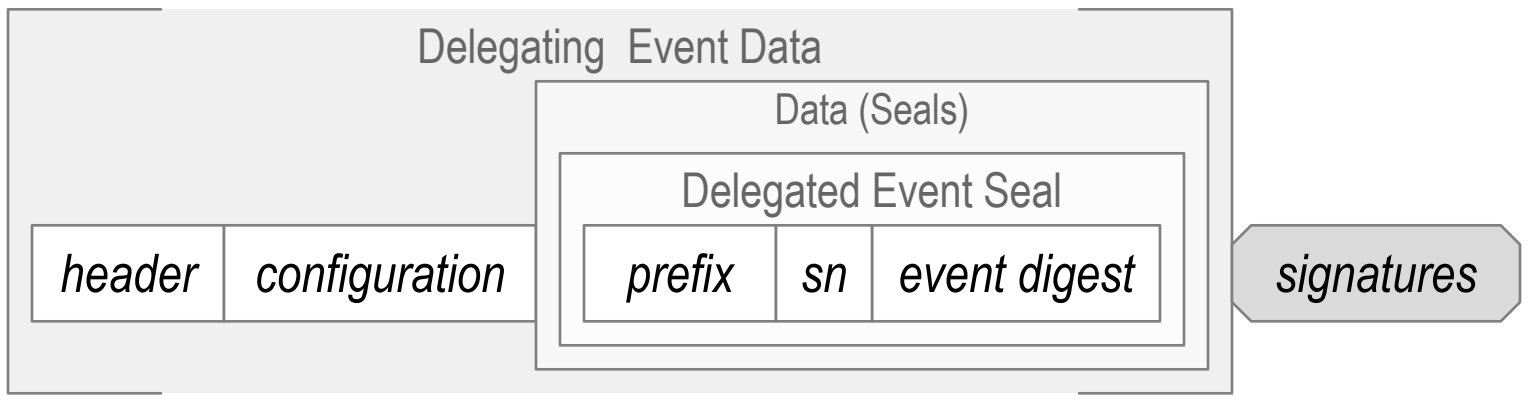

Figure 7.20. Delegating Event Data.

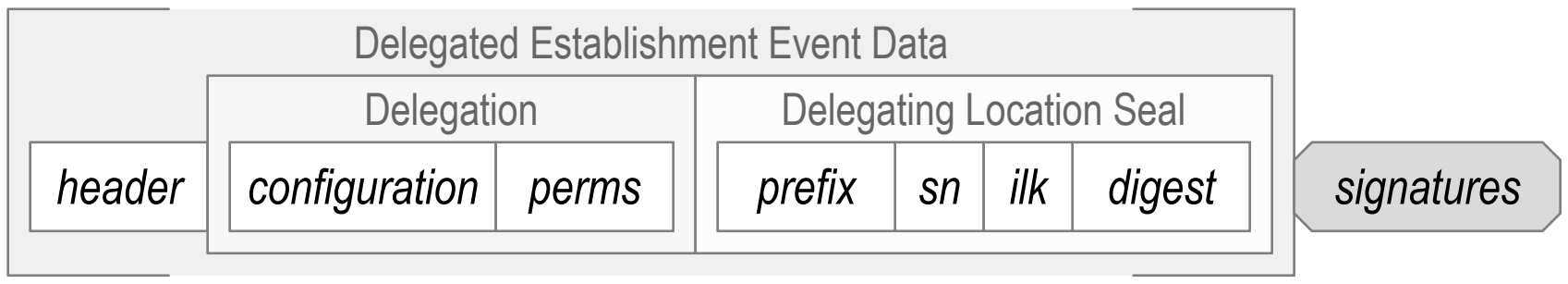

Figure 7.21. Delegated Event Data.

A delegation operation directly delegates an establishment event. Either an inception or rotation. Thus, a delegation operation may either delegate an inception or delegate a rotation that respectively may create and rotate the authoritative keys for the delegated identifier prefix (AID). The identifier prefix for a delegatee (delegated identifier prefix) must be a fully qualified digest of its inception event. This cryptographically binds the delegatee's identifier prefix to the delegator's identifier prefix.

The Delegator (controller) retains establishment control authority over the delegated identifier in that the new delegated identifier may only authorize non-establishment events with respect to itself. Delegation, therefore, authorizes revokable signing authority to some other AID. The delegated identifier has a delegated key event sequence where the inception event is a delegated inception, and any rotation events are delegated rotation events. Control authority for the delegated identifier, therefore, requires verification of a given delegated establishment event, which in turn requires verification of the delegating identifier's establishment event.

To reiterate, because the delegating event seal includes a digest of the full delegated event, it thereby provides a cryptographic commitment to the delegated event and all of its configuration data.

A common use case of delegation would be to delegate signing authority to a new identifier prefix. The signing authority may be exercised by a sequence of revokable signing keys distinct from the keys used for the delegating identifier. This enables horizontal scalability of signing operations. The other major benefit of a cooperative delegation is that any exploiter that merely compromises only the delegate's authoritative keys may not thereby capture the control authority of the delegate. A successful exploiter must also compromise the delegator's authoritative keys. Any exploit of the delegatee is recoverable by the delegator. Conversely, merely compromising the delegator's signing keys may not enable a delegated rotation without also compromising the delegatee's pre-rotated keys. Both sets of keys must be compromised simultaneously. This joint compromise requirement is a distinctive security feature of cooperative delegation. Likewise, as explained later, this cooperative feature also enables recovery of a joint compromise of a delegation at any set of delegation levels by a recovery at the next higher delegation level.

## 7.26 Delegating Interaction Event

A *delegating interaction event* is a type of *interaction event* that includes in its *data payload* a *seal* of a *delegation* operation i.e. a *delegating event seal* that includes a digest of the *delegated event*. The delegating interaction event data is shown below:

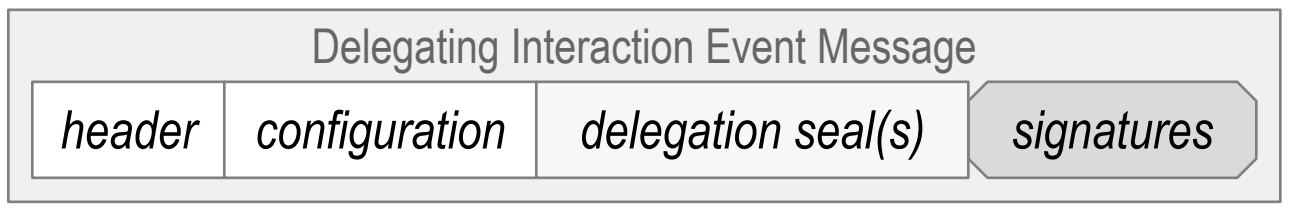

Figure 7.22. Delegating Interaction Event. Event data payload is a delegation seal of data authorizing revokable establishment of some other identifier.

A delegation operation directly delegates an establishment event. Either an inception or rotation.

Figure 7.23. Delegated Establishment Event

The following diagram shows a sequence of delegations by Delegator C that create delegated establishment events for Delegatee D.

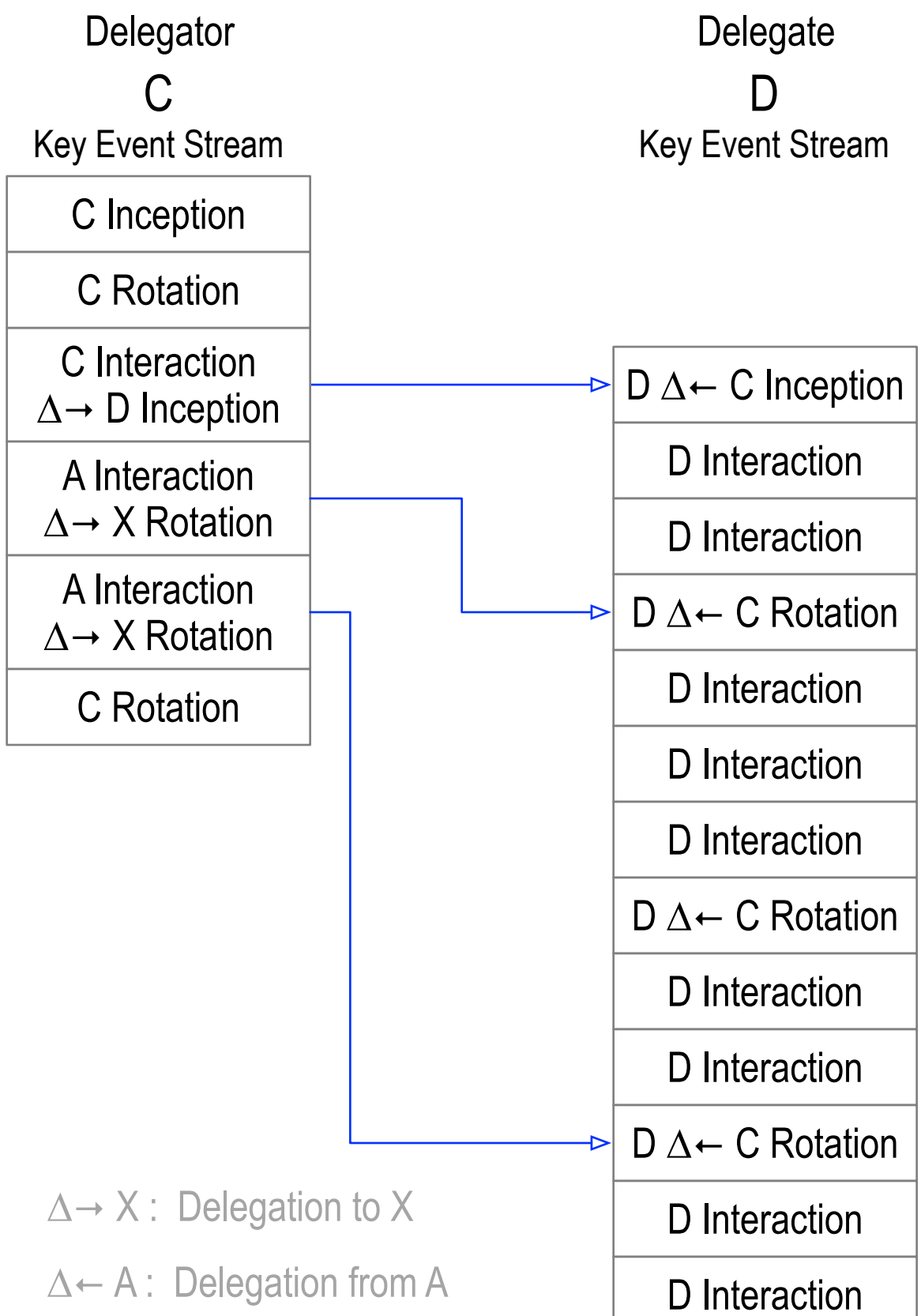

Figure 7.24. Simple Interaction Event Delegated Key Event Stream

The delegation operation may also authorize the delegated identifier to make delegations of its own. This would enable a hierarchy of delegated identifiers that may provide a generic architecture for decentralized key management infrastructure (DKMI). When applied recursively, delegation may be used to compose arbitrarily complex trees of hierarchical (delegative) verifiable key event streams.

## 7.27 Extended Rotation Event

An *extended rotation event* is a type of *rotation event* that also includes a *data payload* as an array of *seals*. This enables the rotation operation to either provide authenticity to any other non-establishment operation on its identifier or an authorized establishment operation on a delegated identifier. In this case the *extended rotation event* operation conveys both a *rotation event* operation and authenticity for *interaction event seals* in one event. This approach provides enhanced security to the interaction *seals* as the effective signing keys are only used once before being rotated. To clarify, an *extended rotation event* that includes an interaction data payload effectively making the event's signing keys one-time usage keys that rotate with each usage. Thus the interaction signing keys are protected from repeated use.

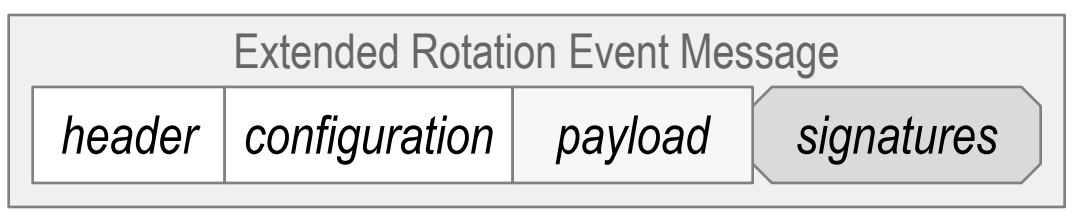

Figure 7.25. Extended Rotation Event. Event payload may contain interaction event data for enhanced security of interaction event.

To clarify, an extended rotation event include an event specific interaction data payload in the form of an array of one or more seals. Each seal acts a an anchor that provides evidence of authenticity and proof of control authority for the associated data. Each entry in the payload array may be a digest of the data for a single interaction, or the Merkle root of a sparse hash tree of the data from a block of interactions. All the associated interaction data is thereby anchored to the event sequence at the location of the rotation event [104]. Applications that use KERI may provide discovery of a digest in an extended rotation event via a hash table (mapping) where the indexes (hash keys) to the table are the digests and the values in the table are the location of the digest at a specific event, at the offset of the entry in the seal array and where applicable the path to its leaf in the associated hash tree.

## 7.28 Delegating Rotation Event

A *delegating rotation event* or *delegating event* is a type of *extended rotation event* that includes in its *data payload* a *delegation seal*. By virtue of mixing the *delegation seal* in a rotation operation, the effective delegation signing keys may be better protected from exposure by their one-time first-time use in the rotation operation. Otherwise a *delegating rotation event* is identical to a *delegating inception event* as described above.

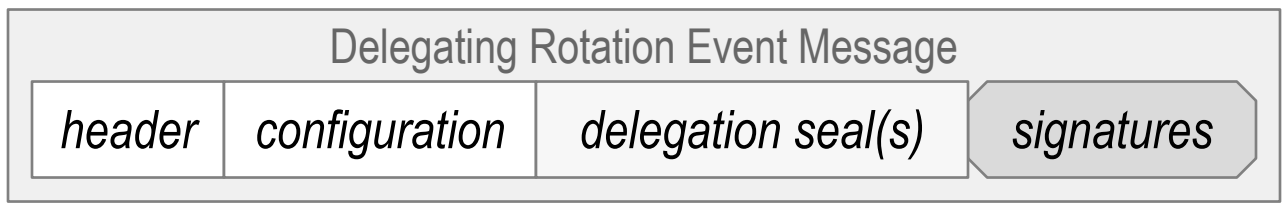

Figure 7.26. Delegating Rotation Event. Event data payload is a delegation seal of data authorizing revokable establishment of some other identifier.

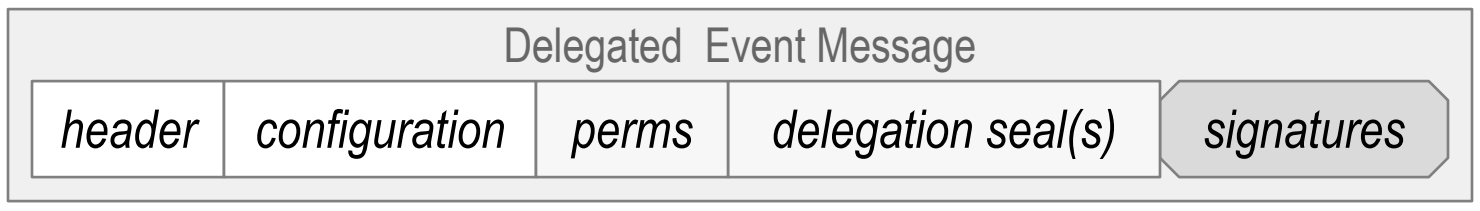

Figure 7.27. Delegated Establishment Event. Event data payload is a delegation authorizing revokable establishment of some other identifier.

The following diagram shows a sequence of delegations operations via pairs of delegating rotation events by Delegator C that create for delegated establishment events for Delegatee D.

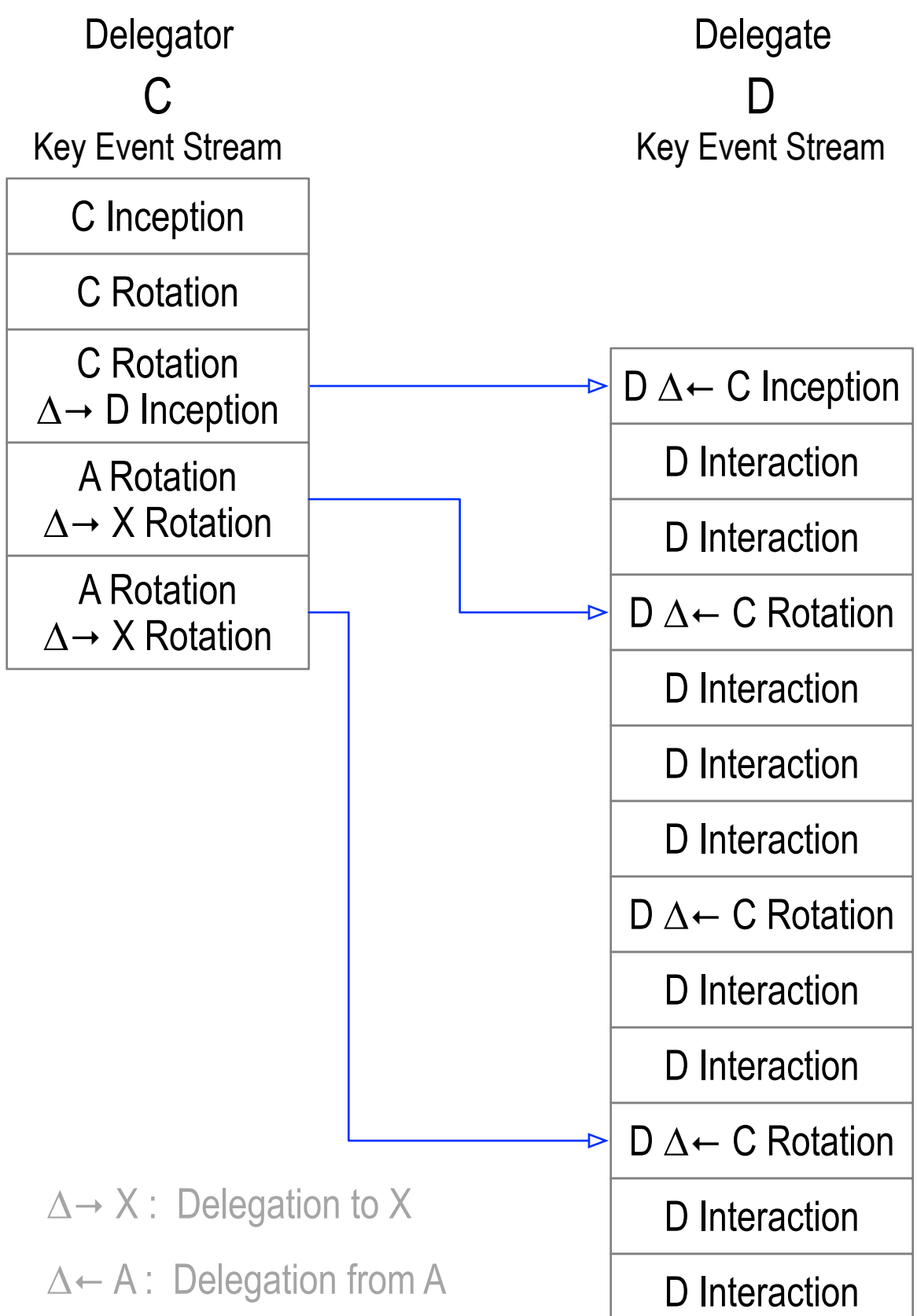

Figure 7.28. Simple Rotation Event Delegated Key Event Stream

Using rotation instead of interaction to provide the delegation operation provides more security at the cost of convenience. The demands of a particular application may determine which approach is more suitable.

One reason to use rotations for delegation is for enhanced security. A rotation event is a first time use of the pre-rotated keys to sign an event. The distinction between an interaction event rotating a delegated identifier's keys and a rotation event rotating a delegated identifier's keys is that the latter enables nested recovery of a compromise of the delegate's keys, even its pre-rotated keys. Recovery is described in more detail in section 11.6, but in summary, a rotation event may be used to supersede an interaction event. When this happens the key event log forks at the superseding rotation event. With delegated events this means recovery is enabled even in the event of the joint compromise of a delegating identifier's signing keys and the delegated identifiers pre-rotated keys. The delegating identifier merely needs to perform a rotation event that provides a superseding rotation of the interaction event used to delegate a rotation of the delegate. This superseding rotation or a subsequent interaction may be used to perform a superseding rotation of the delegate's rotation. This nested recovery may be applied to multiple levels of delegation. A rotation at the next higher level of delegation may be used to recover from key compromise across any set of lower levels of delegation.

## 7.29 Multi-Signature Key-Pair Labeling

The examples in the key-pair labeling descriptions above assume that current control authority over the *identifier* at any point in time is based on a single key-pair. One way to increase security over a given identifier is to use a multi-signature approval scheme multiple simultaneous key-pairs jointly control the identifier. This may require multiple controllers as signers to authorize or approve key events. Multiple signatures may make exploit significantly more difficult be-cause multiple keys must be compromised. Thresholded multiple signatures may mitigate the risk of loss of one or more private keys thereby making key recovery more robust. Multiple signature rotation, however, may be more complex. To simplify and clarify the descriptions of both the single and multiple signature versions of the protocols we introduce appropriate terminology conventions below. For the efficiency of expression all establishment event operations such as inception may be considered a special case of rotation and denoted as such.

Recall that each member of a sequence of key-pairs is labeled with whole number index *j*. In this notation the key-pair $\left(C^j, c^j\right)$ represents the $j^{th}$ key-pair in the sequence of key-pairs controlled by C. Furthermore let each member of the subsequence of rotation events (including inception) from a key event sequence be denoted by whole number index *l*. Recall as well that the sequence of key events may be indexed by *k*, such as $\varepsilon_k$. When only rotation operations (inception inclusive) appear in the key event sequence then the subsequence of rotation operations equals the sequence of key events and $l = k$. Each rotation operation may change the set of controlling key-pairs for the identifier. When the new controlling key-pair set is singular, then each rotation consumes one new key-pair from the key-pair sequence. In this case, the indices of the key-pair sequence and rotation subsequence will have the same value, that is, $j = l$. We may therefore equivalently use *l* as a superscript to represent the $l^{th}$ key-pair, $\left(C^l, c^l\right)$ in the sequence of key-pairs controlled by *C*.

When the controlling key-pair is not singular, however, (i.e. multi-signature) then each rotation operation may consume more than one key-pair from the sequence of controlling key-pairs. In this case then the indices *j* and *l* may not always be equal, i.e. $j \neq l$. Let the number of controlling key-pairs for the $l^{th}$ rotation be denoted $L_l$. The $l^{th}$ rotation thereby consumes $L_l$ key-pairs from the sequence of controlling key-pairs. The key-pairs consumed by the $l^{th}$ rotation form a subsequence of length $L_l$. The subsequence of indices *j* for the subsequence of key-pairs for the rotation *l* of length $L_l$ may be denoted as follows.

$$\left[r_l, r_l+1, r_l+2, \ldots, r_l+L_l-1\right] \tag{7.17}$$

where $r_l$ is the value of index *j* for the first key-pair in the subsequence. This means that

$$r_{l+1} = r_l + L_l\,. \tag{7.18}$$

The set of originating key-pairs are declared by an inception operation that creates the identifier *C*. The inception operation may be thought of as a special case of rotation that is the zeroth rotation operation, that is, $l = 0$. Also let $r_0 = 0$ be the value of the first index from the inception or zeroth rotation. Given that $r_0 = 0$ then the value of the index $r_l$ for $l > 0$ may be computed by summing the lengths, $L_l$, of subsequences of key-pairs consumed by all the prior rotations. This may be expressed as follows:

$$r_l = \sum_{i=0}^{l-1} L_i \,\Big|_{l>0}\,. \tag{7.19}$$

where $r_0 = 0$.

The subsequence of public keys for the $l^{th}$ rotation may be denoted as follows:

$$\left[C^{r_l}, \ldots, C^{r_l+L_l-1}\right]_l, \tag{7.20}$$

with

$$\left[C^{r_0}\right]_0 = \left[C^0\right]. \tag{7.21}$$

Suppose for example that the inception operation uses one key-pair, that is, $L_0 = 1$ and the following three rotations use three, three, and four key-pairs respectively, that is, $L_1 = 3$, $L_2 = 3$, and $L_3 = 4$. The resultant starting indices for each rotation subsequence are as follows:

$$
\begin{aligned}
r^0 &= 0 \\
r^1 &= \sum_{i=0}^{0} L_i = L_0 = 1 \\
r^2 &= \sum_{i=0}^{1} L_i = L_0 + L_1 = 4 \\
r^3 &= \sum_{i=0}^{2} L_i = L_0 + L_1 + L_2 = 7 \\
r^4 &= \sum_{i=0}^{3} L_i = L_0 + L_1 + L_2 + L_3 = 11
\end{aligned}
\quad . \tag{7.22}
$$

Furthermore the resultant subsequences of public-keys for each rotation in order are as follows:

$$
\begin{aligned}
&\left[ C^{r_0} \right]_0 = \left[ C^0 \right] \\
&\left[ C^{r_1}, C^{r_1+1}, C^{r_2+2} \right]_1 = \left[ C^1, C^2, C^3 \right] \\
&\left[ C^{r_2}, C^{r_2+1}, C^{r_2+3} \right]_2 = \left[ C^4, C^5, C^6 \right] \\
&\left[ C^{r_3}, C^{r_3+1}, C^{r_3+2}, C^{r_3+4} \right]_3 = \left[ C^7, C^8, C^9, C^{10} \right]
\end{aligned}
\tag{7.23}
$$

With the nomenclature presented above we may thereby describe more efficiently the use of multi-signature schemes in the protocol.

## 7.30 Verifier

A *verifier* is an *entity* or *component* that cryptographically verifies the signature(s) on an event message. In order to verify a signature, a *verifier* must first determine which set of keys are or were the controlling set for an identifier when an event was issued. In other words a *verifier* must first establish control authority for an identifier. For identifiers that are declared as non-transferable at inception this control establishment merely requires a copy of the inception event for the identifier. For identifiers that are declared transferable at inception this control establishment requires a complete copy of the sequence of key operation events (inception and all rotations) for the identifier up to the time at which the statement was issued.

## 7.31 Validator

A *validator* is an *entity* or *component* that determines that a given signed statement associated with an identifier was valid at the time of its issuance. Validation first requires that the statement is verifiable, that is, has a verifiable signature from the current controlling key-pair(s) at the time of its issuance. Therefore a validator must first act as a verifier in order to establish the root authoritative set of keys. Once verified, the validator may apply other criteria or constraints to the statement in order to determine its validity for a given use case. This use-case specific validation logic may be associated with *interaction* event statements.

## 7.32 Event Location and Version

The *location* of an event in its key event sequence is determined by its previous event digest sequence number and ilk. The *version* of an event of the same *class* at given *location* in the key event sequence is different or inconsistent with some other event of the same *class* the same *location* if any of its other content differs or is inconsistent with that other event of the same *class* and *location*. To clarify, the event version includes the *class* of event, *establishment* or *non-establishment*. The special case where *class* matters is that an *establishment* event (such as rota-tion) at the same *location* may supersede a *non-establishment* event (such as interaction) . This enables recovery of live exploit of the exposed current set of authoritative keys used to sign non-establishment events via a rotation establishment event to the unexposed next set of authoritative keys. The specific details of this recovery are explained later (see Section 11.6). In general, the witnessing policy is that the first seen *version* of an event always wins, that is, the first verified version is *witnessed* (signed, stored, acknowledged and maybe disseminated) and all other versions are discarded. The exception to this general rule is that an establishment event may recover following a set of exploited non-establishment events. The recovery process may fork off a branch from the recovered trunk. This *disputed* branch has the disputed exploited events and the main trunk has the recovered events. The *operational mode* (see Section 10.) and the *threshold of accountable duplicity* determine which events in the disputed branch are *accountable* to the controller (see Section 11.6).

## 7.33 Witness

A *witness* is an *entity* or *component* designated (trusted) by the *controller* of an *identifier*. The primary role of a witness is to verify, sign, and keep events associated with an identifier. A *witness* is the *controller* of its own self-referential *identifier* which may or may not be the same as the *identifier* to which it is a *witness*. As a special case a *controller* may serve as its own *witness*. Witness designations are included in *key (establishment) events*. As a result the role of a witness may be verified using the identifier's rotation history. When designated, a *witness* becomes part of the supporting infrastructure establishing and maintaining control authority over an identifier. An identifier *witness* therefore is part of its *trust basis* and may be controlled (but not necessarily so) by its controller. The purpose of a pool of witnesses is to protect the controller from external exploit of its identifier. A *witness* may use the controlling key-pairs of its own self-referential identifier to create digital *signatures* on *event messages* it has received but are associated with *identifiers* not necessarily under its control. To clarify, a *witness* controls its own self-referential identifier and acts as a witness of *event message*s for some *identifier* not necessarily under its control. A *witness* may receive, verify, and store an *event* on an *identifier*. Verify means verify the signature attached to the event using the current controlling key-pairs for the event at the time of event issuance. Thus a witness first acts as an event verifier. It determines current control authority of the event's identifier with respect to the sequence of key (establishment) events it has so far received for that identifier. The witness follows a policy explained in more detail later for how it treats different versions of an event it may receive. Simply, it always gives priority to the first version of an event it receives (first seen). The witness signifies this by only signing and keeping the first successfully verified version of an event it receives. To restate a witness will never sign any other conflicting version of the same event in an event sequence. The event sequence kept by a witness for an identifier must therefore be internally consistent.

## 7.34 Receipt

A *receipt* or *event receipt* is a special type of *event* conveyed by a *message* that includes one or more signatures on some other event message. A receipt also includes either a copy of the other *event message* or else a reference to that message. Each signature may be created by a *witness*. Indeed the primary purpose of a *witness* is to generate, store, and disseminate an event receipt for the first verified version of an event the *witness* receives. The witness receipt creation policy is described in more details later. Other components such as *validators* and *watchers* (see below) may create receipts that are denoted *validator* or *watcher* receipts respectively. In general the unqualified term *receipt* means *witness receipt* unless clear from the context.

A simple *witness receipt message* conveys information about the receipt. It may include information to uniquely identify the receipted event, such as either the event itself or a reference to the event consisting of the identifier prefix, sequence number and a hash of the event contents together with a set of one or more labeled witness signatures of the receipted event. The witness label might be the identifier of the witness or some other alias uniquely associated with the witness within the context of the witnessed identifier.

Recall that because a signature is a type of cryptographic digest, the signature in the receipt binds the receipt to the contents of the event. Likewise the recipient of the receipt may know the identity of the sending witness and therefore its identifier prefix and hence its controlling key-pair. Nonetheless, in addition to the signature it may be convenient to include in the receipt message the prefix of the witness to make association easier. Typically the recipient of a *key event receipt message* already has a copy of the associated *key event message*. Consequently there may be no need to retransmit a copy of original key event in the receipt message but merely is sequence number and digest for lookup and confirmation. This makes the message bigger but easier to track and manage especially on an asynchronous communications channel. For compactness it may be more convenient to attach prefixes and signatures from multiple witnesses. A diagram of a receipt message with attached witness prefixes and signatures is shown below,

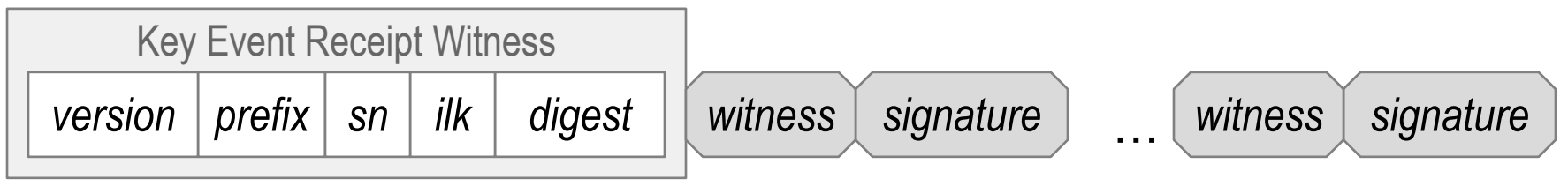

Figure 7.29. Key Event Receipt Message

For a witness of identifier prefix $C$ with identifier prefix $W_0^C$ that is bound to $\left(W_0^C, w_0^C\right)$ its key event receipt message may be expressed as follows:

$$\left\langle \varepsilon_k^C \right\rangle W_0^C \sigma_{W_0^C}^C \tag{7.24}$$

where $\varepsilon_k^C$ is either a copy or a label of the associated key event with sequence number $k$, $W_0^C$ is the witness' identifier, and $\sigma_{W_0^C}^C$ is the signature of the associated key event by the witness. A receipt with two witnessed signatures may be represented as follows:

$$\left\langle \varepsilon_k^C \right\rangle W_0^C \sigma_{W_0^C}^C W_1^C \sigma_{W_1^C}^C \tag{7.25}$$

where $\varepsilon_k^C$ represents the event, $W_0^C$ is the identifier of the first witness, $\sigma_{W_0^C}^C$ is the signature of the first witness, $W_1^C$ is the identifier of the second witness, and $\sigma_{W_1^C}^C$ is the signature of the second witness. This may be extended to any number of witnesses.

The receipt message itself does not need to be signed by the witness because the only important information the message conveys is the witness' signature that indicates that the witness "witnessed" the associated key event. The witness' signature is non-repudiable so a verified copy of the signature is sufficient to convey that information. In order to verify the signature the recipient of the message needs to have a copy of the original key event message and know the signature's verification (public) key. The event label and witness identifier in the receipt message enable convenient lookup of that information.

In general a witness may use a unique identifier for itself for each KERL (key event receipt log) it witnesses. Should the keys for the witness' identifier become compromised, then if the witness identifier is transferable, it may recover from the compromise rotating the compromised key and resuming operation as a witness under the same identifier. If instead the witness' identifier is not transferable then it must cease acting as a witness using the compromised identifier. In the later approach the witness may resume acting as a witness but under a new identifier. Incon-sistent receipts from the same witness are detectable as long as a copy of the original receipt is preserved. Although either approach may be made to work, the later approach avoids recursive rotation validation of KERL entries (where the KERL entries of witnesses to a controller are validated and then the KERL entries of witnesses of the witnesses are validated and so on). Recursive validation complicates the implementation. Consequently, without loss of generality, in this work witnesses may only use non-transferable (non-rotatable) identifier keys. They must create a new identifier when exploited.

## 7.35 Key Event Log

A *key event log* (KEL) is an ordered chained record of the *key event* messages created by the controller(s) of an identifier. It is an append only log. The key events are chained in that each event message besides the *inception* event includes a cryptographic strength content digest of the preceding event message. It must include all the *interleaved key event* messages both *establishment* and *non-establishment*. A verifier may establish current control authority by replaying the subsequence of *establishment events* in a valid KEL. A KEL may be part of the trust basis for an identifier and may act as a secondary root-of-trust.

## 7.36 Key Event Receipt Log

A *key event receipt log* (KERL) is a KEL that also includes all the consistent *key event receipt* messages created by the associated set of witnesses. Witnesses keep a separate KERL for each of the identifiers for which they create receipts. Each witnesses's KERL for an identifier is an append only event log of first seen verified consistent events. The events are chained in that each event message besides the inception event includes a cryptographic strength content digest of the preceding event message. A witness will only keep consistent events and receipts of those consistent events. A witness may cache out of order events or incompletely signed events until such time as they may be verified as consistent or inconsistent. Because a proper event message includes the signature(s) of the controller(s) it may be thought of as a special type of self-signed receipt. In this same way a KEL may be thought of as a special type of KERL. A verifier may establish current control authority by replaying the subsequence of *establishment events* in a valid KERL. A KERL may be part of the trust basis for an identifier and may act as a secondary root-of-trust.

## 7.37 Watcher

A *watcher* is an *entity* or *component* that keeps a copy of a KERL for an identifier but is not designated by the *controller* thereof as one of its *witnesses*. To clarify, a *watcher* is not designated in the associated identifier's key events. A *watcher* is the *controller* of its own self-referential *identifier* which may not be the same as the *identifier* to which it is a *watcher*. An identifier *watcher* may be part of the trust basis of a validator and may also be controlled (but not necessarily so) by the validator's controlling entity. A watcher may sign copies of its KERL or parts of its KERL but because a watcher is not a designated *witness* these are not *witnessed receipts*. They may be considered *watcher receipts* or *ersatz receipts*.

## 7.38 Juror

A *juror* is an *entity* or *component* that performs *duplicity detection* on events and event receipts. A *juror* is the *controller* of its own self-referential *identifier* which may or may not be the same as the *identifier* to which it is a *juror*. The *juror* may thereby create digital signatures on statements about duplicity it has detected. A *juror* detects duplicity by keeping a copy of any mutually inconsistent versions of any events its sees. It may then provide as proof of duplicity the set of mutually inconsistent versions of an event. A *juror* uses the KERL (KEL) from a con-troller, watcher, or witness as the reference for comparison to determine duplicity. Duplicity takes two forms. In the first form, a *controller* may be deemed duplicitous whenever it produces an event message that is inconsistent with another event message it previously produced. In the second form, a *witness* may be deemed duplicitous when it produces an event receipt that is inconsistent with another event receipt it previously produced. The detailed rules for determining inconsistency are described later. By construction, in this protocol, exploits of a controller or its witnesses exhibit themselves as duplicitous statements. Duplicity then becomes a basis for distrust in a controller or its witnesses. To summarize, main role of a *juror* is to provide duplicity detection of a controller and/or its witnesses to validators so that a validator may be protected from exploits of the controller and/or its witnesses. Duplicity is provable by any *juror* to any validator because the validator may itself verify a set of inconsistent statements with respect to a KERL. Duplicity is therefore end-verifiable by any validator at any time given inconsistent statements referenced to a KERL. In other words, a set of jurors may provide ambient duplicity protection to validators. A juror may be part of the trust basis of a validator and may be under the control (but not necessarily so) of that validator.

## 7.39 Duplicitous Event Log

A *duplicitous event log* (DEL) is record of inconsistent event messages produced by a given controller or witness with respect to a given KERL. The *duplicitous events* are indexed to the corresponding event in a KERL. A duplicitous event is represented by a set of two or more provably mutually inconsistent event messages with respect to a KERL. Each juror keeps a *duplicitous event log* (DEL) for each controller and all designated witness with respect to a KERL. Any validator may confirm duplicity by examining a DEL.

## 7.40 Judge

A *judge* is an *entity* or *component* that examines the entries of one or more KERLs and DELs of a given identifier to validate that the event history is from a non-duplicitous controller and has been witnessed by a sufficient number of non-duplicitous witnesses such that it may be trusted or conversely not-trusted by a validator. In this sense a *judge* is a validator of a *controller* and its *witness pool*. A *judge* is the controller of its own self-referential *identifier* which may or may not the same as the identifier to which it is a *judge*. A *judge* may thereby create digital *signatures* on statements about validations it has performed on KERLs and DELS. A *judge* may obtains KERLS from one or more *witnesses* or *watchers* and may obtain DELs from one or more *jurors*. A *judge* may be part of the trust basis of a validator and may also be under the control of a validator. A *judge* may be a second party involved in a transaction with a first party *controller* or a *judge* may be a trusted third party in a multi-party transaction that includes a *controller* and other *validator* parties. A given entity may act in multiple roles such as both *witness* and *juror* or both *juror* and *judge*. A *validator* might perform its function by acting as all of a *watcher*, *juror*, and *judge* or by trusting other *witnesses, watchers, jurors* and *judges*.

## 7.41 Resolver

A *resolver* is an *entity* or *component* that provides discovery for *identifiers*. A *resolver* is the *controller* of its own self-referential *identifier* which may not be the same as the *identifier* to which it is a *resolver*. A *resolver* primarily maps identifiers to the URLs or IP addresses of components of the trust bases for identifiers. These components include *controllers*, *witnesses*, *watchers*, *jurors* and *judges*. Given the URL or IP address of a component, a user may there from obtain or be directed to the associated event histories (KELs, KERLs, and DELs) in order that the user may establish current (root) control authority for the identifier. The resolver may cache these event histories or key event subsequences as end verifiable proofs of root control authority.

A suitable architecture for a KERI resolver may be based on a distributed hash table (DHT) algorithm such as Kademlia [50; 101]. An example of a generic DHT system that could be used for KERI discovery is IPFS [80]. A DHT provides two types of the discovery. The first type is the discovery of the nodes that host portions of the DHT itself. The second type is the discovery of the target data of the DHT. With respect to KERI, the target data for discovery is different for the two operative classes of identifier in KERI, that are, transferable and non-transferable identifier prefixes. In the case of a non-transferable identifier prefix, such as that of a witness or watcher, the target data may include a mapping from the non-transferable identifier prefix to a the service endpoint URL or directly to the IP address of the witness or watcher key event receipt log (KERL) service. In this case a validator could query the resultant IP address for a copy of the full KERL for the transferable identifier prefix to which the witness or watcher is entrained. In the case of transferable identifiers, discovery may provide a mapping of the identifier prefix to a cached copy of either its full key event log (KEL) or a copy of its inception event plus the latest rotation event or equivalently the latest key event state. From this copy, one may extract the identifier prefixes of the current witness set and then use discovery to access the KERLs for those witnesses.

## 7.42 Date-time Stamps

It may be convenient to include absolute (real, astronomical) date-time stamps in messages or log entries. One well known date-time format is the ISO-8601 standard [85; 86]. An ISO-8601 time zone aware UTC date time stamp with microsecond resolution has the following form:

YYYY-MM-DDTHH:MM:SS.mmmmmm+00:00 (7.26)

An example is

`2000-01-01T00:00:00.000000+00:00` (7.27)

# 8 Security Concerns

Every operation in this protocol is expressed via cryptographically verifiable events. Successful exploit therefore must attack and compromise the availability and/or consistency of events. Security analysis therefore is focused on characterizing the nature and timing of these attacks and how well the protocol preserves the availability and consistency of events when subject to attack. We therefore describe potential exploits in terms of these characteristics.

The first characteristic is a *live* versus *dead* event exploit. A *live* exploit involves attacks on current or recent events. Protection from *live* exploit is essential to maintaining the operational security in the present. Protection from *live* exploit focuses on providing both sufficient availability of current events as well as ensuring their consistency (non-duplicity). A *dead* exploit, in contrast, involves attacks on past events. Protection from *dead* exploits is primarily provided by duplicity detection (consistency). One verifiable copy of a KEL (KERL) is enough to detect duplicity in any other verifiable but inconsistent copy. Attacks on the availability of pasts events are relatively easily mitigated by archiving redundant copies. The eventuality of *dead* exploits of compromised signing keys must be mitigated because as computing and cryptographic technology advance over time, (quantum or otherwise), digital signatures may become less secure. Eventually their keys may become compromised via direct attack on their cryptographic scheme.

The second characteristic is a *direct* versus *indirect* operational mode exploit. The protocol may operate in two basic modes, called *direct* and *indirect*. The availability and consistency at-tack surfaces are different for the two modes and hence the mitigation properties of the protocol are likewise mode specific.

The third characteristic is a *malicious third party* versus a *malicious controller* exploit. In the former the attack comes from an external malicious attacker but the controller is honest. In the latter the controller may also be malicious and in some ways may be indistinguishable from a successful malicious third party. The incentive structure for the two exploit types is somewhat different and this affects the mitigation properties of the protocol. We find it helpful in both the design and analysis of protection to consider separately these two kinds of attacks.

The main participants in the protocol are controllers and validators. The other participants, such as, witnesses, watchers, jurors, judges, and resolvers provide support to and may be under the the control of either or both of the two main participants.

The analysis of protection to an attack can be further decomposed into three properties of each protection mechanism with respect to an attack, these are: susceptibility to an attack, vulnerability to harmfulness given an attack, and recoverability given a harmful attack. Security design involves making trade-offs between these three properties of protection mechanisms. Harm from successful exploit may arise in either or both of the following two cases:

A controller may suffer harm due to the loss or encumberment of some or all of its control authority such that the malicious entity may produce consistent verifiable events contrary to the desires of the controller and/or impede the ability of the controller to promulgate new key events.

A validator may suffer harm due to its acceptance of inconsistent verifiable events produced by a malicious entity (controller and/or third party).

Protection consists of either prevention or mitigation of both of the harm cases. The primary protection mechanisms to the controller include best practice key management techniques for maintaining root control authority, redundant confirmation of events by supporting components and duplicity detection on the behavior of designated supporting components. The primary protection mechanism to the validator is duplicity detection on the behavior of supporting components. We describe below, in detail, the properties of the protocol protection mechanisms in terms of the attack characteristics.

# 9 Pre-rotation

## 9.1 Introduction

As described previously, the main purpose of key rotation it to either prevent or recover from a successful compromise of one or more private keys by an exploiter. The main *exposure* attack, *live* or *dead*, is to capture a private key either while it is being used to sign events on some computing device or while it is in transit from some secure storage device to the computing device. Another *dead* exposure attack may be launched some time long after the creation of the key-pair. This attack may then employ advanced computing techniques such as quantum computation to invert the one-way function used to derive the public key.

Given a potentially compromised private key, a secure rotation operation must be authorized using a different private key or else the compromised key may be used by the exploiter to rotate to a key-pair under its control thereby effectively allowing it to capture control over the identifier. One common mitigation is to designate a special key-pair just for rotation operations. This still suffers from the vulnerability that over time the rotation key-pair will also eventually become exposed and thereby at risk of compromise. The rotation key itself may then need to be rotated by yet another special rotation key authorized to rotate the original rotation key and so on. This approach results in a complex multi-layered key management infrastructure.

In contrast, the approach presented here, called *pre-rotation*, is much simpler [136]. As mentioned above one of the primary protection mechanisms for a controller are best practice key management mechanisms for maintaining root control authority. This novel adaptable *pre-rotation* scheme is one of those best practices.

The *pre-rotation* scheme provides secure verifiable rotation that mitigates successful exploit of a given set of signing private keys from a set of (public, private) key-pairs when that exploit happens sometime after its creation and its first use to issue a self-certifying identifier. In other words, it assumes that the private keys remains private until after issuance of the associated identifier. To elaborate, this protocol does not address the cryptographic security of a set of private keys in the face of side channel or other attacks that may capture the private keys in the time period after creation of the key-pairs but before or at issuance of the associated identifier. Instead, well known best practices for key management and signing infrastructure (not covered herein) may be employed to prevent or mitigate attacks that capture the private keys at the time of creation of the key-pairs or anytime before or at issuance of the identifier. The protocol generally assumes that private keys remain private at least until after first exposure. This protocol also generally assumes the non-existence of *live* brute force or other attacks that may break a private key given only knowledge of the public key and/or signed statements until such time that a controller may prophylactically rotate to a stronger signing scheme. It also assumes a cryptographic strength of no less than 128 bits for all cryptographic operations including signing schemes. This protocol is meant to solve the problem of secure transfer of authority from one set of key-pairs to another via verifiable statements not all the problems of keeping private keys private.

As mentioned previously, the primary risk of exposure mitigated by this protocol comes after issuance through the use of a private key to sign statements. A notable protection mechanism of pre-rotation enables recovery from a compromised exposed signing key via a set of unexposed rotation keys that had been previously designated in a cryptographic commitment i.e. a *pre-rotation*. The pre-rotation commitment also provides some degree of post-quantum security to the pre-rotations.

## 9.2 Post-Quantum Security

Post-quantum cryptography deals with techniques that maintain their cryptographic strength despite attack from quantum computers [109; 150]. Because it is currently assumed that practical quantum computers do not yet exist, post-quantum techniques are forward looking to some future time when they do exist. A one-way function that is post quantum secure will not be any less secure (resistant to inversion) in the event that practical quantum computers suddenly or unexpectedly become available. One class of post-quantum secure one-way functions are some cryptographic strength hashes. The analysis of D.J. Bernstein with regards the collision resistance of cryptographic one-way hashing functions concludes that quantum computation provides no advantage over non-quantum techniques [25]. Consequently one way to provide some degree of post-quantum security is to hide cryptographic material behind digests of that material created by such hashing functions [141]. This directly applies to the public keys declared in the pre-rotations. Instead of a pre-rotation making a cryptographic pre-commitment to a public key, it makes a pre-commitment to a digest of that public key. The digest may be verified once the public key is disclosed (unhidden) in a later rotation operation. Because the digest is the output of a one-way hash function, the digest is uniquely strongly bound to the public key. When the unexposed public keys of a pre-rotation are hidden in a digest, the associated private keys are protected from a post-quantum brute force inversion attack on those public keys.

To elaborate, a post-quantum attack that may practically invert the one-way public key generation (ECC scalar multiplication) function using quantum computation must first invert the digest of the public key using non-quantum computation. Pre-quantum cryptographic strength is therefore not weakened post-quantum. A surprise quantum capability may no longer be a vulnerability. Strong one-way hash functions, such as 256 bit (32 byte) Blake2, Blake3 and SHA3, with 128 bits of pre-quantum strength maintain that strength post-quantum. Furthermore, hiding the pre-rotation public keys does not impose any additional storage burden on the controller because the controller must always be able to reproduce or recover the associated private keys to sign the associated rotation operation. When a pre-rotation hides its public keys, however, the length of the subsequent rotation event message increases because it must provide the actual unhidden public keys. The length of the prior rotation event, however, decreases because it must only provide a single digest of the set of public keys. Whereas when a pre-rotation does not hide its set of public keys behind a digest then the subsequent rotation may simply use the unhidden public keys from the prior rotation's pre-rotation set. Hidden public keys may be compactly expressed as Base64 encoded hidden qualified public keys where the hashing function is indicated in the derivation code.

## 9.3 Basic Pre-Rotation

The following diagrams help illustrate how pre-rotation works. For the sake of simplicity and ease of understanding, in this exposition, event configuration is simplified to only include the most relevant details. Likewise all key sets are simplified to have only a single key. Fully explicated event configuration including key sets with multiple keys is described in complete detail later.

Each inception operation involves two sets of keys that each play a role in the event. The two roles are labeled *initial current* or *initial* for short, and *next*. The inception operation creates both sets of key-pairs. The first set consists of the *initial* (original incepting) key-pairs bound to the identifier prefix as the *current* root control authority. The second set consists of the *next* (ensuing) pre-rotated key-pairs. The inception event itself includes the public keys from the *initial* set of key-pairs and a cryptographic digest the public keys from the *next* set of key-pairs. They are all qualified. Recall from fig. 7.3 above the basic format of a key event message. The set of *next* (ensuing) pre-rotated public keys is in *hidden* form as a *qualified digest* of the set of keys. A simplified inception event with a single key per set and a *hidden* next key as a digest is diagrammed as follows:

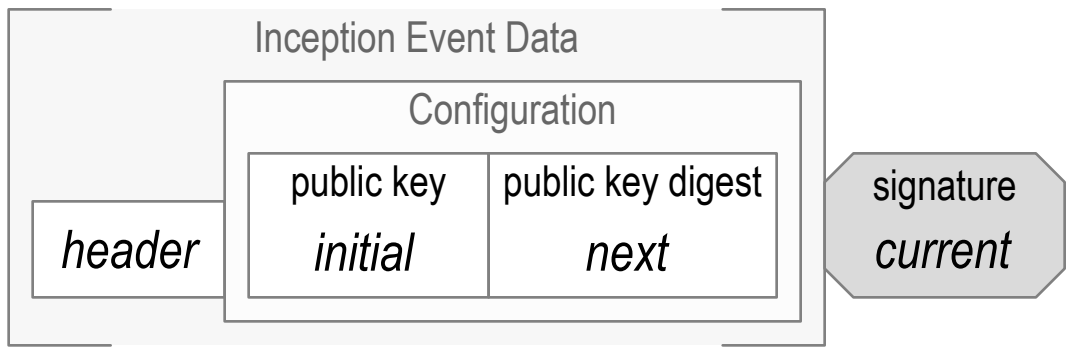

Figure 9.1. Simple Inception Event. The configuration includes the *initial* (original incepting) public key and a digest of the *next* (ensuing) public key (pre-rotation). The *next* public key is therefore hidden. Event is signed with *initial* current private key.

Upon emittance of the inception event the *initial* key-pairs become the *current* authoritative signing key-pairs for the identifier. Emittance *issues* the identifier. Moreover, to be verifiably authoritative, the inception event must be signed by the *initial* private keys. The inception event may be verified against the attached signatures given the included *initial* public keys. The inception event provides the incepting information (including the initial public keys) from which the identifier prefix was originally derived.

Inclusion of the *next* set of public keys in the event effectively performs a *pre-rotation operation* on that set and the set itself may be called a *pre-rotation set* or *pre-rotation* for short. One important function of an *inception event* when initializing a *pre-rotated* self-certifying identifier is to declare, designate and perform a *pre-rotation* on the *next* set of keys. Besides confirming that the controller of the private keys authorized the inception event, verifying the signatures also proves control over the identifier prefix. Furthermore, the inception event also makes a verifiable commitment to its designated set of pre-rotated *next* public keys.

Each rotation operation involves two sets of public keys that each play a role in the event. The two roles are labeled *current*, and *next*. These are all qualified. The current set was created and declared in the most recent prior rotation or inception. A new *next* set is created and declared by the rotation. These key sets may be described as follows:

- The now *current* set of signing keys (formerly *next* before the rotation).
- The new *next* (ensuing) set of signing keys (newly created for the rotation).

To be authoritative, a rotation operation must be signed by the *current* set of authoritative keys. Consequently, only the current set is exposed and the *next* set remains unexposed. By virtue of the rotation, the former set of *next* pre-rotated keys from the more recent prior rotation becomes the newly *current* set of signing keys, and the new *next* set of newly created pre-rotated keys will become the ensuing set of *current* signing keys upon the next rotation. To clarify, as a result of the rotation, the former *next* set of pre-rotated keys now become the authoritative *current* set signing keys. Each rotation operation makes a signed commitment to its set of *next* pre-rotated key-pairs. This commitment is realized by the following rotation that is signed by the newly rotated as *current* controlling sets of key-pairs. Each subsequent rotation operation in turn creates a new *next* set of pre-rotated key-pairs and includes the associated set of public keys (hidden or not) in its rotation event.

In essence each key set follows a rotation lifecycle where it changes its role with each rotation event. A key set first starts as the *next* set, then on the following rotation becomes the *current* set, and then on the following rotation may be discarded. The lifecycle is slightly different for the *initial* key set from an inception event. The *initial* key set first starts as the *current* set and then on the following rotation may be discarded.

Because former set of *next* (ensuing) pre-rotation keys is hidden behind a digest the following rotation must include the associated unhidden public keys. By virtue of the rotation, these have become the newly *current* public keys and are included in the configuration with that label. The rotation event also hides its next (ensuing) rotated public key(s) behind a qualified digest. This is diagrammed as follows:

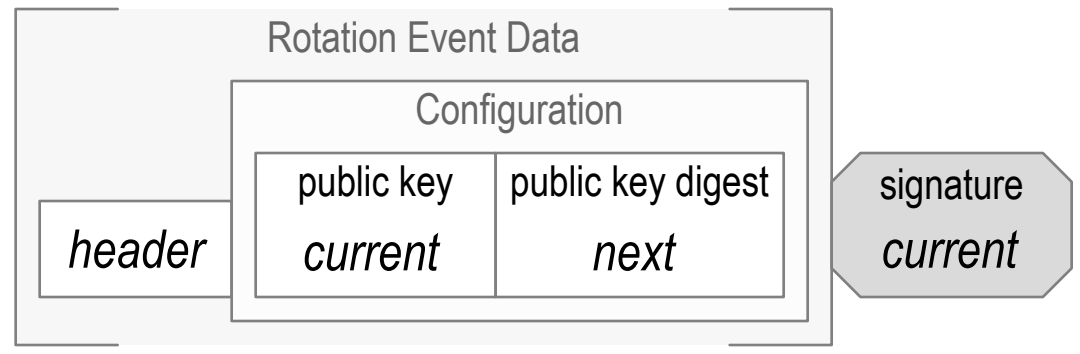

Figure 9.2. Simplified Rotation Event. The configuration includes both the current public key (formerly next) and the digest of the new next (ensuing) public key (pre-rotation). Next public key is therefore hidden. Event is signed with the new current private key (formerly next).

To help clarify an example rotation history is diagrammed below. In the diagram all the keys are taken from to a single indexed key sequence. Each each key labeled with its index and each key element is labeled by role.

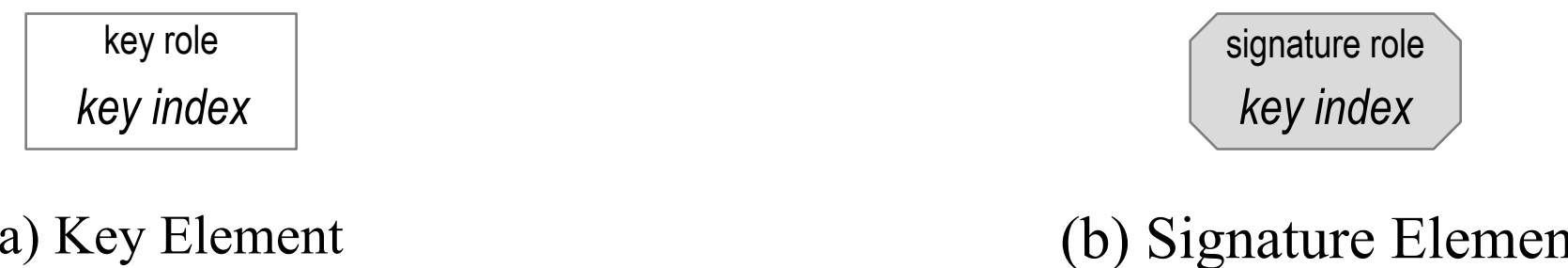

(a) Key Element (b) Signature Element

Figure 9.3. Graphic Labels for Key and Signature Elements. Key element labeled with key role and key index. Signature element labeled with signature role and key index.

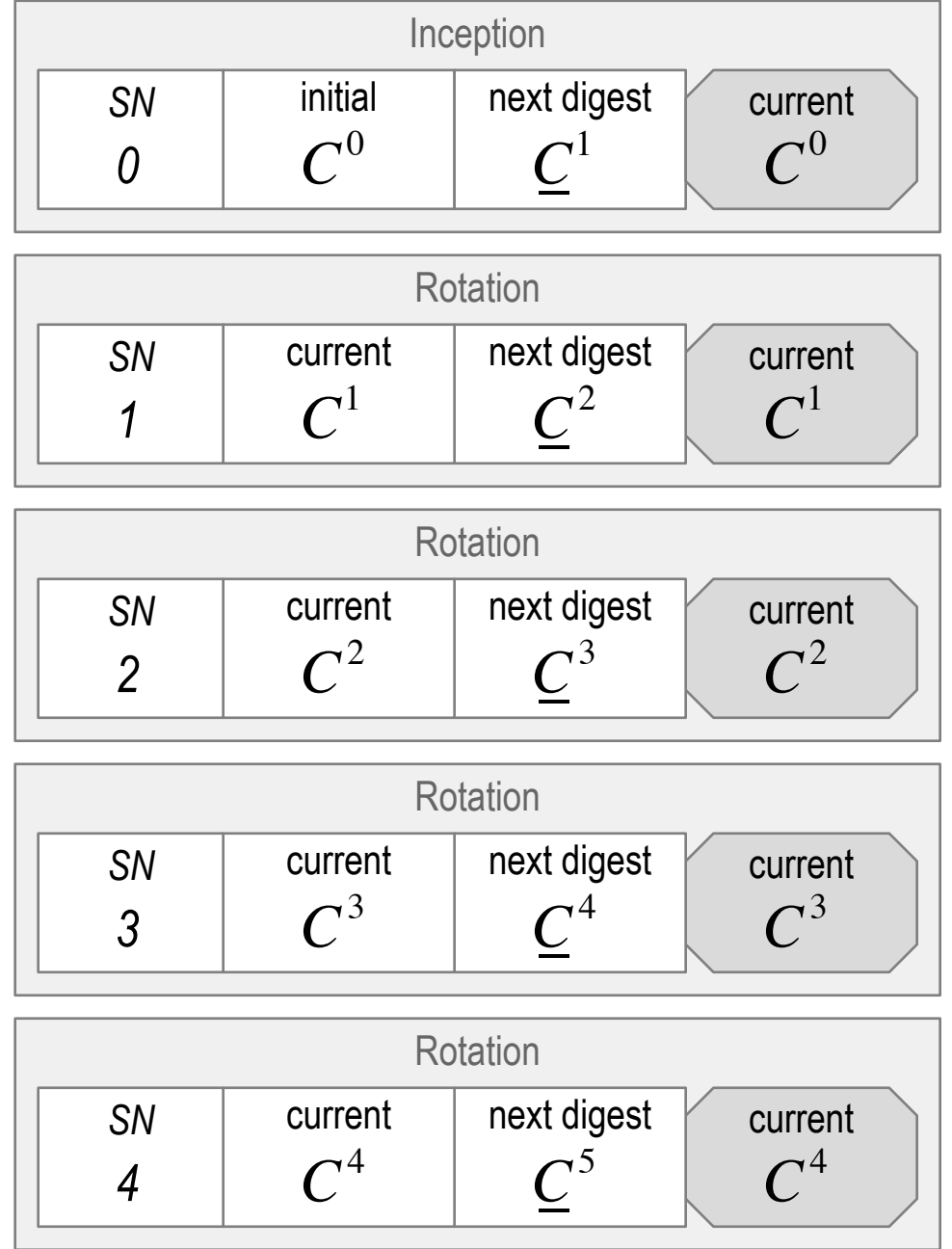

Figure 9.4. Key Event Sequence. Simplified events showing only sequence number (SN) element from header. Inception and rotations showing key-pair sequence indexing where $C^1$ represent key 1 and $\underline{C}^1$ represents hidden key 1.

### 9.3.1 Security

For many exploits, the likelihood of success is a function of exposure to continued monitoring or probing. Narrowly restricting the exposure opportunities for exploit in terms of time, place, and method, especially if the time and place happens only once, makes exploit extremely difficult. The exploiter has to either predict the one time and place of that exposure or has to have continuous universal monitoring of all exposures. By declaring the very first pre-rotation in the inception event, the window for its exploit is as narrow as possible. Likewise, each subsequent rotation event is a one time and place signing exposure of the former next (pre-rotated) rotation key.

Because each pre-rotation makes a cryptographic future commitment to a set of one-time first-time rotation keys, later exploit of the current authoritative signing key(s) may not capture key rotation authority as it has already been transferred via the pre-commitment to new unexposed set of keys. To elaborate, The next (ensuing) pre-rotated key-pairs in an inception event serve as one-time first-time rotation keys in the next rotation operation. Thereafter those key-pairs may be activated as the new current (root) authoritative signing key(s) but no longer have rotation authority. Likewise the next (ensuing) pre-rotated key-pairs in each rotation event serve as one-time first-time rotation keys in the next rotation operation. Thereafter those key-pairs may be activated as the new current (root) authoritative signing key(s) but likewise not longer have rotation authority.

In administrative identity systems the binding between keys, controller, and identifier may be established by administrative fiat. As a result administrative fiat may be used as well as a recovery mechanism for compromised administrative keys. This may permit more exposure through multiple use of each key. In contrast when the binding between keys, controller, and identifier is purely cryptographic (decentralized) such as is the case with this protocol, there is no recovery mechanism once the keys for the root control authority have been fully captured. Therefore security over those keys is more critical. As a result in this protocol administrative (establishment operation) keys are one-time first-time use as administrative keys.

By definition a *dead* attack occurs sometime after creation and propagation of a given rotation event. A successful *dead* attack must first compromise the set of current signing keys for some past rotation event and then create an alternate verifiable version of that past rotation event and then propagate this alternate event to a given validator before the original event has had time to propagate to that validator or any other component the validator may access. The pre-commitment to the *next* set of keys means that no other successful *dead* exploit is possible. A subsequent rotation event that was not signed with the pre-committed next keys from the previous rotation would not be verifiable. But compromising a set of keys, after first-use, given best practices for key storage and key signing, may still be very difficult. Nonetheless, some time later should an attack succeed in compromising a set of keys and thereby creating an alternate but verifiable event, a validator or other component may still be protected as long as the original version of the event has had time to propagate to that validator or other component (such as witness, watcher, juror, judge) that the validator may access. In order to successfully detect duplicity and thereby be protected, any validator need merely compare any later copy of the event with any copy of the original event as propagated to any component it may access. The attacker, therefore must get ahead of the propagation of a past rotation event. Recall that the original version of the event is the one that first exposes the keys to be compromised. This may only allow a very narrow window of time for an attacker to get ahead of propagation. In other words, in order for a dead attack to be successful it must completely avoid detection as duplicitous. To do this it must either prevent the validator from gaining access to any original copy of the key event history or equivalently must first destroy all extant copies of the original key event history accessible to the validator. This may be very difficult. Moreover a controller merely needs to receive confirmation of receipt by a validator of its last rotation event to ensure that that validator is protected from future *dead* exploit.

To summarize, an alternate but verifiable version of a rotation event would be detectably inconsistent with the original version of the event stored in any copy of the original key event history (KEL/KERL). Consequently any validator (or other component or entity) that has access to the original key event history is protected from harm due to a later successful compromise of the keys of any event already in that history.

To further protect the initial key-pairs in an inception event from a dead exploit, a controller may coincidently create both the inception event and an immediately following rotation event and then emitting them together as one. The initial (original incepting) key-pairs may therefore be discarded (including removing all traces from signing infrastructure) after creation but before emission of the coincident events thereby minimizing the exposure of these initial key-pairs.

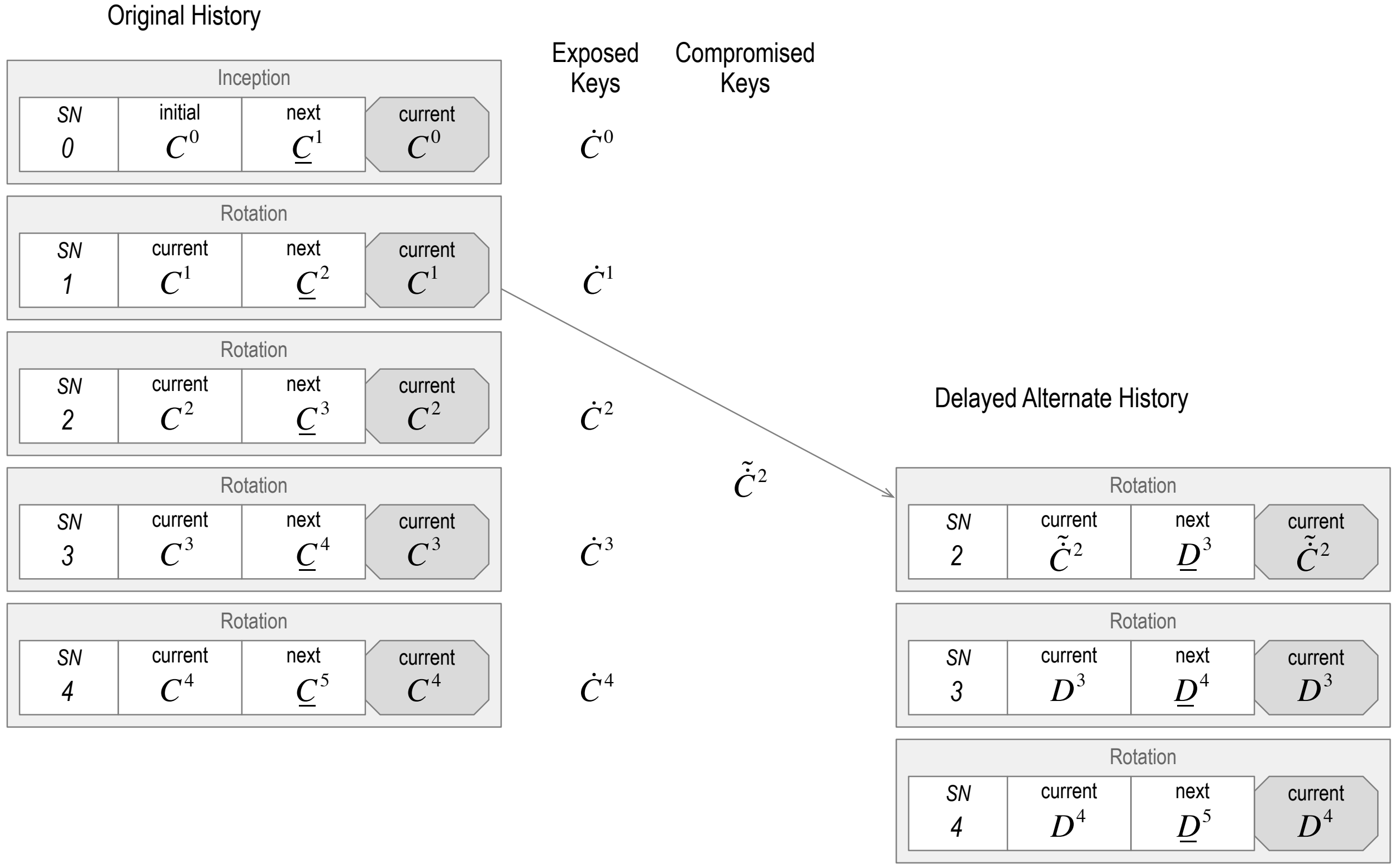

Figure 9.5. Dead Exploit. Simplified key events.

In fig. 9.5 the event key elements are labeled with their key-pair sequence index, such as, $C^2$ for key 2, $\dot{C}^2$ for exposed key 2, and $\tilde{\dot{C}}^2$ for compromised exposed key 2. In a *dead* attack key $C^2$ is compromised some time after and a result of its first exposure in event 2. Time of exposure is shown in diagram by appearance of $\dot{C}^2$ and time of compromise is shown by appearance of $\tilde{\dot{C}}^2$. Given compromised $\tilde{\dot{C}}^2$, attacker may create verifiable but alternate version of event 2 with new next key $D^3$ under attacker's control. But original event 2 has already been emitted, therefore a validator may already have access to a copy of original event 2 and will detect alternate event 2 as duplicitous. This prevents dead attack success. In order to be successful attacker must prevent a validator from first seeing event 2 until after attacker has first seen event 2 and then compromised key $C^2$ and then created alternate version of event 2. By emitting event 2 directly to validator or other component validator may access such as witness, controller may thereby prevent a successful dead exploit at any later time.

In contrast, a *live* attack is defined as an attack that somehow compromises the unexposed next (pre-rotated) set of keys from the latest rotation event before that event has been propagated. This means compromise must occur at or before the time of first use of the keys to sign the event itself. Such a compromise is extremely difficult and the primary reason for employing *pre-rotation* as a *live* exploit protection mechanism.

Moreover, assuming such a compromise, duplicity detection may not protect against a resulting live attack. This is because such a live attack would be able to create a new verifiable rotation event with next keys under the exploiter's control and propagate that event in advance of a new rotation event created by the original controller. Such a successful live exploit may effectively and irreversibly capture control of the identifier. Moreover, in the case of successful live exploit new rotation events from the original controller would appear as duplicitous to any val-idator or other component that already received the exploited rotation event. Protection from *live* exploit comes exclusively from the difficulty of compromising a set of keys at or before the time of first-use.

To elaborate, a successful live exploit must compromise the unexposed next set of private keys from the public-keys declared in the latest rotation. Assuming the private keys remain secret, compromise must come either by brute force inversion or by a side channel attack at the first-use of the private keys to sign the rotation event. By construction, no earlier signing use side-channel attack is possible. This makes successful live exploit from such side channel attacks extremely difficult.

Given the cryptographic strength of the key generation algorithm, a successful brute fore live attack may be practically impossible. Hiding the unexposed next (pre-rotated) public keys behind digests, provides additional protection not merely from pre-quantum brute force attacks but from unexpected post-quantum brute force attacks as well. In this case, a brute force attack would first have to invert the post-quantum resistant one-way hashing function used to create the digest before it may attempt to invert the one-way public key generation algorithm. Moreover as computation capability increases the controller need merely rotate to correspondingly strong cryptographic one-way functions including post-quantum hashing functions. This makes brute force live attack practically impossible indefinitely.

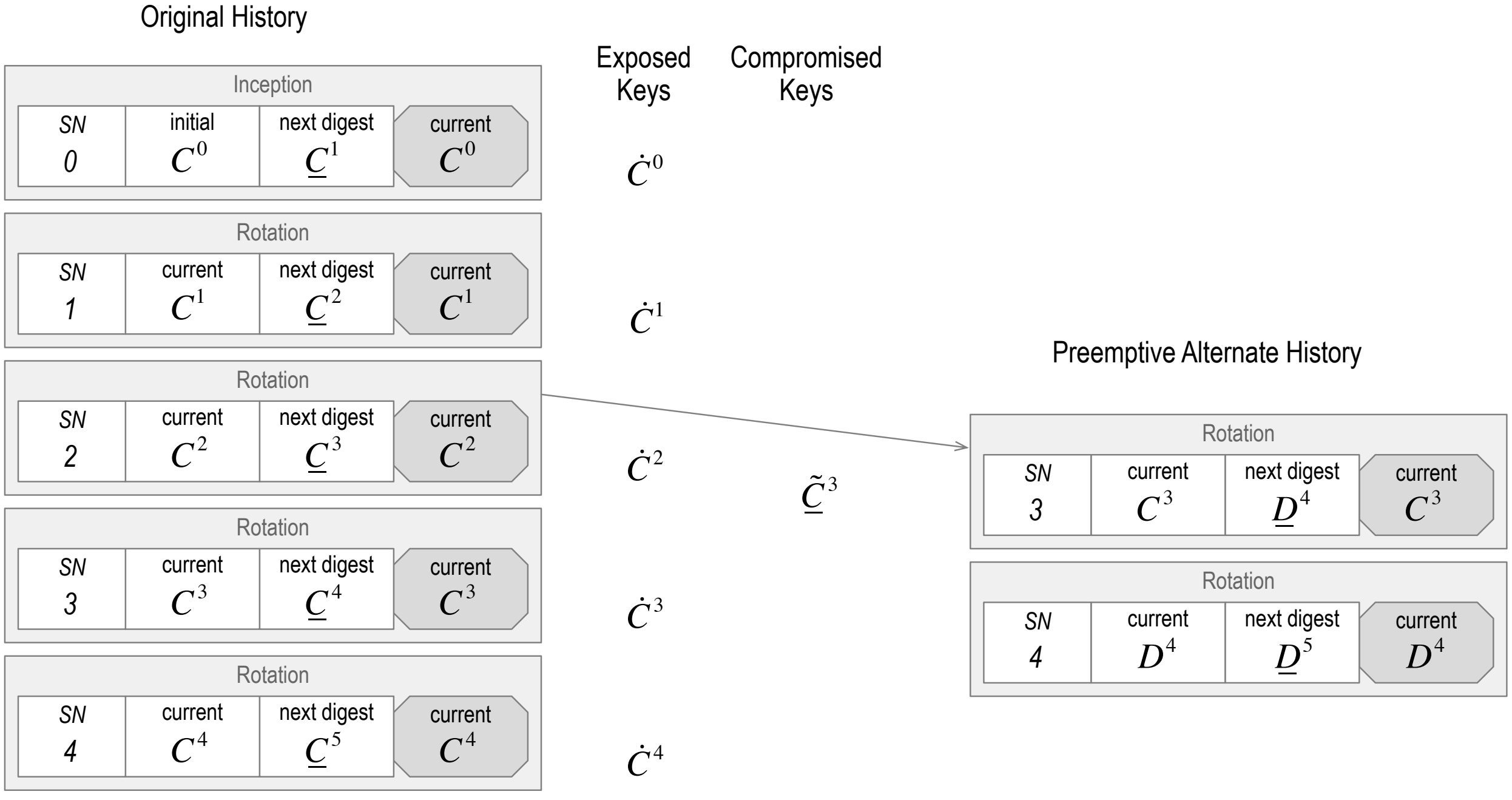

Figure 9.6. Live Exploit. Simplified key events.

In fig. 9.6 the event key elements are labeled with their key-pair sequence index, such as, $C^3$ for key 3, $\dot{C}^3$ for exposed key 2, $\underline{C}^3$ for hidden key 3, and $\underline{\tilde{C}}^3$ for hidden, unexposed, but compromised key 3. Suppose a live attack. Recall that a *live* attack requires that somehow key $\underline{C}^3$ be compromised sometime after it is declared in original event 2 but before it is first exposed in original event 3. Indeed before the existence of event 3. Time of exposure is shown in diagram by appearance of $\dot{C}^3$ and time of compromise is shown by appearance of $\underline{\tilde{C}}^3$. Given compromised $\underline{\tilde{C}}^3$, attacker may create verifiable but alternate version of event 3 with new next hidden key $\underline{D}^4$ under attacker's control. As a result of the pre-exposure compromise of $\underline{\tilde{C}}^3$, the alternate version of event 3 may be created before the original and therefore preempt it.

Nonetheless, given that the private key for $C^3$ remains private prior to its exposure, the only way to compromise $C^3$ is to invert $\underline{C}^3$, a practical impossibility. This prevents live attack success.

In summary, pre-rotation with duplicity detection protects a validator from harm due to both *dead* and *live* attacks. With *dead* exploits, the caveat is that in order for a validating party to able to detect and discard any later forged alternate rotation events it must needs have access to a copy of the original key event history. But with pre-rotation, any verifiable alternate rotation event must undetectably replace a specific preexistent rotation event in every copy a validator may access from any other component. This makes successful *dead* exploit difficult. Replay of any original copy allows the validator to verify the provenance of the original chain of rotations with respect to any alternate copy. In other words, a successful *dead* attack must not only compromise a given set of rotation keys in order to create a specific alternate but verifiable version of the rotation but must also ensure its version is the first version propagated to every component the validator may access. Furthermore a controller merely needs to directly propagate any new rotation event to a validator or other component the validator may access such as a witness to protect that validator from any later *(dead)* attack on that rotation. On the other hand, a successful *live* exploit must compromise a set of pre-rotated keys at or before first use of those keys. This constrains the time and place of first use to the controller's choosing and makes exploit extremely difficult. This difficulty protects against *live* exploits.

### 9.3.2 Other Features

Besides enhanced protection (security), the pre-rotation approach has some other useful features. In its simplest form, pre-rotation does not require any special rotation key infrastructure. This makes the approach self-contained. As discussed earlier, a common approach to key rotation management is to have a special purpose management (administrative) rotation key-pair whose function is to authorize rotation operations on a signing key-pair. This poses the problem that after multiple rotations the management rotation key becomes exposed and may be vulnerable to exploit. Consequently another higher tier of management rotation key may be needed to authorize rotation's of the lower tiers management rotation key and so on. In contrast, in it simplest form pre-rotation uses just one tier of rotation keys. In essence a pre-rotated key set is effectively a one-time first-time use rotation key set on the next (ensuing) rotation. When that ensuing rotation is executed, a new pre-rotation key set is created and the old rotation key set may be repurposed as the current signing key for operations other than rotation. This thereby avoids the infinite regress of successively higher tier rotation keys.

The design aesthetic behind this approach is that of minimally sufficient means. A key state verification engine may employ this single tier or solo architecture as a compositional primitive. Using the primitive, rotation keys are one-time first-tim use but then may be optionally repurposed as signing keys. More complex hierarchical architectures may then be supported via composition of this primitive into multi-tier designs as opposed to building a bespoke primitive for each architecture. The following section will discuss how to compose systems with different features using the core primitive. For the sake of simplicity the examples in the following subsections use single-signature versions of the events but without loss of generality may be extended to thresholded multiple-signature versions as well.

## 9.4 Repurposed Key Mode

Recall that the purpose of the generic interaction event is to enable interactions between a controller and validator that are not expressly establishment (key management) events such as key rotation. These interaction events are signed with the current signing key. Often numerous interaction events will occur between key rotations. Although this exposes the signing key, the pre-rotated ensuing signing key is not exposed and may be used to recover via a rotation from a compromise of the signing key.

To elaborate, a key event history may be characterized as maintaining a single sequence of key-pairs. Each key-pair may serve two different roles in its life-cycle. Each key starts it life-cycle as key authorized for a single (one-time first time) use to sign an establishment (inception or rotation) operation. It then may optionally continue its life-cycle as a repurposed key authorized for multiple use to sign any number of non-establishment (interaction) operations. To better understand how this works consider the sequence of key-pairs as represented by the sequence of associated public keys as follows:

$$C^j \Big|_{j=0,1,\ldots} , \tag{9.1}$$

e.g. $C^0, C^1, C^2, \ldots$. Recall the convention that when a key-pair has been previously exposed in the sense that its private key has already been used sometime in the past to create a signature, then its exposure may be indicated by a dotted key-pair label, e.g, $\dot{C}^j$ . Further recall the convention that a key-pair's role or usage in signing an event may be represented by an uppercase letter subscript e.g $C_I^j$ , $C_R^j$, and $C_X^j$ for inception, rotation and interaction events respectively. Although usually we may treat inception as a special case of rotation. Using this notation a simplified key event history may be diagrammed as follows:

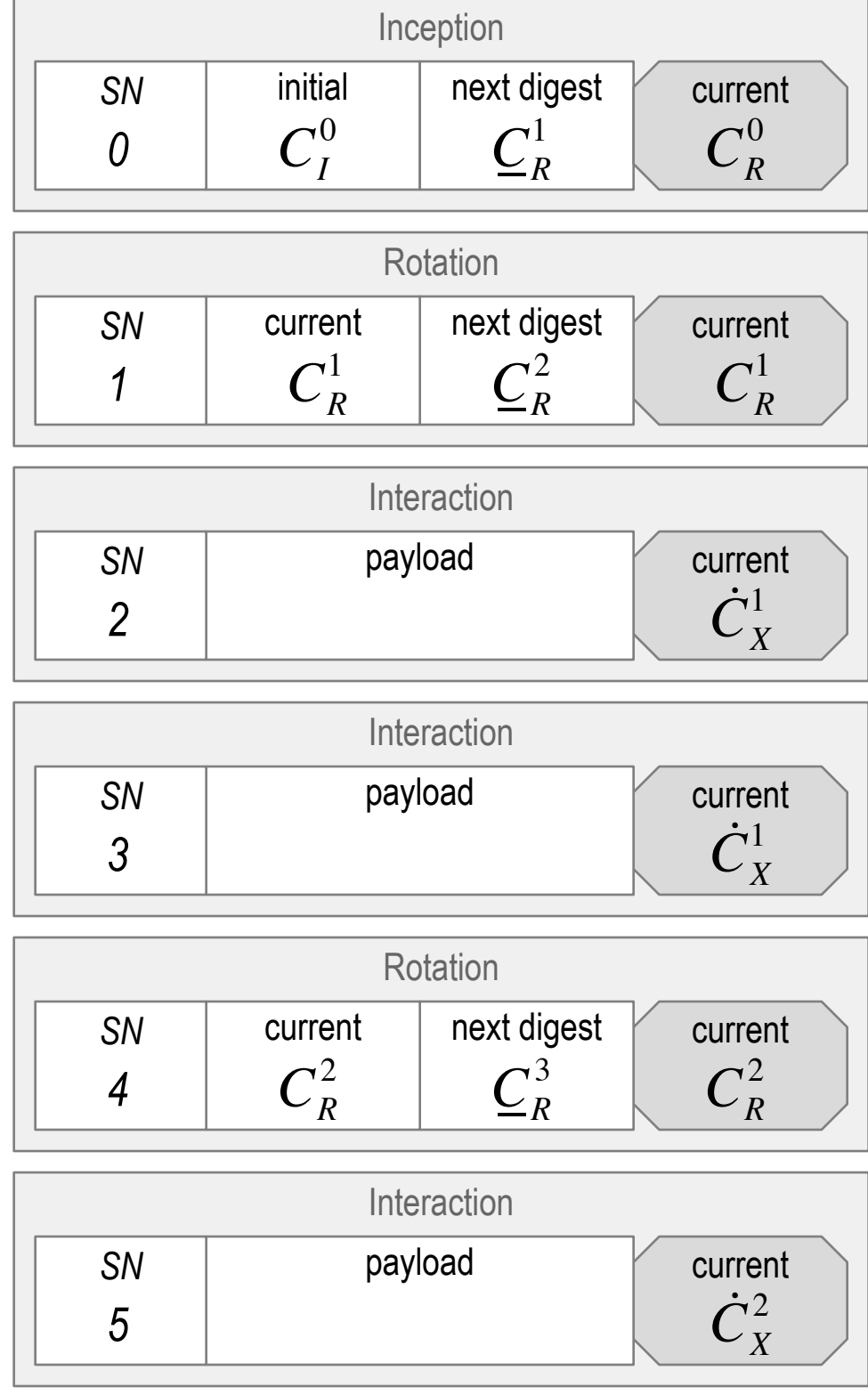

Figure 9.7. Repurposed Key Sequence. Simplified key event history showing rotation signing keys being repurposed for interaction signing.

The inception (event 0) uses $C^0$ to sign as $C_R^0$ after which $\dot{C}^0$ becomes exposed. Key $C^1$ has not yet been exposed. The first rotation (event 1) rotates to $C^1$ and uses it to sign as $C_R^1$, after which $\dot{C}^1$ becomes exposed. Key $\dot{C}^1$ can never be used again for a rotation. Key $C^2$ has not yet been exposed. The next two interactions (events 2 and 3) both use the previously exposed and repurposed $\dot{C}^1$ to sign as $\dot{C}^1_X$. Should $\dot{C}^1$ become compromised the controller may recover control authority by rotating to the unexposed $C^2$. The second rotation (event 4) rotates to $C^2$ and uses it to sign as $C^2_R$ after which $\dot{C}^2$ becomes exposed. Key $C^3$ has not yet been exposed. The following interaction (event 5) uses the previously exposed $\dot{C}^2$ to sign as $\dot{C}^2_X$.

To clarify suppose there are a total of $I$ successive interaction events after a rotation event. Each interaction event is signed with the same signing key. Suppose that following the $I^{th}$ interaction event a new rotation event occurs. After this rotation the key previously used to sign interaction events must be discarded and a new signing key that is the newly repurposed rotation key may be used to sign another set of interaction events.

One major purpose, of this protocol is to support data streaming applications that may benefit from more performant key support infrastructure that inhabits the same computing infrastructure as the streaming application. This approach is supported by the repurposed mode of pre-rotation. Only one key support infrastructure is needed for both the administrative (rotation) and signing keys. Repurposed pre-rotated keys may provide sufficient security for many data streaming applications. The restriction that rotation keys may be only used one-time first-time as rotation keys, provides protection from a compromised signing key via a rotation to the next unexposed rotation key. This approach trades off security for convenience for applications where exposure to exploit of an interaction event is an acceptable trade-off for the convenience of reusing signing keys.

### 9.4.1 Extended Rotations.

In some critical interactions, however, the trade-off between convenience and security made by reusing signing keys for multiple interaction events may not be acceptable. In this case, a more secure approach is to combine an interaction event with a rotation event by adding a payload data structure to the rotation event. The interaction event details are embedded in the rotation event payload. As defined previously this is called an *extended rotation* event. This does not change the semantics of the rotation event verification but merely enables an interaction event to be included in a rotation event. As a result the signing keys are only used once and are rotated after each use. Thus the signing keys are not exposed to repeated use. This better protects each interaction operation with a simultaneous rotation at the cost of using a new set of keys for each interaction operation. A given application may mix extended rotations with non extended rotations followed by interaction events to support different levels of secure interaction workflows with the some key event sequence. In contrast, for stricter security an application may declare in the interaction event configuration that only establishment events are allowed in the key event stream. An simplified key event history the uses extended rotation operations is shown below.

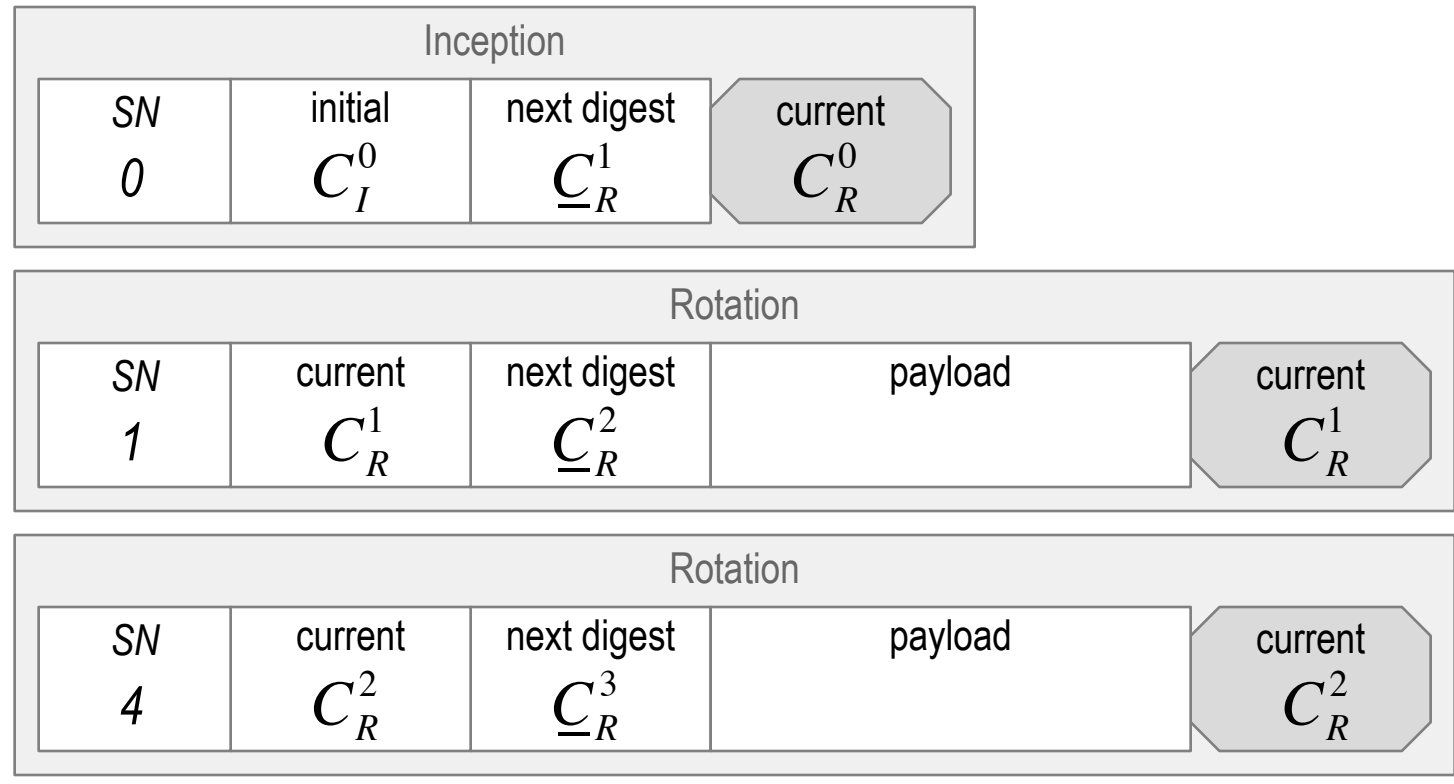

Figure 9.8. Key Sequence with Extended Rotation Events.

In some key management applications, a single key support infrastructure for both administrative (rotation) keys and signing keys. may be preferred. This may be for either performance or cost reasons. These applications may include support infrastructure composed of internet of things (IoT) devices, micro-service servers, or streaming data processing workflows. In these applications depending on the security concerns the *repurposed key mode* with or without *extended rotation operations* may be appropriate.

## 9.5 Delegation Mode

In some key management architectures for administrative identity system, two different key support infrastructure may be used. One for keys to sign administrative operations and another for keys to sign interactions (transactions). Administrative operations such as rotation or delegation may be more security critical but happen much less frequently than interactions. Consequently a more secure albeit less convenient architecture may use special computing devices or components in its infrastructure to store private keys and/or create signatures for administrative actions. These may include one or more of a secure enclave, a hardware security module (HSM), a trusted platform module (TPM) or some type of trusted execution environment (TEE).

### 9.5.1 Key Management Infrastructure Valence

When all storage and signing operations both administrative and not are supported by one computing device or component we call this a *univalent* architecture or infrastructure. When storage and signing operations are split between two computing devices or components, we call this a *bivalent* architecture or infrastructure. In general when storage and signing are split between two or more key computing devices or components, we call this a *multivalent* architecture or infrastructure. Unfortunately, a *bivalent* or *multivalent* key support infrastructure may pose a problem for the repurposed pre-rotation approach described above. With disparate key storage and signing devices for each of rotation events and interaction events, repurposed mode would require that a rotation key be moved from one computing device to the other in order to be repurposed as an interaction key. This may be both inconvenient and insecure. The following diagrams show the different valences for key support infrastructure.

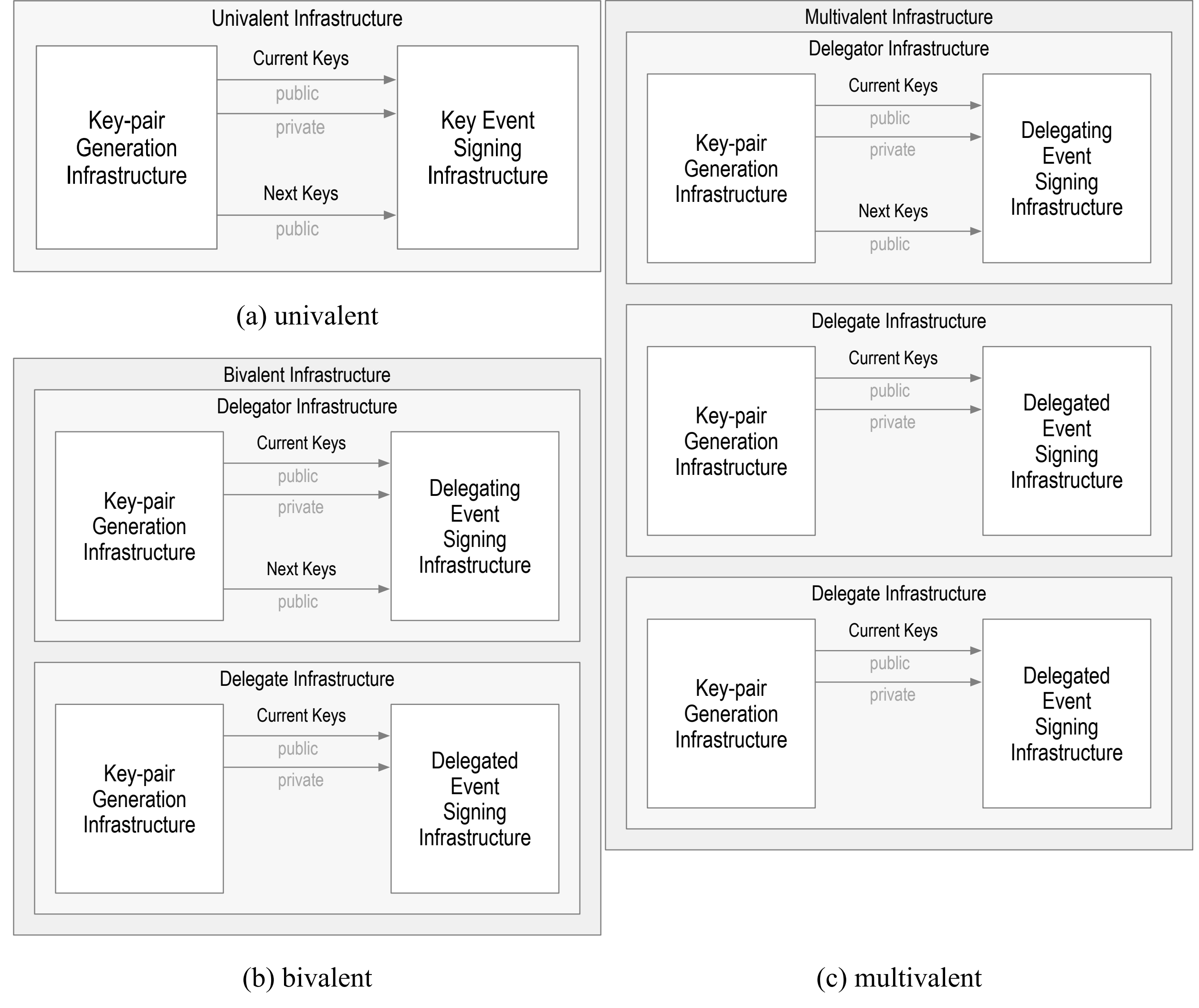

Figure 9.9. Key Support Infrastructure Valence. (a) univalent, (b) bivalent, and (c) multivalent.

### 9.5.2 Delegation and Multivalence

Support for *bivalent* and *multivalent* key support architectures is provided with *delegation mode*. This was introduced earlier but is elaborated here. This is somewhat novel form of delegation that may be described as cooperative delegation (see 7.25). The delegation involves a pair a events in each of the key event logs of both the delegator and delegate. Besides horizontal scalability, a major benefit of a cooperative delegation is that any exploiter that compromises only the delegate's authoritative keys may not capture control authority of the delegate. Any exploit of the delegate only is recoverable by the delegator. A successful exploiter must also compromise the delegator's authoritative keys. A nested set of delegation levels may be used to form a delegation chain or a delegation tree. Each higher level wraps all lower levels with compromised key recovery protection. This maintains the security of the root level for key compromise recovery all the way out to the leaves in spite of the leaves using less secure key management methods.

With delegation mode any number of delegated sequences of interaction signing keys may be managed by one delegating sequence of administrative (establishment) keys. Each key sequence belongs to its own key event sequence. A diagram of key support infrastructures with different valences is shown in the diagram below.More formally, a controller, the *delegate*, receives its control authority over its identifier from another controller, the *delegator*. The *delegate's prefix* is a type of self-addressing self-certifying prefix (see section 2.3.4). This binds the *delegate's* *prefix* to its *delegator's prefix*. Let the identifier of the root controller, the *delegator*, be *C*. Likewise let the identifier of the delegated controller, the *delegate* be *D*. The delegating controller has one key-pair sequence this is represented by the associated public keys as follows:

$$C^j \Big|_{j=0,1,\ldots} , \tag{9.2}$$

e.g. $C^0, C^1, C^2, \ldots$. Each controlled delegate also has a key-pair sequence that is represented by the associated public keys as follows:

$$D^j \Big|_{j=0,1,\ldots} , \tag{9.3}$$

e.g. $D^0, D^1, D^2, \ldots$. Controller *C* uses a special delegating key event to delegate interaction signing authority identifier *D*. Delegation creates a dedicated disjoint key event sequence and disjoint key sequence for D. Because they are disjoint, the keys in $C^j$ may reside in and use completely different key storage and signing components from the keys in $D^j$. A delegator may use successive delegation operations to authorize any number of delegates. Because the two key sequences are disjoint not only are the bound to different identifiers but may be bound to different controllers.

As described previously but clarified here, delegation may be performed either with delegating interaction operations or with delegating (extended) rotation operations. When the architecture is bivalent, that is the delegating key event stream inhabits a dedicated secure computing infrastructure such as a secure enclave or TPM, then using a delegating interaction operation may provide sufficient security. In this case each rotation key set may be repurposed to perform a delegating interaction, but no other interactions. All the other interactions are performed in the delegated key event stream on a different computing platform. The delegating key events are still protected from an exploit of the repurposed delegating interaction event key set because recovery from the exploit may be performed via a subsequent rotation event with the unexposed next keys.

On the other hand using a delegating rotation operation instead may provide enhanced security with either a bivalent or univalent architecture. In this case each delegating operation uses a set of keys only once as part of a dedicated rotation operation for that delegation. This comes at the inconvenience of more rotations.

A set of delegated keys (subsequence from the delegates's key sequence) may be used to sign interaction or other non-establishment events in the delegate's event sequence. In this way, delegation may authorize some other identifier prefix and its associated controlling keys to be authoritative for interaction events. Thus besides merely enabling bivalent key support infrastructure, applications may benefit from other delegative properties such as the ability to attenuate control authority.

As previously described, delegating operations may delegate an establishment event (inception or rotation) in the delegated event stream. Delegating operations may be executed via either a delegating interaction or a delegating rotation operation in the delegating event stream. In the latter case, a delegating rotation operation is a special case of an extended rotation operation. Using an extended rotation operation for each delegating establishment operation may provide better protection to the delegation because the associated keys are only exposed once.

In either case the event payload contains the delegated operation data. The delegated operations are the establishment operations such as inception and rotation for the delegated identifier. Non-establishment operations for the delegated identifier appear only in the delegated event sequence. In the delegating event stream, each delegating event include a commitment, called a delegation seal, to the delegated event. The delegation seal includes the qualified prefix of the delegated identifier and a qualified digest of the delegated event. The delegated *event seal* provides the needed information for a verifier to look up and verify the delegated key event in its event stream (history, log). This is shown below.

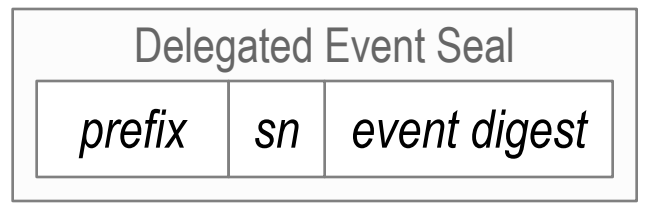

Figure 9.10. Delegated Event Seal.

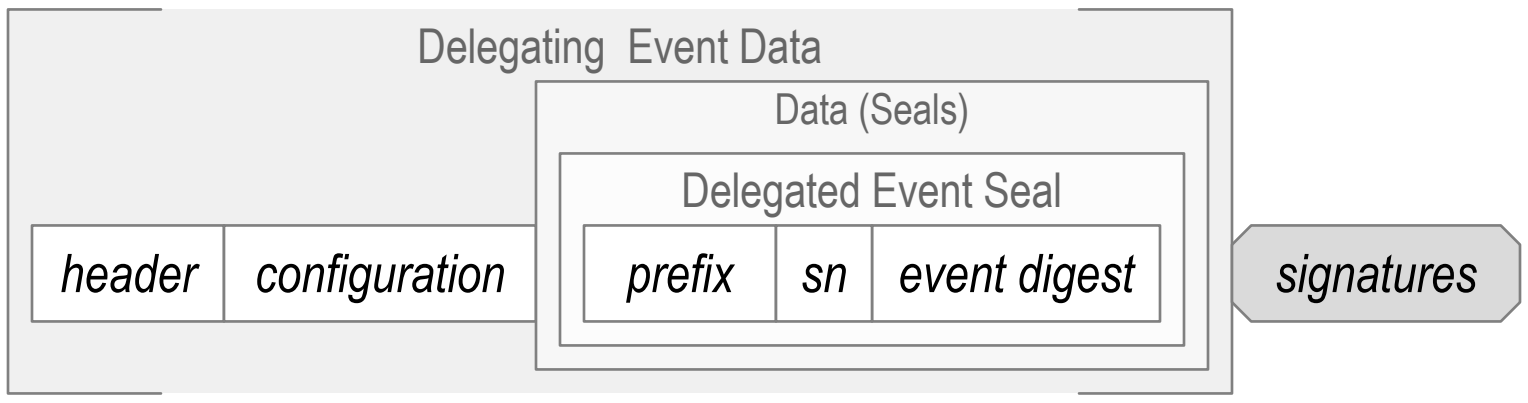

Figure 9.11. Key Event Delegating Inception or Rotation.

In the delegated event stream the delegated establishment events also includes a *delegation seal*, to the unique location of the delegating event. This delegating event *location seal* includes the qualified prefix of the delegating identifier, the sequence number of the delegating event in its event sequence, the *ilk* of the event, and a qualified *digest* of the prior event. Only one event in any event stream may have a given unique combination of these four values. The seal provides the needed information for a verifier to look up the delegating event in its key event stream (history, log). This is shown below.

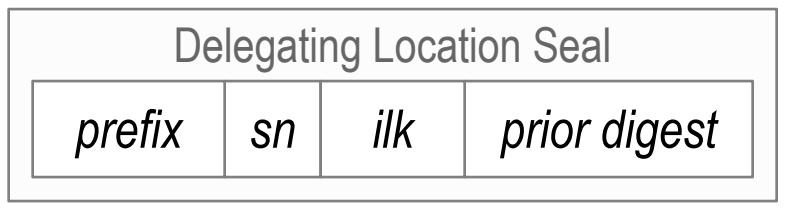

Figure 9.12. Delegating Event Location Seal.

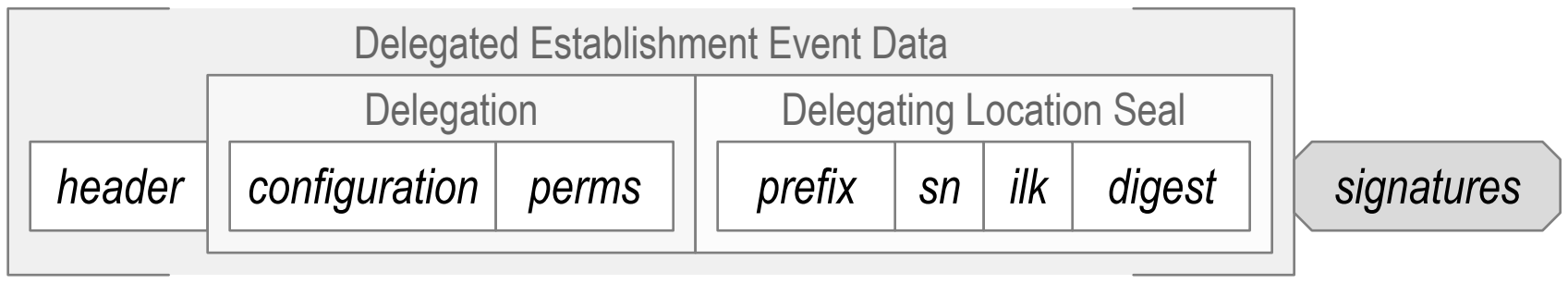

Figure 9.13. Delegated Establishment Event.

A delegated interaction event, however, does not include the seal as it inherits its control authority from the most recent delegated establishment event in its event stream. This is shown below.

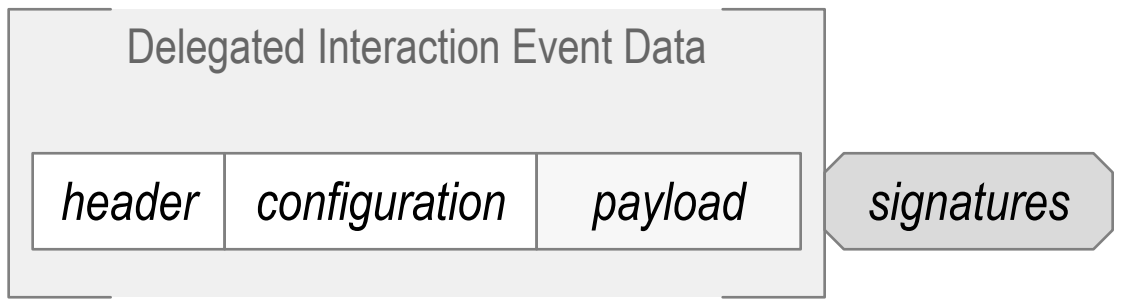

Figure 9.14. Delegated Interaction Event.

From the perspective of support for event streaming applications, delegation mode supports horizontally scalable key management infrastructure. Think of an event sequence as an event stream. In this light, a master controller (identifier) key event stream may delegate signing authority to one or more slave identifier key event streams thereby forming a chained tree of key event streams. A verifier of a delegated key event must also lookup and verify the delegator's control authority from a key event log of the delegator's event stream. Using seals, the delegator's and delegate's event streams are cross linked together. The digest of the delegated's event is included in the delegator's event to cryptographically chain the two streams together. Multiple delegates from the same delegator would form a tree. The following diagram shows a delegating (master) identifier with its associated sequence of controlling events and several delegate (slave) identifier with their sequence of controlled events.

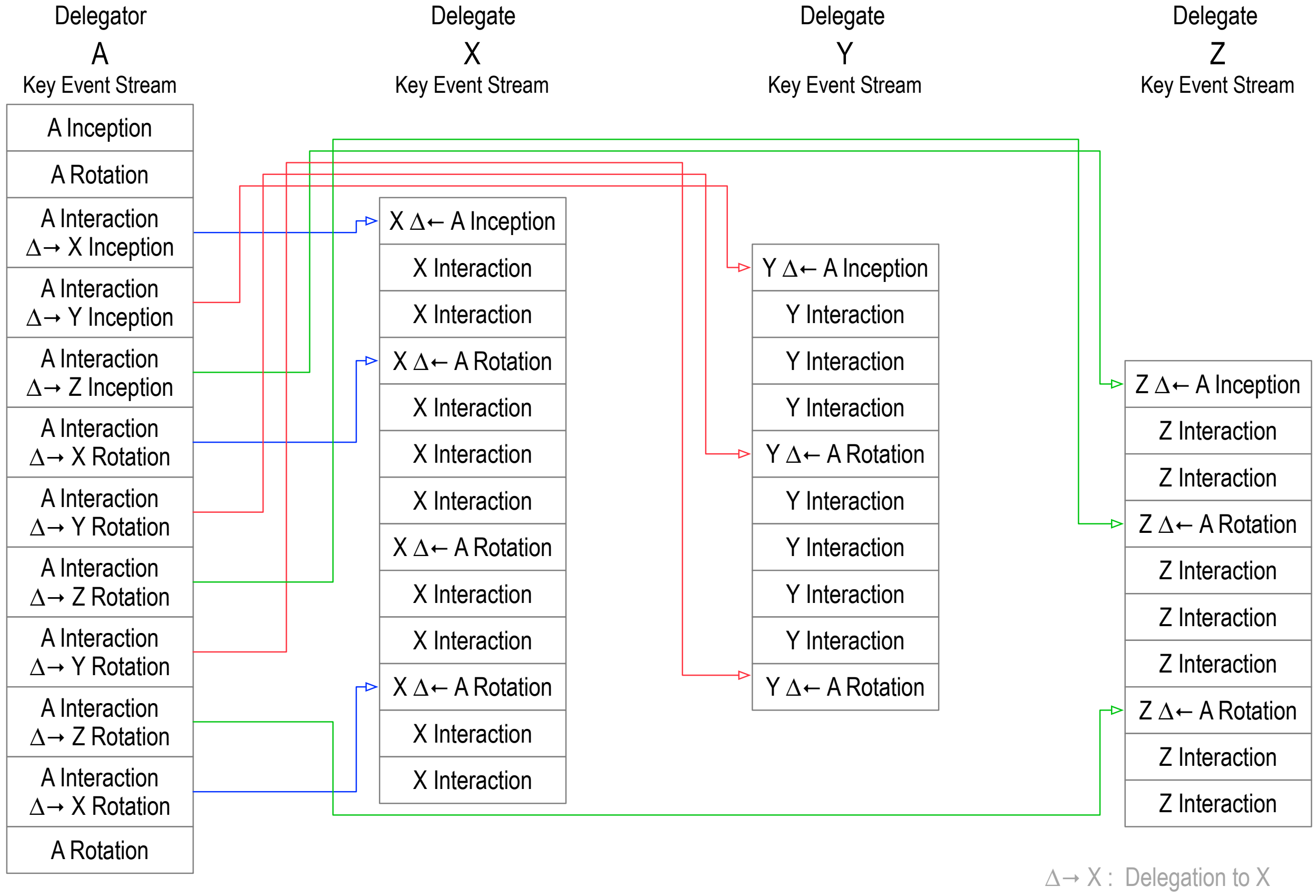

Figure 9.15. Multiple Inception Event Delegated Key Event Streams

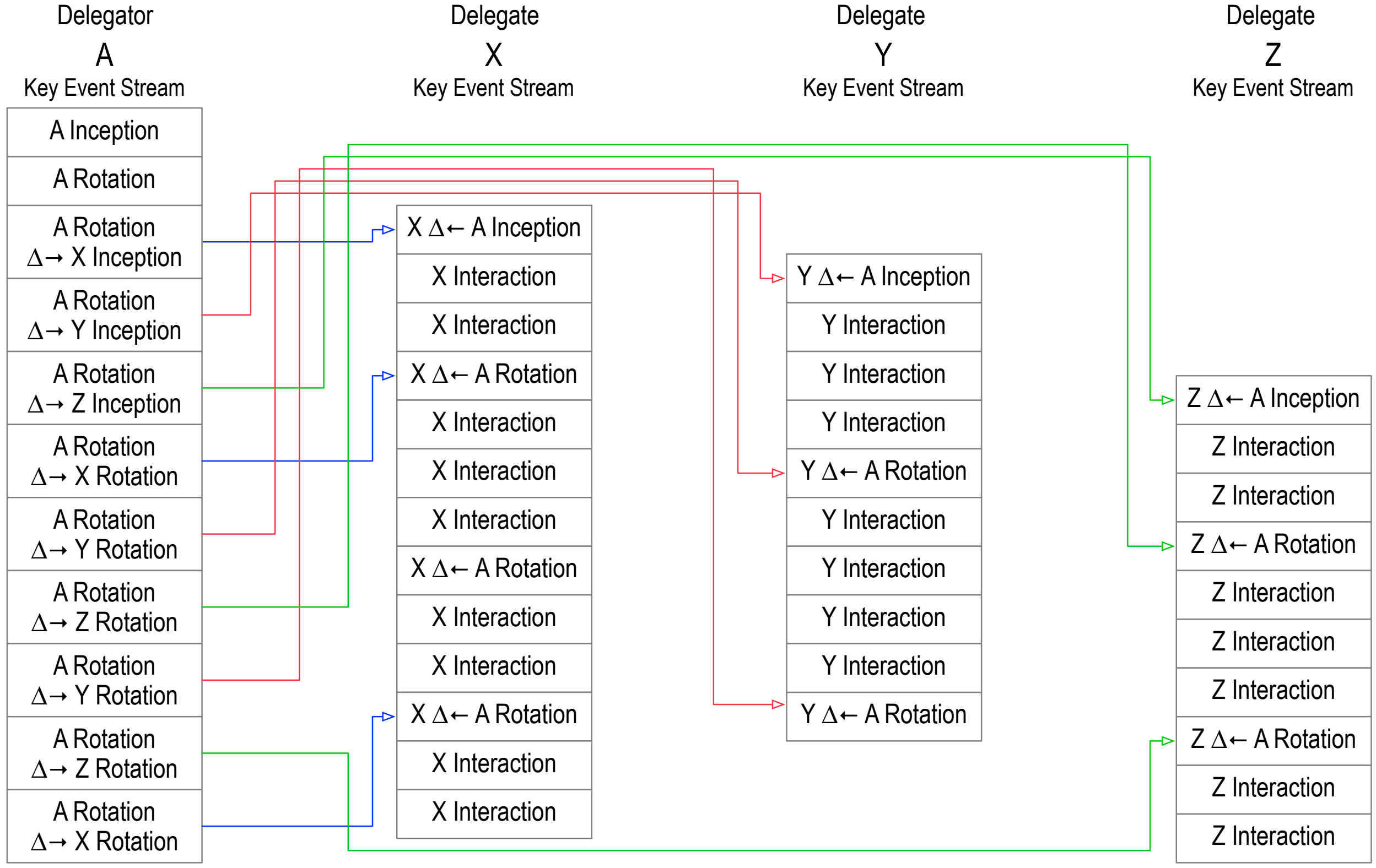

Figure 9.16. Multiple Rotation Event Delegated Key Event Streams

Delegation in this way, semantically creates a key hierarchy. This provides a powerful new building block that allows varied compositions. The only needed additional infrastructure is a key event state verification engine that supports delegated inception and rotation events. One advantage of this approach is that each delegated key event stream is a dedicated chain of events with a dedicated sequence numbering set and dedicated digest chain instead of an interleaved sequence number and digest chain. This makes it easier to manage each stream independently of the others. This makes it more scalable as well and easier to adapt to different key support infrastructures for each delegated signing key event stream.

Delegation may be applied recursively to enable multiple levels of delegation thereby creating a chained tree of key event streams. With this approach any number of layers or levels of management keys may be supported via composition of delegation events. Moreover, in addition to inception and rotation, delegation may be used to manage other types of authorizations.

### 9.5.3 Custodial Key Management

KERI's prerotation mechanism together with delegation enables multivalent custodial key management. In this approach the pre-rotated keys of the delegate are held or controlled by the delegator. The delegate only has control over the current set of signing keys. The delegator may unilaterally revoke that signing capability at any time by issuing a rotation with the delegated pre-rotated keys. To re-assign signing capability the delegator and new delegate perform two successive cooperative delegated rotations. The first to rotate to pre-rotated keys controlled by the new delegate and the second to rotate to pre-rotated keys controlled by the delegator. Because both rotation are not verifiable until the delegator makes a sealed commitment in its event log, the delegator can wait until both rotations have been formulated before committing to the first one. The important point is that in this approach the delegator and delegate never share private keys. Each maintains their own sets of private keys, they merely make successive commitments to each other's associated public keys. This maintains separation between their key management infrastructures.

### 9.5.4 Summary

In summary, the KERI design approach is to build composable primitives instead of custom functionality that is so typical of other DKMI approaches. Consequently when applied recursively, delegation may be used to compose arbitrarily complex trees of hierarchical (delegative) key management event streams. This is a most powerful capability that may provide an essential building block for a generic universal decentralized key management infrastructure (DKMI) that is also compatible with the demands of generic event streaming applications.

## 10 Protocol Operational Modes

This section describes the operational mode of the protocol (conveyance of key events) from controller to validator. In many cases we may make a reasonable assumption of an honest controller. This means the the validator may choose to trust the controller. It it usually in the best interests of the controller to act in a trustworthy manner with respect to its own identifiers i.e. identifiers it control. Especially those with persistently high value. Moreover because each key event must be non-repudiably signed by the controller any inconsistent alternate version of a key event may be eventually detectable by any validator once given access to both versions. This makes inconsistent behavior, that is, *detectable duplicity*, on the part of the controller eminently risky given even a small likelihood of successful detection. Detectable duplicity simultaneously destroys both the trustworthiness of the controller with respect to that identifier and any value it may have built up in the identifier. Because the main purpose of key rotation is to allow a controller to maintain persistent control over an identifier, detectably inconsistent behavior removes any benefit of persistent control of a given identifier. Nevertheless the protocol must still protect against the edge cases where duplicity provides some temporary benefit to either a dishonest controller or equivalently a fully compromised controller.

In this protocol, all the primary activities that protect a validator when engaging with some other controller's identifier, be it verification, control authority establishment, or duplicity detection, are based on an ability to replay the sequence of key events (key event history or log) of that identifier. There are two main operational modes for providing this replay capability that are distinguished by the degree of availability of the identifier's controller when creating and promulgating the key events. The first is direct replay mode and the other is indirect mode.

With direct mode, the promulgation of events to a validator does not happen unless the controller is attached to the network and able to communicate directly with a validator. Direct mode assumes that the controller may have intermittent network availability. This does not preclude the use of network buffers, caches, and other such mechanisms that mitigate temporary communications connectivity issues, but it does assume that these mechanism may not be trusted in any persistent sense to promulgate key events. Nonetheless, direct mode is important as it is compatible with the use of mobile internet devices such as cell-phones. The assumption of intermittent availability means that in order for a validator to access the key event history of an identifier (not its own) that validator must directly receive those events from the identifier's controller. Direct mode is compatible with identifiers for one-to-one exchanges or pair-wise relationships (one identifier per relationship). A single direct mode identifier may be re-used in multiple one-to-one relationships as part of a select group. The assumption of direct communication with in-termittent availability simplifies the set of trusted support infrastructure needed to secure the identifier. The details of direct mode are provided in the next section.

With indirect mode, the promulgation of events to a validator may happen even when the controller is not attached to the network and therefore not able to communicate directly with a validator. Indirect mode supports high (nearly continuous) availability of the key event history to any validator. This means that other components must be trusted to promulgate key events when the controller is not attached to the network. Indirect mode is compatible with identifiers for one-to-many exchanges or any-wise relationships (a controller with any others). A single indirect mode identifier may be used for a public service or business or otherwise when building brand and reputation in that identifier is important. An indirect mode identifier may also be used for private one-to-one or select groups but where intermittent availability is not tolerable. The assumption of high availability complicates the set of trusted support infrastructure needed to secure the identifier. The details of indirect mode are provided in a following section.

## 10.1 Direct Replay Mode

With direct replay mode, the first party is a *controller* of a given identifier and wishes to interact with a second party, the *validator*, using that given identifier. The second party must first validate that the given identifier is under the control of the first party i.e. establish control authority. Because the identifier (self-certifying) is bound to one or more key-pairs via the corresponding public keys, the controller may establish control by sending a verifiable *inception* operation event message signed with the corresponding private keys. From the perspective of the validator, this *inception* event establishes the controller's initial authority over the identifier at issuance. This control authority is subsequently maintained via successive verifiable rotation operation events signed with unexposed pre-rotated keys (recall the description of pre-rotation above).

The controller also establishes the order of events. Because only the controller may create verifiable key events, the controller alone is sufficient to establish an authoritative order of events, i.e. it is the sole source of truth for events. No other source of truth with respect to ordering is necessary. The controller primarily establishes an event sequence by including in each event, except the inception event, a backward cryptographic commitment (digest) to the contents of the previous event. Secondarily each event also includes a monotonically increasing sequence number (non-wrapping whole number counter). The counter simplifies application programmer interfaces (APIs) for managing and reasoning about events. The inclusion of a commitment (digest) to the content of the previous event effectively backward chains the events together in an immutable sequence.

The validator needs access to a record or log of the key event messages in order to initially verify the provenance of the current controlling key-pair for the identifier. This is a *key event log* (KEL). Therefore, as long as the validator maintains a copy of the original event sequence as it first receives it, the validator will be able to detect any later exploit that attempts to change any of the events in that sequence. Moreover, with a backward chained key event sequence, possession of only the latest digest from all the events allows the detection of tampering of any earlier event in any other copy of the key event sequence that is later presented to the validator i.e. duplicity detection. This means a validator does not need to maintain strict custody over its full copy of the KEL but merely the final event in order to detect tampering in any other full copy. To clarify, the validator still needs access initially to a copy of all the events in order to verify and establish control authority but should the validator lose custody of the full event log after validation and only maintain thereafter custody of the last event it has seen it may re-establish the validity of some other copy by backwards validation via the digest from its preserved event.

In this case the sequence number allows a validator to more easily establish if a different event log has been tampered with prior to its last event by examining the event with the same sequence number in the alternate sequence to see if its digest is the same and validates against its previous event and so forth.

To elaborate, as long as the validator maintains a copy the KEL, an exploiter may not establish control of the identifier due to compromise from an exposure exploit of some earlier event in the KEL. For example, a later compromise of the original key-pair could be used to forge a different inception event. As long as the validator has a copy of the original inception event it could detect the forged inception event and ignore it. Likewise later compromise of any of the exposed keys could be used to forge different rotation events. As long as the validator at one time had access to a copy of the original chained key rotation event sequence starting with the original inception event it could detect the exploited rotation events and ignore them. Absent any other infrastructure, in order that the validator obtain a complete event log, the controller must ensure that the validator has received each and every rotation event in sequence. This requires an acknowledged transfer of each new rotation event. In order to ensure that this occurs the controller and validator must both be communicating directly to each other, thus online, at the time of the transfer. If either party becomes unavailable the interaction pauses and the other party must wait until both parties are online to resume the interaction. Consequently this case is only useful for interactions where pausing and resuming (i.e. intermittent availability) is acceptable behavior.

Upon reception of an event the validator sends a receipt message as an acknowledgment. The receipt message includes a signature by the validator of the associated key event message, in other words, an event receipt. The controller now has a signed receipt that attests that the validator received and verified the key event message. The validator is thereby in this narrow sense also acts like a *witness* of the event. The controller can keep the receipt in a key event receipt log (KERL). This provides a trust basis for transactions with the validator. The validator may not later repudiate its signed receipts nor refer to a different key event history in interactions with the controller without detection by the controller. By virtue of choosing to directly communicate to a validator, the controller is implicitly designating that validator as a *witness* of the key event stream.

Each party could establish its own identifier for use with the other in this pair-wise interaction. Each party would thereby in turn be the controller for its own identifier and the validator and witness for the other's identifier. Each could maintain a log of the key events for the other's identifier and key event receipts from the other for its own identifier thereby allowing each to be invulnerable to subsequent exploit of the associated keys with respect to the pair-wise interaction. A log that includes both the signed key events (signed by the controller) and signed key event receipts (signed by the validator) is a log of doubly signed receipts of key events. Each is implicitly designating the other as a witness of its own key events. Any transactions conducted with the associated keys within the time-frame maintained by the logged key event histories may thereby be verified with respect to the keys without the need for other infrastructure such as a distributed consensus ledger. A discussion of how to verify associated transaction events is provided later. Furthermore the validator could keep a log of duplicitous events sent to it directly from the controller e.g. a duplicitous event log (DEL). With this the validator could prove that the controller has been exploited.

Of particular concern with this approach is the original exchange of the inception event. In a direct interaction, however, the controller may create a unique identifier for use with the associated validator and thereby a unique inception event for that identifier. This inception event is therefore a one time, place, and use event. Consequently, as long as the validator retains a copy of the original inception event, (or a copy of any later event), the inception event itself is not subject to later exploit due to exposure and compromise from subsequent usage of the originating key-pair. Another way of looking at this approach is that each pair-wise relationship gets a unique set of identifiers and associated key-pairs for each party.

The exchange of the inception event message must also be made invulnerable to man-in-the-middle attacks (for example by using multi-factor authentication) otherwise an imposter (man-in-the-middle) could create a different identifier under its control and confuse the validator about the correct identifier to use in interactions with the genuine controller. A diagram of the direct replay mode infrastructure is shown below.

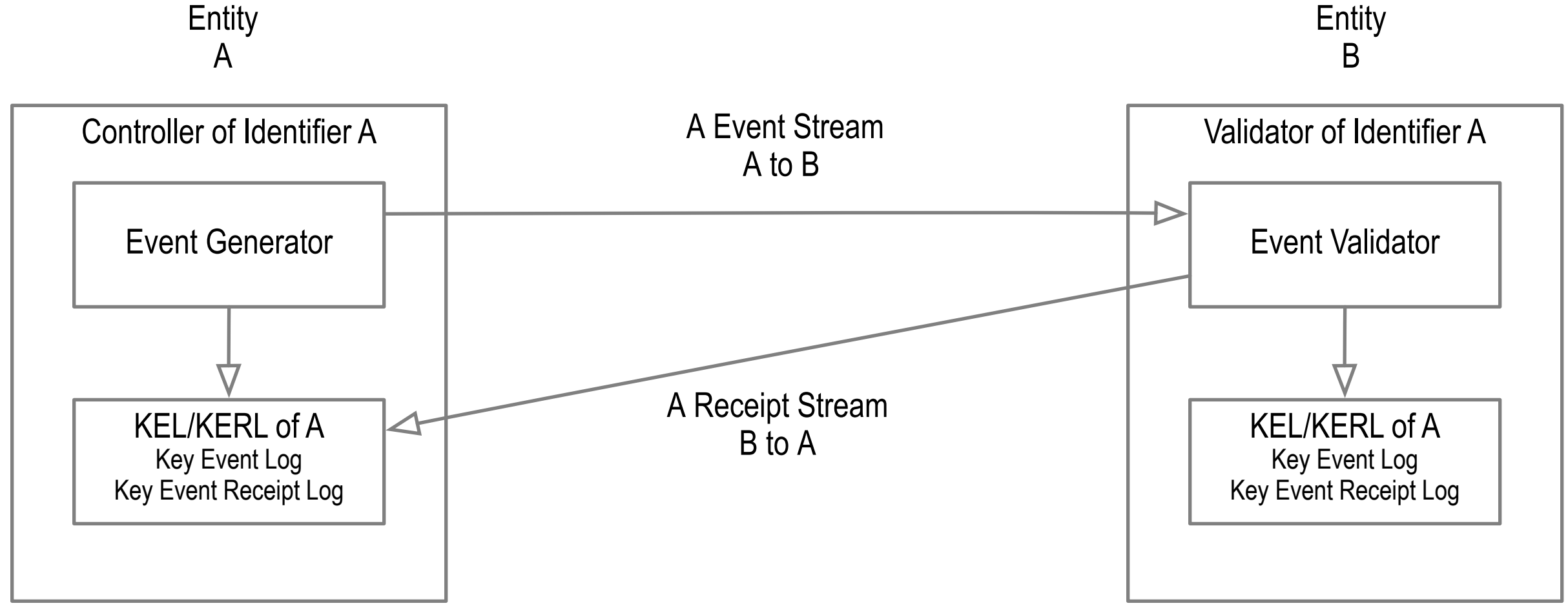

Figure 10.1. Direct Replay Mode: A to B

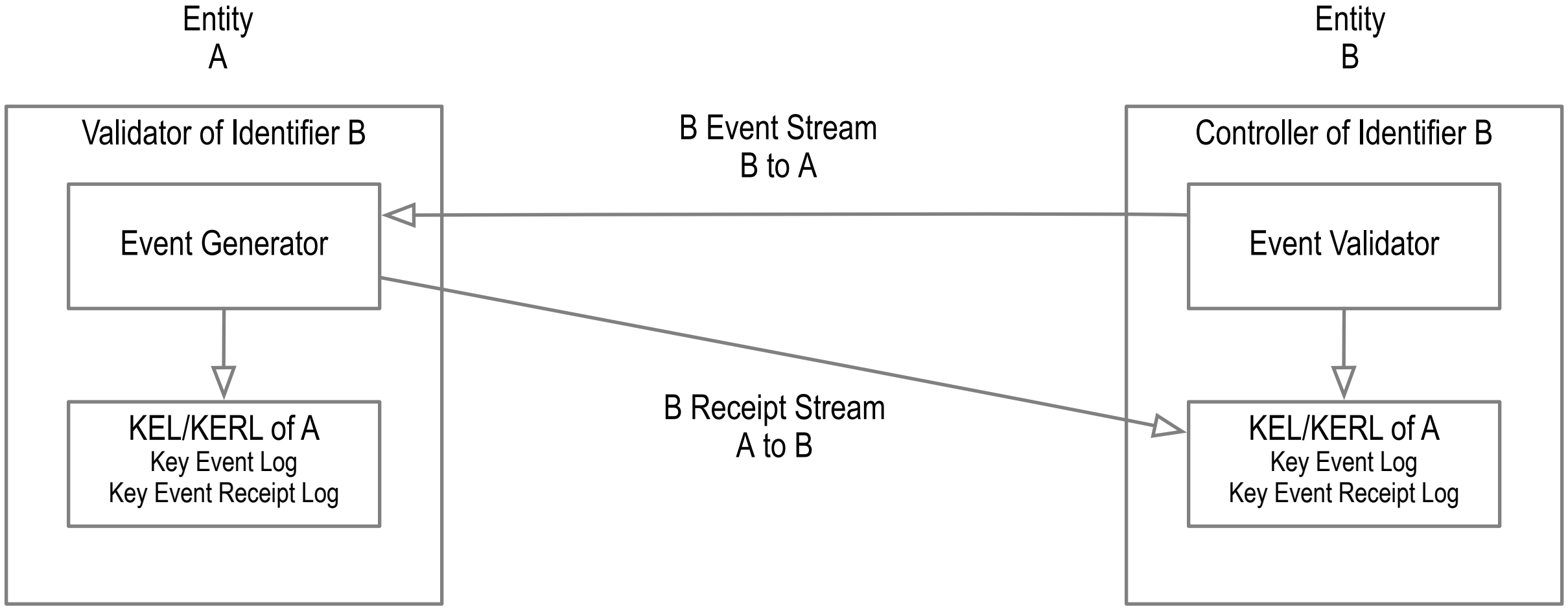

Figure 10.2. Direct Replay Mode: B to A

## 10.2 Indirect Replay Mode

With indirect replay mode a validator may not be available either at the time of creation to receive and acknowledge an inception event or at some later time to receive and acknowledge any other key event. Consequently a later exploit of the associated key-pairs might allow an exploiter to establish an alternate key event history and provide that instead to a validator thereby preempting the original unexploited key event history. The exploiter could thereby more easily capture control of the identifier from the perspective of such a validator. The purpose of indirect mode is to provide a highly available trustworthy service for an identifier that thereby ensures that any validator may have access to a copy of the original version of the key event history. With indirect mode, the controller may designate a set of witnesses (replicas) that essentially store and forward  key events to any requestor. This function of the witnesses may be called a key event promulgation service. It may provide fault tolerant, security, and availability of key events from the standpoint of the controller. A simplified diagram of such a service is shown as follows:

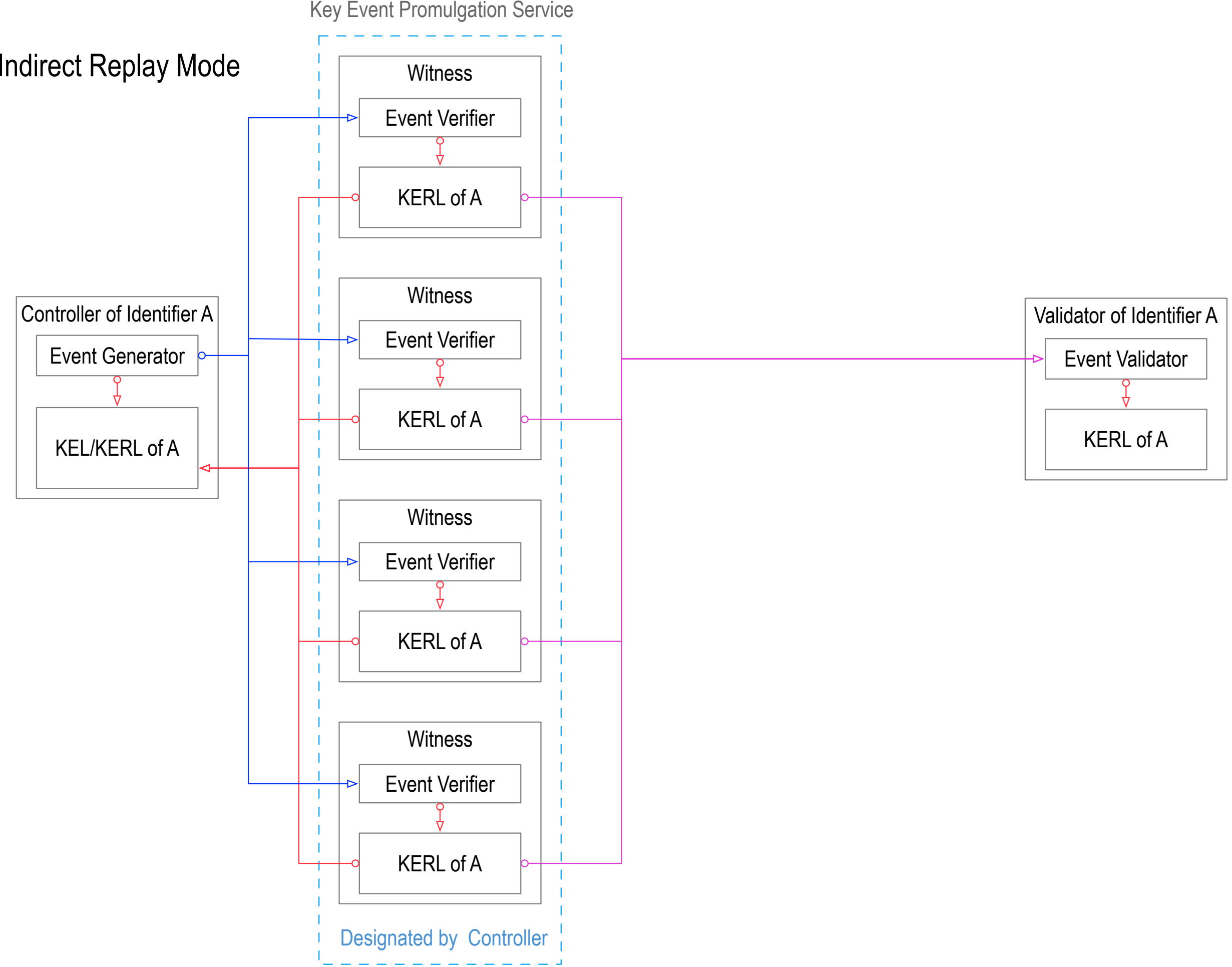

Figure 10.3. Simplified Key Event Promulgation Service via a set of Witnesses.

Although the set of witnesses, as a key event promulgation service, may provide guarantees of fault tolerant availability and security from the controller's perspective, because they are designated by the controller they may not be trusted by a validator. The fact that events are end verifiable, however, means that the validator is free to designate its own set of watchers that may provide high availability and security from the validator's perspective. This watcher service may be called a key event confirmation service. A simplified diagram with both services is shown as follows:

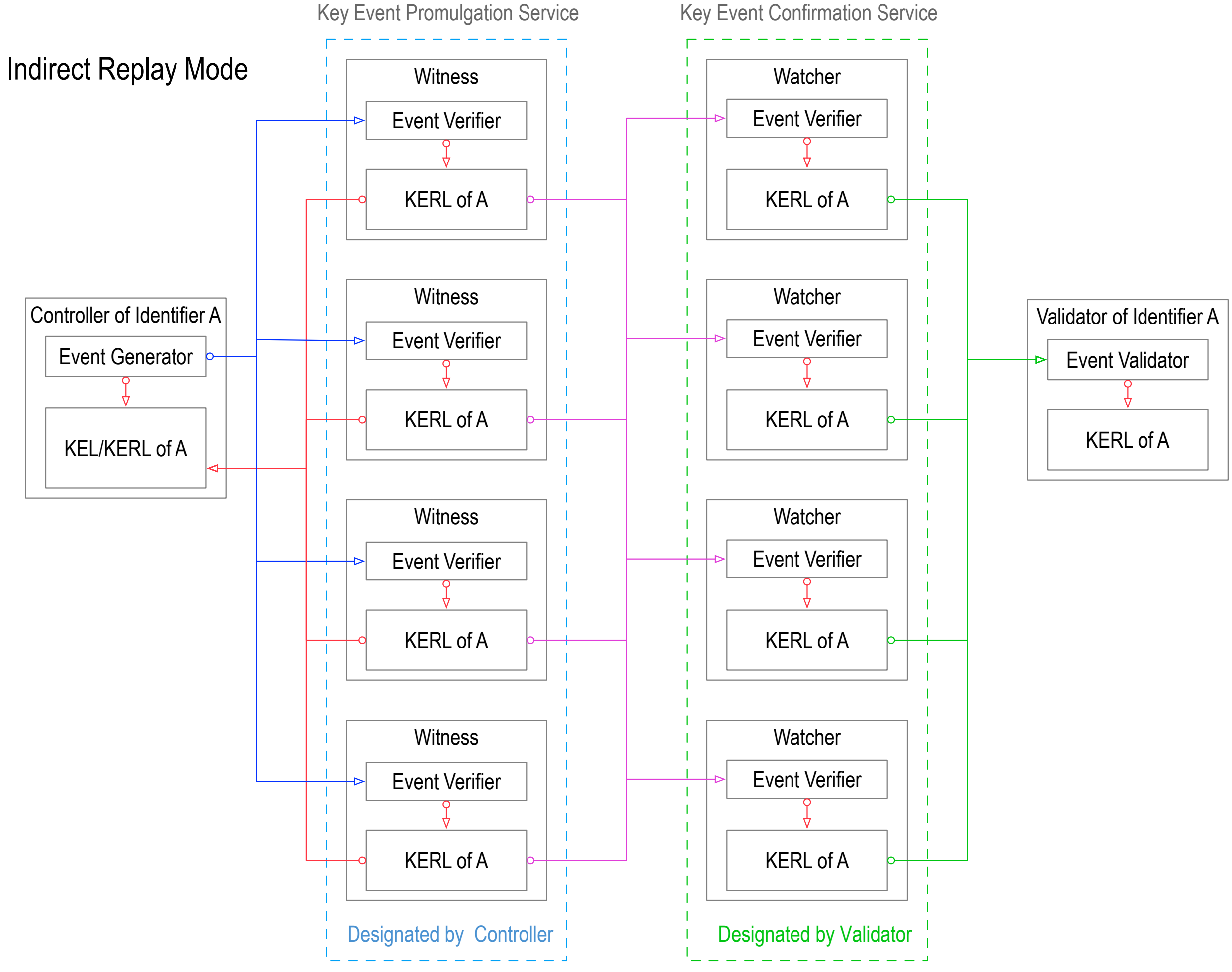

Figure 10.4. Simplified Key Event Promulgation and Confirmation Services.

While a decentralized total ordering distributed consensus ledger may provide such highly available trustworthy service (key event history) by combining the promulgation and conformation into one set of nodes, it may not be the minimally sufficient means for doing so. As discussed previously, total ordering distributed consensus algorithms either suffer from relatively high latency and low throughput or may not scale well with an increasing number of nodes. They may be relatively complex and/or expensive to setup and operate.

This work describes an alternative approach that uses redundant immutable key event receipt logs (KERLS) to provide such a highly available trustworthy service using minimally sufficient means. The separation of promulgation and confirmation into two separate loci-of-control, one the controller's, and the other the validator's simplifies the interaction space between these two parties. This architecturally decentralizes the system and provides for better scalability and performance. As a result the service function may have relatively higher throughput, lower latency, better scalability, lower cost and lower complexity than a totally ordered distributed consensus ledger. Notwithstanding the fact that this work removes the need for a totally ordered distributed consensus ledger, it may still use one when other factors or constraints make it desirable. Indeed, although KERI may utilize simple witnesses, its portable designation support means that when appropriate or desirable more sophisticated witnesses may be used such as witnesses that are oracles to a distributed consensus ledger (i.e. blockchain). In this case the pool of nodes supporting the ledger may appear as one witness from the perspective of KERI. But that one witness may exhibit sufficiently high security and availability for the purposes of both the controller and val-idator. The controller may choose at any time to move to a different ledger or set of witnesses as desired. This means that the identifier is not ledger locked. An example with a ledger oracle as a witness and a ledger oracle as a watcher is shown in the diagram below.

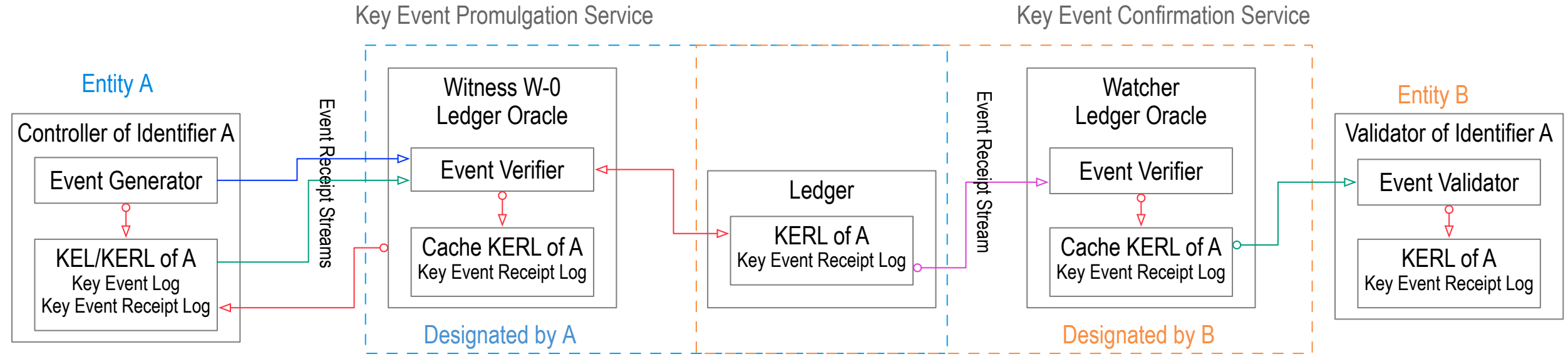

Figure 10.5. Indirect Mode with Ledger Oracles.

To elaborate, the design principle of separating the loci-of-control between controllers and validators removes one of the major drawbacks of total ordered distributed consensus algorithms, that is, shared governance over the pool of nodes that provide the consensus algorithm. Removing the constraint of forced shared governance allows each party, controller and validator, to select the level of security, availability, performance specific to their needs. The validator's confirmation service may be further enhanced to provide duplicity detection and other protections as needed. This is shown in the following diagram.

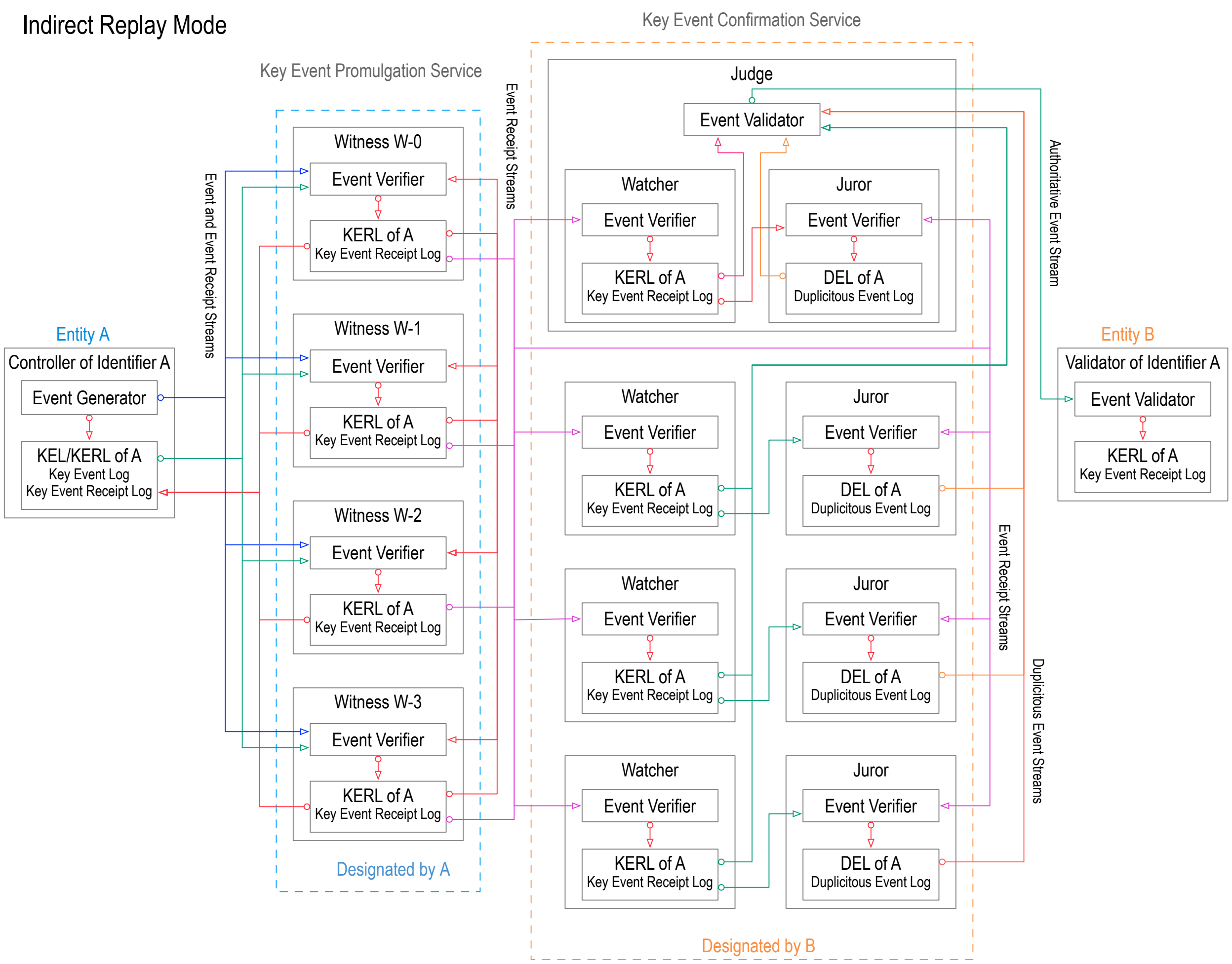

Indirect Replay Mode

Key Event Confirmation Service

Key Event Promulgation Service

Event Receipt Streams

Judge

Event Validator

Watcher

Event Verifier

KERL of A

Key Event Receipt Log

Juror

Event Verifier

DEL of A

Duplicitous Event Log

Authoritative Event Stream

Event and Event Receipt Streams

Witness W-0

Event Verifier

KERL of A

Key Event Receipt Log

Witness W-1

Event Verifier

KERL of A

Key Event Receipt Log

Witness W-2

Event Verifier

KERL of A

Key Event Receipt Log

Witness W-3

Event Verifier

KERL of A

Key Event Receipt Log

Designated by A

Entity A

Controller of Identifier A

Event Generator

KEL/KERL of A

Key Event Log

Key Event Receipt Log

Watcher

Event Verifier

KERL of A

Key Event Receipt Log

Juror

Event Verifier

DEL of A

Duplicitous Event Log

Watcher

Event Verifier

KERL of A

Key Event Receipt Log

Juror

Event Verifier

DEL of A

Duplicitous Event Log

Watcher

Event Verifier

KERL of A

Key Event Receipt Log

Juror

Event Verifier

DEL of A

Duplicitous Event Log

Event Receipt Streams

Duplicitous Event Streams

Entity B

Validator of Identifier A

Event Validator

KERL of A

Key Event Receipt Log

Designated by B

Figure 10.6. Indirect Replay Mode: A to Any.

In more detail, the controller's promulgation service is provided by a set of *N* designated *witnesses*. Although the witnesses are explicitly designated by the controller they may or may not be under the control of the controller. The designation is a cryptographic commitment to the witnesses via a verifiable statement included in an establishment event.The purpose of the witness set is to better protect the service from faults including Byzantine faults [36]. Thus the service employs a type of Byzantine Fault Tolerant (BFT) algorithm. We call this *KERI's Algorithm for Witness Agreement (KAWA)* (formerly known as KA$^2$CE). The primary purpose of the KAWA algorithm is to protect the controller's ability to promulgate the authoritative copy of its key event history despite external attack. This includes maintaining a sufficient degree of availability such that any validator may obtain an authoritative copy on demand.

The critical insight is that because the controller is the sole source of truth for the creation of any and all key events, it alone, is sufficient to order its own key events. Indeed, a key event history does not need to provide double spend proofing of an account balance, merely consistency. Key events by in large are idempotent authorization operations as opposed to non-idempotent account balance decrement or increment operations. Total or global ordering may be critical for non-idempotency, whereas local ordering may be sufficient for idempotency especially to merely prove consistency of those operations. The implication of these insights is that fault tolerance may be provided with a single phase agreement by the set of witnesses instead of a much more complex multi-phase commit among a pool of replicants or other total ordering agreement process as is used by popular BFT algorithms [16; 39; 43; 48; 61; 115; 123; 144]. Indeed the security guarantees of an implementation of KAWA may be designed to approach that of other BFT algorithms but without their scalability, cost, throughput, or latency limitations. If those other algorithms may be deemed sufficiently secure then so may be KAWA. Moreover because the controller is the sole source of truth for key events, a validator may hold that controller (whether trusted or not) accountable for those key events. As a result, the algorithm is designed to enable a controller to provide itself with any degree of protection it deems necessary given this accountability.

The reliance on a designated set of witnesses provides several advantages. The first is that the identifier's trust basis is not locked to any given witness or set of witnesses but may be transferred at the controller's choosing. This provides portability. The second is that the number and composition of witnesses is also at the controller's choosing. The controller may change this in order to make trade-offs between performance, scalability, and security. This provides flexibility and adaptability. Thirdly the witnesses need not provide much more than verification and logging. This means that even highly cost or performance constrained applications may take advantage of this approach.

Likewise, given any guarantees of accountability the controller may declare, a validator may provide itself with any degree of protection it deems necessary by designating a set of observers (watchers, jurors, and judges) . Specifically, a validator may be protected by maintaining a copy of the key event history as first seen (received) by the validator or any other component trusted by the validator (watcher, juror, judge). This copy may be used to detect any alternate inconsistent (duplicitous) copies of the key event history. The validator may then choose how to best respond in the event of a detected duplicitous copy to protect itself from harm. A special case is a malicious controller that intentionally produces alternate key event histories. Importantly, observer components that maintain copies of the key event history such as watchers, jurors, and judges, may be under the control of validators not controllers. As a result a malicious alternate (duplicitous) event history may be eminently detectable by any validator. We call this ambient duplicity detection (which stems from ambient verifiability). In this case, a validator may still be protected because it may still hold such a malicious controller accountable given a duplicitous copy (trust or not trust). It is at the validator's discretion whether or not to treat its original copy as the authoritative one with respect to any other copy and thereby continue trusting or not that original copy. A malicious controller may not therefore later substitute with impunity any alternate copy it may produce. Furthermore, as discussed above, a malicious controller that creates an alternative event history imperils any value it may wish to preserve in the associated identifier. It is potentially completely self-destructive with respect to the identifier. A malicious controller producing a detectably duplicitous event history is tantamount to a detectable total exploit of its authoritative keys and the keys of its witness set. This is analogous to a total but detectable exploit of any BFT ledger such as a detectable 51% attack on a proof-of-work ledger. A detectable total exploit destroys any value in that ledger after the point of exploit.

To restate a controller may designate its witness set in such a way as to provide any arbitrary degree of protection from external exploit. Nonetheless in the event of such an exploit a validator may choose either to hold that controller accountable as duplicitous and therefore stop trusting the identifier or to treat the validator's copy of the key event history as authoritative (ignoring the exploited copy) and therefore continue trusting the identifier. This dependence on the validator's choice in the event of detected duplicity both imperils any potential malicious controller and protects the validator. The details of KAWA will be discussed in more detail below.

# 11 KERI's Algorithm for Witness Agreement.

KERI's Algorithm for Witness Agreement (KAWA) or the *algorithm*, (formerly known as KERI's Agreement Algorithm for Consensus Control establishment or $KA^2C^2E$) is run by the *controller* of an *identifier* in concert with a set of *N witnesses* designated by the *controller* to provide as a service the *key event history* of that *identifier* via a KERL (Key Event Receipt Log) in a highly available and fault-tolerant manner. One motivation for using key event logs is that the operation of redundant immutable (deletion proof) event logs may be parallelizable and hence highly scalable. A KERL is an immutable event log that is made deletion proof by virtue of it being provided by the set of *witnesses* of which only a subset of *F* witnesses may at any time be faulty. In addition to designating the witness set, the controller also designates a *threshold* number, *M*, of *witnesses* for accountability. To clarify, the controller accepts accountability for an event when any subset *M* of the *N* witnesses confirms that event. The threshold *M* indicates the minimum number of confirming witnesses the controller deems sufficient given some number *F* of *potentially faulty* witnesses. The objective of the service is to provide a verifiable KERL to any *validator* on demand. Unlike direct mode where a *validator* may be viewed as an implicit witness, with indirect mode, a *validator* may not be one of the *N* explicitly designated *witnesses* that provide the service.

## 11.1 Witness Designation.

The controller designates both the witness *tally* number and the initial set of witnesses in the inception event configuration. The purpose of the *tally* is to provide a *threshold of accountability* for the number of witnesses confirming an event. This is shown below.

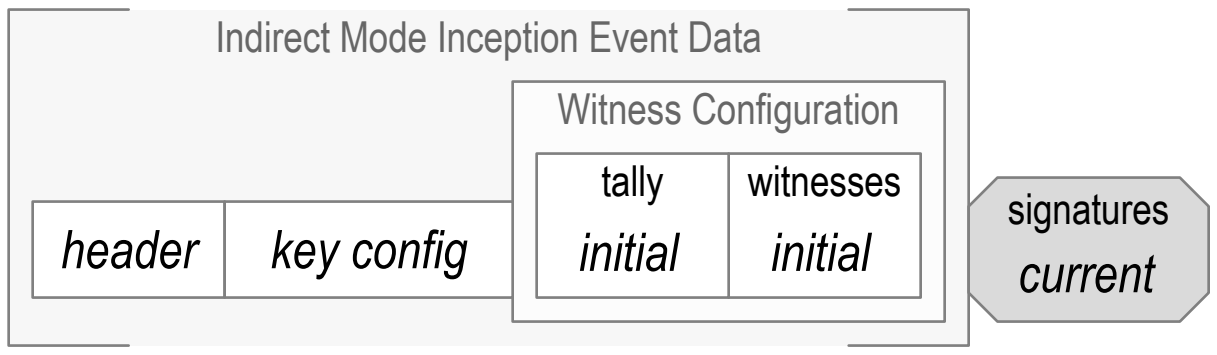

Figure 11.1. Inception Event Witness Configuration.

Subsequent rotation operations may amend the set of witnesses and change the *tally* number. This enables the controller to replace faulty witnesses and/or change the threshold of accountability of the witness set. When a rotation amends the witnesses it includes the new *tally*, the set of *pruned* (removed) witnesses and the set of newly *grafted* (added) witnesses. This is shown below.

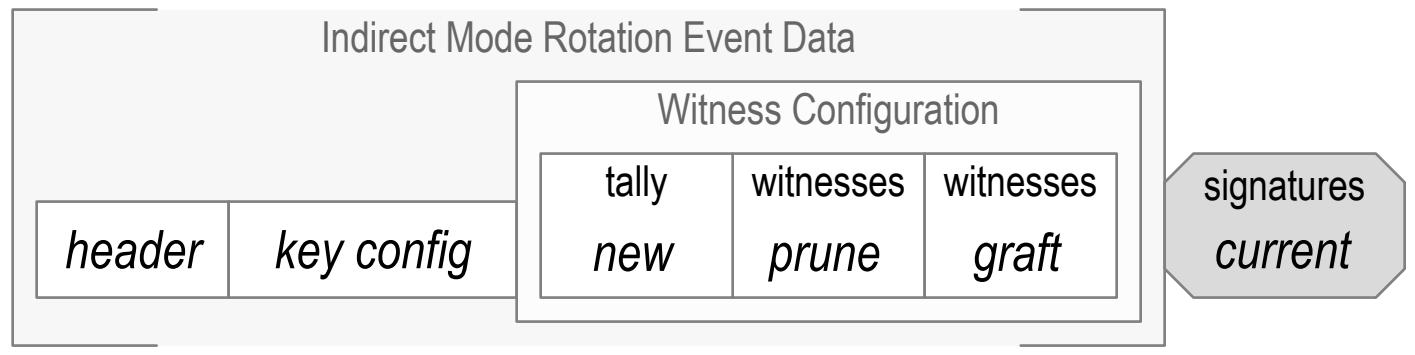

Figure 11.2. Inception Event Witness Configuration.

## 11.2 Witnessing Policy.

In this approach, the *controller* of a given identifier creates and disseminates associated key event messages to the set of *N witnesses*. Each witness, verifies the signatures, content, and consistency of each key event it receives. When a verified key event is also the first *version* of that event the witness has received then it *witnesses* that event by signing the event message to create a *receipt*, storing the *receipt* in its log (KERL), and returning the *receipt* as an acknowledgement to the controller. Depending on its dissemination policy a witness may also send its *receipt* to other witnesses. This might be with a broadcast or gossip protocol or not at all.

Recall as previously defined, the *location* of an event in its key event sequence is determined by its previous event digest and sequence number. Likewise as previously defined, The *version* of an event of the same *class* at given *location* in the key event sequence is different or inconsistent with some other event of the same *class* the same *location* if any of its other content differs or is inconsistent with that other event of the same *class* and *location*. To clarify, the event version includes the *class* of event, *establishment* or *non-establishment*. The special case where *class* matters is that an *establishment* event (such as rotation) at the same *location* may supersede a *non-establishment* event event (such as interaction) in the event of exploit of the non-establishment event. This enables recovery of live exploit of the exposed current set of authoritative keys used to sign non-establishment events via a rotation establishment event to the unexposed next set of authoritative keys. The specific details of this recovery are explained later (see Section 11.6). In general, the witnessing policy is that the first seen *version* of an event always wins, that is, the first verified version is *witnessed* (signed, stored, acknowledged and maybe disseminated) and all other versions are discarded. The exception to this general rule is that an establishment event may recover following a set of exploited non-establishment events. The recovery process may fork off a branch from the recovered trunk. This *disputed* branch has the disputed exploited events and the main trunk has the recovered events. The *operational mode* (see Section 10.) and the *threshold of accountable duplicity* determine which events in the disputed branch are *accountable* to the controller (see Section 11.6).

Later messages or receipts from other witnesses may not change any existing entry in the log (the log is append only i.e. immutable). Each witness also adds to its log any verified signatures from consistent receipts it receives from other witnesses. A consistent receipt is a receipt for the same *version* of the event already in its log. Excepting recovery, inconsistent receipts i.e. for different event *versions* at the same location are discarded (not kept in the log). Although as an option, a controller may choose to run a *juror* (in concert with a witness) that keeps a duplicitous event log (DEL) of the inconsistent or duplicitous receipts that a witness receives. To clarify, a witness' key event receipt log (KERL) is by construction an immutable log. This log includes the events with attached verified signatures, that are the receipts from the controller, the witness itself, and other witnesses.

Initial dissemination of receipts to the $N$ witnesses by the controller may be implemented extremely efficiently with respect to network bandwidth using a round-robin protocol of exchanges between the controller and each of the witnesses in turn. Each time the controller connects to a witness to send new events and collect the new event receipts, it also sends the receipts it has received so far from other witnesses. This round-robin protocol may require the controller to perform at most two passes through the entire set of witnesses in order to fully disseminate a receipt from each witness to every other witness for a given event,This means that at most $2 \cdot N$ acknowledged exchanges are needed for each event in order to create a fully witnessed key event receipt log (KERL) at each and every witness and the controller. Network load therefore scales linearly with the number of witnesses. When network bandwidth is less constrained then a gossip protocol might provide full dissemination with lower latency than a round robin protocol. Gossip protocols are a relatively efficient mechanism and scale with $N \cdot log(N)$ (where $N$ is the number of witnesses) instead of $2 \cdot N$. A directed acyclic graph or other data structure can be used to determine what needs to be gossiped.

## 11.3 Event Escrow

KERI is designed to be highly protocol agnostic, that is, the transport protocol a controller chooses to send events to its witnesses may be implementation dependent. Some of these protocols may benefit, usually for performance reasons,  from the witnesses holding events in escrow. There are two primary use cases: out-of-order events and multi-signature events. Event escrow is an optional implementation specific configurable capability of controllers and witnesses implementations. A witness may employ neither, one or both of an out-of-order event escrow cache and an in-order partial multi-signature event escrow cache. Out-of-order events may be first held in on out-of-order cache before being processed by an in-order but partial multi-signature cache.

### 11.3.1 Out-Of-Order

When the transport protocol is asynchronous such as UTP, then the events may arrive at a witness out of order. An out of order event is not fully verifiable because without a copy of  the immediately preceding event in the sequence, the prior event digest is not verifiable against the prior event. Moreover, for an out-of-order rotation (establishment) event, the newly rotated current set of keys may not be verifiable against the next keys digest from most recent prior establishment event when that prior establishment event has not yet been received and entered into the KEL. Should a witness drop an otherwise verifiable but out of order event, then that event must eventually be retransmitted. This adds network load and latency. Holding otherwise verifiable but out-of-order events in a tunable escrow cache until the missing intervening events show up may be a good way of optimizing network bandwidth and latency in asynchronous protocols. An otherwise verifiable event means that what can be verified does verify. If what can be verified does not verify then there is no reason to hold the event in escrow.

An escrow cache of unverified out-of-order event provides an opportunity for malicious attackers to send forged event that may fill up the cache as a type of denial of service attack. For this reason escrow caches are typically FIFO (first-in-first-out) where older events are flushed to make room for newer events.

### 11.3.2 Multi-Signature

When the identifier prefix established by the KEL uses a multi-signature scheme with a set of authoritative key-pairs there may be a corresponding set of controllers, where each controller holds one or more of the key-pairs from the set. This may be called a multi-controller or collec-tive controller case. This set of controllers may use a collection protocol amongst themselves where one of the controllers first collects enough signatures to reach the multi-signature threshold before sending the event with signatures to the witnesses. In addition to latency, this may make the designated collecting controller a single point of failure.

In this multi-controller case, an alternative approach to collecting signatures is for the witnesses to directly collect events from each of the controllers,, but where each event has only a partial signature set. An event with a partial signature set would not verify against the threshold requirement for the event but may otherwise be verifiable with the provided signatures. In this case the event could be held in escrow until matching events, but with different partial signature sets from other members of the set of controllers, provide enough signatures to reach the threshold. A multi-signature threshold event escrow capability may be more performant and reliable that a two stage protocol where the full set of signatures must first be collected before being sent to the witnesses.

In this case, an in-order but otherwise verifiable event may be held in a multi-signature event escrow cache if the set of attached signatures includes at least one verifiable signature and no unverifiable signatures. This minimizes the exposure of the partial multi-signature escrow cache to a denial of service attack using forged events.

## 11.4 Consensus Process

The purpose of using *N* designated witnesses in the *algorithm* is to ensure both the availability and the security of the service in the event of faulty witnesses. In addition to faults such as unresponsiveness and network interruptions, allowed faults include what are commonly known as Byzantine faults such as malicious or duplicitous behavior (dishonesty) by the witnesses. A faulty controller is a special case to be discussed below. A dishonest witness may promulgate receipts for inconsistent versions of an event or neglect to store or disseminate receipts. An unresponsive but honest witness may be unavailable or otherwise unable to promulgate a receipt. A malicious witness may intentionally be unresponsive. The term Byzantine agreement (BA) is used to describe a type of algorithm that provides some guarantee of consensus agreement despite Byzantine faults, i.e. it's Byzantine Fault Tolerant (BFT) [36]. Unlike more conventional BA algorithms there is no requirement in KAWA that the consensus process provide a total ordering of events. That ordering is already provided to the *algorithm* as an input by the controller. The purpose of the consensus agreement process in KAWA is to produce for each event sourced by the controller a verifiable agreement that includes the events and signatures (receipts) from a sufficient number of witnesses in a secure highly available manner. The *algorithm* merely produces verifiable confirmations of already ordered events. This greatly simplifies the *algorithm*.

Consensus *agreement* guarantees in conventional BA algorithms are often characterized with the terms *correct*, *safe*, and *live* [39]. These guarantees exist for algorithms that undertake multiple phases or rounds in coming to consensus. This complicates the algorithms because agreement at one phase may not propagate to the next phase. In the case of this algorithm, KAWA, the consensus process is different enough that there are no direct analogous definitions of *correct*, *safe*, and *live* albeit some of the related constraints in KAWA are evocative of the constraints in conventional BA algorithms associated with those terms. In order to better understand the differences we first summarize the usual definitions of these terms before providing our terminology. Typically $F$ is defined to be maximum number of faulty nodes at any time (any type of fault). Given at most $F$ faults then any consensus agreement of $F+1$ nodes is consistent because at least one the $F+1$ nodes is honest. Therefore this agreement is called *correct*. But there is no guarantee that any set of $F+1$ nodes will ever be in agreement. A group of nodes that contribute to or come to consensus agreement is called a quorum. In this case $F+1$ is the quorum size for *correct* agreement. Furthermore suppose that $N=2F+1$, where $N$ is total number of nodes. Then a *correct* agreement is guaranteed because at least $F+1$ of the $N$ nodes are honest and they also form a majority of all nodes so no other agreement is possible. Therefore this agreement is called *safe*. But there is no guarantee that the nodes in agreement (quorum) will know they came to agreement in a subsequent round because as many as $F$ of the nodes in agreement in a given round may subsequently become unresponsive in the next round. In a multi-phase or multi-round algorithm, this means that despite being *safe*, the agreement is not *live* because the agreement may not propagate to the next round. Suppose instead that $N=3F+1$, then a correct agreement is guaranteed as before but the agreement will have $2F+1$ nodes (quorum size) and despite some $F$ of those $2F+1$ nodes becoming unresponsive, the remaining $F+1$ honest and responsive nodes will be able to propagate that *correct safe* agreement to the next round. Thus $N=3F+1$ with a quorum size of $2F+1$ guarantees a *correct*, *safe*, and *live* agreement across multiple rounds.

### 11.4.1 Faulty

In KAWA we define $F$ to be the number of *potentially faulty* witnesses at any time. Faulty behavior may be any combination of either unresponsive or responsive but dishonest. This means that no more than $F$ witnesses at any time may exhibit any combination of either unresponsive or responsive but dishonest behavior. Witness faults may be inadvertent or the result of malicious external attack. In either case, a controller may provide fault tolerance by designating a sufficiently redundant set of witnesses. We use the qualification *potentially faulty* because it better describes with how we view faults in a system and clarifies understanding. We may segment the *potentially faulty* witnesses as either honest but unresponsive or malicious. Suppose, for example, that at any time at most $A$ witnesses may be honest but unresponsive and $B$ witnesses may be malicious (either unresponsively or dishonestly), then by our definition:

$$F=A+B\,. \tag{11.1}$$

The malicious, $B$, we may further segment into either maliciously unresponsive or maliciously responsive but dishonest.

### 11.4.2 Agreement

Recall that an *honest witness* (noun) will only *witness* (verb), (i.e. create, store, and acknowledge a receipt for), at most one and only one version of an event. That event version must first be verified. A verified event version must be consistent with prior events and signed by the controller's authoritative keys. The act of witnessing is contingent on a controller first creating a verifiable event version. A receipt has been *promulgated* if it has been witnessed and disseminated to a another witness. Therefore, the controller first creates its own receipt of the event and then promulgates the receipted event to witnesses in order to gather their promulgated receipts. In this *algorithm*, an *agreement* consists of a specific version of an event with verifiable receipts (signatures ) from the controller and a set of witnesses. A state of *agreement* about a version of an event with respect to set of witnesses means that each witness in that set has *witnessed* the same version of that event and each witness' receipt in that set has been promulgated to every other witness in that set. This state of *agreement* may be made known and provable to any validator, watcher, juror, or judge via an *agreement* that is a verifiable fully receipted copy of the event. Such a copy may be obtained from a KERL. The witnesses in a state of *agreement* comprise a set. Each witness in that set is a contributor of a receipt to the associated *agreement*. Likewise the size of an *agreement* is the number of contributor witnesses or equivalently the number of receipts in that *agreement*. An example agreement is diagrammed below.

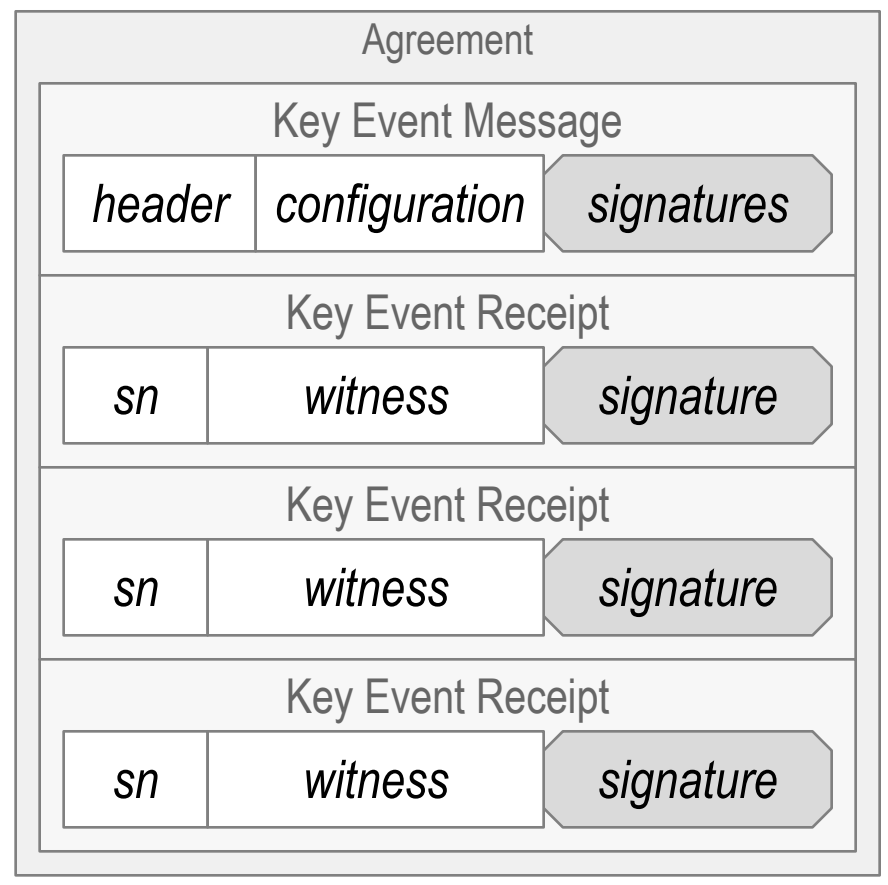

Figure 11.3. Agreement Contents.

Because promulgation of both events and event receipts is the responsibility of the controller, the process of *agreement* is always contingent on a responsive controller (at least initially). From the perspective of the controller, a *sufficient agreement* is an *agreement* proving a state of *agreement* among a large enough set of witnesses that a controller may be held accountable for the *agreement*. Likewise, from the perspective of the validator, a *sufficient agreement* is an *agreement* proving a state of *agreement* among a large enough set of witnesses that a validator may accept the agreement as authoritative. What may be sufficient for a controller may not be sufficient for a validator. To clarify, let $M_C$ denote the threshold size of a *sufficient agreement* from the perspective of a controller and let $M_V$ denote the threshold size of *sufficient agreement* from the perspective of a validator. Typically, $M_V \geq M_C$. The threshold, $M_C$, is the same as the *tally* or *threshold of accountability*, *M*, defined above.

A highly available system needs some degree of fault tolerance. The purpose of the *threshold of accountability (tally)* is enable fault tolerance of the key event service with respect to faulty behavior by either the controller or witnesses. The principal controller fault exhibits as duplicitous behavior in the use of its keys. In this case, the threshold serves as the *threshold of accountable duplicity*. The threshold lets a validator know when it may hold the controller accountable for duplicitous behavior. Without a threshold a validator may choose to hold a controller accountable upon any evidence of duplicity which may make the service fragile in the presence of any degree of such faulty behavior. The primary way that a validator may hold a controller accountable is to stop trusting any use of the associated identifier. This destroys any value in the identifier and does not allow the controller to recover from an exploit. Recall that the one purpose of rotation keys (pre-rotated unexposed) is to enable recovery from compromised interaction signing keys. A compromised interaction signing key may exhibit as duplicitous behavior on the part of the controller. A threshold of accountable duplicity enables a validator to distinguish between potentially recoverable duplicity such as the use of compromised signing key and and non-recoverable duplicity such as a the use of compromised rotation key. This better protects both the validator and the controller and improves the robustness of the service.

The controller is responsible to promulgate a mutually consistent set of receipts, i.e. a *sufficient agreement* of size $M \geq M_C$. Recall that only the controller holds the authoritative keys with which events may be created and witnesses may be designated. Therefore the controller is not replaceable and the overall agreement process is contingent on controller availability, at least initially. As a result the algorithm may pause indefinitely while waiting on the controller. An unresponsive controller may delay agreement indefinitely. Likewise, the algorithm's consensus agreement guarantees may be split into two cases. The first case is an honest controller, the sec-ond case is a dishonest controller. When witnesses also promulgate events then an event may continue to be promulgated after it is witnessed by one non-faulty witness despite a then unresponsive controller. The second case of a dishonest controller which may be further split into two sub-cases. The first sub-case is that a validator may trust that a most $F$ witnesses are potentially faulty despite a dishonest controller. This may arise when the witnesses are not under the direct control of the controller. In addition, a controller whose authoritative interaction signing keys have been exploited may appear to be dishonest. The distinction in this sub-case is that despite a dishonest or equivalently exploited controller, at most $F$ witnesses may have been exploited. The second sub-case is that the witnesses may be under the control of the dishonest or exploited controller and hence more than $F$ of them may be faulty.

Because a dishonest controller may choose to promulgate inconsistent versions of an event, there may be no guarantee that *sufficient agreement* may ever occur. What is more important from the perspective of a validator, however, is that it be protected from any inconsistent or duplicitous agreements despite a dishonest controller and/or some number of dishonest witnesses. Also important is that once *agreement* has been reached, a copy of that agreement may be obtained on demand by any validator (or watcher,  or juror, or judge).

Recall that a successful *live* exploit on the controlling key-pairs may capture the current set of authoritative keys. We expand the definition of *live* exploit to include the witness agreement (consensus) process in the following manner. With respect to the controller, a successful *live* exploit of would prevent *a sufficient agreement*, $M_C$, about any newly promulgated events. With respect to the validator, a successful live exploit would produce undetectably duplicitous but *sufficient agreement*, $M_V$, about any newly promulgated events. Any successful *dead* exploit, in contrast, must happen some time in the future after agreement has occurred and is not concerned with exploiting the process of creating agreement about new events. The primary goal of the consensus algorithm is to provide an agreement process that protects both an honest controller a validator from *live* exploit. Any pre-existing copy of a KERL held by a validator from later dead exploits

11.4.2.1 Proper

Given the assumption of $F$ above any agreement of a set of $F+1$ witnesses means that during the agreement process both the controller was non-faulty and at least one of the witnesses in the agreement was non-faulty (both honest and responsive). Because at least one witness was honest, agreement is deemed *proper*. The event associated with a *proper* agreement may be called a *proper* event. Recall that a witness will only witness the first seen verifiable version of an event. Thus the production of any agreement that includes an honest witness means that all witnesses party to that agreement witnessed the same version of that event. However there is no guarantee that any set of $F+1$ witnesses will ever come to agreement about any version of an event. As many as $F$ may be unresponsive, or else as many as $F$ may be may be responsive but dishonest or the controller may be faulty.

Obviously any usefully sufficient agreement must also be proper. This places a lower bound on the tally, $M$, for a sufficient agreement, that is,

$$M > F, \tag{11.2}$$

where $M$ may be $M_C$ or $M_V$.

11.4.2.2 Utile

An agreement is deemed *utile* if at least one copy of a correct agreement may be obtained by any requestor such as *validator*, *watcher*, *juror, or judge*. In other words the agreement is utile if some copy of the agreement is available to a requestor. By our definition of *agreement*, each wit-ness in that agreement must have provided a receipt to the agreement and received a copy of the agreement (fully receipted copy of the event). Indeed a non-faulty controller will eventually collect a receipt from every witness party to the agreement, all of which receipts, it will disseminate to all the witnesses thereby ensuring that every non-faulty witness will possess copy of the agreement. Furthermore because at least one witness in a *proper* agreement is non-faulty, a copy of that agreement will therefore be available via that witness. Thus any *proper* agreement is also *utile* because at the very least one non-faulty witness will posses a copy and make it available to a requestor. The event associated with a *utile* agreement may be called a *utile* event.

11.4.2.3 Intact

A non-faulty controller desires to select the total number of witnesses, *N*, to ensure or guarantee a *proper* agreement will indeed occur. Recall that at most *F* witnesses are potentially faulty. None of the faulty witnesses, *F*, may ever be depended upon by a non-faulty controller to contribute to any agreement. Excluding potentially faulty witnesses from the set of witnesses contributing to sufficient agreement places an upper bound on the tally *M* with respect to *N*, that is,

$$M \le N - F , \tag{11.3}$$

where *N* is the total number of witnesses and *M* may be $M_C$ or $M_C$. Recall that a *sufficient agreement* must be proper, that is, $M > F$ or equivalently $M \ge F + 1$. This gives a combined constraint on M as follows:

$$F < M \le N - F . \tag{11.4}$$

A *proper* agreement requires that at least $F + 1$ witnesses be in agreement, therefore a guarantee of a proper agreement requires that at least $F + 1$ witnesses be non-faulty. But any of the *N* witnesses in excess of *F* will be non-faulty and may be depended upon by a non-faulty controller to contribute to the agreement. Therefore *N* must be large enough to allow for both $F + 1$ non-faulty witnesses and *F* faulty witnesses. If we substitute $M \ge F + 1$ into eq. 11.3 and solve for *N* we get the following lower bound on *N*, that is,

$$N \ge 2F + 1 . \tag{11.5}$$

Satisfaction of this constraint enables a non-faulty controller to guarantee a *proper* agreement among a set of witnesses. Importantly, because a non-faulty controller will only ever promulgate one version of an event, agreement about one and only one version is guaranteed. Recall that because a non-faulty controller will collect a receipt from every non-faulty witness, a non-faulty controller will eventually possess a copy of the agreement which it will disseminate to all the non-faulty witnesses. Given *N* satisfies this constraint, this guarantee of a *proper* agreement means that the agreement is deemed *intact*. The event associated with an intact agreement may be called an *intact* event. To clarify, *intact* agreement is contingent on a non-faulty controller. Given this contingency, at least $F + 1$ non-faulty witnesses will be guaranteed to possess a copy of the one and only proper agreement which means they all will be available to provide a copy of the agreement to any requestor. Thus an *intact* agreement is also a *utile* agreement.

11.4.2.4 Immune

The assumption of a non-faulty (honest) controller, however, may be unsatisfactory to either a validator or an original controller. Consider the case of a dishonest controller (or exploited controller) but where the witnesses are not under its control such that a validator or original controller may nonetheless trust that there may be no more than *F* faulty witnesses. This is equivalent to the case where the current controlling key set for an identifier may have been compromised (live) but not the key sets for the witnesses, or at most *F* of the key sets for the witnesses. Furthermore, given the practical unfeasibility of *live* exploit of the unexposed next pre-rotated keys, the only practically feasible exploit is a compromise is of exposed current con-troller keys that may be used to sign interaction events. An original controller may recover from such an exploit by promulgating a rotation event using its previously unexposed pre-rotated key set. So establishing a *threshold of accountable duplicity, in this case,* may protect value in the identifier and enable recovery via a rotation. Recall that a controller may be held accountable (liable) for any *sufficient* agreements. Allowing a dishonest or exploited controller to create multiple but inconsistent agreements for which it is liable may be problematic to that controller. Likewise, a validator may desire to protect itself from the harm of such an exploit. In this case, a validator may be harmed if it accepts as authoritative a different version of an event with respect to another validator. When sufficient agreement may occur on more than one version of an event then a validator may not be able to determine which version is authoritative in order to either hold the controller accountable or reconcile that event with respect to other validators.

Protection in both cases for either controller recovery or validator acceptance would be provided if there is a guaranteed agreement on at most one and only version of any event or not at all. This means that only one version could ever be considered authoritative and accepted by any validator. Either agreement happens for only one event version that any validator may accept as authoritative or agreement never happens which means that no version is accepted by any validator. An additional joint constraint on *M*, *N*, and *F* provides this protection.

Consider the case, where given a large enough *N*, a dishonest controller may create two or more *sufficient* agreements by promulgating a different version of an event, each to a disjoint subset of the witnesses where each subset has at least *M* witnesses. This case is prevented when the intersection subset of any two agreement sets is large enough to include at least one honest witness. Recall that an honest witness may never witness two inconsistent versions of the same event. Therefore any two *sufficient* agreements whose intersection includes an honest witness must therefore be in agreement about the same version of that event. This ensures that at most one version of the event will ever have *sufficient* agreement or none at all despite a dishonest (duplicitous) controller with the caveat of at most *F* maliciously faulty (duplicitous) witnesses.

We may derive a minimum constraint on the size of *M* in terms of *F* and *N* that guarantees that either at most one version of an event may have *sufficient agreement* or none at all. In this case, *M* may be used as a *threshold of accountable duplicity*. The number of members in a set is called its cardinality and is given by the $|\,|$ operator. Let the set of all witnesses be denoted $\widehat{N}$, with cardinality,

$$\left|\widehat{N}\right| = N\ . \tag{11.6}$$

Let $\widehat{M}$ represent a *sufficient agreement* set of witnesses. Its cardinality may be denoted $\left|\widehat{M}\right|$. Consider any two agreement sets $\widehat{M}_1$ and $\widehat{M}_2$. The first constraint is that the union of the two agreement sets must include all the witnesses, that is,

$$\widehat{M}_1 \cup \widehat{M}_2 = \widehat{N}\ . \tag{11.7}$$

This ensures that all the witnesses are included in one or the other of the two sets. No witness may be excluded. It places a lower bound on the combined sizes of the agreement sets. From this we may derive a relationship between the cardinalities as follows:

$$\left|\widehat{M}_1 \cup \widehat{M}_2\right| = \left|\widehat{N}\right| = N\ . \tag{11.8}$$

Furthermore, consider the limiting case where the two agreement sets, $\widehat{M}_1$ and $\widehat{M}_2$, each has cardinality equal to the sufficient agreement threshold, M, that is,

$$\left|\widehat{M}_1\right| = \left|\widehat{M}_2\right| = M \ . \tag{11.9}$$

Now recall that given at most $F$ potentially faulty witnesses, any set of size at least $F+1$ is guaranteed to include at least one honest witness. The constraint here is that the intersection must include at least one honest witness. This means the cardinality of the intersection of $\widehat{M}_1$ and $\widehat{M}_2$, must be at least $F+1$, that is,

$$\left|\widehat{M}_1 \cap \widehat{M}_2\right| \geq F+1 \ . \tag{11.10}$$

We will next use the well-known relationship between the sum of the cardinalities of two sets, that is,

$$\left|\widehat{M}_1\right| + \left|\widehat{M}_2\right| = \left|\widehat{M}_1 \cup \widehat{M}_2\right| + \left|\widehat{M}_1 \cap \widehat{M}_2\right| . \tag{11.11}$$

We may substitute and rewrite to form the following equality,

$$2M = N + F + 1 \ . \tag{11.12}$$

After factoring, gives,

$$M = \frac{N+F+1}{2} \ . \tag{11.13}$$

Restoring the inequality gives,

$$M \geq \frac{N+F+1}{2} \ . \tag{11.14}$$

Because *M* must be an integer this is equivalent to,

$$M \geq \left\lceil \frac{N+F+1}{2} \right\rceil \tag{11.15}$$

where $\lceil \ \rceil$ represents the *ceil* or *least integer* function, that is, $\lceil x \rceil$ is defined as the smallest integer that is greater than or equal to *x*. The constraint above implies that the sufficient agreement threshold *M* must also be a majority of *N*. In other words, *M* must be a *sufficient majority* of *N* where *sufficient* is given by the constraint in either eq. 11.14 or eq. 11.15.

The upper and lower constraints on M may be combined as follows:

$$\frac{N+F+1}{2} \leq M \leq N - F \ . \tag{11.16}$$

With the inequality above, there may be more than one solution for *M* for any given *N* and *F*. In that case, we can have an effective number of unavailable witnesses denoted *F** which is given by

$$F^* = N - M \tag{11.17}$$

There are two types of faults we care about with regard *to F* and *F**. One type is unavailability, in which an otherwise honest witness is not available to publish a receipt for a first-seen event. The other type is duplicity, in which a witness publishes a receipt for more than one version of an event given more than one version of an event has been produced by a duplicitous or compromised controller. The inequality means that while no more than *F* witnesses are allowed to be duplicitous, in some cases $F^* \geq F$ honest witnesses may be unavailable.

Satisfaction of this constraint guarantees that at most one *sufficient* agreement occurs or none at all despite a dishonest controller but where at most *F** of the witnesses are potentially unavailable and at most *F* are duplicitous. This guarantee means that the agreement is deemed *immune*.

To elaborate, given at most $F^*$ potentially unavailable or $F$ potentially duplicitous witnesses, an *immune* agreement requires that $M$ be a *sufficient majority* of $N$ and guarantees as a result that the service may either only produce a *sufficient agreement* for one version of each event or none at all despite a dishonest or exploited controller.

This may be applied in combination with the constraint for *intact* agreement from eq. 11.5. As it so happens, whenever $M$, $N$, and $F$ jointly satisfy the *immune* constraint above, (see eq. 11.16) then $N$ and $F$ also satisfy the earlier *intact* constraint above (see eq. 11.5). To see this, rewrite eq. 11.16 without $M$ to find the limiting values. This has the lower bound less than or equal to the upper bound as follows,

$$\frac{N+F+1}{2} \le N-F\ . \tag{11.18}$$

Now set equal as follows,

$$\frac{N+F+1}{2} = N-F\ , \tag{11.19}$$

and solve for $N$ in terms of $F$. This gives $N = 3F+1$ which is clearly greater than $N = 2F+1$ (one can cross-check by substituting in $N = 3F$ to see that anything less does not satisfy the inequality).

This result means that when the controller is honest any agreement deemed *immune* may also be deemed *intact*. Therefore, when the controller is non-faulty, the *immune* constraint implies *intact* which together guarantees that a sufficient agreement of one and only one version of the event will indeed occur. Recall as well that when *intact*, an agreement is also *utile*. Furthermore, in the case of a potentially exploited controller via compromised current authoritative keys but with at most $F^*$ and $F$ potentially faulty witnesses, the *immune* constraint guarantees that recovery from such an exploit may be successful. The details of recovery are provided later.

Some values of $M$, $N$, $F$, and $F^*$ for small $F$ that satisfy the immunity constraint are provided in the table below.

| F | N | 3F+1 | $\left\lceil \frac{N+F+1}{2} \right\rceil$ | N-F | M | F* = N-M |
|---|---|---|---|---|---|---|
| 1 | 4 | 4 | 3 | 3 | 3 | 1 |
| 1 | 5 | 4 | 4 | 4 | 4 | 1 |
| 1 | 6 | 4 | 4 | 5 | 4, 5 | 2,1 |
| 1 | 7 | 4 | 5 | 6 | 5, 6 | 2,1 |
| 1 | 8 | 4 | 5 | 7 | 5, 6, 7 | 3,2,1 |
| 1 | 9 | 4 | 6 | 8 | 6, 7, 8 | 3,2,1 |
| 2 | 7 | 7 | 5 | 5 | 5 | 2 |
| 2 | 8 | 7 | 6 | 6 | 6 | 2 |
| 2 | 9 | 7 | 6 | 7 | 6, 7 | 3,2 |
| 2 | 10 | 7 | 7 | 8 | 7, 8 | 3,2 |
| 2 | 11 | 7 | 7 | 9 | 7, 8, 9 | 4,3,2 |
| 2 | 12 | 7 | 8 | 10 | 8, 9, 10 | 4,3,2 |
| 3 | 10 | 10 | 7 | 7 | 7 | 3 |
| 3 | 11 | 10 | 8 | 8 | 8 | 3 |

| 3 | 12 | 10 | 8 | 9 | 8, 9 | 4,3 |
|---:|---:|---:|---:|---:|---:|---:|
| 3 | 13 | 10 | 9 | 10 | 9, 10 | 4,3 |
| 3 | 14 | 10 | 9 | 11 | 9, 10, 11 | 5,4,3 |
| 3 | 15 | 10 | 10 | 12 | 10, 11, 12 | 5,4,3 |

Figure 11.4. Some values of $M$, $N$, and $F$ that satisfy immunity.

In summary, when $M$, $N$, and F satisfy eq. 11.16 then despite a dishonest or compromised controller, the service may either only produce *proper utile* agreements about events or none at all. Or equivalently, given the *immune* constraint is satisfied, the service may not produce multiple divergent but *proper* key event receipt logs (KERLs). Thus any user of the service, be it validator, watcher, juror, or judge will be able to obtain either a *proper* event agreement on demand from some witness or none at all. Any non-faulty witness with a *proper* agreement will keep that agreement in its KERL and provide it on demand. Moreover, in order to be deemed *proper*, an agreement must have been verified as consistent with all prior events by every non-faulty witness that is party to that *agreement*. Recall that *intact* agreement is a guarantee of *proper* agreement. Thus *intact agreement* on any event is tantamount to *intact agreement* on all prior events. This means that any associated KERL up to and including the *intact* event (guaranteed *proper* event) may be deemed *intact (*guaranteed *proper)*. Consequently, the availability of a *proper* event at a witness is tantamount to the availability of a *proper* log (KERL) of all prior events consistent with that event at that witness, and thereby high availability of the service is assured.

## 11.5 Witness Rotation.

As introduced above, the initial set of designated witnesses is specified in the inception event and then may be changed by a rotation event. This may be thought of as *rotating* the witness set. Specifically, a rotation event may change the tally, the size, and the composition of the witness set. This allows the controller to prune faulty witnesses and graft on new ones. This facility gives the controller the ability with a rotation event to not only rotate authoritative signing keys in order to recover from their compromise but also to replace witnesses that may become unresponsive or compromised. As a result of the rotation event, the witness tally or threshold of accountability may change along with a change to the witnesses in the witness set. For the associated event agreement to be immune it must be witnessed by at least this threshold from the new witness set. In the case that the members of the new witness set are significantly different from the old set, i.e. most of the witnesses have been pruned then a validator may have more difficulty discovering the new event and witness set because the old pruned witnesses may no longer provide the event. One way to ameliorate this effect would be to require that a sufficient number of the old witness set also witness the event. This may make is easier for validators to securely discover and verify changes in the witness set. In other words each witness is required to witness the event that prunes that witness from the witness set.

### 11.5.1 Basic Witness Rotation

The following discussion details this joint agreement condition, that is, sufficient agreement from both the old and new witness sets.

Let the symbol $W_i$ denote a witness from a set of $N$ witnesses where the subscript $i$ indicates a specific witness. Let the symbol $\widehat{W}_l$ denote a set of witnesses where the subscript $l$, denotes the witness set corresponding to the $l^{th}$ establishment event. Thus the initial witness set, $\widehat{W}_0$, designated by the inception event at $l = 0$ may be expressed as follows:

$$\widehat{W}_0 = \left[ W_0 , W_1 , \cdots , W_{N-1} \right]. \tag{11.20}$$

Any subsequent witness set, $\widehat{W}_l$, for $l^{th}$ establishment (rotation) event may be expressed as follows:

$$\widehat{W}_l = \left( \widehat{W}_{l-1} - \widehat{X}_l \right) \bigcap \widehat{Y}_l \tag{11.21}$$

where $\bigcap$ is set intersection, $\widehat{W}_l$ is the set of witnesses generated by the current, $l^{th}$, rotation, $\widehat{W}_{l-1}$ is the set of witnesses from the previous, $(l-1)^{th}$, establishment event (inception or rotation), $\widehat{X}_l$ is the set of newly pruned (excluded) witnesses from $\widehat{W}_{l-1}$, and $\widehat{Y}_l$ is the set of newly grafted (included) witness in $\widehat{W}_l$. The following set valued operations and properties apply:

$$\widehat{X}_l \subseteq \widehat{W}_{l-1}, \ \widehat{Y}_l \not\subset \widehat{W}_{l-1}, \text{ and } \widehat{X}_l \not\subset \widehat{W}_l, \tag{11.22}$$

where $\subseteq$ is a subset, $\not\subset$ is not a subset. The total, $N_l$, of witnesses after the $l^{th}$ rotation may be computed as follows:

$$N_l = N_{l-1} - O_l + P_l \tag{11.23}$$

where $N_{l-1}$ is the total after the $(l-1)^{th}$ rotation, $O_l$ is the number of witnesses newly pruned (excluded) by the $l^{th}$ rotation, and $P_l$ is the number of witnesses newly grafted (included) by the by the $l^{th}$ rotation. The the $l^{th}$ tally, $M_l$ must satisfy:

$$M_l \leq N_l. \tag{11.24}$$

Let $| \, |$ represent the cardinality (number of elements) of a set. Then we have the following:

$$\left| \widehat{X}_l \right| = O_l, \ \left| \widehat{Y}_l \right| = P_l, \text{ and } \left| \widehat{W}_l \right| = N_l. \tag{11.25}$$

As described in more detail in a previous section, the tally $M_l$, indicates the number of sufficient confirming witness receipts from the witness list that a validator must obtain before accepting the associated, $l^{th}$, event.

### 11.5.2 Joint Confirmed Witness Rotation

As an additional protection to a controller or validator, a rotation event that changes the witness set must also have confirming receipts from a sufficient number of the previous, $(l-1)^{th}$, set of witnesses. This number is given by the previous tally, $M_{l-1}$. To clarify a rotation that changes the witness set must satisfy two thresholds for two confirmation sets, the first drawn from the $(l-1)^{th}$ and the second drawn from the $l^{th}$ sets of witnesses. Let $\widehat{U}_{l-1}$ denote the set of confirming witness receipts drawn from first set of witnesses as designated by the $(l-1)^{th}$ rotation event. There must be at least $M_{l-1}$ receipts in $\widehat{U}_{l-1}$. This requirement may be expressed as,

$$\widehat{U}_{l-1} \subseteq \widehat{W}_{l-1}, \text{ and } \left| \widehat{U}_{l-1} \right| \geq M_{l-1}, \tag{11.26}$$

where $| \, |$ is the set cardinality. Likewise let $\widehat{U}_l$ denote the set of confirming witness receipts drawn from second set of witnesses as designated by the $l^{th}$ rotation event. There must be at least $M_l$ receipts in $\widehat{U}_l$. This requirement may be expressed as,

$$\widehat{U}_l \subseteq \widehat{W}_l \text{, and } \left|\widehat{U}_l\right| \geq M_l . \tag{11.27}$$

The two confirmations together satisfy the following:

$$\left|\widehat{U}_{l-1} \cup \widehat{U}_l\right| \leq M_{l-1} + M_l , \tag{11.28}$$

where $\cup$ is set union. The requirement of confirmation from both sets means that an exploiter must compromise a sufficient number of witness from both the previous and current sets of witnesses in order for the the event to be validated. This prevents an exploiter who generates an an alternate version of a rotation event from merely replacing all the witnesses from prior events with compromised witnesses under the exploiter's control. The exploiter must also compromise a sufficient number of witnesses from the previous establishment event event in order to change them in the exploited event. To clarify, a validator may not accept the rotation event unless there are a sufficient number of confirming receipts from both sets of witnesses. This may be an optional feature. It may benefit from adding a unique confirming witness prior threshold to establishment events. It could use as a default the prior threshold from the previous establishment event. However a any validator could choose their own prior confirming threshold. The only requirement is that old honest witnesses are required to also witness the event that removes them as a witness. The problem with this joint confirmation is that if too many witnesses are compromised then the prior witness set confirmation will never reach the threshold and recovery will not be accepted by the validator. The identifier becomes dead to the validator, i.e. unrecoverable by the controller from the validator's perspective. This is equivalent to exceeding the $F$ threshold for allowed faulty witnesses at any given time. This additional set of witness receipts could aid a validator in reconciling any key compromises that the associated rotation is meant to recover from. It could also aid in reconciling dead attacks as a dead attack would have to compromise two sets of keys not just one set.

## 11.6 Recovery

Recall that one of the main purposes of key rotation is to enable recovery from a *live* exploit that has compromised the current set of authoritative keys. Because the current set of authoritative keys have been exposed via their use to sign one or more events there may be some likelihood of such compromise. Recovery is performed via a rotation to a previously unexposed set of keys, i.e. the *next* or pre-rotated set. Compromise of unexposed keys may be made practically infeasible. The rules for how witnesses process verify, store, and promulgate events in the case of compromise are explained here. Further recall that a controller may be only held accountable for an event agreement when that agreement includes a sufficient number of receipts, $M$.

The policy for a witness to *witness*, that is, verify, store, and promulgate an event are different when a rotation event may used as a recovery event promulgated by the original controller in response to its detection of *exploited* interaction events created by a compromised set of keys. This applies not only to indirect mode but to direct mode where the validator is an ersatz witness/watcher. This case is diagrammed below:

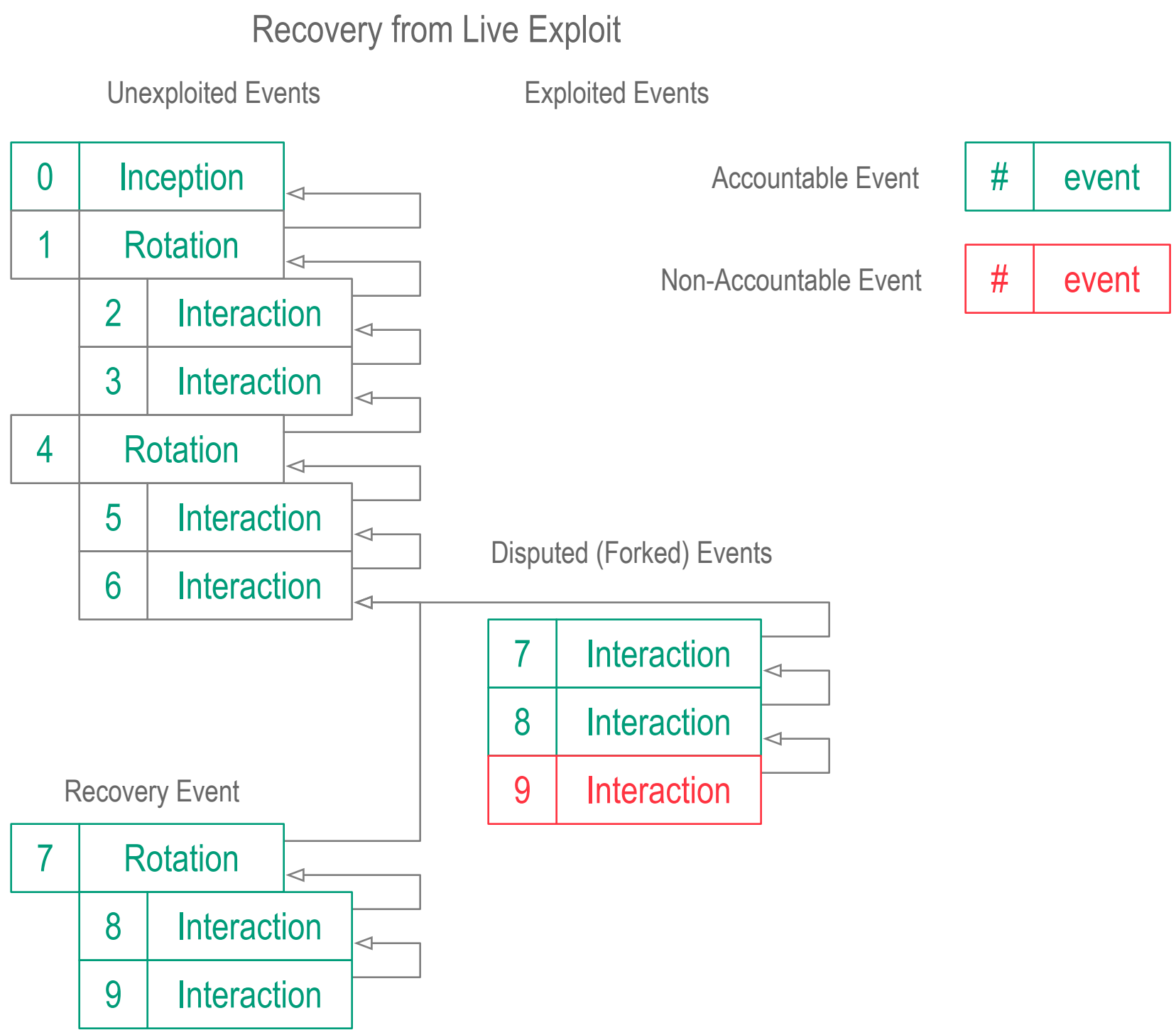

Figure 11.5. KERL for a Recovery Rotation Event.

We will walk through the example in the KERL diagram above in order to motivate the rules as they are presented. Each key event in the diagram is labeled with its sequence number. Each key event also has an arrow showing the digest chain to the previous event in the sequence. There are three types of events, inception events, rotation events and interaction events. The inception event may be considered a special case of a rotation event. There may be only one inception event. Moreover because an inception event does not participate in recovery from exploit no special recovery rules apply to it. The events are also labeled with a color. Events in green are entries in the log that have attached to them enough receipts to reach or exceed the threshold of accountability *M*. Events in red do not have enough receipts attached to reach the threshold of accountability. The diagram show the events in two columns. The first column contains unexploited events, that is, events signed with un-compromised keys. The second column contains exploited events, that is, events signed with compromised keys.

Events labeled 0 through 6 are unexceptional. Note that Event 4 is a rotation event, so the current authoritative keys for signing interaction events 5 and 6 are first exposed in event 4. Event 4 also commits to a next set of keys that may be not exposed until a subsequent rotation event. Suppose that the current authoritative keys exposed in event 4 are compromised by an attacker sometime after event 6. This attacker chooses to use those compromised create verifiably consistent interaction events 7,8, and 9 as part of its exploit. These events are promulgated to the witnesses which create witness receipts. The particular witness to whom this log belongs verifies event signatures and consistency for exploited events 7, 8, and 9 as the first seen versions of these events and therefore enters them into its log. This witness also receives and logs a sufficient number of receipts for exploited events 7 and 8 to reach the accountability threshold *M*. The original (true) controller notices the exploited events in a witness log and decides to recover using a rotation. This controller creates its own version of event 7 in the sequence as a rotation event and promulgates it to this witness. This effectively acts to dispute interaction events 7, 8, and 9 already in the log.

The special recovery rule described here is that a rotation event overrides or supersedes an interaction event with the same event sequence number and digest. This rule allows recovery of a compromised set of current signing keys used to sign interaction events with a superseding rotation event. Recall that the definition of same event version at a location also includes the event class. To clarify event version is defined not only by the event's sequence number and the event's digest of the previous event (chaining) but also but also by the event's class (establishment or non-establishment). The rule is that a witness (or validator as ersatz witness) will accept a rotation event with the same sequence number and chaining digest of a pre-existing interaction event (i.e. at the same location). In doing so the witness creates a fork or set of duplicate entries in its log at the location of the superseding or overriding rotation event. One branch contains the rotation event and subsequent entries the other branch contains the overridden interaction event and and subsequent interaction events chained from it. The former branch is the *recovered* (unexploited) branch, the latter branch is the *disputed* (exploited) branch.

Another special recovery rule described here is that once the witness (validator as ersatz witness) has accepted a superseding rotation event, the witness will no longer accept any new events of any type into the *disputed* branch.

Disputed events may only be interaction events and may be in one of two states, *accountable* or *non-accountable*. An event is deemed *accountable* if the witness has already received enough receipts to reach or exceed the threshold of accountability *M*. In the case of direct mode where the validator is an ersatz witness, the threshold of accountability is 1. This means that in direct mode all *disputed* events are also *accountable*.

Another special rule introduced here is that an *accountable* but *disputed* (interaction) event may be accepted as valid. In other words *accountable* events (*disputed* or not) are otherwise treated normally. Whereas disputed but not yet accountable events, however, are treated specially. The special case is when the dissemination policy for the witness allows it to initiate dissemination of receipts directly to other witnesses (gossip). The witness treats disputed but not yet accountable events specially by ceasing to initiate any dissemination (gossip) of the associated receipts to other witnesses. The witness will however provide upon request any receipts it already has logged for disputed events. The witness will also continue to accept receipts from other witnesses for disputed events. Should the witness subsequently receive enough receipts for a disputed event to reach the threshold of accountability then the event becomes an accountable but disputed event and the witness may resume its normal dissemination policy with regards that event.

The purpose of treating disputed but not yet accountable events specially by not allowing witness initiated dissemination of receipts is to provide more opportunity for the original true controller to propagate the recovery rotation event to all the witnesses before they may first see a disputed event. Recall that as soon as a witness receives a superseding recovery rotation event it ceases to accept any new events into the newly disputed branch created by that rotation event. This also gives the controller more of an opportunity to have the recovery rotation event be seen first by any non-responsive or offline witnesses. Stopping witness initiated dissemination (gossip) or receipts creates an exponential die off of gossiped receipts of disputed events. Recovery may create an unavoidable race condition but the special rule minimizes the extent of that race condition.

The rule allowing the acceptance as valid of accountable but disputed events prevents a dishonest but unexploited controller from later maliciously disputing events it previously created. Once the event receipts have reached the threshold of accountability then their validity becomes immutable. As a result, a controller may recover control of compromised keys but may not repudiate already accountable interaction events. This protects validators from dishonest controllers. This capability is a tradeoff between security and convenience. This type of exploit may be avoided entirely by using extended rotation events to provide interaction events. In this case the current authoritative signing keys are rotated to unexposed keys after each and every use. This comes at the cost of more rotations which may inconvenient in many applications.

### 11.6.1 Nested Delegation Recovery

Delegated identifiers introduce additional rules for recovery. In KERI delegations are cooperative, this means that both the delegator and delegate must contribute to a delegation. The delegator creates a cryptographic commitment in either a rotation or interaction event via a seal a a delegated establishment event. The delegate creates a cryptographic commitment in its establishment event via a seal to the delegating event. Each commitment is signed respectively by the committer. Delegator signs its delegating event and delegate signs its delegated event. Consequently, in order to forge a delegation, an attacker must compromise both the pre-rotated signing keys of the delegate and either the current signing keys of the delegator when the delegating event is an interaction event and the pre-rotated signing keys of the delegator when the delegating event is a rotation. This cooperative delegation together with special superseding recovery rules for events enables cooperative recovery.

The rules are as follows: A rotation event may supersede an interaction event in the key event log of the delegator. The superseding delegating rotation or a subsequent delegating interaction may provide a cryptographic commitment to a superseding rotation event in the delegate's key event log that recovers control over the delegate's KEL. This superseding delegated rotation may supersede a rotation event not merely an interaction event. Thus in a delegated key event log a fork may occur at a rotation event not merely at an interaction event. The delegation chain of events determines if such a superseding event is valid. The special rule is that the delegator's keys in its KEL used to delegate the superseding event in the delegate's KEL must have been rotated in the delegator's KEL by a later rotation or interaction event in the delegator's KEL than the keys in the delegator's KEL used to delegate the superseded event in the delegate's KEL.

This superseding rule may be recursively applied to multiple levels of delegation, thereby enabling recovery of any set of keys signing or pre-rotated in any lower levels by a superseding rotation delegation at the next higher level. This cascades the security of the key management infrastructure of higher levels to lower levels. This is a distinctive security feature of the cooperative delegation of identifiers in KERI.

### 11.6.2 DDOS

An unresponsive controller may not promulgate a new event to any of the witnesses thereby delaying agreement on its own key events. Unresponsiveness on the part of the controller maybe either by choice or as the result of an attack on the controller. Because the non-responsiveness of a controller, however honest or not, is primarily a threat to the controller not to a validator there may be little incentive for self-inflicted unresponsiveness. However there may be significant incentive for an external attacker to delay the promulgation of new key events especially to prevent recovery via a new rotation event. An attacker may intercept the controller's network access by capturing control of the controller's gateways or routers. One way to mitigate this attack is for the controller to use multiple unpredictable network access points. Recall that a controller is self-authenticating against its authoritative key-pairs not its domain name or IP address. Thus any network access mechanism may be verifiable by witnesses. Similarly an attacker may interfere with the controller's network access using a distributed denial of service (DDOS) attack. This is only viable if the controller uses a public or known resolvable domain name or IP address to target. Using dynamically selected network access points may likewise mitigate a DDOS attack on the controller. Generally speaking, mechanisms to mitigate network access attacks are well known and are not, therefore, provided directly by this algorithm. In any case, this *algo-* *rithm* assumes that a controller may follow best practices to ensure a sufficient level of responsiveness for its needs.

Because the service provided by the witnesses as a group may be public, they each may require a publicly resolvable domain name or IP address. Consequently a distributed denial of service attack against a witness may not be so easy to mitigate. Recall however, that a witness is self-authenticating to its public key-pair(s) not its IP address. Consequently a controller make spin up clones of any witnesses under DDOS attack but at different IP addresses. The controller may then publicize these new addresses to any resolvers. The DDOS attack must then attack not only the clones but the resolvers. A controller may provision new clones much faster than a robotized DDOS attack may be able to shift its attack to the new clones. A controller may designate only a handful of witnesses but dozens of clones of each witness. At worst this slows down the witnessing but does not stop it. A DDOS attack must to take out all the clones of a majority of witnesses to be successful. Mitigation of DDOS may be achieved through horizontal scaling of each witness. This provides DDOS attack resistance to KERI.

Suppose for example, a controller designates seven witnesses with a threshold of five witnesses with at most two faulty witnesses for a properly witnessed event. The controller also spins up ten clone instances of each witness. A successful DDOS attack on one witness requires a successful attack on all ten of its clones. A successful DDOS attack on of the event requires a successful DDOS attack on all the clones of a least 3 witnesses. Given cloned witnesses, increasing the number of witnesses by even a small amount to allow for more faulty witnesses increases the degree of difficulty exponentially.

Moreover, the controller may create clones whose addresses are not publicly resolvable. These witnesses may then broadcast their receipts to ambient observers (validators, watchers, jurors, judges) (for example IPFS) that merely record receipted logs. The number of observers may be very large. Validators that use observers to access the logs may not then be impeded at all by a DDOS attack unless the DDOS attack also targets the observers. The number of potential watchers is unconstrained in KERI and contributes to ambient verifiability. This may make it prohibitively expensive to mount a successful DDOS attack.

## 11.7 Security Concerns.

As described in the previous section, the *algorithm* assumes a sufficiently responsive controller. Lack of responsiveness is primarily a threat to the controller not a validator. Consequently providing sufficient controller responsiveness is the responsibility of the controller not the *algorithm*. In contrast, a responsive but dishonest (or compromised) controller may pose a *live* threat to a validator with respect to new events never before seen by the validator. The *algorithm* must provide means for the validator to protect itself from such threats. When the controller is responsive but dishonest it may create inconsistent versions of an event that are first seen by different subsets of the witnesses. In the case where only $F$ of the witnesses are faulty despite a dishonest controller, the validator may protect itself by requiring a large enough *sufficient agreement* or *threshold of accountable duplicity*, $M_V$, that guarantees that either only one satisfying agreement or none at all, e.g. makes the service *immune*. To restate, the validator may select $M_V$ to ensure the the service is *immune* such that the service will either provide one and only one *proper* key event receipt log (KERL) or none at all. This protects the validator.

A greater threat to a validator may be that of a dishonest controller that may collude with its witnesses to promulgate alternative (divergent) event version agreements each with sufficient agreement. But this would violate the assumption of at most $F$ faulty witnesses. In this case, the witness consensus process i.e the *algorithm*, may not protect the validator. Protection must come from some other process under the validator's control. In this case, a validator may protect itself with duplicity detection via a set of observers (validators, watchers, jurors, judges). In such a case, in order to undetectably promulgate alternate but sufficiently accountable event version agreements, a dishonest controller with dishonest witnesses must prevent any validator from communicating with any other observer that may have seen any alternate event version agreement. This attack may be made practically unfeasible given a large and diverse enough set of observers. Once duplicity is detected that identifier loses all its value to any detecting validator. This imperils any dishonest controller who attempts such an attack.

The final threat is the threat of *dead* exploit where some time in the future the exposed key-pairs used to sign past events in a KERL may be compromised. The compromised keys may then be used to create an alternate or divergent verifiable event history. Recall, however, that a proper KERL enables validation of the controlling keys of the associated identifier over the time-frame of the events in the log. Once produced, a proper KERL may be provided by any observer (validator, watcher, juror, or judge) that has retained a copy of it not merely the witnesses. Subsequent compromise of a controller's keys and a compromise of witnesses may not invalidate any of the events in a pre-existent proper KERL.

Therefore, in order to fool a validator into accepting an erroneous or compromised divergent key event history, a successful exploiter must forge a *proper* KERL but with a different sequence of key events. To do this the exploiter must not only exploit the controller's signing keys that were authoritative at some event but also exploit $M$ of the $N$ designated witness's keys at that event as well. The exploiter must also prevent that validator from accessing any other but alternate proper KERL from any other observer (validator, watcher, juror, judger) that may have a copy as a check against such an attack. The combination of these tasks make such an exploit extremely difficult to achieve.

Consequently, even in the extreme case that some time in the future a complete and total *dead* exploit of the controller keys and at least $M$ of the witness' keys occurs such that they forge a seemingly proper but divergent KERL, any prior copy of a proper KERL will enable detection and proof of accountable duplicity of that *dead* exploit. In this case the validator may choose to use its prior copy or the prior copy from some set of *jurors* it trusts to determine which of the divergent KERLs is authoritative. This is similar to how certificate transparency works [70; 95; 97]. In order for such a *dead* attack to succeed the attacker must prevent a targeted validator from accessing any other copies of an alternate KERL. The idea of *ambient verifiability* mentioned above comes from the fact that the original KERL may be distributed among any number of watchers from whom a validator may obtain a copy. At some point the degree of accessibility to an original copy becomes essentially ubiquitous at which point verifiability may be considered ambient. Given *ambient verifiability* then duplicity detection becomes likewise ambient. To elaborate, a successful *dead* attack requires isolation of an validator from ambient sources of the KERL. In general, isolation from ambient sources may be prohibitively expensive. Consequently, *ambient verifiability* provides asymmetry between attacker and defender in the favor of the defender. Indeed the end goal of this autonomic identity system is to achieve ambient security in the sense that nearly anyone, anywhere, at anytime can become a verifiable controller of a verifiable identity that is protected by ambient verifiability and hence duplicity detection of the associated KERL.

Furthermore, any mutual interaction events between a validator and controller may provide proof of priority. In a mutual interaction, the validator includes a copy or digest of an interaction event sourced by the controller in an event sourced by the validator. A total compromise of the controller and all witnesses would not be able to forge the validator's signature on the mutual interaction event. Thus the existence of any mutual interaction events may then be used to prove priority even in the extremely unlikely case of a complete and total *dead* exploit of a controller and all of its witnesses.

Alternatively, in the case of a complete and total *dead* exploit, the validator and controller may jointly agree to use some other more formal mechanism to resolve the priority of divergent KERLs. This may be the median of the astronomical time of original reception of a receipt by a mutually trusted set of observers. Or this may be through the use of anchor transactions on a distributed consensus ledger. This later approach would only require minimal use of a distributed consensus ledger in order to resolve the most extreme and unlikely case of total *dead* exploit.

Finally, however unlikely, subsequent improvements in cryptographic attack mechanisms such as quantum computing may enable at some future time complete compromise all exposed key-pairs. One solution would be for the market to operate a trusted set of jurors that archive KERLs just in case of some such future total compromise. These trusted jurors may secure their archives with post quantum cryptography. Thus any post quantum attack may be detectable merely by appeal to one or more of these archives.

# 12 Event Semantics and Syntax

The following sections map the various options for key events onto a single universal syntax and associated semantics. This allows a single implementation for both servers and clients while yet accommodating application specific tuning of the features. In general, indirect replay mode with multiple signatures, witnesses and an interaction data payload comprises the most generic set of syntax and semantics. All the other variations are special cases. The convention is that missing elements have defined default values. A given variation may elide or leave empty or more of the elements in the associated events but the default semantics provide consistent interpretation. Thereby all the syntax variations for any event can be mapped to a single equivalent universal syntactical expression with defined semantics. This enables a single semantic implementation of the underlying verification and validation logic. The following sections describe these mappings.

Formally, the events defined below reference entities with identifiers that are bound to key-pairs. For example, a controller has an identifier prefix labeled $C$ that is bound to one or more key-pairs, depending on the prefix derivation, from a sequence of key-pairs starting at $\left(C^0, c^0\right)$. Likewise a validator has an identifier prefix labeled $V$ that is bound to one or more key-pairs, depending on the prefix derivation, starting at $\left(V^0, v^0\right)$. A set of witnesses indexed by subscript $i$ each have an identifier prefix $W_i\big|_{i=0,1,2,\ldots}$ that is bound to one or more key-pairs, depending on the derivation, starting with $\left(W_i^0, w_i^0\right)$. When clear from the context the identifier binding to a key-pair may simply represented by the non-superscripted symbol, such as, $C = C^0$, $V = V^0$, and $W_i = W_i^0$.

In the following event descriptions, the index $k$ indexes all key events of any *ilk*. The index $l$ indexes the subsequence of establishment (inception and rotation) events taken from the super sequence indexed by $k$. This means that in general the $k^{th}$ event may not be the same as the $l^{th}$ event. Only when the all events are establishment events will $k = l$. The inception event is special it always has $k = l = 0$. The inception event may be considered a special case of a rotation event. Non-establishment events may be interleaved between establishment events.

All events except the inception event include a digest of the previous event. The inception event has no digest because it is the first or zeroth event in the sequence. The digest backward chains the events and reinforces the sequence ordering provided by the sequence number. The sequence of digests and the sequence numbers must correspond.

## 12.1 General Inception

A diagram of the general inception event is shown as follows:

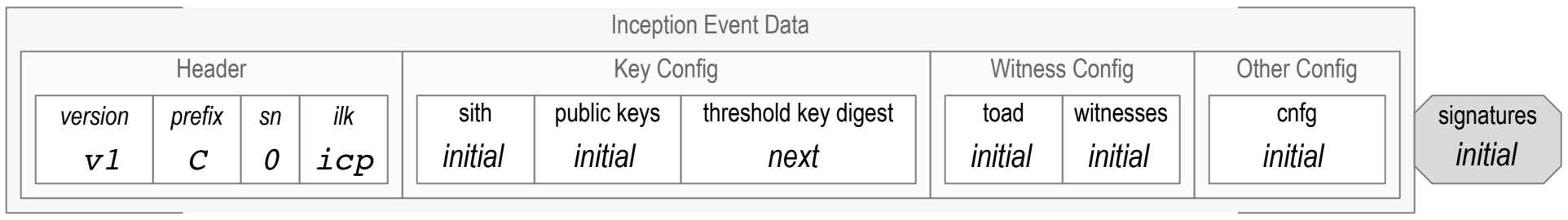

Figure 12.1. General Inception Event

The general symbolic expression for inception event appearing as the zeroth key event in the key event sequence for identifier prefix labeled *C* is as follows:

$$\varepsilon_0^C = \left\langle v_0^C, C, t_0^C, \mathtt{icp}, K_0^C, \widehat{C}_0^C, \eta_0^C\left(\left\langle K_1^C, \widehat{C}_1^C\right\rangle\right), M_0^C, \widehat{W}_0^C, [cnfg]\right\rangle \widehat{\sigma}_0^C, \tag{12.1}$$

where, $\varepsilon_0^C$ is the zeroth event in the event stream of identifier prefix *C*, $v_0^C$ is the protocol version (including encoding), *C* is the qualified identifier prefix, $t_0^C$ with value $t_0^C = 0$ is the unique monotonically increasing sequence number for this event , `icp` is the event *ilk* representing an inception event, $K_0^C$ is the required signing threshold (*sith*) of $L_0^C$ signers from the initial set of authoritative key-pairs, $\widehat{C}_0^C$, given by,

$$\widehat{C}_0^C = \left[C^0, \ldots, C^{L_0^C - 1}\right]_0^C, \tag{12.2}$$

where $\left[C^0, \ldots, C^{L_0^C - 1}\right]_0^C$ is the set of associated qualified public keys, these public keys are included in the derivation of prefix *C*, $K_0^C$ may be a number or when using a weighted threshold scheme, it may be a list or list of lists (see Section 15.), $\eta_0^C\left(\left\langle K_1^C, \widehat{C}_1^C\right\rangle\right)$ is the digest of the serialized data $\left\langle K_1^C, \widehat{C}_1^C\right\rangle$ for the next set of authoritative keys, where $K_1$, is the required the next threshold out of $L_1^C$ signers from , $\widehat{C}_1^C$, given by,

$$\widehat{C}_1^C = \left[C^{r_1}, \ldots, C^{r_1 + L_1^C - 1}\right]_1^C, \tag{12.3}$$

where $\left[C^{r_1}, \ldots, C^{r_1 + L_1^C - 1}\right]_1^C$ is the set of associated list of qualified public keys, $M_0^C$ is the *tally* or threshold of accountable duplicity (toad) for the number of witness receipts to the event,

$$\widehat{W}_0^C = \left[W_0^C, \ldots, W_{N_0^C - 1}^C\right]_0^C \tag{12.4}$$

is the list of the identifier prefixes for the total $N_0^C$ designated witnesses, $[cnfg]$ is an optional initial configuration data element that is represented as an ordered list of ordered mappings, depending on the serialization encoding this may be expressed as an ordered list of lists of (label, value) pairs that provides any additional confirmation data used in the derivation of *C*, each value set mush be unique, one class of mappings are unique *trait* values. for example when an element of $[cnfg]$ includes the mapping `("trait", "EstOnly")` then only establishment events are allowed in the key event stream, this may provide enhanced security, and finally $\widehat{\sigma}_0^C$ is the set of digital signatures each computed on the contents of the brackets,

$\langle\ \rangle$, that is,

$$\widehat{\sigma}_0^C = \sigma_{C^{s_0}} \ldots \sigma_{C^{s_{s_0^C-1}}} \,, \qquad (12.5)$$

where $\sigma_{C^{s_0}} \ldots \sigma_{C^{s_{s_0^C-1}}}$ is a set of attached signatures.

In a sense, an inception operation may be considered as a special case of rotation, that is, the zeroth rotation. The inception operation pre-rotates (declares) the subsequence $\widehat{C}_1^C$ to be the ensuing set of key-pairs. The inception event demonstrates control over the identifier prefix $C$ via the set of signatures $\widehat{\sigma}_0^C$ that must satisfy signing threshold $K_0^C$ out of total $L_0^C$ signers.

As a clarifying example, suppose the inception event is multi-signature, that is, $K_0^C = 2$ and $L_0^C = 3$. The next set of keys are also 2 of 3, that is, $K_1^C = 2$ and $L_1^C = 3$ and . Further, suppose also that sufficient witness receipts are 3 out of a total of 4, that is, $M_0^C = 3$ of $N_0^C = 4$ . Suppose the attached signatures are from $C^0$ and $C^2$. The corresponding inception event may be denoted as follows:

$$\left\langle \texttt{K1}, C, 0, \texttt{icp}, 2, \left[ C^0, C^1, C^2 \right], \eta_0^C, 3, \left[ W_0, W_1, W_2, W_3 \right], [\ ] \right\rangle \sigma_0 \sigma_2 \,, \qquad (12.6)$$

where $\eta_0^C\left(\left\langle 2, \left[ C^3, C^4, C^5 \right] \right\rangle\right)$ is the digest of the next set threshold and keys.

## 12.2 General Rotation

A diagram of the general rotation event is shown as follows:

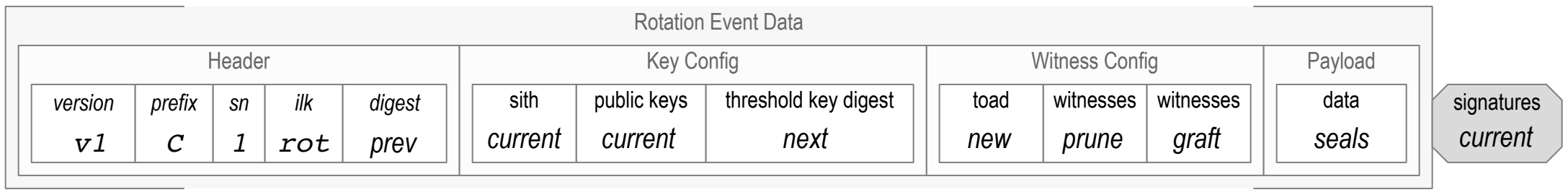

Figure 12.2. General Rotation Event

The general expression for a rotation event appearing as the $l^{th}$ event from the subsequence of establishment events that is also the $k^{th}$ key event from the key event sequence for identifier prefix labeled $C$, is as follows:

$$\varepsilon_k^C = \left\langle v_k^C, C, t_k^C, \eta_k^C\left(\varepsilon_{k-1}^C\right), \texttt{rot}, K_l^C, \widehat{C}_l^C, \eta_l^C\left(\left\langle K_{l+1}^C, \widehat{C}_{l+1}^C \right\rangle\right), M_l^C, \widehat{X}_l^C, \widehat{Y}_l^C, [seals] \right\rangle \widehat{\sigma}_{kl}^C \,, \qquad (12.7)$$

where, $\varepsilon_k^C$ is the $k^{th}$ event in the event stream of identifier prefix $C$ that is also the $l^{th}$ event in the subsequence of establishment events, $v_k^C$ is the protocol version (including encoding), $C$ is the qualified identifier prefix, $t_k^C$ with value $t_k^C = k$ is the unique monotonically increasing sequence number for this event , $\eta_k^C\left(\varepsilon_{k-1}^C\right)$ is the digest of the previous. $(k-1)^{th}$, key event , in the sequence, `rot` is the event *ilk* representing a rotation event, $K_l^C$ is the required signing threshold (*sith*) out of $L_l^C$ signers of the set of current authoritative keys $\widehat{C}_l^C$ given by,

$$\widehat{C}_l^C = \left[ C^{r_l^C}, \ldots, C^{r_l^C + L_l^C - 1} \right]_l^C \,, \qquad (12.8)$$

$r_l^C$ is the starting index of the subsequence of controlling key-pairs for this the $l^{th}$ rotation, $K_l^C$ may be a number or when using a weighted threshold scheme, it may be a list or list of lists (see

Section 15.), $\eta_l^C\left(\left\langle K_{l+1}^C, \widehat{C}_{l+1}^C \right\rangle\right)$ is the digest of the serialized data, $\left\langle K_{l+1}^C, \widehat{C}_{l+1}^C \right\rangle$ for the next set of authoritative threshold and keys where $K_{l+1}^C$ is the required threshold of $L_{l+1}^C$ signers from the next set of authoritative key-pairs, $\widehat{C}_{l+1}^C$, given by,

$$\widehat{C}_{l+1}^C = \left[ C^{r_{l+1}^C}, \ldots, C^{r_{l+1}^C + L_{l+1}^C - 1} \right]_{l+1}^C, \tag{12.9}$$

where $\left[ C^{r_{l+1}^C}, \ldots, C^{r_{l+1}^C + L_{l+1}^C - 1} \right]_{l+1}^C$ is the set of associated list of qualified public keys, $r_{l+1}^C$ is the starting index of the subsequence of controlling key-pairs for this the $(l+1)^{th}$ rotation, $M_l^C$ is the *tally* or threshold of accountable duplicity (toad) out of a total of $N_l^C$ of witness receipts to the event, $\widehat{X}_l^C$ is the $l^{th}$ prune list of $O_l^C$ witnesses to remove from the witness list, given by,

$$\widehat{X}_l^C = \left[ X_0^C, \ldots, X_{O_l^C - 1}^C \right]_l^C \tag{12.10}$$

where $\left[ X_0^C, \ldots, X_{O_l^C - 1}^C \right]_l^C$ is a list qualified identifier prefixes of the pruned witnesses, $\widehat{Y}_l^C$ is the $l^{th}$ graft list of $P_l^C$ witnesses to append to the witness list, given by,

$$\widehat{Y}_l^C = \left[ Y_0^C, \ldots, Y_{P_l^C - 1}^C \right]_l^C, \tag{12.11}$$

where where $\left[ Y_0^C, \ldots, Y_{P_l^C - 1}^C \right]_l^C$ is a list qualified identifier prefixes of the grafted witnesses, $[seals]$ is an array of zero or more seals, each seal is an ordered mapping that depending on the serialization encoding may be expressed as a list of lists of (label, value) pairs that may be used to anchor an interaction event data or a delegation (details of interaction or delegation are provided later), and finally $\widehat{\sigma}_{kl}^C$ is the set of digital signatures attached to the $k^{th}$ event using the keys drawn from the $l^{th}$ set of authoritative keys $\widehat{C}_l^C$, each computed on the contents of the brackets, $\langle \; \rangle$,that is,

$$\widehat{\sigma}_{kl}^C = \sigma_{C^{r_l^C + s_0}} \ldots \sigma_{C^{r_l^C + s_{S_{kl}^C - 1}}}, \tag{12.12}$$

where $\sigma_{C^{r_l^C + s_0}} \ldots \sigma_{C^{r_l^C + s_{S_{kl}^C - 1}}}$ is a set of attached signatures, $r_l^C$ is the starting index of the subsequence of controlling key-pairs for this the $l^{th}$ rotation.

As a clarifying example, lets reuse the inception event example above, that is,

$$\left\langle \texttt{K1}, C, 0, \texttt{icp}, 2, \left[ C^0, C^1, C^2 \right], \eta_0^C, 3, \left[ W_0, W_1, W_2, W_3 \right], [\;] \right\rangle \sigma_0 \sigma_2, \tag{12.13}$$

where the next set of keys was committed to in the digest, $\eta_0^C\left(\left\langle 2, \left[ C^3, C^4, C^5 \right] \right\rangle\right)$, this is followed by the rotation event,

$$\left\langle \texttt{K1}, C, 1, \texttt{rot}, 2, \left[ C^3, C^4, C^5 \right], \eta_l^C, 3, \left[ X_0 \right], \left[ Y_0 \right], [\;] \right\rangle \sigma_4 \sigma_5 . \tag{12.14}$$

where $\eta_l^C\left(\left\langle 2, \left[ C^6, C^7, C^8 \right] \right\rangle\right)$ is the digest of the next set of keys. The inception event, $\varepsilon_0$, always implies that $K_0 = 1$, $L_0 = 1$, and $r_0 = 0$. Given that the inception event initial key set has 3 keys we can compute $r_1 = 3$. The inception event also declares that $K_1 = 2$ and the length of next keys list is $L_1 = 3$. The rotation event, $\varepsilon_1$, has $\left[C^3, C^4, C^5\right]$ for the current set of keys. The rotation event declares that $K_2 = 2$ and the length of the next keys list is $L_2 = 3$. The tally is still 3. It has one pruned witness whose public key is $X_0 = W_1$ and one grafted witness whose public key is $Y_0 = W_4$ leaving a total of 4 witnesses.

## 12.3 General Interaction

Interaction events may be conveyed in three different ways. These are as a distinct interaction event message, as a payload in an extended rotation event message, and as a distinct delegation event message. The payload may express a single interaction or transaction, a block of interactions, or a Merkle root of a hash tree of a block of interactions that are thereby anchored to the event sequence at the location of the interaction event [104].

A diagram of the general interaction event is shown as follows:

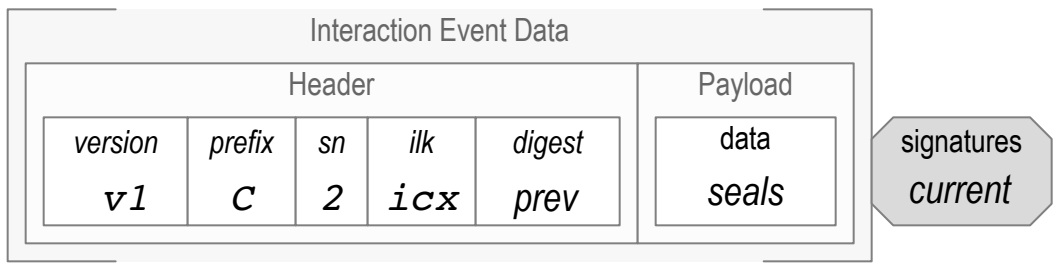

Figure 12.3. General Interaction Event

The general expression for the generic interaction event is as follows:

$$\varepsilon_k^C = \left\langle v_k^C, C, t_k^C, \eta_k^C\left(\varepsilon_{k-1}^C\right), \mathtt{ixn}, \left[seals\right] \right\rangle \widehat{\sigma}_{kl}^C, \tag{12.15}$$

where, $\varepsilon_k^C$ is the $k^{th}$ event in the event stream of identifier prefix *C* with the most recent rotation being the $l^{th}$ event in the subsequence of establishment events, $v_k^C$ is the protocol version (including encoding), *C* is the qualified identifier prefix, $t_k^C$ with value $t_k^C = k$ is the unique monotonically increasing sequence number for this event, $\eta_k^C\left(\varepsilon_{k-1}^C\right)$ is the digest of the previous key event , $\left(k-1\right)^{th}$, in the sequence, `ixn` is the event *ilk* representing an interaction event, $\left[seals\right]$ is an array of one or more seals, each seal is an ordered mapping data structure element, that depending on the serialization encoding this may be expressed as a list of lists of (label, value) pairs, each seal provides a cryptographic commitment to the details of data associated with the actual interaction(s) conveyed by this event, although not supplied explicitly in this event, the set of current authoritative keys, $\widehat{C}_l^C$, of length $L_l^C$ was supplied by the most recent, that is, $l^{th}$, establishment event and is given by,

$$\widehat{C}_l^C = \left[C^{r_l^C}, \ldots, C^{r_l^C + L_l^C - 1}\right]_l^C, \tag{12.16}$$

where $\left[C^{r_l^C}, \ldots, C^{r_l^C + L_l^C - 1}\right]_l^C$ is a list of qualified public keys, the current threshold, $K_l^C$ out of $L_l^C$ signers was also supplied by the most recent, that is, $l^{th}$, establishment event , $r_l^C$ is the starting index of the subsequence of controlling key-pairs from the $l^{th}$ rotation, and finally $\widehat{\sigma}_{kl}^C$ is the set of digital signatures attached to the $k^{th}$ event using the keys indexed by $\widehat{s}_{kl}^C$ and drawn from the $l^{th}$ set of authoritative keys $\widehat{C}_l^C$, each computed on the contents of the brackets, $\langle\ \rangle$,that is,

$$\widehat{\sigma}_{kl}^C = \sigma_{C^{r_l^C + s_0}} \ldots \sigma_{C^{r_l^C + s_{S_{kl}^C - 1}}}, \tag{12.17}$$

where $\sigma_{C^{r_l^C+s_0}} \ldots \sigma_{C^{r_l^C+s_{S_{kl}^C-1}}}$ is a set of attached signatures, $r_l^C$ is the starting index of the subsequence of controlling key-pairs from the $l^{th}$ rotation.

## 12.4 Delegation

The section defines delegative inception and rotation operations for managing a different set of signing keys. The examples on show one level of delegation is supported. These may be extended recursively to support multiple levels of delegation. The controlling keys are used to establish ( incept and rotate) other identified key event streams that provide signing keys. In these definitions, the identifier prefix labels are $C$ for the delegator and $D$ for the delegate. The event indices from the two key event sequences, one for delegator and one for delegate are labeled with the respective prefix. Recall that delegation may be performed in the delegating key event sequence with either a delegating interaction event for a delegating extended rotation event. Diagrams of the associated delegating events are shown as follows:

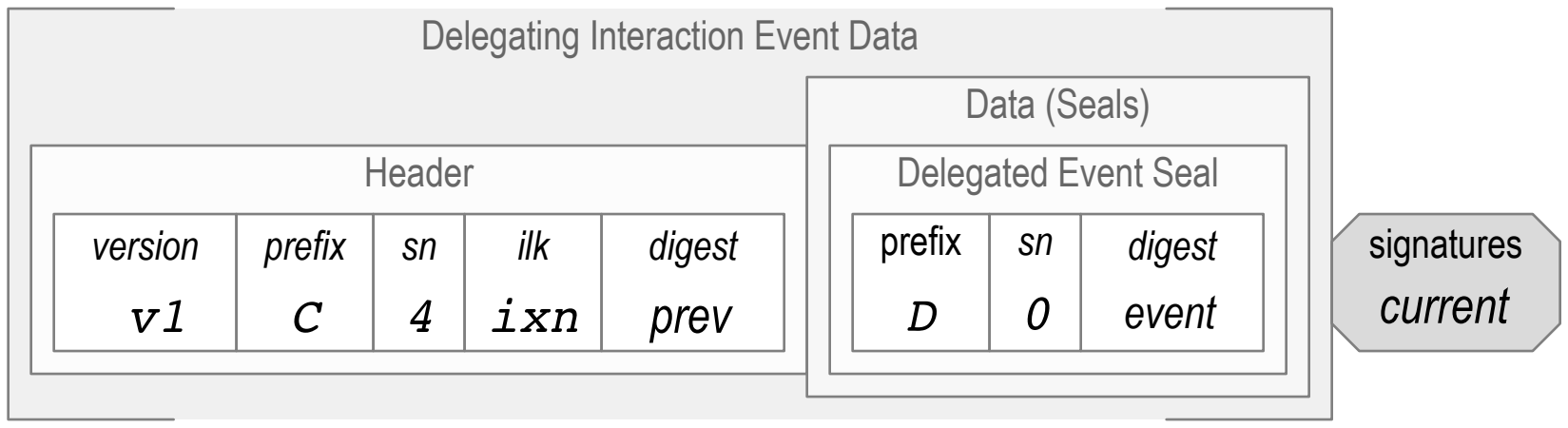

Figure 12.4. Delegating Interaction Event.

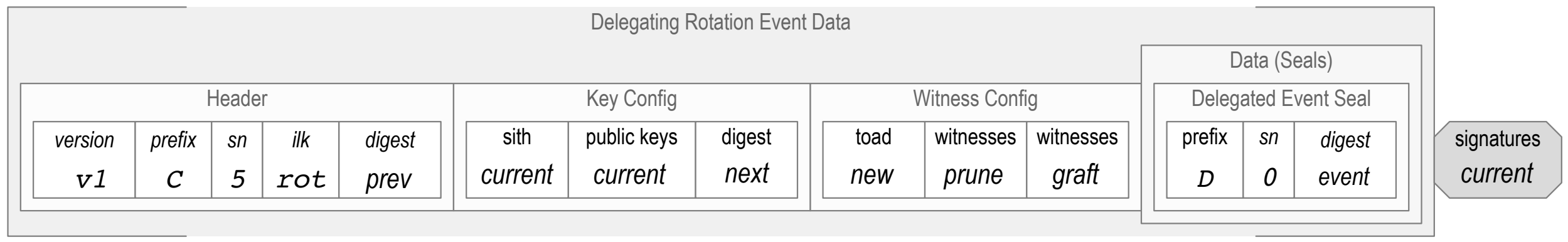

Figure 12.5. Delegating Rotation Event.

Events associated with delegation include what is called a delegation seal. A pair of seals cross anchor a delegating and delegated event pair. The format of the seals is diagrammed as follows:

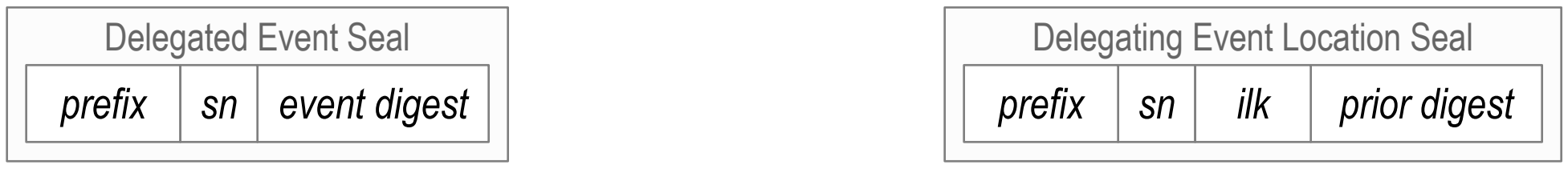

Figure 12.6. Delegation Seals.

The delegated event seal digest is computed on the complete serialized delegated event. The delegating event location seal indicates the unique location of the delegating event.

### 12.4.1 Inception Delegation

Delegation of inception is performed with a delegating event in the delegator's event sequence. In the delegating event, the value of its [*seals*] element is an array of one or more *delegation seals*, each seal is an ordered mapping data structure that depending on the serialization encoding this may be expressed as a list of lists of (label, value) pairs, that is, $seal = \widehat{\Delta}_0^D$, where $\widehat{\Delta}_0^D$, is called the delegator's incepting delegating seal that is delegating event 0 in the delegate's event stream. This may be denoted as follows:

$$\widehat{\Delta}_0^D = \left\{ D, t_0^D, \eta_k^C\left(\varepsilon_0^D\right) \right\}, \tag{12.18}$$

where $\hat{\Delta}_0^D$ delegates the inception of $D$ in the zeroth event of $D$ from the event in $C$ where it appears, $t_0^D$ is the sequence number of the delegated inception event, that is, 0, and, $\eta_k^C\left(\varepsilon_0^D\right)$ is a digest the inception (zeroth) event of $D$ from the $k^{th}$ event of $C$, $\varepsilon_0^D$ is the complete serialized inception event of $D$.

## 12.4.2 Delegated Inception

The associated inception event in the key event sequence for D may be denoted as follows;

$$\varepsilon_0^D = \left\langle v_0^D, D, t_0^D, \texttt{dip}, K_0^D, \hat{D}_0^D, M_0^D, \hat{W}_0^D, \left[cnfg\right], \hat{\Delta}_k^C \right\rangle \hat{\sigma}_0^D \tag{12.19}$$

where $\varepsilon_0^D$ is the zeroth event in the event stream of D, $v_0^D$ is the version of the incepting event, $D$ is the quantified identifier prefix, $t_0^D$ is the sequence number, `dip` is the event *ilk* for delegated inception, $K_0^D$ is the required threshold of $L_0^D$ signers from the delegated set of authoritative key-pairs, $\hat{D}_0^D$, given by,

$$\hat{D}_0^D = \left[ D^0, \ldots, D^{L_0^D - 1} \right]_0^D, \tag{12.20}$$

where $\left[ D^0, \ldots, D^{L_0^D - 1} \right]_0^D$ is the set of associated list of qualified public keys, $K_0^D$ may be a number or when using a weighted threshold scheme, it may be a list or list of lists (see Section 15.), , $M_0^D$ is the *tally* or threshold of accountability out of a total of $N_0^D$ for the number of witness receipts to the event, $\hat{W}_0^D$ is the list of witnesses denoted,

$$\hat{W}_0^C = \left[ W_0^C, \ldots, W_{N_0^C - 1}^C \right]_0^C \tag{12.21}$$

where $\left[ W_0^D, \ldots, W_{N_0^D - 1}^D \right]_o^D$ is the list qualified identifier prefixes for the $N_0^D$ designated witnesses, $\left[cnfg\right]$ is a list strings of configuration traits or options, each configuration string value must be unique., and finally, $\hat{\Delta}_k^C$ is the delegate's incepting delegated seal that refers to the $k^{th}$ event of $C$ as the delegating event. $\hat{\Delta}_k^C$ references the delegating inception of $D$ from the $k^{th}$ key event of $C$ and may be denoted as follows:

$$\hat{\Delta}_k^C = \left\{ C, t_k^C, ilk, \eta_k^C\left(\varepsilon_{k-1}^C\right) \right\} \tag{12.22}$$

where $C$ is the qualified prefix for the delegating event stream, in other words $C$ refers to the delegator, $t_k^C$ is the sequence number of the $k^{th}$ event in the key event stream of delegator $C$, *ilk* is the delegator's $k^{th}$ event ilk (`ixn`, or `rot`), and $\eta_k^C\left(\varepsilon_{k-1}^C\right)$ is the prior event digest appearing in the delegator's $k^{th}$ event that is the digest of the prior, $\left(k-1\right)^{th}$, event $\varepsilon_{k-1}^C$ of $C$, and finally $\hat{\sigma}_0^D$ is the set of digital signatures computed on the contents of the brackets, $\langle \; \rangle$, that is,

$$\hat{\sigma}_0^D = \sigma_{D^{s_0}} \ldots \sigma_{D^{s_{S_0^D - 1}}}, \tag{12.23}$$

where $\sigma_{D^{s_0}} \ldots \sigma_{D^{s_{S_0^D - 1}}}$ is a set of attached signatures. A diagram of the the delegated inception event is shown below:

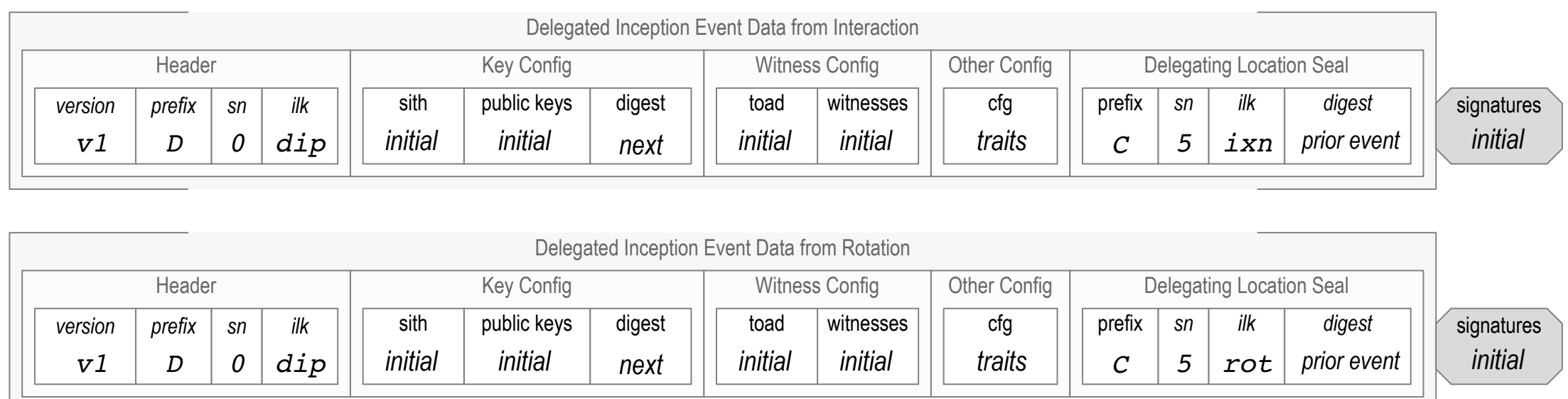

Figure 12.7. Delegated Inception Events.

### 12.4.3 Rotation Delegation

Delegation of rotation is performed with an extended rotation event that is the delegating event. In the delegating event, the value of its $[seals]$ element is an array of one or more *delegation seals*, each seal is an ordered mapping data structure that depending on the serialization encoding this may be expressed as a list of lists of (label, value) pairs, that is, $seal = \widehat{\Delta}_k^D$, where $\widehat{\Delta}_k^D$, is called the delegator's rotating delegating seal that delegates the $k^{th}$ event in D and may be denoted as follows:

$$\widehat{\Delta}_k^D = \left\{D, t_k^D, \eta_k^C\left(\varepsilon_k^D\right)\right\}, \qquad (12.24)$$

where $\widehat{\Delta}_k^D$ delegates the rotation of $D$ in the $k^{th}$ event of D from the event of $C$ where it appears, $t_k^D$ is the sequence number of the $k^{th}$ event of $D$, and $\eta_k^C\left(\varepsilon_k^D\right)$ is a digest of the complete serialized $k^{th}$ event of D appearing in the $k^{th}$ event of $C$ where the the two $k$s may not be the same.

### 12.4.4 Delegated Rotation

The associated rotation event in the key event sequence for D may be denoted as follows;

$$\varepsilon_k^D = \left\langle v_k^D, D, t_k^D, \eta_k^D\left(\varepsilon_{k-1}^D\right), \texttt{drt}, K_l^D, \widehat{D}_l^D, M_l^D, \widehat{X}_l^D, \widehat{Y}_l^D, [seals], \widehat{\Delta}_k^C \right\rangle \widehat{\sigma}_{kl}^D \qquad (12.25)$$

where $\varepsilon_k^D$ is the $k^{th}$ event in the event stream of D, $v_k^D$ is the version of the rotating event, $D$ is the quantified identifier prefix, $t_k^D$ is the sequence number, $\eta_k^D\left(\varepsilon_{k-1}^D\right)$ is the digest of the previous key event , $(k-1)^{th}$, in the sequence, `drt` is the event *ilk* for delegated rotation, $K_l^D$ is the required threshold of $L_l^D$ signers in the from the $l^{th}$ delegated set of authoritative key-pairs, $\widehat{D}_l^D$ for $D$, given by,

$$\widehat{D}_l^D = \left[D^{r_l^D}, \ldots, D^{r_l^D+L_l^D-1}\right]_l^D, \qquad (12.26)$$

where $\left[D^{r_l^D}, \ldots, D^{r_l^D+L_l^D-1}\right]_l^D$ is the set of associated list of qualified public keys, $K_l^D$ may be a number or when using a weighted threshold scheme, it may be a list or list of lists (see Section 15.), $M_l^D$ is the *tally* or threshold of accountability out of a total of $N_l^D$ for the number of witness receipts to the event, $\widehat{X}_l^D$ is the $l^{th}$ prune list of $O_l^D$ witnesses to remove from the witness list, given by,

$$\widehat{X}_l^D = \left[X_0^D, \ldots, X_{O_l^D-1}^D\right]_l^D \qquad (12.27)$$

where $\left[X_0^D,\ldots,X_{O_l^D-1}^D\right]_l^D$ is a list qualified identifier prefixes of the pruned witnesses, $\widehat{Y}_l^D$ is the $l^{th}$ graft list of $P_l^D$ witnesses to append to the witness list, given by,

$$\widehat{Y}_l^D=\left[Y_0^D,\ldots,Y_{P_l^D-1}^D\right]_l^D, \tag{12.28}$$

where where $\left[Y_0^D,\ldots,Y_{P_l^D-1}^D\right]_l^D$ is a list qualified identifier prefixes of the grafted witnesses, $[seals]$ is a list seals which each an ordered mapping, and finally, $\widehat{\Delta}_k^C$ is the delegate's rotating delegated seal that refers to the $k^{th}$ event of *C* as the delegating event. $\widehat{\Delta}_k^C$ references the delegating rotation of *D* from the $k^{th}$ key event of *C* and may be denoted as follows:

$$\widehat{\Delta}_k^C=\left\{C,t_k^C,ilk,\eta_k^C\left(\varepsilon_{k-1}^C\right)\right\} \tag{12.29}$$

where *C* is the qualified prefix for the delegating event stream, in other words *C* refers to the delegator, $t_k^C$ is the sequence number of the $k^{th}$ event in the key event stream of delegator *C*, *ilk* is the delegator's $k^{th}$ event ilk (`ixn`, or `rot`), and $\eta_k^C\left(\varepsilon_{k-1}^C\right)$ is the prior event digest appearing in the delegator's $k^{th}$ event that is the digest of the prior, $\left(k-1\right)^{th}$, event $\varepsilon_{k-1}^C$ of *C*, and finally $\widehat{\sigma}_{kl}^D$ is the set of digital signatures computed on the contents of the brackets, $\langle\ \rangle$, that is,

$$\widehat{\sigma}_{kl}=\sigma_{C^{+r_l^D+s_0}}\ldots\sigma_{C^{r_l^D+s_{S_{kl}^D}-1}}, \tag{12.30}$$

where $\sigma_{C^{+r_l^D+s_0}}\ldots\sigma_{C^{r_l^D+s_{S_{kl}^D}-1}}$ is a set of attached signatures.

A diagram of the the delegated rotation event is shown below:

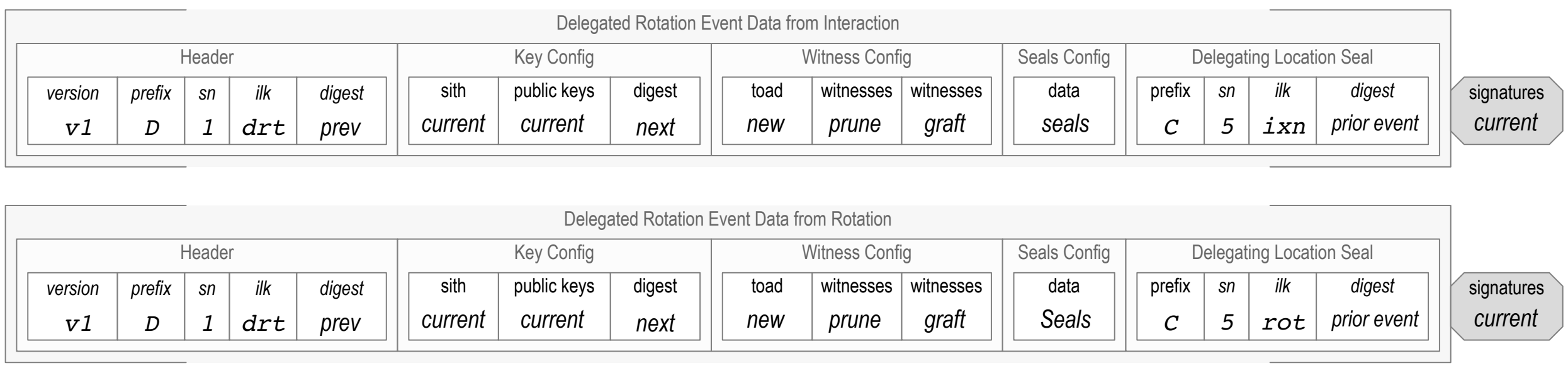

Figure 12.8. Delegated Rotation Event.

### 12.4.5 Delegated Interaction

The delegated interaction event is in every way identical to the non-delegated generic interaction event with the exception that the identifier prefix belongs to a delegated event sequence. Let D be the label for the delegated identifier prefix. The general expression for the delegated generic interaction event is as follows:

$$\varepsilon_k^D=\left\langle v_k^D,D,t_k^D,\eta_k^D\left(\varepsilon_{k-1}^D\right),\texttt{ixn},\left[seals\right]\right\rangle\widehat{\sigma}_{kl}^D, \tag{12.31}$$

where the elements are defined as above. This event is diagrammed as follows:

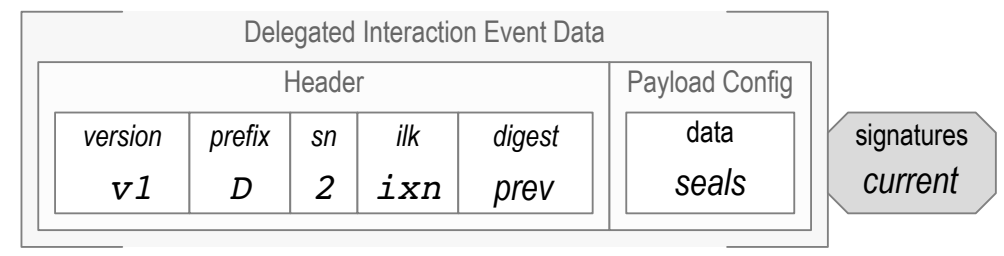

Figure 12.9. Delegated Interaction Event.

### 12.4.6 Algorithm for Generating Delegation Events

Given the complexity of the cross-anchoring process using delegation seals between the pair of delegation events (delegating and delegated), a description of the steps to generate the delegation events is provided below for clarity.

1. Create a partial delegating event, either interaction or rotation as appropriate.
2. Fill in the *prefix*, *sn*, *ilk* and prior *digest* fields in the header of the event.
3. Create a delegating event location seal using the delegating *prefix*, *sn*, *ilk* and prior *digest* fields from the header of the delegating event above.
4. Create a complete delegated event setting its *seal* field to the location seal created above.
5. Compute the digest of complete serialized delegated event.
6. Create a delegated event seal of the delegated event using delegated *prefix*, *sn* and and computed *digest* of complete delegated event
7. Add the delegated event seal to data field list (seals) of the delegating event.
8. Repeat steps 4 through 7 above for any other delegated events.
9. Complete the creation of the delegating event.

## 12.5 Event Receipt Messages

The structure of the event receipt message depends on whether or not the signers of the receipt use a transferable or non-transferable identifier prefix. Because witnesses and watchers use non-transferable prefixes, the public key for verifying their receipt signature may be extracted from their prefix. This simplifies the receipt. This also applies to validators that use non-transferable prefixes. In the case of a validator with a non-transferable prefix then a different event receipt message is needed.

### 12.5.1 Event Receipt Message from Non-Transferable Prefix

In general, when using non-transferable identifier prefixes, a validator receipt, a witness receipt, or a watcher receipt, all have the same format, the only difference is the identifier prefix of the respective entity that creates the receipt. Thus the same receipt message may be used for all. Moreover, communications between witnesses or watchers may benefit from batch sending of multiple receipts for a given event. Batch receipts may be more efficiently transmitted with a receipt message that allows multiple prefixes and signatures in one message for a given event rather than sending multiple messages for that same given event. Furthermore, a multi-receipt message need only be slightly more verbose than the single receipt version so that it is also usable when there is only one receipt. Therefore we only need to implement a multi-receipt version.

Because derivation codes allow the parsing of concatenated identifier prefixes and signatures, a compact multi-receipt message may use multiple attached couplets, each with a witness identifier and a signature. A sequence of these couplets may follow the receipt message that identifies the event. The following diagram illustrates the format.

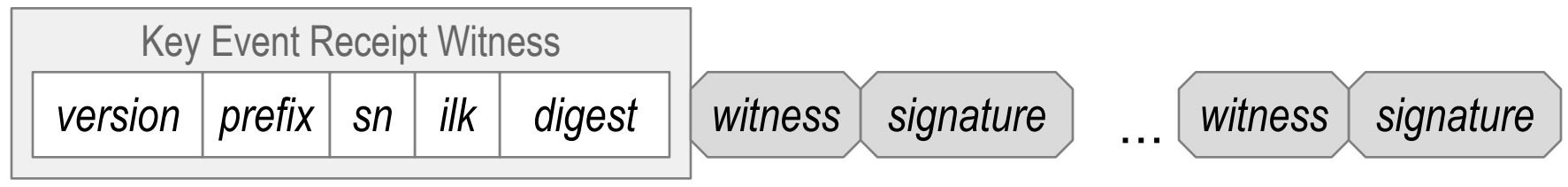

Figure 12.10. Multi-Witnessed Receipt.

Because the controller has a copy of the event message, a compact receipt message created by a witness need not include a copy of the event message but may include only the protocol version, the identifier prefix of the controller, the event ilk, `rct`, the identifier of the witness and the signature of the witness. A compact streaming multi-receipt message for identifier prefix *C* may be denoted as follows:

$$\rho^{C}_{\tilde{W}^{C}_{ls}}\left(\varepsilon^{C}_{k}\right)=\left\langle \nu^{C}_{k}, C, t^{C}_{k}, \texttt{rct}, \eta^{C}_{k}\left(\varepsilon^{C}_{k}\right)\right\rangle W^{C}_{l0}\sigma^{C}_{W^{C}_{l0}},\ldots,W^{C}_{lN^{C}_{s}-1}\sigma^{C}_{W^{C}_{lN_s-1}} \tag{12.32}$$

where $\rho^{C}_{\tilde{W}^{C}_{lk}}\left(\varepsilon^{C}_{k}\right)$ is the multi-receipt on event $\varepsilon^{C}_{k}$, $\nu^{C}_{k}$ is the protocol version, *C* is the identifier of the controller, $t^{C}_{k}$ is the sequence number of the event $\varepsilon^{C}_{k}$, `rct` is the event *ilk* representing an event receipt, $\eta^{C}_{k}\left(\varepsilon^{C}_{k}\right)$ is the digest of the event to enable lookup, $W^{C}_{l0}\sigma^{C}_{W^{C}_{l0}},\ldots,W^{C}_{lN^{C}_{s}-1}\sigma^{C}_{W^{C}_{lN_s-1}}$ is a sequence of couplets of length $N^{C}_{s}$ for whom signatures are included in the receipt, each couplet consists of a fully qualified witness identifier and a fully qualified signature from that witness, *l* represents the establishment event witness set and *s* represents the set of witnesses in this receipt.

To clarify, a witness, $W^{C}_{li}$, from the witness set of an identifier prefix *C*, where *l* denotes the establishment event for the witness set and *i* is index of the the witness within the set, may create an event receipt by signing with its associated key-pair, $\left(W^{C}_{li}, w^{C}_{li}\right)$. Recall that witness identifier prefixes are non-transferable and derived from a single key-pair. The receipt signature may be denoted as follows:

$$\sigma^{C}_{W^{C}_{li}}\left(\left\langle \varepsilon^{C}_{k}\right\rangle\right), \tag{12.33}$$

where $\varepsilon^{C}_{k}$ is the event. The specific event details are dependent on the event itself.

### 12.5.2 Event Receipt Message from Transferable Prefix

A validator, *V*, may create an event receipt by signing with its associated key-pair(s). Given the simple case of a non-transferable prefix with a single key, the the validator receipt is the same as a witness receipt above. However when the validator is using a transferable identifier then the receipt needs to include a seal of the latest establishment event for the validator in order that the correct set of signing keys may be used to verify the receipt signatures. The identifier prefix for a validator may make use of multiple keys and therefore require multiple signatures.

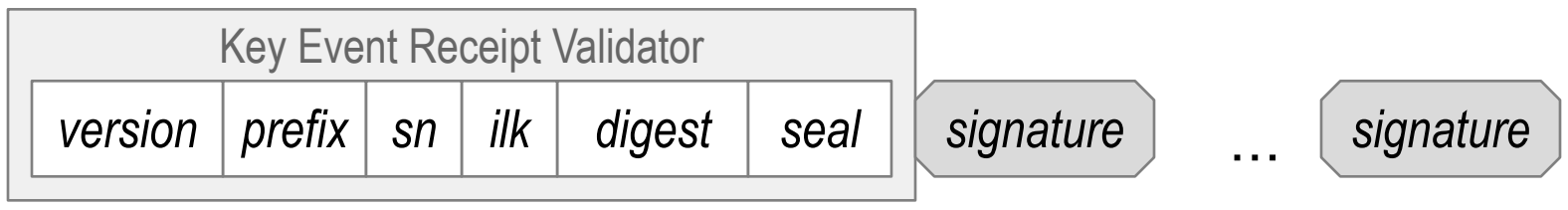

Figure 12.11. Multi-signature Validator Receipt

When using only a single signature, the receipt signature of on key event $\varepsilon^{C}_{k}$ from *V* may be denoted as follows:

$$\sigma^{C}_{V}\left(\left\langle \varepsilon^{C}_{k}\right\rangle\right). \tag{12.34}$$

where $\varepsilon_k^C$ is the event. The specific event details are dependent on the event itself. When the validator is using a multi-signature scheme then a set of signatures will be needed. This set may be represented as follows:

$$\widehat{\sigma}_{V_l}^C \tag{12.35}$$

Because the controller has a copy of the event message, a compact receipt message created by the validator need not include a copy of the event message but may include only the protocol version, the identifier prefix of the controller, the event ilk, `vrc`, a digest of the event, and an event seal of the latest establishment event of the validator and the signature(s) of the validator. Such a compact event receipt message, $\rho_V^C\left(\varepsilon_k^C\right)$, may be denoted as follows:

$$\rho_V^C\left(\varepsilon_k^C\right)=\left\langle v_k^C, C, t_k^C, \mathtt{vrc}, \eta_k^C\left(\varepsilon_k^C\right), \widehat{\Delta}_k^V\right\rangle\widehat{\sigma}_{V_l}^C \tag{12.36}$$

where $\rho_V^C\left(\varepsilon_k^C\right)$ is the receipt on event $\varepsilon_k^C$, $v_k^C$ is the protocol version, *C* is the identifier of the controller, $t_k^C$ is the sequence number of the event $\varepsilon_k^C$, `vrc` is the event *ilk* representing a validator event receipt, $\eta_k^C\left(\varepsilon_k^C\right)$ is the digest of the event to enable lookup, $\widehat{\Delta}_k^V$ is the event seal of the latest establishment event of the validator, and $\widehat{\sigma}_{V_l}^C$ is the set of signatures of the validator on the key event message labeled $\varepsilon_k^C$. The event seal, $\widehat{\Delta}_k^V$, is denoted as follows:

$$\widehat{\Delta}_k^V=\left\{V, \eta_k^V\left(\varepsilon_k^V\right)\right\}, \tag{12.37}$$

where *V* is the identifier prefix of the validator and $\eta_k^V\left(\varepsilon_k^V\right)$ is the digest of the latest establishment event of the validator, the *k* in this case may be different than the *k* of the event in the controller's, *C*, event sequence.

# 13 Implementation

## 13.1 State Verification Engine

Using the general representations provided above for the three events kinds, namely, inception, rotation, and interaction a generic state verification algorithm or engine can be formulated. This becomes the KERI verifier core and would be portable to all applications. The state verifier engine maintains the current verified state. The state engine process events from an event stream and updates the state from the current to the next state as a function of the current state and a verified event message. Processing is largely signature verification. The state maintains the list of designated witnesses but does not perform any validation against the number of event receipts. A witness, watcher, judge, juror, or validator application would include the KERI state engine to provide event message verification. A controller application also maintains its copy of the state with a complementary state generator engine that instead of updating state based on external events, updates state based on internally generated events. A diagram of the KERI event message state verification or state engine is shown below:

KERI Core — State Verifier Engine

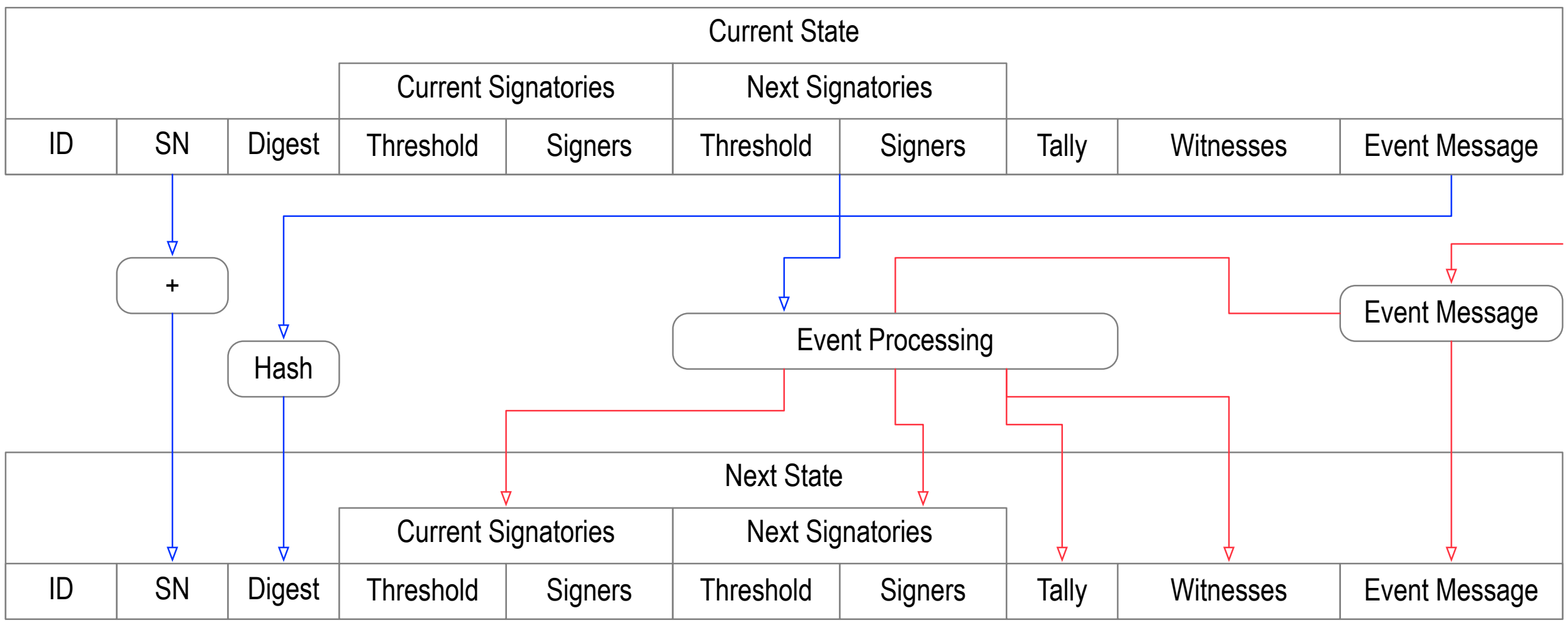

Figure 13.1. KERI Core State Verification Engine

## 13.2 Delegated State Verification Engine

A delegated key event stream has a slightly different verification engine. It includes delegation data and requires verification of the delegator's delegating event. Its diagram is shown below:

KERI Delegated Core — State Verifier Engine

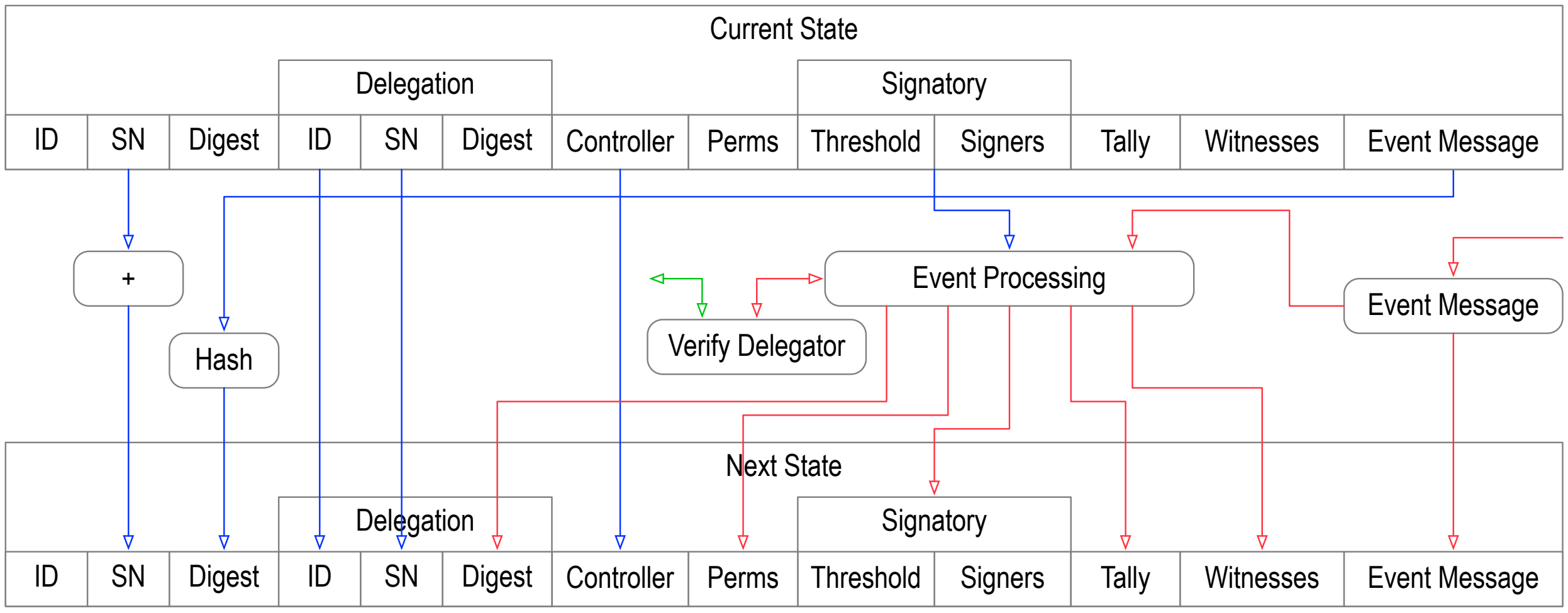

Figure 13.2. KERI Delegated Core State Verifier Engine

## 13.3 Implementation Choices

Because KERI is designed to be compatible with event streaming applications, its design lends itself to a simple state verification engine with compact and efficient syntax. Furthermore KERI has advanced key management features. KERI provides for reconfigurable thresholded multiple signature schemes where both the threshold total number of signatures may change are each rotation. The pre-rotation makes a forward commitment to an unexposed key-pair(s) that provides security that may not be undone via exploit of any exposed key-pairs. This allows for reconfigurability without sacrificing security. These advanced key management features alone make KERI desirable even in non-data streaming applications where scalability and performance or not so important. Likewise many applications for other reasons require a distributed consensus ledger. In that case the best approach might be to use KERI but without witnesses and leverage the trust provided by the distributed consensus nodes. A smart contract system like Ethereum (public or private) has the capacity to support the semantics of the KERI Core state verification engine on chain. Alternatively, KERI could be implemented as a side state channel with a Judge or Validator periodically anchoring the current state to a distributed ledger such as Ethereum or Bitcoin. This side channel approach will work with most distributed ledgers.

As a specific comparison, consider other smart contract based systems such as identifiers created with the ERC-1056 standard [60; 62] or control of tokenized assets using the Gnosis MuliSigWallet on Ethereum [106]. Both are vulnerable to capture via exploit of the exposed signing key-pairs. In the case of ERC-1056 only one key-air need be exploited or a threshold number of key-pairs for MultiSig. The Gnosis MultiSig wallet is a smart contract with advanced features for multiple signatures. It allows changes to the threshold and number of signatures as well as revoking and replacing signatures. In this sense it is of comparable complexity to the KERI engine. The crucial limitation, however, with the Gnosis MultiSig wallet is that an exploiter of a threshold number of exposed signing key-pairs may undo or permanently capture the wallet. Whereas KERI's pre-rotation scheme makes a forward commitment to unexposed key-pairs that may not be undone via exploit of any exposed signing key-pairs. This means that recovery from an exploit in KERI may be overcome (recaptured) by performing a rotation but not with the Gnosis multi-sig wallet. At least the Gnosis MultiSig Wallet has the advantage of multiple-signatures which ERC-1056 does not. But neither benefit from the in-stride pre-rotation security of KERI.

Non distributed consensus ledger implementations uniquely benefit from KERI's event streaming design. An example implementation target may be Apache Kafka [12] or Apache Flink [11]. Both Apache Kafka and Flink provide libraries for building scalable event processing streams for data intensive applications. The features and semantics may differ somewhat. Nonetheless, the KERI core stater verification engine can be implemented as either a Kafka streams application or a Flink ProcessFunction. This allows implementation of KERI witnesses and validators as Kafka or Flink streams. The witness receipt validation function of a KERI Judge may also be implemented as a Kafka or Flink stream. Thus the whole KERI DKMI may be hosted on scalable Kafka or Flink clusters. Similarly KERI may be easily implemented using asynchronous flow based processing frameworks such as Ioflo [122].

# 14 Derivation Codes

As previously described, the purpose of the *derivation code* is to compactly encode the derivation information needed to extract and verify the cryptographic material (encoded data) in the identifier prefix. By including the derivation code in the prefix we cryptographically bind the derivation to the identifier prefix. Different derivation results in a different identifier. The cryptographic material is encoded with Base64-URLSafe [87]. The BASE64 encoding employs either zero, one, or two pad characters to account for the difference in binary and Base64 lengths of the encoded cryptographic material. The prefix coding takes advantage of the pad characters when present by replacing them with a derivation code of the appropriate length. Many of the associated cryptographic material formats are 32 bytes in length which encode to 44 Base64 characters including1 pad character. In this case replacing the pad character with a one character derivation code provides the derivation information with a net zero increase in length of the Base64 encoding. Likewise Base64 encoded cryptographic material that includes 2 pad characters uses a two character derivation code. This is the case for cryptographic material that is 64 bytes in length. The rare cases where the cryptographic material length results in 0 pad characters (divisible by 3) then a 4 characters derivation code is used. This preserves evenly matched Base64 encoding lengths where 4 Base64 characters evenly encode 3 binary bytes.

As previously described, in order to properly extract and use the public key embedded in a self-certifying identifier we need to know the derivation process surrounding the associated key-pairs and other configuration material. This includes the cypher-suite and operation. In general this provides the process used to derive the self-certifying identifier. Because this derivation information is needed to correctly parse the encoded public key and the convention is to parse from left-to-right, we replace the pad characters by prepending the *derivation code* and then deleting the pad character(s) from the end. Because the string is valid Base-64 it may be converted to binary before being parsed with binary operations or it may be parsed before conversion. If parsed before conversion, the derivation code characters must be extracted from the front of the string. The first character tells how to parse the remaining characters if any. The corresponding number of pad characters must be re-appended before converting back to binary the remaining characters that comprise the encoded public key. Once converted the binary version of the public key may be used  in cryptographic operations. The proposed set of derivation codes for KERI are provided in the table below. Each is either 1, 2 or 4 characters long to replace 1, 2 or 0 pad characters respectively. This ensures that each self-contained identifier string is compliant with Base-64 specification which must be a multiple of 4 base-64 characters. This allows conversion before parsing which may be more efficient.

There are two classes of derivation codes. The first class encodes cryptographic material for elements associated with identifier prefixes and key events this may be identifier prefix, public keys, or digest. The second class encodes only attached signatures.

As an example of the first class, suppose the cryptographic material is an Ed25519 public key using the basic derivation of a self-certifying identifier prefix. An Ed25519 public key is 32 bytes in length. This encodes to 44 Base64 characters with 1 pad character inclusive. This then requires a 1 character derivation code. This has derivation code `B` in the one character derivation table. Let the the public key encoded in Base64 including pad character be as follows:

`DKrJxkcR9m5u1xs33F5pxRJP6T7hJEbhpHrUtlDdhh0=`,

then the encoding with derivation prefixed is as follows:

`BDKrJxkcR9m5u1xs33F5pxRJP6T7hJEbhpHrUtlDdhh0`.

## 14.1 Cryptographic Material Derivation Codes

For cryptographic material, the first 12 entries in the 1 character derivation table are selector characters to enable parsing of multiple length tables. This provides support for 1, 2 and 4 character derivation codes as well as future extensibility to longer codes. A few of the characters are reserved as selector codes to select the appropriate table. Provided below are suggested notional values for some of the entries in the associated tables.

## 14.2 Cryptographic Material Derivation CodeTables

| Selection Code | Select Description |
|---|---|
| 0 | Two character derivation code. Use two character table. |
| 1 | Four character derivation code. Use four character table. |
| 2 | Five character derivation code. Use  five character table. |
| 3 | Six character derivation code. Use  six character table. |
| 4 | Eight character derivation code. Use eight character table. |
| 5 | Nine character derivation code. Use nine character table. |
| 6 | Ten character derivation code. Use ten character table. |
| | |
| - | Count of attached receipt couplet code. Use four character attached receipt couplet count table |

TABLE 14.1 One character prefix table select characters.

| Derivation Code | Prefix Description | Data Length Bytes | Pad Length | Derivation Code Length | Prefix Length Base64 | Prefix Length Bytes |
|---|---|---|---|---|---|---|
| -A*XX* | Count of Attached Qualified Base64 Receipts | 0 | 0 | 4 | 4 | 3 |
| -B*XX* | Count of Attached Qualified Base2 Receipts | 0 | 0 | 4 | 4 | 3 |

TABLE 14.2 Four character attached receipt couplet count code. Partial table.

| Derivation Code | Prefix Description | Data Length Bytes | Pad Length | Derivation Code Length | Prefix Length Base64 | Prefix Length Bytes |
|---|---|---|---|---|---|---|
| A | Random seed of Ed25519 private key of length 256 bits | 32 | 1 | 1 | 44 | 33 |
| B | Non-transferable prefix using Ed25519 public signing verification key. Basic derivation. | 32 | 1 | 1 | 44 | 33 |
| C | X25519 public encryption key. May be converted from Ed25519 public signing verification key. | 32 | 1 | 1 | 44 | 33 |
| D | Ed25519 public signing verification key. Basic derivation. | 32 | 1 | 1 | 44 | 33 |
| E | Blake3-256 Digest. Self-addressing derivation. | 32 | 1 | 1 | 44 | 33 |
| F | Blake2b-256 Digest. Self-addressing derivation. | 32 | 1 | 1 | 44 | 33 |
| G | Blake2s-256 Digest. Self-addressing derivation. | 32 | 1 | 1 | 44 | 33 |
| H | SHA3-256 Digest. Self-addressing derivation. | 32 | 1 | 1 | 44 | 33 |
| I | SHA2-256 Digest. Self-addressing derivation. | 32 | 1 | 1 | 44 | 33 |
| J | ECDSA secp256k1 seed for private key of length 256 bits | 32 | 1 | 1 | 44 | 33 |
| K | Random seed of Ed448 private key of length 448 bits | 56 | 1 | 1 | 76 | 57 |
| L | X448 public encryption key | 56 | 1 | 1 | 76 | 57 |

TABLE 14.3 One character prefix table encoding. Partial table.

| Derivation Code | Prefix Description | Data Length Bytes | Pad Length | Derivation Code Length | Prefix Length Base64 | Prefix Length Bytes |
|---|---|---|---|---|---|---|
| 0A | Random seed or private key of length 128 bits | 16 | 2 | 2 | 24 | 18 |
| 0B | Ed25519 signature. Self-signing derivation. | 64 | 2 | 2 | 88 | 66 |
| 0C | ECDSA secp256k1 signature. Self-signing derivation. | 64 | 2 | 2 | 88 | 66 |
| 0D | Blake3-512 Digest. Self-addressing derivation. | 64 | 2 | 2 | 88 | 66 |
| 0E | SHA3-512 Digest. Self-addressing derivation. | 64 | 2 | 2 | 88 | 66 |
| 0F | Blake2b-512 Digest. Self-addressing derivation. | 64 | 2 | 2 | 88 | 66 |
| 0G | SHA2-512 Digest. Self-addressing derivation. | 64 | 2 | 2 | 88 | 66 |

TABLE 14.4 Two character prefix table encoding. Partial table.

| Derivation Code | Prefix Description | Data Length Bytes | Pad Length | Derivation Code Length | Prefix Length Base64 | Prefix Length Bytes |
|---|---|---|---|---|---|---|
| 1AAA | ECDSA secp256k1 non-transferable prefix public signing verification key. Basic derivation. | 33 | 0 | 4 | 48 | 36 |
| 1AAB | ECDSA secp256k1 public signing verification or encryption key. Basic derivation. | 33 | 0 | 4 | 48 | 36 |
| 1AAC | Ed448 non-transferable prefix public signing verification key. Basic derivation. | 57 | 0 | 4 | 80 | 60 |
| 1AAD | Ed448 public signing verification key. Basic derivation. | 57 | 0 | 4 | 80 | 60 |
| 1AAE | Ed448 signature. Self-signing derivation. | 114 | 0 | 4 | 156 | 117 |

TABLE 14.5 Four character prefix table encoding. Partial table.

## 14.3 Attached Signature Derivation Codes

The primary distinction of attached signature derivation codes is that the derivation code includes an index or offset into the list of current signing public keys. This provides a self-contained way of matching the signature to the public key needed to verify the signature. This means that the minimum length attached signature derivation code is 2 characters, the first character for the signature scheme and the second character for the index. One Base64 character may encode up to 64 values (0-63). So at most there may be 64 attached signatures using this scheme. In the highly unlikely case of more than 64 attached signatures then a longer derivation code may be used. The most commonly used ECC signature schemes use 64 byte signatures which, after conversion to Base64, have two pad characters. This is perfect for a two character derivation code with one index character included. Other signature schemes may have signatures with different number of pad bytes thereby requiring a different derivation code table. A few of the characters are reserved as selector codes to select the appropriate table.

One desirable feature of KERI is that it be transmission protocol agnostic. There are two main types of internet protocols. These are framed and unframed (streaming). In a framed protocol each packet has a known boundary. This means that a serialized event may be followed by any number of attached signatures as long as they all fit within one maximum sized packet frame. Parsing the attached signatures may proceed until the end of the frame is reached. The packet parser does not need to know in advance the number of attached signatures as it will discover that once it reaches the end of the frame. In a unframed protocol, on the other hand, there is no well defined packet boundary so a parser does not know when to stop parsing for attached signatures and start parsing for the next serialized event. In this case it would be helpful to have an indication of the count of attached signatures that appears after the event but before the attached signatures. This would make each event plus attached signatures self-framing. This feature makes KERI protocol agnostic with respect to framing. Of the two primary low level Internet protocols, UDP is framed and TCP is unframed. Whereas HTTP adds framing to TCP and RTSP effectively removes framing from UDP. KERI provides this self-framing capability with a special attached signature code that merely provides the count of attached signatures. KERI also anticipates more complex attached signature processing with a special code that allows for an operation code or op code table. Provided below are suggested notional values for some of the entries in the associated tables.

## 14.4 Attached Signature Derivation CodeTables

<table>
<tr><th>Selection Code</th><th>Select Description</th></tr>
<tr><td>0</td><td>Four character attached signature code. Use four character table</td></tr>
<tr><td>1</td><td>Five character attached signature code. Use five character table</td></tr>
<tr><td>2</td><td>Six character attached signature code. Use six character table</td></tr>
<tr><td></td><td></td></tr>
<tr><td>-</td><td>Count of attached signatures code. Use four character attached signature count table</td></tr>
<tr><td>-</td><td>Op-code for attached signatures . Use op-code table. | _ | opcode | opmod | # quadlets |</td></tr>
</table>

TABLE 14.6 Select characters from two character code. Partial table.

| Derivation Code | Prefix Description | Data Length Bytes | Pad Length | Derivation Code Length | Prefix Length Base64 | Prefix Length Bytes |
|---|---|---|---|---|---|---|
| -A*XX* | Count of Attached Qualified Base64 Signatures | 0 | 0 | 4 | 4 | 3 |

| -B*XX* | Count of Attached Qualified Base2 Signatures | 0 | 0 | 4 | 4 | 3 |
|---|---|---|---|---|---|---|

TABLE 14.7 Attached signature scheme and count from four character count code. Partial table.

| Derivation Code | Prefix Description | Data Length Bytes | Pad Length | Derivation Code Length | Prefix Length Base64 | Prefix Length Bytes |
|---|---|---|---|---|---|---|
| A*X* | Ed25519 signature plus index offset | 64 | 2 | 2 | 88 | 66 |
| B*X* | ECDSA secp256k1 signature plus index offset | 64 | 2 | 2 | 88 | 66 |

TABLE 14.8 Signature scheme character from two character attached signature code. Partial table.

| Derivation Code | Prefix Description | Data Length Bytes | Pad Length | Derivation Code Length | Prefix Length Base64 | Prefix Length Bytes |
|---|---|---|---|---|---|---|
| 0A*XX* | Ed448 signature plus index offset | 114 | 0 | 4 | 156 | 117 |

TABLE 14.9 Signature scheme character from four character attached signature code. Partial table.

# 15 Weighted Threshold Multi-Signature Scheme

The threshold multi-signature schemes presented elsewhere in this work may be modified to support a fractional weighted threshold by changing the threshold number (integer) into a list of fractional values or a list of lists of fractional values. In the simpler case of a single list, each value in the list is between 0 and 1. In this fractional weighted scheme, a valid signature set is any subset of signatures whose corresponding weights sum to 1 or greater. For example a threshold signature scheme for the following ordered list of signers:

$$\hat{C} = \left[ C^1, C^2, C^3 \right], \tag{15.1}$$

where any 2 of 3 signatures is valid may be represented equivalently by the following ordered list of fractional weights,

$$\hat{K} = \left[ \tfrac{1}{2}, \tfrac{1}{2}, \tfrac{1}{2} \right], \tag{15.2}$$

where any combination of two or more weights would sum to at least 1. To generalize let the $l^{th}$ ordered set of controlling signers be denoted as follows:

$$\hat{C}_l = \left[ C_l^1, \ldots, C_l^{L_l} \right]_l, \tag{15.3}$$

where $\hat{C}_l$ represents the $l^{th}$ list, $L_l$ is the number of signers in the list, and each $C_l^j$ in the list is the public key from a signing key-pair. The corresponding $l^{th}$ ordered list of fractional weights (one-to-one for each signer) may be denoted as follows:

$$\hat{K}_l = \left[ U_l^1, \ldots, U_l^{L_l} \right]_l, \tag{15.4}$$

where $\hat{K}_l$ represents the $l^{th}$ list, $L_l$ is the number of weights in the list, and each $U_l^j$ in the list is the weight for the corresponding signer. Each weight satisfies:

$$0 < U_l^j \leq 1. \tag{15.5}$$

A subset of the full list of signatures may be attached to some event. Let it be the $k^{th}$ event. This subset may be represented by a list of zero based indexes (offsets) into the $l^{th}$ list of signers and weights given by eq. 15.3 and eq. 15.4 The indexed list may denoted as follows:

$$\widehat{s}_k^l = \left[ s_0, \ldots, s_{S_k^l - 1} \right]_k^l , \tag{15.6}$$

where $\widehat{s}_k^l$ represents the ordered indexed list of offsets into the $l^{th}$ list of signers attached to the $k^{th}$ event, $S_k^l$ is the number of attached signers, and each $s_i$ is a zero based offset into both $\widehat{C}_l$ and $\widehat{K}_l$. A set of signatures is valid when the associated weights sum to greater than or equal 1, such as:

$$\overline{U}_l = \sum_{i=s_0}^{s_{S_k-1}} U_l^i \geq 1 , \tag{15.7}$$

where $\overline{U}_l$ represents the sum, *i* in the summation is assigned to successive values from the list of offsets, $\widehat{s}_k^l = \left[ s_0, \ldots, s_{S_k^l - 1} \right]_k^l$, that is, $i \in \widehat{s}_k^l$, and $U_l^i$ is the weight at the offset *i*. Care must be taken when using floating point representations to account for floating point rounding errors in the summation. One way to avoid this problem is to use rational fractional number computations. Some programming libraries, the *fractions* module in Python, support explicit rational fractional calculations.

In its simplest form, as the example above showed, this weighted threshold scheme may be made to act equivalently to the trivial *K* of *L* threshold scheme by assigning equal weights as follows:

$$U_l^j = {}^1\!/_{K_l} , \tag{15.8}$$

where $K_l$ is the threshold count of signatures in a conventional *K* of *L* threshold scheme. The real power of the weighted threshold scheme, however, is realized by unequal weights. This allows different combinations of signers to reach a valid signature set.

Suppose for example that the $l^{th}$ weight list is as follows:

$$\widehat{K}_l = \left[ {}^1\!/_2, {}^1\!/_2, {}^1\!/_4, {}^1\!/_4, {}^1\!/_4, {}^1\!/_4 \right]_l . \tag{15.9}$$

In this case, a valid signature set occurs either when both the first two signers sign, or when any one of the first two sign and any two or more of the last four sign, or when all four of the last four sign. This allows different degrees of signing strength or signing authority to be assigned to signers as a way to reflect different degrees of trust in the signers. This allows the assignment of role based signing authority and hierarchical authority. Effectively, in the example above, the first two signers have the same authority together as the last four have together. The first two could be higher level managers and the last four lower level managers.

This approach may be extended further to support a more complex logical AND combination of the satisfaction truth value of multiple fractional weighted thresholds. In this case the threshold is expressed as a list of weight lists where each individual weight list in the list of lists must satisfy a threshold of 1 for its satisfaction truth value to be considered *true* else it is *false*. These truth values, one from each weighted list, are then combined with a logical AND to determine the truth value of the overall threshold. Suppose for example, the complex list of weight lists as follows:

$$\widehat{K}_l = \left[ \left[ {}^1\!/_2, {}^1\!/_2, {}^1\!/_4, {}^1\!/_4, {}^1\!/_4, {}^1\!/_4 \right], \left[ {}^1\!/_2, {}^1\!/_2, {}^1\!/_2, {}^1\!/_2 \right], \left[ 1, 1, 1, 1 \right] \right] \tag{15.10}$$

For this complex list of weight lists, one example of a valid signature set may be expressed as follows: from the first list any combination of signers that adds to 1 or more such as any one of the first two signers plus any two of the last four signers, AND from the second list any two or more of the four signers AND from the third list, any one or more of the four signers.

Using delegation further extends the capability. Multiple distinct delegated sets of weighted multi-signature schemes provides extreme flexibility for managing signing authority arrangements. As described above, simply replacing the integer valued $K_l$ with the list valued $\widehat{K}_l$ in any of the aforementioned multiple signature capable event definitions may enable weighted multiple signatures. For example the inception event, $\varepsilon_0 = \left\langle C, t_0, \mathtt{icp}, C^0, K_1, \widehat{C}_1, M_0, \widehat{W}_0 \right\rangle \sigma_{C^0}$ becomes $\varepsilon_0 = \left\langle C, t_0, \mathtt{icp}, C^0, \widehat{K}_1, \widehat{C}_1, M_0, \widehat{W}_0 \right\rangle \sigma_{C^0}$ where $\widehat{K}_1$ is a weight list. Furthermore the delegated inception $D_{\varepsilon_0} = \left\langle D, t_0, \mathtt{dip}, \widehat{\Delta}_k^C, D^0, perms, K_0^D, \widehat{D}_0^D, M_0^D, \widehat{W}_0^D \right\rangle \sigma_{D^0}$ becomes $D_{\varepsilon_0} = \left\langle D, t_0, \mathtt{dip}, \widehat{\Delta}_k^C, D^0, perms, \widehat{K}_0^D, \widehat{D}_0^D, M_0^D, \widehat{W}_0^D \right\rangle \sigma_{D^0}$ where $\widehat{K}_0^D$ is a weight list.

# 16 Serialization

Recall that each event includes a *protocol version code string*. The protocol version code also includes the serialization coding (such as JSON, MessagePack, CBOR, etc) used to encode the associated event. This allows a parser to determine how to parse the message.

There are two uses for serialization. The first is the serialization of a full event the second is serialization of set of elements extracted from a full event. The elements of the former do not need to be in any canonical order. The elements of the later must be in a canonical order in order that a digest or signature of the extracted set may be reproduced from a full event.

## 16.1 Extracted Element Canonical Form

The data structure that provides the extracted elements must have a canonical order so that it may be reproduced in a digest of elements of a event. A canonical form for extracted elements that is universal across different types of serialization encodings is a a concatenation of all the values. The order of the concatenation is a depth first traversal of the elements in a given order. The order of appearance is standardized including nested elements.

The extraction sets are as follows:

- Inception event elements used to derive digest of self-addressing prefix.
- Inception event elements used to derive signature of self-signing prefix.
- Event elements used to derive digest of next key set with threshold
- Event elements used to derive digest in seal of delegated prefix inception event.
- Seal elements.

## 16.2 Serialization Algorithms

There are two serialization algorithms needed for KERI. The first is the serialization of full events. The second is the serialization of data sets extracted from a full event.

In KERI, complete key events need to be serialized in order to create cryptographic signatures of each event. Complete events also need to be serialized in order to create cryptographic digests of each event. Complete events are propagated over the network in signed serialized form. Complete events also need to be deserialized after network propagation such that the resultant deserialization preserves the semantic structure of the data elements in the event. Because the cryptographic material in the events may be of variable length, a fixed field length serialization is not a viable approach. Consequently KERI must support variable length field serialization. Moreover serialization encodings that support arbitrary nested self-describing data structures provide fu-ture proofing to event composition. Notably, JSON, CBOR, and MsgPack are well known broadly supported serialization encodings that support arbitrary nested self-describing data structures. JSON is an ascii text encoding that is human readable, albeit somewhat verbose, that is commonly used in web applications. On the other hand both CBOR and MsgPack are binary encodings that are much more compact. Supporting all three encodings (and possibly others) would satisfy a wide range of application, transport, and resource constraints. Importantly, because only the controller of the associated identifier prefix may compose KERI events, the ordering of elements in the event serialization may be determined solely by that controller. Other entities may sign the event serialization provided by the controller but do not need to provide a serialization of their own.

Within the KERI protocol cryptographic digests and signatures are created of specified subsets of the data elements extracted from complete KERI key events. Consequently these extracted data sets also need to be serialized. Importantly, however, the extracted data set serialization do not need to be propagated over the network. This means that the serializations do not need to preserve the semantic structure of the data. This allows extracted data serialization encoding to simplified. These digests of extracted data, however, need to be reproducible by any other entity. This means that the ordering of the data elements in the serialization must be exactly specified.

In summary, the necessary constraint for complete event serialization is support for arbitrary data structures with variable length fields that are serializable and deserializable in multiple formats. Reproducible ordering is not a necessary constraint. The necessary constraint for extracted data element sets is reproducible ordering but there is no need to deserialize.

Beneficial if not necessary constraints are ease and convenience of development and maintenance of the associated code libraries. One possibility that was considered and rejected would be to use same serializations for both the complete event serialization and the extracted data set serialization. The problem with that approach is that on an event by event basis the serialization encoding may change, that is, one event may be encoded using JSON and another using CBOR. This would require keeping track of which encoding was used on the event in order to reproducibly perform an extracted data serialization. More problematic is that for delegated events, the extracted data set digest included in one event may be from data extracted from an event belonging to a different identifier. In such a situation, keeping track of the encoding provides an inconvenience that obviates the advantages of using the same encodings for both events and extracted data sets. Given this, then a simplified but repeatable encoding algorithm may be used for the extracted data set serializations. Simplified because the extracted data sets do not need to be deserialized.

The one constraint for extracted data set serializations is that the ordering of elements be exactly specified. One way to simplify that specification is to use the ordering of data elements in the complete event serialization. But this means imposing an ordering constraint on the complete event serialization. The advantage is that ordering is only expressed once per event not once per extracted data set. This better future proofs the protocol as there is always only one place for ordering and that is the complete event element ordering.

Another benefit of ordering the complete event serialization is that automated self-contained discovery of the serialization encoding used in an event becomes trivialized by requiring that the first element in the serialization be the version string element. A deserializer may then unambiguously determine the encoding by inspecting the first few bytes of any serialized event. This better supports a wider variety of transport protocols and databases.

Until recently native JSON serialization libraries in Javascript did not preserve a logical ordering of elements from a Mapping (JavaScript Object) data structure. The only ordering possible was lexicographic by sorting the fields in the mapping by their labels. This meant that arbitrary logical ordering of mapping fields was not possible and no semantic meaning could be imposed on the serialization based on order of appearance of fields. One could approximate some ordering by imposing lexicographic constraints on the field names but that makes the field names less usable. Ordered mappings support a logical ordering by preserving the order of creation of fields in the mapping. Any arbitrary order may be imposed by changing the order of creation of the fields in the mapping.

Fortunately Javascript ES6 (ES2015) and later now provide a mechanism to impose property (field) creation order on JSON.stringify() serializations. The latest versions of Javascript now natively preserve property creation order when serializing. This means that deserialization will automatically recreate the properties in the same order as serialized. For those implementations of Javascript ES6 or later that do not yet support native JSON.stringify property creation ordering, a JSON.stringify parameter replacer function that uses the [[ownPropertyKeys]] internal method, namely, Reflect.ownKeys(object) may be employed to ensure property creation order. This means that native Javascript support may be easily and broadly provided. All major JavaScript implementations now support JavaScript ES6. Example JavaScript code is provided later in this document.

Other languages like Python, Ruby, Rust etc. have long supported native creation order preserving serializations of ordered mappings. These preserve field element (property equivalent) creation order. This in combination with the recent Javascript support means that KERI may impose an ordering constraint on complete event data elements that may be used for canonical ordering of both complete events and extracted data element sets.

### 16.2.1 Event Serialization Algorithm

The ordering of elements in each mapping in each event element is specified (see below). This includes nested elements. Serialization using the standard JSON, CBOR, and MsgPack libraries follow a recursive depth first traversal of nested elements.

Complete events are serialized/deserialized from/to ordered mappings using the JSON, CBOR or MsgPack functions for serializing/deserializing. Each programming language may name the functions differently for serialization/deserialization. In JavaScript they are JSON.stringify/JSON.parse. In Python they are json.dumps/json.loads. Likewise the CBOR and MsgPack libraries in each language may use different names for these functions. The JSON serialization is without whitespace.

### 16.2.2 Extracted Data Set Serialization Algorithm

The extracted data set serialization algorithm serializes the elements in each set using the same order of appearance as the associated event. Nested elements are serialized using a depth first traversal. The serialization appends (concatenates) the serialized value of each element as it encounters it. Each element value is serialized as a UTF-8 encoded string of characters. Each implementation must perform the recursive depth first traversal.

## 16.3 JSON Encoding

One popular encoding used for serializing data structures or tuples is the JSON (JavaScript Object Notation) standard. One of the historical limitations of JSON, however, is that the field order of a serialized JavaScript object is not normative, that is, a valid JSON serialization of a JSON object does not guarantee the order of appearance of the fields within a JavaScript object (or equivalent such as a Python dict) that results from the deserialization. In addition whitespace in a JSON serialization is not normative. Also there is some variation in how JSON serializers deal with unicode strings. Consequently round trip serializations and deserializations may not be identical across all implementation of JSON serializers and therefore would not verify against the same cryptographic signature. This is the so called canonicalization problem. There are a couple of ways to address this problem. The ES6 version of JavaScript has an internal method [[ ownPropertyKeys]] that preserves the creation order of JavaScript object properties. This may be used to serialize Javascript objects in property creation order. Most is not all JSON serializers support a mode where no whitespace is used. This is the most compact JSON encoding. Thus ES6 or later implementations of JavaScript may be canonically serialized using object property creation order. Given ordering and control over whitespace the serializations can be made repeatable. Another simple solution to this problem when using JSON is that the data associated with a signature may only be serialized once by the signer. Users of the data may deserialize but never re-serialize unless they also re-sign. Any compliant JSON deserialization will produce an equivalent Javascript object (same field names and values but order and whitespace are ignored). But to ensure repeatability both field order and whitespace must be constrained.

In many protocols the signatures are attached to the serialized data in one data string. This encounters another limitation of JSON, that is, many JSON implementations raise an error if a deserialization attempt on a string does not consume all the characters in the string. Thus a hybrid data string that consists of a serialized JSON object followed by a signature string might be difficult to deserialize with some JSON implementations. A portable approach, however, is to concatenate the signature but separate it from the JSON serialization with a unique string of characters that will not be produced by a JSON serializer. A parser first searches for the separator string and then separately extracts both the JSON serialization and the signature. One such human friendly separator string is the 4 character sequence of whitespace characters, CR LF CR LF (in ascii notation), where CR represents the CarriageReturn character (ASCII 13) and LF represents the LineFeed character (ASCII 10). In escaped notation this string is "\r\n\r\n". If these separator characters were to appear within a quoted string they would be doubled escaped by the JSON serializer. Furthermore, neither of these characters appear in a Base64 serialization. Consequently this same separator or merely a CarriageReturn or LineFeed character may be used to separate multiple signatures. This approach makes it easy for a parser to separate, extract, and verify the serialized data with the attached signature(s) without deserializing the JSON. One approach is to serialize a nested JavaScript object with JSON followed by the separator "\r\n\r\n” then a JSON string delimited (double quoted) Base64 serialization of a signature. Additional signatures are each separated by the separator. An example follows:

```
{
  "id": "Xq5YqaL6L48pf0fu7IUhL0JRaU2_RxFP0AL43wYn148=",
  "sn": 1,
  "kind": "ek",
  "signer": "Xq5YqaL6L48pf0fu7IUhL0JRaU2_RxFP0AL43wYn148=",
  "ensuer": "Qt27fThWoNZsa88VrTkep6H-4HA8tr54sHON1vWl6FE="
}
\r\n\r\n
"AeYbsHot0pmdWAcgTo5sD8iAuSQAfnH5U6wiIGpVNJQQoYKBYrPPxAoIc1i5SHC
IDS8KFFgf8i0tDq8XGizaCg=="
\r\n\r\n
"AuSQAfnH5U6wiIGpVNJQQAeYbsHot0pmdWAcgTo5sD8ioYKBYrPPxAoIc1i5SHC
IDS8KFFgf8i0tDq8XGizaCg=="
```

In KERI the key events are transmitted as strings of UTF-8 characters. The derivation code prepended to signatures enables a slightly more compact encoding for JSON. The JSON encod-ed data appears first in the form of a string of UTF-8 characters. This is followed by the "\r\n\r\n" separator and then one or more signatures as qualified Base64 strings each separated by a "\n" (LineFeed) character. An example follows: (whitespace inserted for readability only)

```
{
  "vs"   : "KERI_json_1.0",
  "id"   : "AaU6JR2nmwyZ-i0d8JZAoTNZH3ULvYAfSVPzhzS6b5CM",
  "sn"   : "0",
  "ilk"  : "icp",
  "sith" : "1",
  "keys" : ["AaU6JR2nmwyZ-i0d8JZAoTNZH3ULvYAfSVPzhzS6b5CM"],
  "next" : "DZ-i0d8JZAoTNZH3ULvaU6JR2nmwyYAfSVPzhzS6b5CM",
  "toad" : "1",
  "wits" : [],
  "data" : [],
  "sigs" : ["0"]
}
\r\n\r\n
0AAeYbsHot0pmdWAcgTo5sD8iAuSQAfnH5U6wiIGpVNJQQoYKBYrPPxAoIc1i5SH
CIDS8KFFgf8i0tDq8XGizaCg
\n
0AAuSQAfnH5U6wiIGpVNJQQAeYbsHot0pmdWAcgTo5sD8ioYKBYrPPxAoIc1i5SH
CIDS8KFFgf8i0tDq8XGizaCg
```

## 16.4 HTTP Signature Header

When using HTTP as the transport it may be more convenient to attach the signatures in an HTTP Header. One way to do this is with a custom header. The format of the custom *Signature* header follows the conventions of RFC7230 [64]. The header format is as follows:

```
Signature: HeaderValue

HeaderValue:
  tag = "value"
or
  tag = "value"; tag = "value"  ...
```

where the each `tag` in each clause of the Signature HeaderValue is a name in the form of a unique string that identifies the signature that follows as the value of the tagged clause in the HeaderValue. Each tagged clause's signature value is a doubly quoted string "" that contains the actual signature in Base64 URL-safe format.

An optional clause with a tag named *kind* may be present of the form:

```
  kind = "value"
```

where the value in double quotes of the tagged clause is a string that specifies the cipher suite type used to create the signature clauses within that header. All signatures within a header must be of the same cypher suite type. An example header follows where the tag names are `current`, `next` and `kind` . The first two values are a Base64 URL-safe signatures and the last value is a cipher suite type string.

```
Signature:
current="Y5xTb0_jTzZYrf5SSEK2f3LSLwIwhOX7GEj6YfRWmGViKAesa08UkNW
ukUkPGuKuu-EAH5U-sdFPPboBAsjRBw=="; next="Xhh6WWGJGgjU5V-
e57gj4HcJ87LLOhQr2Sqg5VToTSg-
SI1W3A8lgISxOjAI5pa2qnonyz3tpGvC2cmf1VTpBg==";
kind="Ed25519VerificationKey2018";
```

Multiple *Signature* headers may be present and the set of signatures is the union of all the signatures from all the clauses from all the *Signature* headers. If the same clause tag appears multiple times in this union then only the last occurrence is used.

Different conventions may be used to map the signature to a public key. One is that each clause tag name besides the tag name *kind* must equal a field name in the JSON body. The associated JSON field value is the public key or prefix from which the public key may be derived that may be used to verify the signature provided in the corresponding header clause with the same name (tag).

The convention used for KERI key events is the each tag name is the string equivalent of the integer offset into the associated public signing key list. The signatures are also fully qualified in that they include a prepended derivation code that replaces the pad characters so that the type of signature algorithm is given by the derivation code as well as the offset. As a result, a "*kind*" tag is not needed. The following is an example of a signature header for a KERI key event.

```
Signature:
0="AAY5xTb0_jTzZYrf5SSEK2f3LSLwIwhOX7GEj6YfRWmGViKAesa08UkNWukUk
PGuKuu-EAH5U-sdFPPboBAsjRBw"; 1="ABXhh6WWGJGgjU5V-
e57gj4HcJ87LLOhQr2Sqg5VToTSg-
SI1W3A8lgISxOjAI5pa2qnonyz3tpGvC2cmf1VTpBg";
```

# 17 Conclusion

KERI is a decentralized key management infrastructure (DKMI) based on the principle of minimally sufficient means. KERI is designed to be compatible with event streaming applications but may be employed in distributed ledger systems as well. The event streaming design lends itself to a simple state verification engine. The syntax is compact and efficient. Nonetheless KERI has advanced key management features. The principle key management operation is key rotation via a novel in-stride key pre-rotation scheme. This utilizes a single sequence of controlling key-pairs for easier management as well as advanced features such as post-quantum security. KERI provides for reconfigurable thresholded multiple signature schemes where both the threshold and total number of signatures may change at each rotation. KERI also provides for fractional weighted multiple signatures schemes. The pre-rotation makes a forward commitment to unexposed key-pair(s) that provide security that may not be undone via exploit of any exposed key-pairs. This allows for reconfigurability without sacrificing security. KERI also provides for reconfigurable designation of witnesses and quorum (tally) sizes where both the total number of witnesses and quorum size may change from rotation to rotation. Once again the pre-rotation forward commitment to unexposed key-pairs means that the witness configuration may not be undone via exploit of any exposed key-pairs. KERI is the only event-streaming capable system that we know of that provides this combination of advanced features.

A delegated version of KERI is also provided that enables hierarchical key management where a master controller (identifier) key event stream may delegate signing authority to one or more slave identifier key event streams thereby forming a chained tree of key event streams.

KERI scalable design supports multiple use cases. Two primary trust modalities motivated the design, these are a direct (one-to-one) mode and an indirect (one-to-any) mode. In the direct mode the identity controller establishes control via verified signatures of the controlling key-pair. The indirect mode extends that trust basis with witnessed key event receipt logs (KERL) for validating events. This gives rise to the acronym KERI for key event receipt infrastructure. The The security and accountability guarantees of indirect mode are provided by KAWAor KERI's Algorithm for Witness Agreement among a set of witnesses. The KAWA approach may be much more performant and scalable than more complex approaches that depend on a total ordering distributed consensus ledger. Nevertheless KERI may employ a distributed consensus ledger when other considerations make it the best choice. In other words KERI may be augmented with distributed consensus ledgers but does not require them. KERI is applicable to DKMI in data streaming, web 3.0, and IoT applications where performance and scalability are important. KERI's core services are identifier independent. This makes KERI a simple universal portable DKMI.

## ACKNOWLEDGMENTS

The author wishes to thank all those that provided helpful feedback. This include the participants in the associated sessions on KERI at the Spring and Fall 2019 Internet Identity Workshops, side discussions at the face-to-face meeting of the W3C working group on DIDs in early 2020 and recognition by DIF's Identity and Discovery project. Specifically the author wishes to thank the following for their insightful comments as well as moral support for this effort (not in any order): Alan Karp, Daniel Hardman, Drummond Reed, Christian Lundkvist, Rouven Heck, Orie Steele, Markus Sabadello, Dave Huseby, Mike Lodder, and Carsten Stocker. Many of the improvements in this version of KERI were a direct result of their suggestions.

## AUTHOR

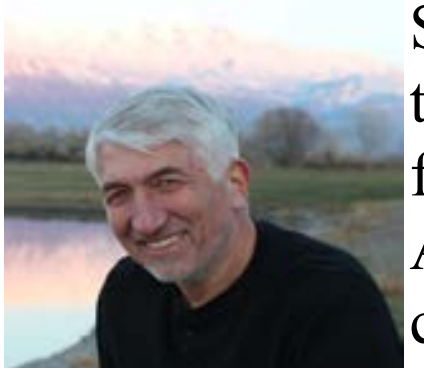

Samuel M. Smith Ph.D. has a deep interest in decentralized identity and reputation systems. Samuel received a Ph.D. in Electrical and Computer Engineering from Brigham Young University in 1991. He then spent 10 years at Florida Atlantic University, eventually reaching full professor status. In addition to decentralized identity and reputation, he has performed pioneering research in automated reasoning, machine learning, and autonomous vehicle systems. He has over 100 refereed publications in these areas and was principal investigator on numerous federally funded research projects. Dr. Smith has been an active participant in open standards development for networking protocols, and decentralized identity. He is also a serial entrepreneur.

## REFERENCES

[1] "Ed25519: high-speed high-security signatures,"
http://ed25519.cr.yp.to

[2] "locus of control," APA Dictionary of Psychology,

https://dictionary.apa.org/locus-of-control

[3] "802.1AR-2018 - IEEE Standard for Local and Metropolitan Area Networks - Secure Device Identity," IEEE, 2018/08/02

https://ieeexplore.ieee.org/document/8423794

[4] "A Deep Dive on the Recent Widespread DNS Hijacking Attacks," KrebsonSecurity, 2019/02/19

https://krebsonsecurity.com/2019/02/a-deep-dive-on-the-recent-widespread-dns-hijacking-attacks/

[5] "A Universally Unique Identifier (UUID) URN Namespace," IETF RFC-4122, 2005/07/01

https://tools.ietf.org/html/rfc4122

[6] Agoulmine, N., "Introduction to Autonomic Concepts Applied to Future Self-Managed Networks," Academic Press, 2011.

https://www.sciencedirect.com/science/article/pii/B9780123821904000012

[7] Akhshabi, S. and Dovrolis, C., "The Evolution of Layered Protocol Stacks Leads to an Hourglass-Shaped Architecture," SIGCOMM'11, 2011/08/15-19

http://conferences.sigcomm.org/sigcomm/2011/papers/sigcomm/p206.pdf

[8] Alexander, T. P. and Coleman, J. H., "Authoritative Source of Truth," OMG MBSE Wiki, 2018/12/14

https://www.omgwiki.org/MBSE/doku.php?id=mbse:authoritative_source_of_truth

[9] Allen, C. and Applecline, S., "Hierarchical Deterministic Keys: BIP32 & Beyond,"

https://github.com/WebOfTrustInfo/rwot1-sf/blob/master/topics-and-advance-readings/hierarchical-deterministic-keys--bip32-and-beyond.md

[10] Allen, C., "The Path to Self-Sovereign Identity," Life With Alacrity, 2016/04/25

http://www.lifewithalacrity.com/2016/04/the-path-to-self-sovereign-identity.html

[11] "Apache Flink," Apache Software Foundation,

https://flink.apache.org

[12] "Apache Kafka," Apache Software Foundation,

https://kafka.apache.org

[13] "Arbitrary-precision arithmetic," Wikipedia,

https://en.wikipedia.org/wiki/Arbitrary-precision_arithmetic

[14] Arciszewski, S., "No Way, JOSE! Javascript Object Signing and Encryption is a Bad Standard That Everyone Should Avoid," Paragon Initiative, 2017/03/14

https://paragonie.com/blog/2017/03/jwt-json-web-tokens-is-bad-standard-that-everyone-should-avoid

[15] "Argon2," GitHub,

https://github.com/p-h-c/phc-winner-argon2

[16] Aublin, P.-L., Mokhtar, S. B. and Quéma, V., "Rbft: Redundant byzantine fault tolerance," vol. Distributed Computing Systems, no. ICDCS, pp. 297-306, 2013

http://pakupaku.me/plaublin/rbft/report.pdf

[17] Aumasson, J.-P., "BLAKE2 — fast secure hashing,"

https://blake2.net

[18] "Autonomic," Dictionary.com,

https://www.dictionary.com/browse/autonomic?s=t

[19] "Autonomic Computing," Wikipedia,

https://en.wikipedia.org/wiki/Autonomic_computing

[20] "Autonomic Nervous System," Wikipedia,

https://en.wikipedia.org/wiki/Autonomic_nervous_system

[21] "Autonomous," Dictionary.com,

https://www.dictionary.com/browse/autonomous

[22] Barker, E. and Barker, W. C., “Recommendation for Key Management:
Part 2 – Best Practices for Key Management Organizations,” NIST Special Publication 800-57 Part 2 Revision 1, 2019/05/01

https://nvlpubs.nist.gov/nistpubs/SpecialPublications/NIST.SP.800-57pt2r1.pdf

[23] Beck, M., “On The Hourglass Model, The End-to-End Principle and Deployment Scalability,” Computer Science, 2016

https://www.semanticscholar.org/paper/On-The-Hourglass-Model%2C-The-End-to-End-Principle-Beck/564ddb378077feb171650a8a0d56bce860bfee3e

[24] Beck, M., “On The Hourglass Model,” Communications of the ACM, vol. Vol. 62 No. 7, pp. 48-57, 2019/07/01

https://cacm.acm.org/magazines/2019/7/237714-on-the-hourglass-model/fulltext

[25] Bernstein, D. J., “Cost analysis of hash collisions: Will quantum computers make SHARCS obsolete?,” Cr.yp.to, 2009/08/23

https://cr.yp.to/hash/collisioncost-20090823.pdf

[26] Bernstein, D. J., Duif, N., Lange, T., Schwabe, P. and Yang, B.-Y., “High-speed high-security signatures,” NSF, 2011

http://ed25519.cr.yp.to/ed25519-20110926.pdf

[27] “BIP32 recommends a 256 bit seed. Why do most Bitcoin wallets only use a 128 bit seed?,” StackExchange,

https://bitcoin.stackexchange.com/questions/72612/bip32-recommends-a-256-bit-seed-why-do-most-bitcoin-wallets-only-use-a-128-bit

[28] Birge-Lee, H., Sun, Y., Edmundson, A., Rexford, J. and Mittal, P., “Using BGP to acquire bogus TLS certificates,” vol. Workshop on Hot Topics in Privacy Enhancing Technologies, no. HotPETs 2017, 2017

[29] Birge-Lee, H., Sun, Y., Edmundson, A., Rexford, J. and Mittal, P., “Bamboozling certificate authorities with {BGP},” vol. 27th {USENIX} Security Symposium, no. {USENIX} Security 18, pp. 833-849, 2018

https://www.usenix.org/conference/usenixsecurity18/presentation/birge-lee

[30] Birge-Lee, H., Wang, L., Rexford, J. and Mittal, P., “Sico: Surgical interception attacks by manipulating bgp communities,” vol. Proceedings of the 2019 ACM SIGSAC Conference on Computer and Communications Security, pp. 431-448, 2019

https://dl.acm.org/doi/abs/10.1145/3319535.3363197

[31] “Blake3,” Github,

https://github.com/BLAKE3-team/BLAKE3

[32] “BLAKE3 Is an Extremely Fast, Parallel Cryptographic Hash,” InfoQ, 2020/01/12

https://www.infoq.com/news/2020/01/blake3-fast-crypto-hash/

[33] “blake3.pdf,” Github Blake3,

https://github.com/BLAKE3-team/BLAKE3-specs/blob/master/blake3.pdf

[34] Bly, J., “Mobile Edge Of The Internet Is Rapidly Moving To IPv6,” TeamARIN, 2020/01/16

https://teamarin.net/2020/01/16/mobile-edge-of-the-internet-is-rapidly-moving-to-ipv6/

[35] Brown, D. R. L., “ECDSA Secp256k1,” Certicom Research, vol. Version 2.0, 2010/01/27

https://en.bitcoin.it/wiki/Secp256k1

[36] “Byzantine fault tolerance,” Wikipedia,

https://en.wikipedia.org/wiki/Byzantine_fault_tolerance

[37] Cameron, K., “The Laws of Identity,” Microsoft, vol. 2005/05/11,

http://www.identityblog.com/stories/2005/05/13/TheLawsOfIdentity.pdf

[38] Cameron, K., “Microsoft’s Vision for an Identity Metasystem,” Microsoft Whitepaper, 2006/01/08

http://www.identityblog.com/stories/2005/10/06/IdentityMetasystem.pdf

[39] Castro, M. and Liskov, B., “Practical Byzantine fault tolerance,” vol. OSDI 99, pp. 173-186, 1999

http://www.pmg.lcs.mit.edu/papers/osdi99.pdf

[40] "Certificate authority," Wikipedia,

https://en.wikipedia.org/wiki/Certificate_authority

[41] "Certificate Transparency Ecosystem,"

https://certificate.transparency.dev

[42] Clark, D. D., "Interoperation, Open Interfaces, and Protocol Architecture," The National Academies Press OpenBook Chap 16, vol. The Unpredictable Certainty: White Papers, no. 1997, pp. 133, 1997

https://www.nap.edu/read/6062/chapter/17#138

[43] Clement, A., Wong, E. L., Alvisi, L., Dahlin, M. and Marchetti, M., "Making Byzantine Fault Tolerant Systems Tolerate Byzantine Faults.," vol. NSDI 9, pp. 153-168, 2009

http://static.usenix.org/events/nsdi09/tech/full_papers/clement/clement.pdf

[44] "Convergence (SSL)," Wikipedia,

https://en.wikipedia.org/wiki/Convergence_(SSL)

[45] Conway, S., Hughes, A., Ma, M. et al., "A DID for Everything," Rebooting the Web of Trust RWOT 7, 2018/09/26

https://github.com/SmithSamuelM/Papers/blob/master/whitepapers/A_DID_for_everything.pdf

[46] "Cryptographically secure pseudorandom number generator," Wikipedia,

https://en.wikipedia.org/wiki/Cryptographically_secure_pseudorandom_number_generator

[47] Dahlberg, R., Pulls, T. and Peeters, R., "Efficient sparse merkle trees," vol. Nordic Conference on Secure IT Systems, pp. 199-215, 2016

hhttps://eprint.iacr.org/2016/683.pdf

[48] Mazieres, D., "The Stellar Consensus Protocol:

A Federated Model for Internet-level Consensus," 2016/02/26

https://www.stellar.org/papers/stellar-consensus-protocol.pdf

[49] "Digital Signatures," PyNaCL readthdocs.org,

https://pynacl.readthedocs.io/en/stable/signing/

[50] "Distributed hash table," Wikipedia,

https://en.wikipedia.org/wiki/Distributed_hash_table

[51] Ditizio, F. B., Hoyle, S. B. and Pruitt, C. H. L., "Autonomic Ship Concept," Naval Engineers Journal, 1995/09/01

https://onlinelibrary.wiley.com/doi/abs/10.1111/j.1559-3584.1995.tb03098.x

[52] "DNS Certification Authority Authorization,"

https://en.wikipedia.org/wiki/DNS_Certification_Authority_Authorization

[53] "Do you need more than 128-bit entropy?," StackExchange,

https://security.stackexchange.com/questions/102157/do-you-need-more-than-128-bit-entropy

[54] "Domain Name System," Wikipedia,

https://en.wikipedia.org/wiki/Domain_Name_System

[55] "Domain Name System Security Extensions (DNSSEC_,"

https://en.wikipedia.org/wiki/Domain_Name_System_Security_Extensions

[56] "Ed25519 to Curve25519," LibSodium Documentation,

https://libsodium.gitbook.io/doc/advanced/ed25519-curve25519

[57] Eijdenberg, A., Laurie, B. and Cutter, A., "Verifiable Data Structures," 2015/11/01

https://github.com/google/trillian/blob/master/docs/papers/VerifiableDataStructures.pdf

[58] "Elliptic Curve Digital Signature Algorithm (ECDSA)," Wikipedia,

https://en.wikipedia.org/wiki/Elliptic_Curve_Digital_Signature_Algorithm

[59] "Elliptic-curve cryptography," Wikipedia,
https://en.wikipedia.org/wiki/Elliptic-curve_cryptography

[60] "ERC-1056 Lightweight Identity," Ethereum Foundation EIP,
https://github.com/ethereum/EIPs/issues/1056

[61] Buchman, E., Kwon, J. and Milosevic, Z., "The latest gossip on BFT consensus," 2018/07/13
https://tendermint.com/docs/tendermint.pdf

[62] "Ethr-DID Library," uPort.me,
https://github.com/uport-project/ethr-did

[63] "Exascale computing," WIkipedia,
https://en.wikipedia.org/wiki/Exascale_computing

[64] Fielding, R. and Reschke, J., "RFC-7320: Hypertext Transfer Protocol (HTTP/1.1): Message Syntax and Routing," Internet Engineering Task Force (IETF), 2014/06/01
https://tools.ietf.org/html/rfc7230#section-3.2.1

[65] Freed, N., Klensin, J. and Hansen, T., "Media Type Specifications and Registration Procedures
draft-ietf-appsawg-media-type-regs-14," IETF, 2017/06/20
https://trac.tools.ietf.org/html/draft-ietf-appsawg-media-type-regs-14

[66] Gavrichenkov, A., "Breaking HTTPS with BGP Hijacking," BlackHat, 2015
https://www.blackhat.com/docs/us-15/materials/us-15-Gavrichenkov-Breaking-HTTPS-With-BGP-Hijacking-wp.pdf

[67] "General Data Protection Regulation (GDPR)," Wikipedia,
https://en.wikipedia.org/wiki/General_Data_Protection_Regulation

[68] Girault, M., "Self-certified public keys," EUROCRYPT 1991: Advances in Cryptology, pp. 490-497, 1991
https://link.springer.com/content/pdf/10.1007%2F3-540-46416-6_42.pdf

[69] Goodin, D., "A DNS hijacking wave is targeting companies at an almost unprecedented scale," Ars Technica, 2019/01/10
https://arstechnica.com/information-technology/2019/01/a-dns-hijacking-wave-is-targeting-companies-at-an-almost-unprecedented-scale/

[70] Google, "Certificate Transparency,"
http://www.certificate-transparency.org/home

[71] Grant, A. C., "Search for Trust: An Analysis and Comparison of CA System Alternatives and Enhancements," Dartmouth Computer Science Technical Report TR2012-716, 2012
https://pdfs.semanticscholar.org/7876/380d71dd718a22546664b7fcc5b413c1fa49.pdf

[72] Hansen, T., "Additional Media Type Structured Syntax Suffixes
draft-ietf-appsawg-media-type-suffix-regs-02," IETF, 2012/07/17
https://trac.tools.ietf.org/html/draft-ietf-appsawg-media-type-suffix-regs-02

[73] "HD Protocol, HD Wallet, BiP32," Bitcoin Glosssary,
https://bitcoin.org/en/glossary/hd-protocol

[74] "How Cybercrime Exploits Digital Certificates," InfoSecInstitute, 2014/07/28
https://resources.infosecinstitute.com/cybercrime-exploits-digital-certificates/#gref

[75] "How many bits of entropy does an elliptic curve key of length n provide?," StackExchange,
https://crypto.stackexchange.com/questions/26791/how-many-bits-of-entropy-does-an-elliptic-curve-key-of-length-n-provide

[76] "HTTP Public Key Pinning," Wikipedia,
https://en.wikipedia.org/wiki/HTTP_Public_Key_Pinning

[77] "IANA,"

https://www.iana.org

[78] "Information Theory," Wikipedia,
https://en.wikipedia.org/wiki/Information_theory

[79] "Information-theoretic security," Wikipedia,
https://en.wikipedia.org/wiki/Information-theoretic_security

[80] "Interplanetary File System IPFS,"
https://docs.ipfs.io

[81] "IPv4," Wikipedia,
https://en.wikipedia.org/wiki/IPv4

[82] "IPv4 address exhaustion," Wikipedia,
https://en.wikipedia.org/wiki/IPv4_address_exhaustion

[83] "IPv6," Wikipedia,
https://en.wikipedia.org/wiki/IPv6

[84] "IPv6 deployment," Wikipedia,
https://en.wikipedia.org/wiki/IPv6_deployment

[85] "ISO 8601," Wikipedia,
https://en.wikipedia.org/wiki/ISO_8601

[86] "ISO 8601 and Nanosecond Precision Across Languages," nbsoft solutions, 2016/06/14
https://nbsoftsolutions.com/blog/iso-8601-and-nanosecond-precision-across-languages

[87] Josefsson, S., "The Base16, Base32, and Base64 Data Encodings," IETF 4648, 20016/10/01
https://tools.ietf.org/html/rfc4648#section-5

[88] Josefsson, S., "RFC-3548: The Base16, Base32, and Base64 Data Encodings," IETF, 2006-10-01
https://tools.ietf.org/html/rfc4648

[89] Josefsson, S. L., I., "Edwards-Curve Digital Signature Algorithm (EdDSA)," IRTF RFC-8032, 2017/01/01
https://tools.ietf.org/html/rfc8032

[90] Kaminsky, M. and Banks, E., "SFS-HTTP: Securing the Web with Self-Certifying URLs," MIT, 1999
https://pdos.csail.mit.edu/~kaminsky/sfs-http.ps

[91] "Key Derivation," LibSodium Documentation,
https://libsodium.gitbook.io/doc/key_derivation

[92] "Key stretching," Wikipedia,
https://en.wikipedia.org/wiki/Key_stretching

[93] "Key Transparency," Google,
https://github.com/google/keytransparen…

[94] Kudva, S., "The Difference Between System of Record and Source of Truth," LinkedIn, pp. 2016/03//31,
https://www.linkedin.com/pulse/difference-between-system-record-source-truth-santosh-kudva/

[95] Laurie, B. and Kasper, E., "Revocation Transparency,"
https://www.links.org/files/RevocationTransparency.pdf

[96] Laurie, B., Langley, A. and Kasper, E., "RFC6962: Certificate Transparency," IETF, 2013
https://tools.ietf.org/html/rfc6962

[97] Laurie, B., "Certificate Transparency: Public, verifieable, append-only logs," ACMQueue, vol. Vol 12, Issue 9, 2014/09/08
https://queue.acm.org/detail.cfm?id=2668154

[98] "Linked Data," W3C,
https://www.w3.org/standards/semanticweb/data

[99] “List of arbitrary-precision arithmetic software,” Wikipedia,

https://en.wikipedia.org/wiki/List_of_arbitrary-precision_arithmetic_software

[100]
“Locus of Control,” Wkipedia,

https://en.wikipedia.org/wiki/Locus_of_control

[101]
Maymounkov, P. and Mazieres, D., “Kademlia: A peer-to-peer information system based on the xor metric,” vol. International Workshop on Peer-to-Peer Systems, pp. 53-65, 2002

https://link.springer.com/chapter/10.1007/3-540-45748-8_5

[102]
Mazieres, D. and Kaashoek, M. F., “Escaping the Evils of Centralized Control with self-certifying pathnames,” MIT Laboratory for Computer Science, 2000

http://www.sigops.org/ew-history/1998/papers/mazieres.ps

[103]
Mazieres, D., “Self-certifying File System,” MIT Ph.D. Dissertation, 2000/06/01

https://pdos.csail.mit.edu/~ericp/doc/sfs-thesis.ps

[104]
“Merkle tree,” Wikipedia,

https://en.wikipedia.org/wiki/Merkle_tree

[105]
“multicodec,” github multiformats/multicodec,

https://github.com/multiformats/multicodec#multicodec-table

[106]
“MultiSigWallet,” Gnosis,

https://github.com/gnosis/MultiSigWallet

[107]
Nakov, S., “ECDSA: Elliptic Curve Signatures,” Cryptobook, 2018/11/01

https://cryptobook.nakov.com/digital-signatures/ecdsa-sign-verify-messages

[108]
“Namespace,” Wikipedia,

https://en.wikipedia.org/wiki/Namespace

[109]
Niederhagen, R. and Vaidner, M., “Practical Post-Quantum Cryptography,” Fraunhofer White Paper, vol. ISSN 2192-8169, 2017/08/18

https://www.sit.fraunhofer.de/fileadmin/dokumente/studien_und_technical_reports/
Practical.PostQuantum.Cryptography_WP_FraunhoferSIT.pdf?_=1503992279

[110]
“Non-repudiation,” Wikipedia,

https://en.wikipedia.org/wiki/Non-repudiation

[111]
“One-way function,” Wikipedia,

https://en.wikipedia.org/wiki/One-way_function

[112]
“One-way Function,” Crypto-IT,

http://www.crypto-it.net/eng/theory/one-way-function.html

[113]
“OpenID Connect,” OpenID,

https://openid.net/connect/

[114]
“Password Hashing Competition: and our recommendation for hashing passwords: Argon2,”

https://password-hashing.net/#argon2

[115]
“Plenum Byzantine Fault Tolerant Protocol,” HyperLedger Indy,

https://github.com/hyperledger/indy-plenum/wiki

[116]
“Project Haystacks: TagModel,”

https://project-haystack.org/doc/TagModel

[117]
“Public-key Cryptography,” Wikipedia,

https://en.wikipedia.org/wiki/Public-key_cryptography

[118]
“Remote ATtestation proceduresS (RATS) WG,” IETF,

https://datatracker.ietf.org/wg/rats/about/

[119]
“Roots of Trust,” NIST,

https://csrc.nist.gov/Projects/Hardware-Roots-of-Trust

[120]
Rytter, M. and Jørgensen, B. N., “Independently extensibile contexts,” vol. European Conference on Software Architecture, pp. 327-334, 2010

https://link.springer.com/chapter/10.1007/978-3-642-15114-9_25

[121]
Saarinen, M.-J. and Aumasson, J.-P., “The BLAKE2 Cryptographic Hash and Message Authentication Code (MAC) IETF RFC-7693,” IETF RFC-7693, 2015/11/01

https://tools.ietf.org/html/rfc7693

[122]
Smith, S. M., “Ioflo,”

https://github.com/ioflo/ioflo

[123]
Nakamoto, S., “Bitcoin: A Peer-to-Peer Electronic Cash System,” 2008/10/31

https://bitcoin.org/bitcoin.pdf

[124]
“IEEE Std 802.1AR-2018: IEEE Standard for Local and metropolitan area networks–Secure Device Identity,” IEEE Std 802.1AR-2018, 2018/08/02

https://1.ieee802.org/security/802-1ar/

[125]
“Secp256k1 (ECDSA),” Bitcoin Wiki,

https://en.bitcoin.it/wiki/Secp256k1

[126]
“Securing End-to-End Communications,” CISA Alert (TA15-120A), 2016/09/26

https://www.us-cert.gov/ncas/alerts/TA15-120A

[127]
“self-sovereign-identity,” GitHub,

https://github.com/WebOfTrustInfo/self-sovereign-identity

[128]
Serrano, N., Hadan, H. and Camp, L. J., “A complete study of PKI (PKI’s Known Incidents),” Available at SSRN 3425554, 2019

https://papers.ssrn.com/sol3/papers.cfm?abstract_id=3425554

[129]
Shae, M., Smith, S. M. and Stocker, C., “Decentralized Identity as a Meta-platform: How Cooperation Beats Aggregation,” Rebooting the Web of Trust, vol. RWOT 9, 2019/11/19

https://github.com/SmithSamuelM/Papers/blob/master/whitepapers/CooperationBeatsAggregation.pdf

[130]
“Side-channel attack,” Wikipedia,

https://en.wikipedia.org/wiki/Side-channel_attack

[131]
“Single source of truth,” Wikipedia,

https://en.wikipedia.org/wiki/Single_source_of_truth

[132]
Smith, S. M., “Open Reputation Framework,” vol. Version 1.2, 2015/05/13

https://github.com/SmithSamuelM/Papers/blob/master/whitepapers/open-reputation-low-level-whitepaper.pdf

[133]
Smith, S. M. and Khovratovich, D., “Identity System Essentials,” 2016/03/29

https://github.com/SmithSamuelM/Papers/blob/master/whitepapers/Identity-System-Essentials.pdf

[134]
Smith, S. M., “Meta-Platforms and Cooperative Network-of-Networks Effects: Why Decentralized Platforms Will Eat Centralized Platforms,” SelfRule, 2018/04/25

https://medium.com/selfrule/meta-platforms-and-cooperative-network-of-networks-effects-6e61eb15c586

[135]
Smith, S. M., “Key Event Receipt Infrastructure (KERI) Design and Build,” arXiv, 2019/07/03

https://arxiv.org/abs/1907.02143

[136]
Smith, S. M., “Decentralized Autonomic Data (DAD) and the three R’s of Key Management,” Rebooting the Web of Trust RWOT 6, Spring 2018

https://github.com/SmithSamuelM/Papers/blob/master/whitepapers/DecentralizedAutonomicData.pdf

[137]
“Social Engineering,” imperva,

https://www.imperva.com/learn/application-security/social-engineering-attack/

[138]
Staff, P. I. E., “One Login To Rule them All - Seamless and Secure Cross-Domain Authentication,” Paragon Initiative, 2016/02/22

https://paragonie.com/blog/2016/02/one-login-rule-them-all-seamless-and-secure-cross-domain-authentication

[139]
Staff, P. I. E., “Split Tokens: Token-Based Authentication Protocols without Side-Channels,” Paragon Initiative, 2017/02/28

https://paragonie.com/blog/2017/02/split-tokens-token-based-authentication-protocols-without-side-channels

[140]
Stevens, G., “DNS Poisoning Attacks: A Guide for Website Admins,” HashedOut, 2020/01/21

https://www.thesslstore.com/blog/dns-poisoning-attacks-a-guide-for-website-admins/

[141]
Stocker, C., Smith, S. and Caballero, J., “Quantum Secure DIDs,” RWOT10, 2020/07/09

https://github.com/WebOfTrustInfo/rwot10-buenosaires/blob/master/final-documents/quantum-secure-dids.pdf

[142]
Szabo, N., “Secure Property Titles with Owner Authority,” 1998

https://nakamotoinstitute.org/secure-property-titles/

[143]
TCG, “Implicit Identity Based Device Attestation,” Trusted Computing Group, vol. Version 1.0, 2018/03/05

https://trustedcomputinggroup.org/wp-content/uploads/TCG-DICE-Arch-Implicit-Identity-Based-Device-Attestation-v1-rev93.pdf

[144]
"The Ethash Algorithm," Ethereum Foundation,

http://www.ethdocs.org/en/latest/mining.html#the-algorithm

[145]
"Top 500: The List,"

https://www.top500.org

[146]
"Trusted Platform Module," Wikipedia,

https://en.wikipedia.org/wiki/Trusted_Platform_Module

[147]
"Uniform Resource Identifier (URI): Generic Syntax," IETF RFC-3986, 2005/01/01

https://tools.ietf.org/html/rfc3986

[148]
"URL Uniform Resource Locators," Wikipedia,

https://en.wikipedia.org/wiki/URL

[149]
Valsorda, F., "USING ED25519 SIGNING KEYS FOR ENCRYPTION," Fillipo.io, 2019/05/18

https://blog.filippo.io/using-ed25519-keys-for-encryption/

[150]
Various, "Status Report on the First Round of the NIST Post-Quantum Cryptography Standardization Process," NISTIR 8240, 2019/01/01

https://nvlpubs.nist.gov/nistpubs/ir/2019/NIST.IR.8240.pdf

[151]
W3C, "Decentralized Identifiers (DIDs)," W3C Draft Community Group Report,

https://www.w3.org/TR/did-core/

[152]
Windley, P. J., "Soverign-Source Identity, Autonomy, and Learning," Technometria, 2016/01/19

http://www.windley.com/archives/2016/01/soverign-source_identity_autonomy_and_learning.shtml

[153]
Windley, P. J., "Life-Like Identity: Why the Internet Needs an Identity Metasystem," Technometria, 2019/08/20

http://www.windley.com/archives/2019/08/life-like_identity_why_the_internet_needs_an_identity_metasystem.shtml

[154]
Working Group, R. A. T. S., "Network Device Attestation Workflow," IETF Internet-Draft, 2019/12/03

https://tools.ietf.org/html/draft-fedorkow-rats-network-device-attestation-01

[155]
"Zooko's Triange," Wikipedia,

https://en.wikipedia.org/wiki/Zooko%27s_triangle